\documentclass{article}
\usepackage{amsmath}
\usepackage{amssymb}

\input{tcilatex}
\begin{document}

\begin{center}
{\Huge The duality-character solution-information-carrying unitary
propagators}$%
\begin{array}{c}
\text{\QQfnmark{%
The term \textit{information-carrying (}$IC$\textit{)} used in the previous
paper [$1$] is modified to the present term \textit{%
solution-information-carrying (}$SIC$\textit{)} in this paper\textit{\ }as
the latter reflects better the mathematical-logical principle of
unstructured search problem.}} \\ 
\\ 
\end{array}%
\QQfntext{0}{
The term \textit{information-carrying (}$IC$\textit{)} used in the previous
paper [$1$] is modified to the present term \textit{%
solution-information-carrying (}$SIC$\textit{)} in this paper\textit{\ }as
the latter reflects better the mathematical-logical principle of
unstructured search problem.}$%
\begin{equation*}
\end{equation*}

{\large Xijia Miao}

{\large Somerville, Massachusetts}

{\large Dated: December 2020}%
\begin{equation*}
\end{equation*}

{\Large Abstract}
\end{center}

The $HSSS$ quantum search process owns the dual character that it obeys both
the unitary quantum dynamics and the mathematical-logical principle of the
unstructured search problem. It is essentially different from a conventional
quantum search algorithm. It is constructed with the duality-character
oracle operations of unstructured search problem and the relevant
quantum-mechanical\ unitary operators. It consists of the two consecutive
steps that the first step is the search-space dynamical reduction and the
second step the dynamical quantum-state-difference amplification (QUANSDAM).
At the second step the QUANSDAM process is directly constructed with the $%
SIC $ unitary propagators, while the latter each are prepared with the basic 
$SIC $ unitary operators. All these $SIC$ unitary operators and propagators
each own the dual character. Here the preparation for the $SIC$ unitary
propagators of a typical quantum system (i.e., a single-atom system) is
concretely carried out by starting from the basic $SIC$\ unitary operators.
The $SIC$ unitary propagator of a quantum system may reflect the quantum
symmetry of the quantum system, while the basic $SIC$ unitary operators may
not. The quantum symmetry is considered as the fundamental
quantum-computing-speedup resource in the quantum-computing speedup theory.
Therefore, the purpose for the preparation of the $SIC$ unitary propagators
of a quantum system is ultimately to employ the quantum symmetry of the
quantum system to simplify the construction and realization of the QUANSDAM
process and hence speed up the QUANSDAM\ process. The preparation process is
a solution-information transfer process from the original quantum subsystem
(e.g., an $n-$qubit spin system) to the final quantum subsystem (e.g., a
single-atom system). It is unitary and deterministic. It obeys the
information conservation law. In methodology the preparation is based on the
energy eigenfunction expansion principle and the multiple-quantum operator
algebra space. Furthermore, a general theory mainly based on the Feynman
path integration technique and also the energy eigenfunction expansion
method is established to treat theoretically and calculate a $SIC$\ unitary
propagator of any quantum system in the coordinate representation, which may
be further used to construct theoretically an exponential QUANSDAM process
in future.\newline
\newline

\begin{center}
{\LARGE Contents}
\end{center}

\begin{flushleft}
$%
\begin{array}{l}
{\large 1.\ Introduction}%
\end{array}%
\begin{array}{c}
........................................................................................%
\end{array}%
\begin{array}{r}
{\large 2}%
\end{array}%
$

$%
\begin{array}{l}
{\large 2.\ The\ duality-character\ oracle\ operations\ of\ the\ HSSS\ }%
\end{array}%
$

$%
\begin{array}{c}
\text{ \ \ \ }{\large quantum\ search\ process}%
\end{array}%
\begin{array}{c}
....................................................................%
\end{array}%
\begin{array}{r}
{\large 6}%
\end{array}%
$

$%
\begin{array}{l}
{\large 3.\ Preparation\ and\ realization\ for\ the\ SIC\ unitary\
propagators}%
\end{array}%
\begin{array}{c}
\text{ \ \ }%
\end{array}%
\begin{array}{r}
{\large 15}%
\end{array}%
$

$%
\begin{array}{l}
{\large 3.1.\ Existence\ of\ the\ SIC\ unitary\ propagators}%
\end{array}%
\begin{array}{c}
.................................%
\end{array}%
\begin{array}{r}
{\large 24}%
\end{array}%
$

$%
\begin{array}{l}
{\large 3.2.\ The\ practical\ SIC\ unitary\ propagators}%
\end{array}%
\begin{array}{c}
.......................................%
\end{array}%
\begin{array}{r}
{\large 46}%
\end{array}%
$

$%
\begin{array}{l}
{\large 4.\ The\ SIC\ unitary\ propagators\ in\ the\ coordinate\
representation}%
\end{array}%
\begin{array}{c}
\text{ \ }%
\end{array}%
\begin{array}{r}
{\large 62}%
\end{array}%
$

$%
\begin{array}{l}
{\large 5.\ Discussion\newline
}%
\end{array}%
\begin{array}{c}
\text{ \ \ \ \ \ \ \ \ }%
\end{array}%
\begin{array}{c}
..........................................................................%
\end{array}%
\begin{array}{r}
{\large 85}%
\end{array}%
$

$%
\begin{array}{l}
{\large References}%
\end{array}%
\begin{array}{c}
\text{ \ \ \ \ \ \ \ \ \ \ \ }%
\end{array}%
\begin{array}{c}
..........................................................................%
\end{array}%
\begin{array}{r}
{\large 91}%
\end{array}%
$

$%
\begin{array}{l}
{\large Appendix\ A.\ The\ perturbation\ equations\ for\ a\ SIC\ unitary}%
\end{array}%
$

$%
\begin{array}{c}
\text{\ \ \ \ \ \ \ \ \ \ \ \ \ \ \ \ \ }{\Large \ }{\large propagator}\text{%
\ }{\large and\ a\ SIC\ Green\ function}%
\end{array}%
\begin{array}{c}
.........................%
\end{array}%
\begin{array}{r}
{\large 93}%
\end{array}%
$\newline
\end{flushleft}

\begin{equation*}
\end{equation*}%
{\LARGE 1. Introduction}

A quantum-computing speedup process [$1,2$] obeys not only the fundamental
quantum-physical laws (i.e., the unitary quantum dynamics and the
quantum-mechanical symmetry \footnote{%
In quantum mechanics [$10$] a symmetry operation is generally described by a
unitary operator. An exception is the time-reversal symmetry operation which
is described by an anti-unitary operator. But time reversal symmetry
operation still may be effectively realized (or simulated) unitarily. In
this sense unitary quantum dynamics may be thought to be more fundamental in
quantum mechanics, although quantum-mechanical symmetry is a fundamental
atribute of the natural world and independent of unitary quantum dynamics.})
but also the mathematical-logical principle of a computational problem to be
solved [$3$]. This dual character of quantum-computing speedup reflects the
interaction between the unitary quantum dynamics and the
mathematical-logical principle in the quantum-computing speedup process. It
is essential in the quantum-computing speedup theory [$2$]. In the theory,
as an independent contributor to the quantum-computing speedup, the
mathematical-logical principle should be described in the picture of the
theoretical (or mathematical) reversible computation [$4$] in computational
science. But the key to achieving an essential quantum-computing speedup is
the interaction picture of reversible computation to describe the
mathematical-logical principle [$1,2$]. The mathematical-logical principle
was not considered as an independent principle (independent of
quantum-physical laws) to make contribution to the quantum-computing speedup
until early 2001 $[3].$ It is this mathematical-logical principle that makes
the mathematical (math) Hilbert space necessary to describe completely any
quantum-computing speedup process together with the physical Hilbert space [$%
1,2$].

In the quantum-computing speedup theory symmetrical structure and property
of the Hilbert space of a quantum system (or briefly quantum symmetry of a
quantum system) is thought of as the fundamental quantum-computing-speedup
resource that is responsible for an essential quantum-computing speedup [$%
2,1,3$]. Generally, in quantum mechanics quantum symmetry of a quantum
system may be characterized by the quantum symmetry group which induces
unitary transformations on the Hilbert space of the quantum system. However,
in the quantum-computing speedup theory the fundamental
quantum-computing-speedup resource, i.e., quantum symmetry of a quantum
system, may be completely characterized by the quantum symmetry group which
induces the unitary transformations on the physical Hilbert space and the
one which induces the unitary transformations on the corresponding math
Hilbert space.

Therefore, in the quantum-computing speedup theory both the unitary quantum
dynamics and the fundamental quantum-computing-speedup resource are
considered as the two pillars to build an efficient quantum-computing
process which has an exponential quantum-computing speedup over its
classical counterpart.

The quantum-computing speedup theory therefore puts forward a
duality-character quantum search process, i.e., the $HSSS$ quantum search
process, which is based on the \textit{H}ilbert-\textit{s}pace \textit{s}%
ymmetrical \textit{s}tructure ($HSSS$) and the unitary quantum dynamics and
works on both the physical Hilbert space and its corresponding math Hilbert
space. The $HSSS$ quantum search process that solves an unstructured search
problem consists of the two successive steps that the first step is the
search-space dynamical reduction [$2,3$] and the second the dynamical
quantum-state-difference amplification [$2$]. The search-space dynamical
reduction is realized efficiently in Ref. [$5$]. It has to employ the
specific tensor-product symmetric structure of a composite quantum system
that this\ composite quantum system has an exponentially large Hilbert space
and its component systems each have a polynomially large Hilbert subspace.
For example, a composite quantum spin system that consists of $n$ spins
(e.g., nuclear spins, electronic spins, etc.) with spin quantum number $%
I=1/2 $ owns this specific tensor-product (i.e., direct-product) symmetric
structure. This tensor-product symmetric structure is the special form of
the fundamental quantum-computing-speedup resource. It is necessary for the
search-space dynamical reduction [$3,2,5$]. It is known from the magnetic
resonance theory that the spin energy for such a composite spin system does
not grow exponentially as the spin number of the system. But evidently this
(energy) property is not considered as the fundamental
quantum-computing-speedup resource in the quantum-computing speedup theory.
Beside this tensor-product symmetrical structure the fundamental
quantum-computing-speedup resource also includes any other symmetrical
structures and properties of a quantum system including those of the
multiple-quantum operator algebra space [$6,7,3$]. They still play important
role in the quantum-computing speedup. Especially they play important role
in constructing and realizing a dynamical \textit{quan}tum-\textit{s}tate-%
\textit{d}ifference-\textit{am}plification ($QUANSDAM$) process even in a
simple quantum system (e.g., a single-atom system) that does not own this
tensor-product symmetric structure. Although a simple quantum system does
not have the specific tensor-product symmetric structure, it still may own
other symmetric structures and properties and a QUANSDAM process could
employ the symmetric structures and properties to achieve a significant
quantum-computing speedup for the $HSSS$ quantum search process.

The key criterion for the duality-character description of quantum-computing
speedup to be correct is whether this duality-character description can lead
to that the $HSSS$ quantum search process has a super-square or even
exponential quantum-searching speedup. It is well-known that a conventional
quantum search algorithm [$8$] can achieve a square quantum-searching
speedup and moreover, this square quantum-searching speedup is the limit
speedup [$9$]. If the $HSSS$ quantum search process can achieve a
super-square quantum-searching speedup at least, then that means that this
square speedup limit is broken down.

A quantum-state-difference-amplification process is the inverse of a \textit{%
uni}tary \textit{dy}namical \textit{s}tate-\textit{lock}ing ($UNIDYSLOCK$)
process $[1,2,34]$. A UNIDYSLOCK process is defined as a unitary process
that can transform simultaneously two or more orthogonal quantum states to
their corresponding non-orthogonal states whose quantum-state differences
may be arbitrarily small. It makes sense only in the quantum-computing
speedup theory. It is not a conventional quantum computation (algorithm)
which is essentially a purely quantum-physical process (See Ref. [$1$] and
also the last section `Discussion'). Then a QUANSDAM\ process can transform
a pair of highly overlapping non-orthogonal quantum states to a pair of
orthogonal states at the same time. It is unitary. It is essentially
different from a $QM$\ unitary dynamical process, because the latter is not
able to change simultaneously a pair of non-orthogonal states to their
corresponding orthogonal states. It is also essentially different from a
non-equilibrium irreversible process, because although the latter is also
able to change a pair of non-orthogonal states to a pair of orthogonal
states, it is not unitary. The characteristic feature for a UNIDYSLOCK (or
QUANSDAM)\ process is that a UNIDYSLOCK\ (or QUANSDAM) process is a unitary
quantum dynamical process, but it can transform a pair of orthogonal quantum
states to their corresponding non-orthogonal states (or vice versa) at the
same time, that is, it can change the quantum-state difference of a pair of
quantum states. Furthermore, it is common for any quantum-computing speedup
processes in the quantum-computing speedup theory. Then a quantum-computing
speedup process is not a conventional quantum computation (algorithm) due to
that the latter is essentially a purely quantum-physical process and does
not change unitarily quantum-state difference of a pair of quantum states [$%
1 $]. From the point of view of pure quantum mechanics a QUANSDAM (or
UNIDYSLOCK) process does not exist at all. One therefore is forced to employ
the duality-character description of quantum-computing speedup [$2,1$], in
order that a UNIDYSLOCK (or QUANSDAM)\ process can be described correctly
and reasonably. This is the first example that \textit{the dual character
that a quantum-computing speedup process obeys both the unitary quantum
dynamics and the mathematical-logical principle of a computational problem
to be solved} is shown to be nontrivial. A QUANSDAM process must be
comprehended completely from both the aspect of the unitary quantum dynamics
and the one of the mathematical-logical principle of the unstructured search
problem as well as their interaction. This is required by the dual
character. The dual character is universal for any quantum-computing speedup
processes including the QUANSDAM\ (or UNIDYSLOCK) process.

In principle a QUANSDAM process is built out of the basic $SIC$\ unitary
operators [$1,2,3$] that carry information of the solution to the
unstructured search problem. But generally the basic $SIC$\ unitary
operators are not able to reflect the symmetrical structure and property of
the quantum system used to perform the QUANSDAM process. In contrast, the $%
SIC$\ unitary propagators of the quantum system, which also carry
information of the solution to the unstructured search problem, tend to be
able to reflect the symmetrical structure and property of the quantum
system. Both the basic $SIC$ unitary operators and the $SIC$\ unitary
propagators each own the dual character and therefore are the
duality-character unitary operators. Then in order to make use of the
symmetric structure and property to speed up the QUANSDAM\ process one
should choose the $SIC$\ unitary propagators as the building blocks of the
QUANSDAM process. Therefore, the first task to construct a QUANSDAM process
is to prepare the suitable $SIC$\ unitary propagators that can reflect the
symmetric structure and property of a quantum system by starting from the
basic $SIC$ unitary operators. This is also one main purpose of the present
work in this paper.

It has been proposed [$34,2,15,1$] that a UNIDYSLOCK (or QUANSDAM) process
is constructed and realized in a single-atom system, where the unitary
manipulation of a single atom in time and space [$30,31$] is of the
fundamental importance.

In this paper it describes in detail how the $SIC$ unitary propagators of a
concrete quantum system, i.e., a single-atom system, is prepared from the
basic $SIC$ unitary operators on the basis of the energy eigenfunction
expansion [$13$, $10$] and the multiple-quantum operator algebra space [$%
6,24,3,7$]. It does not care about how the basic $SIC$ unitary operators are
prepared explicitly via the search-space dynamical reduction. Instead, the
basic $SIC$ unitary operators are directly considered as the basic building
blocks of a $SIC$ unitary propagator and a QUANSDAM process. The
computational complexity for a QUANSDAM process and also for a $SIC$\
unitary propagator then is measured by using directly the basic $SIC$
unitary operators instead of the original duality-character oracle
operations of the unstructured search problem in the $HSSS$ quantum search
process. Moreover, the complexity for a QUANSDAM process also may be
measured by using directly the $SIC$ unitary propagators.

A general theory then is established to treat theoretically and calculate a $%
SIC$\ unitary propagator of any quantum system in the coordinate
representation. It is mainly based on the Feynman path integration technique
[$14$] and also the energy eigenfunction expansion method in quantum
mechanics [$13,10$]. In particular, with the theory the $SIC$ unitary
quadratic propagators in the coordinate representation are determined
exactly in a quantum system such as a single-atom system in time and space.
The theoretical work is basic for further constructing theoretically an
exponential QUANSDAM\ process by starting from the $SIC$ unitary propagators
in a quantum system such as a single-atom system in future.\newline
\newline
\newline
{\Large 2. The duality-character oracle operations of the }${\Large HSSS}$%
{\Large \ quantum search\ process}

The initial and preliminary research work on the $HSSS$ quantum search
process was reported in the early 2001 [$3$]. The historical evolution and
the mechanism for the $HSSS$ quantum search process are described in detail
in Refs. [$1,2$]. Below a simple description is given to the $HSSS$ quantum
search process in mechanism and especially a more detailed description to
the core of the search process, i.e., the duality-character $SIC$\ oracle
operations of the unstructured search problem.

The $HSSS$ quantum search process consists of the two consecutive steps: the
search-space dynamical reduction and the dynamical quantum-state-difference
amplification. From the viewpoint of computation it must be first a quantum
search algorithm that can solve a search problem with exponentially large
unstructured search space (typically $2^{n}-$dimensional unstructured search
space). This has nothing to do with whether it performs the search-space
dynamical reduction or not. In this sense the unstructured search space is
thought to be irreducible in the search-space dynamical reduction.
Therefore, the $HSSS$ quantum search process and the search-space dynamical
reduction each always can be expressed as a sequence of the
duality-character oracle operations of the search problem with the original
unstructured search space and the relevant $QM$ unitary operators [$1,2$].
Then the search-space dynamical reduction makes sense only in unitary
quantum dynamics. Its mechanism is described in Refs. [$1,2$] and its
efficient realization in Ref. [$5$].

Big quantum system and/or high quantum number tend to be related to the
classical-physical world. This is consistent with the spirit of the Bohr's
correspondence principle in quantum physics. Of course, the
classical-physical world is the special and small region of the
quantum-physical world and also obeys the same quantum-physical laws. The
largest contribution of the search-space dynamical reduction to the $HSSS$
quantum search process could be that the search-space dynamical reduction
eliminates dynamically the exponentially large unstructured search space and
purges the classical-physical effect, resulting in that during searching for
the real solution to the unstructured search problem the unitary quantum
dynamics can run effectively in a domain with smaller quantum system, lower
quantum number, and stronger quantum effect. The search-space dynamical
reduction is necessary but not sufficient for the $HSSS$ quantum search
process to achieve an essential quantum-computing speedup. In fact, it is
not free. But this cost could be sufficiently compensated by the powerful
quantum-state-difference amplification at the second step of the $HSSS$
quantum search process. The search-space dynamical reduction plays a key
role in avoiding the $HSSS$ quantum search process falling in analog
classical computation.

The $HSSS$ quantum search process is a duality-character quantum search
algorithm. According to the duality-character description of
quantum-computing speedup it must be described completely in both the
physical Hilbert space of a quantum system and its corresponding math
Hilbert space. In the $HSSS$ quantum search process the search-space
dynamical reduction must be first carried out. It usually may be carried out
in the $2^{n}-$dimensional Hilbert space of an $n-$qubit spin system (or
more generally an $n-$qubit composite quantum system\footnote{%
In conventional quantum computation a qubit means a superposition of two
orthogonal states (e.g., $\left\vert 0\right\rangle $ and $\left\vert
1\right\rangle $) used to treat quantum information. However, in the
quantum-computing speedup theory, a two-level quantum system (or subsystem)
used to treat quantum information is called a qubit as usual.}) and the
corresponding math Hilbert space. Here the unstructured search space of the
search problem is defined in the math Hilbert space. It is formed by all the
candidate solution states, among which only one is the real solution state
to the unstructured search problem. It must be considered as a whole (or as
a single entity). Suppose that $|x_{0}\rangle $ and $|S\rangle $ are the
real solution state and any candidate solution state to the unstructured
search problem, respectively. Theoretically any candidate solution state $%
|S\rangle $ belongs to only the math Hilbert space ($H_{m}$) if it is not
the real solution state $|x_{0}\rangle $, while the real solution state $%
|x_{0}\rangle $ belongs to the physical Hilbert space ($H_{p}$) and also the
math Hilbert space ($H_{m}$). For convenience, here the physical Hilbert
space ($H_{p}$) is taken as the $2^{n}-$dimensional Hilbert space of the $n-$%
qubit spin system which consists of the $n$ spins, each of which has the
spin quantum number $I=1/2$. Correspondingly there is the $2^{n}-$%
dimensional math Hilbert space. Then in both the physical and math Hilbert
spaces the duality-character oracle operation of the unstructured search
problem in the $HSSS$ quantum search process may be apparently defined as [$%
1 $]%
\begin{equation}
C_{S}\left( \theta \right) :|S\rangle \rightarrow \left\{ 
\begin{array}{c}
\exp \left( -i\theta \right) |x_{0}\rangle ,\text{ } \\ 
\exp \left( -i\theta \right) |S\rangle ,\text{ }%
\end{array}%
\begin{array}{ccc}
|x_{0}\rangle \in H_{p},\text{ }|x_{0}\rangle \in H_{m} & if & S=x_{0} \\ 
|S\rangle \notin H_{p},\text{ }|S\rangle \in H_{m} & if & S\neq x_{0}%
\end{array}%
\right.  \tag{2.1}
\end{equation}%
This oracle operation $C_{S}\left( \theta \right) $ is a duality-character
operation. It is a reversible (or unitary) selective diagonal operator. It
may be written in the exponential operator form [$3$]%
\begin{equation}
C_{S}\left( \theta \right) =\exp \left( -i\theta D_{S}\right) ,  \tag{2.2}
\end{equation}%
where the oracle diagonal operator $D_{S}=|S\rangle \langle S|$ is expressed
as%
\begin{equation}
D_{S}=\underset{m=1}{\overset{n}{\dbigotimes }}\left( \frac{1}{2}%
E_{m}+a_{m}^{s}I_{mz}\right)  \tag{2.3a}
\end{equation}%
and the corresponding candidate solution state $\left\vert S\right\rangle $
is written as%
\begin{equation}
\left\vert S\right\rangle =\underset{m=1}{\overset{n}{\dbigotimes }}\left( 
\frac{1}{2}T_{m}+a_{m}^{s}S_{m}\right) ,  \tag{2.3b}
\end{equation}%
here $T_{m}=\left\vert 0_{m}\right\rangle +\left\vert 1_{m}\right\rangle $
and $S_{m}=\frac{1}{2}\left( \left\vert 0_{m}\right\rangle -\left\vert
1_{m}\right\rangle \right) ,$ and $\left\vert 0_{m}\right\rangle $ and $%
\left\vert 1_{m}\right\rangle $ are the two usual computational bases of the 
$m-$th spin$-1/2$ of the $n-$qubit spin system, and $E_{m}$ and $I_{mz}$ are
the unity operator and the $z-$component spin operator of the $m-$th spin$%
-1/2$, respectively. The most important quantity in (2.3a) and (2.3b) is the
duality-character double-valued logical number vector $\left\{
a_{m}^{s}\right\} $ ($a_{m}^{s}=\pm 1$ in value with $1\leq m\leq n$) which
characterizes completely the duality-character oracle operation $C_{S}\left(
\theta \right) .$ A more detailed description for the duality-character
quantity $a_{m}^{s}$ may be seen in the section 3 later. The
duality-character oracle operation $C_{S}\left( \theta \right) $ of (2.2) in
exponential operator form works in the $n-$qubit spin system, but it can be
easily generalized in a general $n-$qubit quantum system\footnote{%
In a general $n-$qubit quantum system one needs only to replace the unity
operator $E_{m}$ and the $z-$component spin operator $I_{mz}$ in (2.3a) with
the diagonal operator $E_{m}^{01}=\left\vert 0_{m}\right\rangle \left\langle
0_{m}\right\vert +\left\vert 1_{m}\right\rangle \left\langle
1_{m}\right\vert $ and the $z-$component pseudospin operator $Q_{mz}^{01}=%
\frac{1}{2}\left( \left\vert 0_{m}\right\rangle \left\langle
0_{m}\right\vert -\left\vert 1_{m}\right\rangle \left\langle
1_{m}\right\vert \right) ,$ respectively, while keeping $T_{m}$ and $S_{m}$
in (2.3b) unchanged, where $\left\vert 0_{m}\right\rangle $ and $\left\vert
1_{m}\right\rangle $ are the two usual computational bases of the $m-$th
qubit of the $n-$qubit quantum system.}. It was first put forward long time
ago [$3$], but in the quantum-computing speedup mechanism it was not fully
understood until recent years [$1,2$].

Apparently the duality-character oracle operation $C_{S}\left( \theta
\right) $ of (2.1) seems to be the unity operation up to a global phase
factor $\exp \left( -i\theta \right) $ in the math Hilbert space alone\ from
the viewpoint of the mathematical reversible computation [$4$]. If this was
true, then the search-space dynamical reduction could not be realized in the
math Hilbert space alone. The theoretical interpretation for the $%
C_{S}\left( \theta \right) $ of (2.1) then is not so simple. However, it is
sure that the search-space dynamical reduction can not be completely
realized in the math Hilbert space alone. It could be completely realized
only when the physical Hilbert space takes part in. Actually, one can not
describe exactly the duality-character oracle operation from the viewpoint
of the mathematical reversible computation alone. Likewise the
duality-character oracle operation also can not be exactly described alone
in the classical-physical reversible computation [$11$]. As shown in (2.1),
the real solution state $|x_{0}\rangle $ is the bridge to connect the
physical Hilbert space with the math Hilbert space, because it belongs to
the physical Hilbert space and also the math Hilbert space. Therefore, the
exact description for the duality-character oracle operation of the $HSSS$
quantum search process is involved in the interaction between the
mathematical-logical principle of the unstructured search problem and the
fundamental quantum-physical laws. \textit{In the quantum-computing speedup
theory this kind of interaction is fundamental to achieve an essential
quantum-computing speedup not only for the }$HSSS$\textit{\ quantum search
process but also for any quantum-computing speedup process.}

From the point of view of the mathematical-logical principle of the
unstructured search problem any candidate solution state could be possibly
considered as the real solution state, before the real solution state is
found in the $HSSS$ quantum search process. However, the real solution state
is distinct from any candidate solution state that is not the real solution
state. \textit{The search principle for the }$HSSS$\textit{\ quantum search
process then is to distinguish unambiguously the real solution state }$%
|x_{0}\rangle $\textit{\ from any candidate solution state }$|S\rangle $%
\textit{\ that is not the real solution state.} As shown in (2.1), the
unique difference between $|x_{0}\rangle $ and $|S\rangle $ ($\neq
|x_{0}\rangle )$ (except the difference ($S\neq x_{0}$) between themselves)
is that the real solution state $|x_{0}\rangle $ is in the physical Hilbert
space $H_{p}$ of the quantum system, while any candidate solution state $%
|S\rangle $ ($\neq |x_{0}\rangle )$ is not. This leads to that any quantum
state of the quantum system such as $|x_{0}\rangle $ that carries the
information of the real solution to the unstructured search problem is
directly observable in the quantum system, while any candidate solution
state $|S\rangle $ ($\neq |x_{0}\rangle )$ or generally any state that
carries the information of any candidate solution different from the real
solution is not observable in the quantum system. The final search result
for the $HSSS$ quantum search process can be output only by measuring the
quantum state that carries the information of the real solution in the
quantum system. The real solution state never leaves the physical Hilbert
space, while any candidate solution state different from the real solution
state is always in the math Hilbert space during the $HSSS$ quantum search
process.

The duality-character oracle operation $C_{S}\left( \theta \right) $ of
(2.1) or (2.2) acts on simultaneously the physical Hilbert space and the
corresponding math Hilbert space in the $HSSS$ quantum search process. Now
consider first the quantum-physical aspect of the duality-character oracle
operation which is related to the physical Hilbert space $H_{p}$. In the
quantum-physical aspect the duality-character oracle operation is really the
unitary (or reversible) selective phase-shift operation:%
\begin{equation}
C_{x_{0}}\left( \theta \right) :|x_{0}\rangle \rightarrow \exp \left(
-i\theta \right) |x_{0}\rangle .  \tag{2.4}
\end{equation}%
This selective phase-shift operation acts on the real solution state $%
|x_{0}\rangle $ alone in the physical Hilbert space. It has nothing to do
with any candidate solution state other than $|x_{0}\rangle $. In the
physical Hilbert space $H_{p}$ the selective phase-shift operation $%
C_{x_{0}}\left( \theta \right) $ also may be fully written as%
\begin{equation}
C_{x_{0}}\left( \theta \right) :|x\rangle \rightarrow \left\{ 
\begin{array}{cc}
\exp \left( -i\theta \right) |x_{0}\rangle & if \\ 
|x\rangle & if%
\end{array}%
\begin{array}{c}
x=x_{0} \\ 
x\neq x_{0}%
\end{array}%
\right.  \tag{2.5}
\end{equation}%
Here both the states $|x\rangle $ and $|x_{0}\rangle $ are the basis states
of the physical Hilbert space of the quantum system, i.e., $|x_{0}\rangle ,$ 
$|x\rangle \in H_{p}$. However, the real solution state $|x_{0}\rangle $ is
essentially different from any basis state $|x\rangle $ ($x\neq x_{0})$ in
that the basis state $|x_{0}\rangle $ also is a member of the unstructured
search space, while $|x\rangle $ is not. Moreover, any candidate solution
state $|S\rangle $ $(\neq |x_{0}\rangle )$ does not appear in the physical
Hilbert space, i.e., $|S\rangle \notin H_{p}.$ Therefore, there is an
essential difference between $|x_{0}\rangle $ and $|S\rangle $ $(\neq
|x_{0}\rangle )$ that the real solution state $|x_{0}\rangle $ is observable
in quantum mechanics, while any candidate solution state $|S\rangle $ $(\neq
|x_{0}\rangle )$ is not.

It must be pointed out that in the $HSSS$ quantum search process the
selective phase-shift operation $C_{x_{0}}\left( \theta \right) $ of (2.4)
alone is not an oracle operation that can characterize completely an
unstructured search problem. The operation $|x\rangle \rightarrow |x\rangle $
for $x\neq x_{0}$ in (2.5) is an identical state transformation. It may have
two different physical interpretations. For the $HSSS$ quantum search
process it may be intuitively explained: Take no action on the state $%
|x\rangle .$ And the state $|x\rangle $ does not act as a member of the
unstructured search space as here the math Hilbert space is not the physical
one. These result in that the unitary (or reversible) operation $%
C_{x_{0}}\left( \theta \right) $ of (2.5) alone is not an oracle operation
in the $HSSS$ quantum search process. For the conventional quantum search
algorithm [$8$] the same identical state transformation may be intuitively
explained: Take action on the state $|x\rangle $ and the net action result
is nothing (or zero). Then in this case the reversible operation $%
C_{x_{0}}\left( \theta \right) $ of (2.5) is not considered as a selective
operation but a nonselective operation [$1$] in the physical Hilbert space.
It alone can be an oracle operation and may act as the oracle operation of
the conventional quantum search algorithm. And the state $|x\rangle $ is a
member of the unstructured search space as here the math Hilbert space is
just the physical Hilbert space. In this paragraph one sees clearly that
there is an essential difference between the $HSSS$ quantum search process
and the conventional quantum search algorithm: The unitary operation $%
C_{x_{0}}\left( \theta \right) $ of (2.5) is not an oracle operation for the
former, while it is for the latter.

Similarly, from the aspect of the mathematical-logical principle (or briefly
the mathematical-logical aspect) which is related to the math Hilbert space $%
H_{m}$, the duality-character oracle operation $C_{S}\left( \theta \right) $
of (2.1) or (2.2) is the unitary (or reversible) selective phase-shift
operation:%
\begin{equation}
C_{s}\left( \theta \right) :|S\rangle \rightarrow \exp \left( -i\theta
\right) |S\rangle  \tag{2.6}
\end{equation}%
with any candidate solution state $|S\rangle \neq |x_{0}\rangle .$ In the
math Hilbert space $H_{m}$ it can be fully expressed as%
\begin{equation}
C_{s}\left( \theta \right) :|y\rangle \rightarrow \left\{ 
\begin{array}{cc}
\exp \left( -i\theta \right) |S\rangle & if \\ 
|y\rangle & if%
\end{array}%
\begin{array}{c}
y=S \\ 
y\neq S%
\end{array}%
\right.  \tag{2.7}
\end{equation}%
Here both the candidate solution state $|S\rangle $ and the state $|y\rangle 
$ ($y\neq S$) are the basis states of the math Hilbert space, i.e., $%
|S\rangle ,$ $|y\rangle \in H_{m}.$ However, $|S\rangle $ is also a member
of the unstructured search space, while $|y\rangle $ is not.

There is the special case that in (2.6) and (2.7) the candidate solution
state $|S\rangle $ is taken as the real solution state $|x_{0}\rangle $ in
the math Hilbert space. It is known from (2.1) that the real solution state $%
|x_{0}\rangle $ is in the physical Hilbert space, but it also belongs to the
math Hilbert space. Though in theory $|x_{0}\rangle $ also belongs to the
math Hilbert space, i.e., $|x_{0}\rangle \in H_{m},$ it is not observable in
the math Hilbert space alone. And it can be observable only in the physical
Hilbert space. Generally, any state that appears in the math Hilbert space
alone is not observable in quantum\ mechanics. This is different from that
case in the physical Hilbert space.

A quantum-computing speedup process is a duality-character
quantum-com-puting process. Its characteristic property is that it can
change unitarily the quantum-state difference of a pair of quantum states.
Below it is shown that the duality-character oracle operation $C_{S}\left(
\theta \right) $ owns this characteristic property. This duality-character
oracle operation is applied simultaneously to both the physical and math
Hilbert spaces of an $n-$qubit spin system (or an $n-$qubit quantum system
generally). Denote $\left\vert \Psi _{0}\right\rangle $ as an arbitrary
state of the system. It can be expanded in terms of the conventional
computational bases $\{\left\vert x\right\rangle \}$ of the system: $%
\left\vert \Psi _{0}\right\rangle =\sum_{x=0}^{N-1}a_{x}\left\vert
x\right\rangle ,$ where $\sum_{x=0}^{N-1}\left\vert a_{x}\right\vert ^{2}=1.$
Now consider the state $\left\vert \Psi _{0}\right\rangle $ as any one of
the initial two states which belong to the physical and math Hilbert spaces,
respectively, when the oracle operation acts on both the Hilbert spaces at
the same time. Obviously, here both the initial states are taken as the same
one $(\left\vert \Psi _{0}\right\rangle )$, that is, they completely overlap
with one another. By acting the oracle operation $C_{S}\left( \theta \right) 
$ of (2.2) on the initial state $\left\vert \Psi _{0}\right\rangle $ one
obtains the final state $\left\vert \Psi \right\rangle =C_{S}\left( \theta
\right) \left\vert \Psi _{0}\right\rangle .$ This final state can be
determined explicitly in the duality-character description [$1$]. On the one
hand, in the physical Hilbert space the final state $\left\vert \Psi
\right\rangle $ is explicitly written as, with the help of the unitary
transformation of (2.5),%
\begin{equation}
\left\vert \Psi _{x_{0}}\right\rangle =a_{x_{0}}\exp \left( -i\theta \right)
\left\vert x_{0}\right\rangle +\sum_{x=0,x\neq x_{0}}^{N-1}a_{x}\left\vert
x\right\rangle  \tag{2.8}
\end{equation}%
On the other hand, in the math Hilbert space the final state $\left\vert
\Psi \right\rangle $ is explicitly expressed as, with the aid of the unitary
transformation of (2.7), 
\begin{equation}
\left\vert \Psi _{S}\right\rangle =a_{S}\exp \left( -i\theta \right)
\left\vert S\right\rangle +\sum_{y=0,y\neq S}^{N-1}a_{y}\left\vert
y\right\rangle  \tag{2.9}
\end{equation}%
where $\left\vert S\right\rangle \neq \left\vert x_{0}\right\rangle .$ The
quantum-state difference between the final two states $\left\vert \Psi
_{x_{0}}\right\rangle $ and $\left\vert \Psi _{S}\right\rangle $ may be
measured by the overlapping integration of these final two states. By using
these final states $\left\vert \Psi _{x_{0}}\right\rangle $ of (2.8) and $%
\left\vert \Psi _{S}\right\rangle $ of (2.9) the overlapping integration can
be calculated explicitly and it is given by%
\begin{equation}
\left\langle \Psi _{x_{0}}|\Psi _{S}\right\rangle =1-\left\vert
a_{x_{0}}\right\vert ^{2}\left( 1-\exp \left( +i\theta \right) \right)
-\left\vert a_{S}\right\vert ^{2}\left( 1-\exp \left( -i\theta \right)
\right)  \tag{2.10}
\end{equation}%
It is known that at the initial time the overlapping integration of the same
initial two states $\{\left\vert \Psi _{0}\right\rangle \}$ is equal to $%
\left\langle \Psi _{0}|\Psi _{0}\right\rangle =1.$ Now it can be seen from
(2.10) that generally the overlapping integral $\left\langle \Psi
_{x_{0}}|\Psi _{S}\right\rangle \neq 1$ at the final time. Therefore, the
overlapping integral $\left\langle \Psi _{0}|\Psi _{0}\right\rangle =1$ at
the initial time can be changed to $\left\langle \Psi _{x_{0}}|\Psi
_{S}\right\rangle \neq 1$ (in absolute value) at the final time, after the
oracle operation $C_{S}\left( \theta \right) $ is applied to the initial
state $\left\vert \Psi _{0}\right\rangle .$ This indicates that the
duality-character oracle operation of (2.1) or (2.2) owns the characteristic
property of quantum-computing speedup.

The experimental implementation for the duality-character oracle operation
may be based on the two inherent properties of the oracle operation: $%
(O_{i}) $ in the quantum-physical aspect the oracle operation owns the
unitary transformation of (2.5) and $(O_{ii})$ in the mathematical-logical
aspect it must obey the mathematical-logical principle of the unstructured
search problem that before the real solution is found, any candidate
solution state may be possibly considered as the real solution state in
searching for the real solution. In addition to these two inherent
properties $(O_{1})$ and $(O_{ii})$ the experimental implementation also may
employ the relations that both the physical Hilbert space and the math
Hilbert space have the same basis set and moreover, the basis set also forms
exactly the unstructured search space. Then, on the one hand, the second
property $(O_{ii})$ can be deduced from (2.4) and (2.6) together that before
the real solution state $\left\vert x_{0}\right\rangle $ is found, any
candidate solution state $\left\vert S\right\rangle $ may be possibly
considered as $\left\vert x_{0}\right\rangle $ in searching for the real
solution. On the other hand, it can be deduced from (2.5) that the second
property $(O_{ii})$ can be made equivalent to that any $QM$ basis state $%
\left\vert x\right\rangle $ may be possibly considered as $\left\vert
x_{0}\right\rangle $ in the physical Hilbert space, before $\left\vert
x_{0}\right\rangle $ is found. Why this equivalent relation is said to be
made? This is because the $QM$ basis state $\left\vert x\right\rangle $ ($%
x\neq x_{0})$ is not a member of the unstructured search space. However, one
may set up the one-to-one corresponding relation between the $QM$ basis
state $\left\vert x\right\rangle $ and the candidate solution state $%
\left\vert S\right\rangle .$ With the help of this one-to-one corresponding
relation the mathematical-logical principle of the unstructured search
problem that governs the oracle operation may be faithfully mapped onto the
physical Hilbert space. This strategy could be generalized to other
computational problems (not limited to an unstructured search problem), when
these problems are solved with the the quantum-computing speedup processes.

These two inherent properties $(O_{i})$ and $(O_{ii})$ of the
duality-character oracle operation have nothing to do with the two different
physical explanations mentioned above for the identical state transformation 
$|x\rangle \rightarrow |x\rangle $ for $x\neq x_{0}$ in the unitary
transformation (2.5). The first property $(O_{i})$ is due to that the
unitary transformation (2.5) itself is not able to distinguish the two
different physical explanations from one another. The second property $%
(O_{ii})$ is obvious.

The experimental implementation is carried out in the physical Hilbert space
and it realizes the quantum-physical aspect of the duality-character oracle
operation, which obeys the mathematical-logical principle of the
unstructured search problem. Apparently it seems not to involve the math
Hilbert space or the mathematical-logical aspect of the oracle operation.
However, for the duality-character oracle operation both the aspects must be
considered completely; no aspect is allowed to neglect; and either aspect
alone does not own the dual character. There is no exception for the
experimental and theoretical implementations of the duality-character oracle
operation.

From the viewpoint of pure quantum mechanics the unitary (or reversible)
transformation (2.5) may be simply realized in any $n-$qubit quantum system.
However, the purely quantum-mechanical unitary transformation (2.5) is not
exactly the quantum-physical aspect of the duality-character oracle
operation. The quantum-physical aspect is exactly the unitary (or
reversible) transformation (2.5) that obeys deterministically the
mathematical-logical principle of the unstructured search problem.

How to realize that the unitary transformation (2.5) obeys the
mathematical-logical principle? One scheme for the experimental
implementation is described as follows. In this scheme the
mathematical-logical principle is realized in the physical Hilbert space.
The faithful mapping for the mathematical-logical principle consists of the
following two steps from the duality-character oracle operation onto the
physical Hilbert space. The first step is to set up the one-to-one
corresponding relation between the candidate solution state $\left\vert
S\right\rangle $ of the unstructured search space in the math Hilbert space
and the $QM$ basis state $\left\vert x\right\rangle $ of the unitary
transformation (2.5). The second step is that the $QM$ basis state $%
|x\rangle $ is replaced with the member $|x\rangle $ of the unstructured
search space in the physical Hilbert space. This scheme could be outlined as%
\begin{equation*}
\text{The duality-character oracle operation (2.2)}
\end{equation*}%
\begin{equation*}
\Rightarrow \text{The unitary (or reversible) transformation (2.5)}
\end{equation*}%
\begin{equation*}
\Rightarrow \text{The unitary (or reversible) transformation (2.5) }
\end{equation*}%
\begin{equation*}
+\text{ The mathematical-logical principle,}
\end{equation*}%
where the mathematical-logical principle means that before the real solution
state $\left\vert x_{0}\right\rangle $ is found, any basis state $\left\vert
x\right\rangle $ in (2.5), which is also a member of the unstructured search
space, may be possibly considered as $\left\vert x_{0}\right\rangle $ in the
physical Hilbert space. Suppose in this paper that this statement that any
candidate solution state $\left\vert x\right\rangle $ may be possibly
considered as the real solution state $\left\vert x_{0}\right\rangle $ means
the deterministic description, while the `inverse' statement that the $%
\left\vert x_{0}\right\rangle $ may be possibly considered as any $%
\left\vert x\right\rangle $ means the indeterministic description. Then here
the unitary (or reversible) transformation (2.5) with the
mathematical-logical principle in the $HSSS$ quantum search process obeys
deterministically the mathematical-logical principle of the unstructured
search problem. In contrast, for the unitary (or reversible) transformation
(2.5) that acts as the oracle operation in the conventional quantum search
algorithm, whether the mathematical-logical principle is obeyed or not is
indeterministic. (See the last section `Discussion')

For the quantum-physical aspect of the duality-character oracle operation
(2.2) in this scheme the mathematical-logical principle of the unstructured
search problem is realized in the physical Hilbert space (i.e., via the
Bennett's reversible computation [$11$,$12$]) and obviously it is not
realized in the math Hilbert space (i.e., via the Lecerf's reversible
computation [$4$]). Particularly it is not realized directly in the hybrid
of the physical and math Hilbert spaces (i.e., via the duality-character
reversible computation or the interaction picture of reversible
computation). Due to this point this scheme may be considered as an indirect
experimental implementation.

A concrete (indirect) experimental implementation is already given in Refs. [%
$1,5,3$]. It consists of the reversible (or unitary) sequence of the
reversible Boolean functional operations [$1,5,3$]: 
\begin{equation*}
BFSEQ=V_{0}U_{f}V(\theta )U_{f}V_{0}
\end{equation*}%
where the $QM$ unitary operators $V\left( \theta \right) $ and $V_{0}$ and
the reversible Boolean functional operation $U_{f}$ which characterizes the
unstructured search problem are described in Ref. [$1$] in detail. This
reversible sequence $BFSEQ$ realizes the unitary (or reversible)
transformation (2.5) and obeys deterministically the mathematical-logical
principle of the unstructured search problem. And hence it may be used for
the indirect experimental implementation of the duality-character oracle
operation (2.2).

The selective phase-shift operation $C_{x_{0}}\left( \theta \right) $ of
(2.4) may be expressed as the unitary (or reversible) exponential
phase-shift operator $C_{x_{0}}\left( \theta \right) =\exp \left( -i\theta
D_{x_{0}}\right) $ with the diagonal operator $D_{x_{0}}=|x_{0}\rangle
\langle x_{0}|.$ Apparently the latter is dependent only upon the real
solution state $|x_{0}\rangle $ and is not involved in any other basis
states including any candidate solution state different from $|x_{0}\rangle
. $ Therefore, for the exponential phase-shift operator $C_{x_{0}}\left(
\theta \right) $ there is not ambiguity about the physical explanation of
the identical state transformation $|x\rangle \rightarrow |x\rangle $ for $%
x\neq x_{0}$ in (2.5) which is mentioned above. Moreover, by starting from
the exponential phase-shift operation $C_{x_{0}}\left( \theta \right) $ it
may be more clear to understand the mechanism for how the duality-character
oracle operation $C_{S}\left( \theta \right) =\exp \left( -i\theta
D_{S}\right) $ of (2.2) is implemented experimentally in the $HSSS$ quantum
search process in the scheme mentioned above. Note that $|x_{0}\rangle $
belongs to the physical Hilbert space and also the math Hilbert space. $(i)$
If any candidate solution state $\left\vert S\right\rangle $ in the math
Hilbert space may be possibly considered as the real solution state $%
|x_{0}\rangle $ in the exponential phase-shift operation $C_{x_{0}}\left(
\theta \right) $, then such exponential phase-shift operation $%
C_{x_{0}}\left( \theta \right) $ is the quantum-physical aspect of the
duality-character oracle operation $C_{S}\left( \theta \right) $. $(ii)$ If
any basis state $\left\vert x\right\rangle $ in the physical Hilbert space
that is not a candidate solution state may be possibly considered as $%
|x_{0}\rangle $ in the $C_{x_{0}}\left( \theta \right) $, then the $%
C_{x_{0}}\left( \theta \right) $ may stand for the quantum-physical aspect
of the oracle operation $C_{S}\left( \theta \right) .$ $(iii)$ The
experimental implementation mentioned above for the quantum-physical aspect:
If any basis state $\left\vert x\right\rangle $ in the physical Hilbert
space that is also a candidate solution state may be possibly considered as $%
|x_{0}\rangle $ in the $C_{x_{0}}\left( \theta \right) $, then the $%
C_{x_{0}}\left( \theta \right) $ may stand for the quantum-physical aspect
of the duality-character oracle operation $C_{S}\left( \theta \right) $ and
can be implemented experimentally. $(iv)$ If in the $C_{x_{0}}\left( \theta
\right) $ the real solution state $|x_{0}\rangle $ may be possibly
considered as any candidate solution state $\left\vert x\right\rangle $ in
the physical Hilbert space, then the $C_{x_{0}}\left( \theta \right) $ is
the oracle operation of the conventional quantum search algorithm.

A direct and native experimental implementation for the duality-character
oracle operation (2.2) should be carried out in the hybrid of the physical
and math Hilbert spaces. This really means that the mathematical-logical
principle of the unstructured search problem should be realized in the
hybrid of the physical and math Hilbert spaces (i.e., via the
duality-character reversible computation). The duality-character oracle
operation (2.2) is an exponential operator and also a unitary (or
reversible) selective diagonal operator. This point could be important for
the direct and native experimental implementation. Ultimately one could need
the duality-character reversible computation (or the interaction picture of
reversible computation) so that with the duality-character reversible
computation the duality-character oracle operation (or more generally any
duality-character functional operation) is constructed and realized directly
and naturally.

The duality-character oracle operation of the $HSSS$ quantum search process
is substantially different from the oracle operation of a conventional
quantum search algorithm, here the latter is essentially a purely
quantum-physical operation (See Ref. [$1$] and also the last section
`Discussion'). They reflect the striking difference: whether quantum
mechanics alone is sufficient to describe completely any quantum-computing
process or not.

The $HSSS$ quantum search process may be constructed and realized concretely
in the three consecutive stages: (1)\ the basic $SIC$ unitary operators are
prepared through the search-space dynamical reduction; (2) the $SIC$ unitary
propagator of a quantum system is constructed by starting from the basic $%
SIC $ unitary operators; (3) the dynamical QUANSDAM process is constructed
by directly using the $SIC$ unitary propagators. One main work in this paper
is devoted to the second stage which connects the first step (i.e., the
search-space dynamical reduction) and the second step (i.e., the dynamical
quantum-state-difference amplification) of the $HSSS$ quantum search
process. \newline
\newline
\newline
{\Large 3. Preparation and realization for the }${\Large SIC}${\Large \
unitary propagators}

The expected result that is finally obtained from the search-space dynamical
reduction is the basic \textit{s}olution-\textit{i}nformation-\textit{c}%
arrying ($SIC$) unitary operators that carry the information\ of the
solution to the unstructured search problem. These basic $SIC$\ unitary
operators were first proposed in the early 2001 [$3$]. They are the
duality-character unitary operators and own the characteristic property of
quantum-computing speedup [$1$]. They are further used to build a QUANSDAM
(or UNIDYSLOCK) process of the $HSSS$ quantum search process to solve the
unstructured search problem. Here for convenience suppose that the
search-space dynamical reduction for the $2^{n}-$dimensional unstructured
search space of the search problem is performed in an $n-$qubit spin $%
(I=1/2) $ system, and the reduced search space obtained finally from the
search-space dynamical reduction is two-dimensional, which is the smallest
reduced search space. Then the basic $SIC$ unitary operator, which is
prepared by the search-space dynamical reduction and has the two-dimensional
reduced search space, may be written as [$3,5,1$]%
\begin{equation}
U_{\lambda }^{sic}(a_{m}^{s},\theta _{m})=\exp (-i\theta
_{m}a_{m}^{s}I_{m\lambda }),\text{ }1\leq m\leq n,\text{ }\lambda =x,y,z 
\tag{3.1}
\end{equation}%
where the Cartesian spin operator $I_{m\lambda }=\frac{1}{2}\sigma
_{m\lambda }$ with the Pauli spin operator $\sigma _{m\lambda }$ of the $m-$%
th spin in the $n-$qubit spin system\footnote{%
In a general $n-$qubit quantum system the basic $SIC$\ unitary operator is
still defined by (3.1) by replacing the spin operator in (3.1) with the
corresponding pseudospin operator of the $n-$qubit quantum system.} and the
exponentially small rotating angle $\theta _{m}$ is proportional to $1/2^{n}$
$[5].$ The angle $\theta _{m}$ is exponentially smaller than that one $%
\left( \theta \right) $ of the original duality-character oracle operation
of (2.2), i.e., $C_{S}\left( \theta \right) =\exp \left( -i\theta
D_{S}\right) $. This is the cost to prepare the basic $SIC$ unitary operator
of (3.1). The most important one in (3.1) is the duality-character quantity $%
a_{m}^{s}$ which takes $a_{m}^{s}=\pm 1$ in value and is called the
double-valued mathematical-logical number (or the unit number [$3$]). It
represents the solution information of the unstructured search problem. It
carries the information of the $m-$th component state $\left( |s_{m}\rangle
\right) $ of the solution state $\left( |S\rangle =|s_{n}\rangle
...|s_{m}\rangle ...|s_{1}\rangle \right) $ of the search problem. Its
transfer process is deterministic [$1,2$] and obeys the information
conservation principle (See below). The basic $SIC$\ unitary operator
effectively acts on the Hilbert space of the $m-$th single spin of the $n-$%
qubit spin system, which is two-dimensional and minimum in quantum mechanics.

The basic $SIC$ unitary operator of (3.1) (up to a global phase factor), on
the one hand, is built out of the duality-character oracle operations of
(2.2) and the relevant $QM$ unitary operators [$3,5,2,1$]. Therefore, it is
still a duality-character unitary operation. Through the duality-character
oracle operations it still obeys the mathematical-logical principle of the
unstructured search problem. Thus, in principle it still can act as a
building block of the $HSSS$ quantum search process to solve the
unstructured search problem, although it belongs to the two-dimensional
reduced search space. On the other hand, dynamically its own reduced search
space is two-dimensional and hence far smaller than the original
exponentially large unstructured search space. This also means that the
basic $SIC$ unitary operators can run effectively in a simple quantum system
with a Hilbert space far smaller than the original exponentially large one
of the $n-$qubit spin system.

Quantum mechanically the basic $SIC$ unitary operator of (3.1) belongs to
the $m-$th spin of the $n-$qubit spin system. But it can be transformed to
another $SIC$ unitary operator of other quantum system than the $n-$qubit
spin system by a $QM$ unitary transformation. Thus, by the $QM$ unitary
transformation the solution information $\left( a_{m}^{s}\right) $ of the
basic $SIC$\ unitary operator may be transferred from the $n-$qubit spin
system to other quantum system such as a single-atom system.\footnote{%
Strictly speaking, this solution-information transfer should be from a
subsystem to another in composite quantum system, as shown later.} This
solution-information transfer obeys the information conservation principle
and is a deterministic process. This means that the information of the real
solution (any candidate solution) is never transferred from the physical
(math) Hilbert space to the math (physical) Hilbert space alone by the $QM$
unitary transformation.

The basic $SIC$ unitary operator is simple, but generally it could not be
used directly in the construction of a powerful QUANSDAM process. It does
not reflect the symmetrical structure and property of the quantum system
that is used to realize a QUANSDAM\ process. In contrast, a $SIC$\ unitary
propagator of the quantum system could be more advantageous in the
construction of a powerful QUANSDAM process, because it can directly take
into account the symmetrical structure and property of the quantum system
which is the fundamental quantum-computing-speedup resource to speed up a
quantum computing process.\textit{\ }This is the main reason why in this
section the author makes a great effort to prepare the $SIC$\ unitary
propagator of a quantum system such as a single-atom system. A $SIC$ unitary
propagator is usually employed directly to build a powerful QUANSDAM\
process. Generally, it is composed of the basic $SIC$ unitary operators and
the relevant $QM$ unitary operators [$1$]. Hence it is still a
duality-character unitary operation.

A $SIC$\ unitary propagator of a quantum system with time-independent
Hamiltonian $H$ may be obtained by loading the solution information ($%
a_{m}^{s}$) onto the propagator of the quantum system. In a simplest form it
may be formally written as [$1$]%
\begin{equation}
U_{H}^{sic}(a_{m}^{s},t_{m})=\exp (-ia_{m}^{s}Ht_{m}/\hslash )  \tag{3.2}
\end{equation}%
where the exponentially small time parameter $t_{m}$ is usually taken as $%
t_{m}\backsim \theta _{m}\backsim 1/2^{n},$ here $\theta _{m}$ is given in
(3.1). This $SIC$ unitary propagator is built out of the basic $SIC$ unitary
operators of (3.1). As proposed in Ref. [$1$], it may be generally created
by a sequence of the basic $SIC$ unitary operators $\left\{ U_{\lambda
_{l}}^{sic}(a_{m}^{s},\theta _{m}^{l})\right\} $ of (3.1) and the relevant $%
QM$ unitary operators $\{V_{l}\}$:%
\begin{equation}
USEQ\left( K\right) =V_{K}U_{\lambda _{K}}^{sic}(a_{m}^{s},\theta
_{m}^{K})V_{k-1}...V_{1}U_{\lambda _{1}}^{sic}(a_{m}^{s},\theta
_{m}^{1})V_{0}.  \tag{3.3}
\end{equation}%
Then theoretically this $SIC$ unitary operator sequence should be equal to
the $SIC$ unitary propagator of (3.2):%
\begin{equation}
U_{H}^{sic}(a_{m}^{s},t_{m})=USEQ\left( K\right) .  \tag{3.4}
\end{equation}%
This is the basic equation to construct theoretically the $SIC$ unitary
propagator. It is pointed out in Ref. [$1$] that a $SIC$\ unitary propagator
should be constructed and realized in a quantum system which owns a discrete
energy spectrum.

Obviously, the $SIC$ unitary propagator of (3.2) is still a
duality-character unitary operation and hence it can be further used to
solve the unstructured search problem, if it can be constructed through the
equation (3.4).

The time evolution process of the quantum system governed by the $SIC$
unitary propagator of (3.2) is written as%
\begin{equation}
|\Psi (a_{m}^{s},t_{m})\rangle =U_{H}^{sic}(a_{m}^{s},t_{m})|\Psi _{0}\rangle
\tag{3.5}
\end{equation}%
where $|\Psi _{0}\rangle $ is arbitrary initial state of the quantum system.
This is an effective time evolution process as any relevant auxiliary
quantum system is hidden here. This time evolution process is really a
UNIDYSLOCK (or QUANSDAM) process [$1$]. Theoretically it may be calculated
by a variety of quantum-mechanical methods. The typical methods include the
Green's function method (or the Feynman path integral technique [$14$]) and
the eigenfunction expansion method [$13,10$]. As pointed out in Ref. [$1$],
it could be more convenient to use the Green's function method [$14$] to
calculate theoretically and the energy eigenfunction expansion method to
realize experimentally the time evolution process of a quantum system which
owns a discrete energy spectrum in time and space.

According to the basic equation of (3.4) and with the help of (3.3) one also
can write the time evolution process of (3.5) as%
\begin{equation}
U_{H}^{sic}(a_{m}^{s},t_{m})|\Psi _{0}\rangle =V_{K}U_{\lambda
_{K}}^{sic}(a_{m}^{s},\theta _{m}^{K})V_{k-1}...V_{1}U_{\lambda
_{1}}^{sic}(a_{m}^{s},\theta _{m}^{1})V_{0}|\Psi _{0}\rangle  \tag{3.6}
\end{equation}%
This time evolution process also may be used to construct (exactly or
approximately) the $SIC$ unitary propagator of (3.2). It could be more
convenient to use this method to construct the $SIC$ unitary propagator of a
quantum system in time and space, when combining with the energy
eigenfunction expansion in quantum mechanics. As far as efficient
construction of the $SIC$ unitary propagator is concerned, this joint
construction method [$1$] based on the equation (3.6) and the energy
eigenfunction expansion is not involved in any quantum effect of the quantum
system with arbitrary state $|\Psi _{0}\rangle $ in (3.6).

The energy eigenfunction expansion principle is the special case of the
eigenfunction expansion principle which is considered as a mathematical, but
not a physical, principle in quantum mechanics [$10,13$].
Quantum-mechanically this general principle (the latter one) may be
generally used to treat a quantum system in time and space whose
eigenfunction expansion is infinite. Therefore, here consider mainly a
quantum system in time and space whose energy eigenfunction expansion is
infinite. This principle (the former one) simply states that any quantum
state of the quantum system can be expanded in terms of all the energy
eigenfunctions which form a complete eigenfunction set in the quantum
system. The energy eigenfunction expansion is theoretical basis for
realizing experimentally the time evolution process governed by a $SIC$
unitary propagator in a quantum system which has a discrete energy spectrum.
The key point in theory to use the energy eigenfunction expansion consists
in that infinite eigenfunction expansion series for any quantum state in a
quantum system is always convergent [$13$]. The energy (or momentum)
eigenfunction expansion has been used in the unitary manipulation of a
single atom with a Gaussian wave-packet state in time and space [$30$].

A further consideration is the fast-convergent energy eigenfunction
expansion for a quantum state with an infinite expansion series. If
convergence of eigenfunction expansion for any quantum state of a quantum
system is mathematical attribute, there is no reason why fast (or slow)
convergence should not be mathematical attribute. The fast convergence is
closely related to efficient construction (or generally computational
complexity of the construction) of a $SIC$ unitary propagator in a quantum
system which owns infinitely many discrete energy eigenfunctions. It has
nothing to do with whether the state that is expanded in a complete
eigenfunction set is quantum or classical [$1$], because a classical state
also may have a fast-convergent eigenfunction expansion. In fact, any state
that may be quantum or classical always can have a fast-convergent
eigenfunction expansion, when the complete eigenfunction set is chosen
suitably. Later in the subsection 3.2 it will be shown how the
fast-convergent eigenfunction expansion is explicitly related to efficient
construction of a $SIC$ unitary propagator.

Fast-convergent eigenfunction expansion of a wave-packet quantum state%
\footnote{%
Here it could be more suitable to name a wave-packet quantum state a shaped
quantum state.} could be related to the wave-packet shape of the quantum
state (See, for example, Refs. $[15,30]$). Intuitively (not strictly)
speaking, more narrow a wave-packet state in shape, faster the convergence
of eigenfunction expansion of the state. For example, if a Gaussian shaped
(wave-packet) momentum state in one-dimensional momentum space has a more
narrow wave-packet spread, then convergence is faster for the momentum
eigenfunction expansion of the state.

Now it is considered concretely that a $SIC$ unitary propagator is
constructed theoretically in a single-atom system in time and space. A
single atom owns the internal-motion degree of freedom and the
external-motion (or center-of-mass (COM) motion) degrees of freedom. In the
absence of the interaction between the atomic internal and COM motions, the
Hamiltonian of the single-atom system is the sum of the two independent
parts, one of which is of the atomic internal motion and another of the
atomic COM motion. Denote $H_{a}$ and $H_{c}$ as the internal-motion and the
external-motion part of the atomic Hamiltonian, respectively. Then the
atomic Hamiltonian is given by $H=H_{a}+H_{c}.$ Here assume that all these
three Hamiltonians $H,$ $H_{a},$ and $H_{c}$ are time-independent. The time
evolution propagator generated by the atomic Hamiltonian $H$ is $%
U_{H}(t)=\exp \left( -iHt/\hslash \right) .$ It describes the time evolution
process of the single-atom system in the absence of the interaction.
Similarly, the time evolution propagators generated by the Hamiltonians $%
H_{c}$ and $H_{a}$ are given by $U_{c}(t)=\exp \left( -iH_{c}t/\hslash
\right) $ and $U_{a}(t)=\exp \left( -iH_{a}t/\hslash \right) ,$
respectively. They describe the COM motion and the internal motion of the
single-atom system, respectively. All these three time evolution propagators
are purely quantum-mechanical. Now according to the expression (3.2) with
the Hamiltonian $H,$ $H_{c},$ or $H_{a}$ there are formally the three $SIC$
unitary propagators $U_{H}^{sic}(t)=\exp \left( -ia_{m}^{s}Ht/\hslash
\right) ,$ $U_{c}^{sic}(t)=\exp \left( -ia_{m}^{s}H_{c}t/\hslash \right) ,$
and $U_{a}^{sic}(t)=\exp \left( -ia_{m}^{s}H_{a}t/\hslash \right) ,$ which
are correspondent to these three $QM$\ unitary propagators $U_{H}\left(
t\right) ,$ $U_{c}\left( t\right) ,$ and $U_{a}\left( t\right) ,$
respectively.

Below it is described mainly how to construct and realize the $SIC$ unitary
propagator: 
\begin{equation*}
U_{at}^{sic}(t_{m})=\exp \left[ -ia_{m}^{s}\left( H_{a}^{\prime
}+H_{c}\right) t_{m}/\hslash \right]
\end{equation*}%
with an appropriate internal-motion Hamiltonian $H_{a}^{\prime }$ in a
single-atom system with external harmonic potential field, one special form
of a general quadratic potential field. This $SIC$\ unitary propagator
corresponds to the $QM$ unitary propagator $U_{at}(t_{m})=\exp \left[
-i\left( H_{a}^{\prime }+H_{c}\right) t_{m}/\hslash \right] .$ Obviously, $%
U_{at}^{sic}(t_{m})=U_{H}^{sic}(t_{m})$ when $H_{a}^{\prime }=H_{a}$ and $%
U_{at}^{sic}(t_{m})=U_{c}^{sic}(t_{m})$ up to a global phase factor $\exp
\left( -ia_{m}^{s}t_{m}/\hslash \right) $ when $H_{a}^{\prime }=E.$ Though
the $SIC$ unitary propagator $U_{at}^{sic}(t_{m})$ is not explicitly
involved in the interaction between the atomic internal and COM motions, it
will be seen below that its theoretical construction and experimental
realization in the single-atom system must employ this interaction from the
perspective of the unitary quantum dynamics of the single-atom system.
Manipulating and controlling unitarily a single-atom system by using this
interaction is pivotal in the unitary manipulation of a single atom in space
and time [$15,30,31$].

Before the $SIC$ unitary propagator is constructed in a single-atom system,
it is needed to know some basic knowledge of a single-atom system in quantum
mechanics. This also is the basic knowledge for the unitary manipulation of
a single atom in space and time. Since a single-atom system is described by
both its internal and external motions independently, its Hilbert space is
the tensor product of the component Hilbert subspace of its internal motion
and the one of its COM motion. Suppose that $\{\varphi _{j}\left( \mathbf{r,x%
}\right) \}$\ is the complete energy eigenfunction orthonormal set of the
Hilbert space of the single-atom system; and $\left\{ u_{k}\left( \mathbf{x}%
\right) \right\} $ and $\left\{ \psi _{k}\left( \mathbf{r}\right) \right\} $
are the complete energy eigenfunction orthonormal sets of the component
Hilbert subspaces of the COM motion and the internal motion of the
single-atom system, respectively. Here $\left\{ \varphi _{j}\left( \mathbf{%
r,x}\right) \right\} $ may be conveniently taken as the tensor-product basis
set $\left\{ \varphi _{kl}\left( \mathbf{r,x}\right) \right\} =\left\{ \psi
_{k}\left( \mathbf{r}\right) \tbigotimes u_{l}\left( \mathbf{x}\right)
\right\} \equiv \left\{ \psi _{k}\left( \mathbf{r}\right) u_{l}\left( 
\mathbf{x}\right) \right\} .$ Then the energy-eigenvalue equations for the
above atomic Hamiltonians $H=H_{a}+H_{c},$ $H_{c},$ and $H_{a}$ are $%
H\varphi _{kl}\left( \mathbf{r,x}\right) =E_{kl}\varphi _{kl}\left( \mathbf{%
r,x}\right) ,$ $H_{c}u_{k}\left( \mathbf{x}\right) =E_{k}^{c}u_{k}\left( 
\mathbf{x}\right) ,$ and $H_{a}\psi _{k}\left( \mathbf{r}\right)
=E_{k}^{a}\psi _{k}\left( \mathbf{r}\right) ,$ respectively. Here $%
E_{kl}=E_{k}^{a}+E_{l}^{c}$ is the total energy (eigenvalue) of the
single-atom system associated with the eigenfunction $\varphi _{kl}\left( 
\mathbf{r,x}\right) $; and $E_{l}^{c}$ and $E_{k}^{a}$ are the energies of
the COM motion and the internal motion of the single-atom system associated
with $u_{l}\left( \mathbf{x}\right) $ and $\psi _{k}\left( \mathbf{r}\right) 
$, respectively. It is known in quantum mechanics [$10$] that these energy
eigenvalues $E_{kl},$ $E_{k}^{c},\ $and $E_{k}^{a}$ all are discrete for a
single atom in external harmonic potential field or in infinite square
potential well.

Different species of atoms may have different internal-motion Hamiltonians,
but their COM-motion Hamiltonians may be the same one which may be
controlled simply by an external potential field. Generally, the
internal-motion Hamiltonian of a single atom may be written as $%
H_{a}=\sum_{k}E_{k}^{a}\left\vert \psi _{k}\right\rangle \left\langle \psi
_{k}\right\vert $ in the energy representation ($\left\vert \psi
_{k}\right\rangle \equiv \psi _{k}\left( \mathbf{r}\right) $). Suppose that
the atomic COM motion is in one-dimensional coordinate space. Then the
time-independent COM-motion Hamiltonian of a single atom in external
harmonic potential field may be written as%
\begin{equation}
H_{c}=\allowbreak \frac{1}{2m}p^{2}+\frac{1}{2}m\omega ^{2}x^{2},  \tag{3.7a}
\end{equation}%
where $m$ is the atomic mass and $\omega $ the COM-motion oscillatory
(angular) frequency of the atom in the harmonic potential field. This type
of quadratic (or harmonic) COM-motion Hamiltonians have been realized
experimentally in an individual (atomic) ion in external quadratic (or
harmonic) potential field [$16,17$] since 1970s and early 1980s. It is known
in quantum mechanics [$10$] that the atomic harmonic COM motion with the
Hamiltonian $H_{c}$ of (3.7a) has the discrete energy eigenvalue $%
E_{k}^{c}=\left( k+\frac{1}{2}\right) \hslash \omega $ with $k=0,1,2,...$
and its associated energy eigenbase: 
\begin{equation}
u_{k}\left( x\right) =\sqrt{\frac{\alpha }{\sqrt{\pi }2^{k}k!}}H_{k}\left(
\alpha x\right) \exp \left( -\frac{1}{2}\alpha ^{2}x^{2}\right)  \tag{3.7b}
\end{equation}%
with $\alpha ^{2}=m\omega /\hslash .$ All these infinitely many discrete
energy eigenbases $\{u_{k}\left( x\right) \}$ form a complete orthonormal
basis set. For a single atom in infinite square potential well ($\left\vert
x\right\vert \leq L$) [$10$] the atomic COM motion has the energy eigenvalue 
$E_{n}^{c}=\pi ^{2}\hslash ^{2}n^{2}/(8mL^{2})$ with $n=1,2,...$ and its
associated energy eigenbase $u_{n}\left( x\right) =\sqrt{\frac{1}{L}}\sin
\left( \frac{n\pi x}{2L}\right) $ with $n=2,4,...$ and $u_{n}\left( x\right)
=\sqrt{\frac{1}{L}}\cos \left( \frac{n\pi x}{2L}\right) $ with $n=1,3,...$
All these infinitely many discrete energy eigenbases $\{u_{n}\left( x\right)
\}$ also form a complete orthonormal basis set.

The eigenfunction expansion principle shows that arbitrary quantum state $%
\Psi \left( \mathbf{x},\mathbf{r,}t_{0}\right) $ of a single-atom system may
be expanded in the complete set of the energy eigenbases $\{\varphi
_{kl}\left( \mathbf{r,x}\right) \},\ i.e.,$ $\Psi \left( \mathbf{x},\mathbf{%
r,}t_{0}\right) =\sum_{kl}C_{kl}(t_{0})\psi _{k}\left( \mathbf{r}\right)
u_{l}\left( \mathbf{x}\right) ,$ where $C_{kl}(t_{0})$ is the expansion
coefficient of the eigenbase $\varphi _{kl}\left( \mathbf{r,x}\right) =\psi
_{k}\left( \mathbf{r}\right) u_{l}\left( \mathbf{x}\right) .$ This expansion
series is infinite at least because $\left\{ u_{k}\left( \mathbf{x}\right)
\right\} $ is an infinite eigenbasis set, but it is always convergent [$13$%
]. The energy eigenfunction expansion may be used to construct the $SIC$
unitary propagator $U_{at}^{sic}(t_{m})$ on the basis of the formula (3.6)
in the single-atom system with external harmonic potential field. This is
the joint construction method [$1$] that is applied to the single-atom
system. Here for simplicity the atomic internal Hamiltonian $H_{a}^{\prime }$
is taken as the simplest form: $H_{a}^{\prime }=E_{0}^{a}\left\vert \psi
_{0}\right\rangle \left\langle \psi _{0}\right\vert $ with the atomic
internal ground state $\left\vert \psi _{0}\right\rangle $ (i.e., $\psi
_{0}\left( \mathbf{r}\right) $)$.$ In this simplest case the $%
U_{at}^{sic}(t_{m})$ is a simplest $SIC$ unitary propagator that acts on the
whole single-atom system. In this section this simplest $SIC$\ unitary
propagator is treated in detail. For a more complex Hamiltonian $%
H_{a}^{\prime }$ this joint construction method is still available, and
hence a similar treatment can be available as well for a more complex $SIC$\
unitary propagator, but it will not be further described in this section.

Now the state $|\Psi _{0}\rangle $ on the right-hand (RH) side of (3.5) or
on the left-hand (LH) side of (3.6) may be chosen as $\Psi \left( \mathbf{x},%
\mathbf{r,}t_{0}\right) =\left\vert \psi _{0}\right\rangle \tbigotimes \Psi
_{c}\left( x,t_{0}\right) $ of the single-atom system, where $\Psi
_{c}\left( x,t_{0}\right) $ is an arbitrary state of the atomic COM motion.
A particularly important case is that the atomic COM-motion state $\Psi
_{c}\left( x,t_{0}\right) $ is taken as a Gaussian wave-packet state [$%
1,2,15,31,30,34$]. Note that all the COM-motion energy eigenbases of the
single-atom system form a complete orthonormal basis set. Then according to
the energy eigenfunction expansion the state $\Psi _{c}\left( x,t_{0}\right) 
$ can be expanded in terms of the COM-motion energy eigenbases $\left\{
u_{k}\left( x\right) \right\} :$%
\begin{equation}
\Psi _{c}\left( x,t_{0}\right) =\sum_{l=0}^{\infty }B_{l}(t_{0})u_{l}\left(
x\right)  \tag{3.8}
\end{equation}%
where $B_{l}(t_{0})$ is an expansion coefficient. This expansion series is
infinite, but it is always convergent. For the single-atom system with
external harmonic potential field the discrete energy eigenbase $u_{k}\left(
x\right) $ in (3.8) is given by (3.7b). The energy eigenfunction expansion
for the state $|\Psi _{0}\rangle $ on the RH side of (3.5) or on the LH side
of (3.6) then may be written in the form%
\begin{equation}
|\Psi _{0}\rangle =\left\vert \psi _{0}\right\rangle \tbigotimes \Psi
_{c}\left( x,t_{0}\right) =\left\vert \psi _{0}\right\rangle \tbigotimes
\sum_{l=0}^{\infty }B_{l}(t_{0})u_{l}\left( x\right)  \tag{3.9}
\end{equation}%
This expansion series is still infinite, but it is always convergent.

The time evolution process (3.5)\ of the single-atom system with the initial
state $|\Psi _{0}\rangle $ of (3.9), when the system is acted on by the $SIC$
unitary propagator $U_{at}^{sic}(t_{m})$ with $H_{a}^{\prime
}=E_{0}^{a}\left\vert \psi _{0}\right\rangle \left\langle \psi
_{0}\right\vert ,$ is explicitly written as [$1$]%
\begin{equation*}
\Psi ^{sic}\left( x\mathbf{,}r\mathbf{,}t_{m}\right)
=U_{at}^{sic}(t_{m})|\Psi _{0}\rangle
\end{equation*}%
\begin{equation}
=\exp \left( -ia_{m}^{s}E_{0}^{a}t_{m}/\hslash \right) \left\vert \psi
_{0}\right\rangle \tbigotimes \sum_{l=0}^{\infty }B_{l}(t_{0})\exp \left(
-ia_{m}^{s}E_{l}^{c}t_{m}/\hslash \right) u_{l}(x).  \tag{3.10}
\end{equation}%
Similarly, the inverse of the time evolution process may be expressed as%
\begin{equation*}
\Psi ^{sic}\left( x\mathbf{,}r\mathbf{,-}t_{m}\right)
=U_{at}^{sic}(t_{m})^{+}|\Psi _{0}\rangle
\end{equation*}%
\begin{equation*}
=\exp \left( ia_{m}^{s}E_{0}^{a}t_{m}/\hslash \right) \left\vert \psi
_{0}\right\rangle \tbigotimes \sum_{l=0}^{\infty }B_{l}(t_{0})\exp \left(
ia_{m}^{s}E_{l}^{c}t_{m}/\hslash \right) u_{l}(x).
\end{equation*}%
Both the time evolution process and its inverse can be calculated and
realized in the same way.

In order to employ the joint construction method to construct the $SIC$
unitary propagator $U_{at}^{sic}(t_{m})$ one needs to calculate explicitly
both sides of the equation (3.6). The time evolution process (3.10) (or its
inverse) may be used to calculate the LH side of (3.6) (or its inverse),
when the state $|\Psi _{0}\rangle $ on the LH side of (3.6) is expressed as
the infinite expansion series of (3.9). Here only the single-atom system is
considered, since the $SIC$ unitary propagator acts on effectively the
single-atom system, as shown in (3.10). However, in the joint construction
method one needs to consider explicitly the whole composite quantum system
whose component systems are the single-atom system and the spin system that
is used to prepare the basic $SIC$ unitary operators of (3.1) as well as the
relevant auxiliary quantum system, when the equation (3.6) is used to
construct theoretically the $SIC$ unitary propagator$\ $and especially when
one treats theoretically the RH side of (3.6).

Because the expansion series (3.9) is infinite, though the $SIC$\ unitary
propagator $U_{at}^{sic}(t_{m})$ could be exactly constructed in theory by
the joint construction method, it could not be realized in practice.
Generally the $SIC$ unitary propagator can be constructed approximately by
the joint construction method based on the equation (3.6) and the infinite
expansion series (3.9). In this case it may be approximately realized in
practice.

Theoretically the energy eigenfunction expansion principle to realize the
time evolution process of (3.10) is to generate the time($t_{m}$)-dependent $%
SIC$ complex exponential functions $\{\exp \left[ -ia_{m}^{s}\left(
E_{0}^{a}+E_{k}^{c}\right) t_{m}/\hslash \right] \}$ for all the energy
eigenbases (or levels) $\{\psi _{0}\left( \mathbf{r}\right) u_{k}(x)\}$ in
(3.10) at the same time. Here, as shown in (3.10), $\exp \left(
-ia_{m}^{s}E_{0}^{a}t_{m}/\hslash \right) $ is a global phase factor and may
be discarded. And the COM-motion energy eigenvalue $E_{k}^{c}$ for the
single-atom system is discrete and dependent on the quantum number $k$. As
one typical example, for the single-atom system with external harmonic
potential field one has $E_{k}^{c}=\left( k+1/2\right) \hslash \omega $ with
the associated energy eigenbase $u_{k}\left( x\right) $ of (3.7b). This
energy eigenvalue $E_{k}^{c}$ is a linear function of the quantum number $k$%
. As another typical example, for the single-atom system with infinite
square potential well ($\left\vert x\right\vert \leq a$) this energy
eigenvalue is given by $E_{k}^{c}=\pi ^{2}\hslash ^{2}k^{2}/(8ma^{2})$ [$10$%
] (See above also). It is proportional to the square quantum number $k^{2}$
or it is a quadratic function of the quantum number $k$. Generally, for any
quantum system with energy eigenvalue set $\left\{ E_{k}\right\} $ the time($%
t_{m}$)-dependent $SIC$ complex exponential function may be formally written
as%
\begin{equation}
\exp \left( -ia_{m}^{s}E_{k}t_{m}/\hslash \right) =\exp \left[
-ia_{m}^{s}\alpha \left( t_{m}\right) f\left( \left\{ k\right\} \right) %
\right]  \tag{3.11}
\end{equation}%
where $f\left( \left\{ k\right\} \right) $ is a real function of the quantum
number set $\left\{ k\right\} $ and $\alpha \left( t_{m}\right) $ is a real
function of the time $t_{m}$ and independent of the quantum number set $%
\left\{ k\right\} $. Obviously, $\exp \left[ -ia_{m}^{s}\left(
E_{0}^{a}+E_{k}^{c}\right) t_{m}/\hslash \right] $ in (3.10) is a special
case of the general $SIC$ complex exponential function of (3.11). Thus, it
can be expressed as (3.11) and it also can, even if the global phase factor $%
\exp \left( -ia_{m}^{s}E_{0}^{a}t_{m}/\hslash \right) $ is omitted from it.
The function $f\left( \left\{ k\right\} \right) $ is characterized
completely by the energy eigenvalues of the quantum system. For the COM\
motion of the single-atom system in the two typical examples above the
quantum number set $\left\{ k\right\} $ may contain only one quantum number $%
k$. But for a complex quantum system such as a composite quantum system it
may contain many independent quantum numbers.

The energy eigenvalue equation sets up the relation between the quantum
number (energy eigenvalue) and its associated eigenfunction, which is
perhaps the oldest number-state relation in quantum mechanics. It is a
theoretical basis for the energy eigenfunction expansion principle to
realize (or calculate) the time evolution process of a quantum system.

So far the LH side of the equation (3.6) is treated clearly. But the much
more important is how to treat clearly the RH side of (3.6) for the
construction of the $SIC$ unitary propagator $U_{at}^{sic}(t_{m})$. It is
known from the equation (3.6) that the RH side of (3.6) consists of the
basic $SIC$ unitary operators and the relevant $QM$ unitary operators, where
only the basic $SIC$ unitary operators carry the solution information $%
\left( a_{m}^{s}\right) $ of the search problem. Then the solution
information carried by the $SIC$ unitary propagator $U_{at}^{sic}(t_{m})$ on
the LH side of (3.6) is uniquely original from the basic $SIC$ unitary
operators $\left\{ U_{\lambda _{l}}^{sic}(a_{m}^{s},\theta _{m}^{l})\right\} 
$ on the RH side of (3.6). However, here the $SIC$ unitary propagator $%
U_{at}^{sic}(t_{m})$ ($i.e.,U_{H}^{sic}(a_{m}^{s},t_{m})$) on the LH side of
(3.6) runs effectively in the single-atom system, while the basic $SIC$
unitary operators on the RH side of (3.6) run effectively in the $n-$qubit
spin system. Therefore, the theoretical treatment for the RH side of (3.6)
must consider the whole composite quantum system that consists of the spin
system which is used to prepare the basic $SIC$ unitary operators, the
single-atom system, and the relevant auxiliary quantum system.

If the $SIC$ unitary propagator $U_{at}^{sic}(t_{m})$ can be constructed
through the equation (3.6), then in the construction the solution
information $\left( a_{m}^{s}\right) $ must be transferred from the
(component) Hilbert subspace of the spin system to the (component) one of
the single-atom system in the composite quantum system; and moreover this
solution-information transfer is realized only through the RH side of (3.6).
Conversely, this solution-information transfer that is realized on the RH
side of (3.6) is necessary to construct the $SIC$ unitary propagator by the
equation (3.6).

Now it is described how the solution-information transfer on the RH side of
(3.6) is realized in the composite quantum system under the condition that
the equation (3.6) holds. As shown in the basic equation (3.4), due to the
basic $SIC$ unitary operators the $SIC$ unitary propagator $%
U_{at}^{sic}(t_{m})$ obeys the mathematical-logical principle of the
unstructured search problem. Consequently this solution-information transfer
obeys the information conservation principle and it is a deterministic
process. It has been proposed that the solution-information transfer is
realized by the two consecutive steps in the composite quantum system [$1,2$%
]. The first step is to transfer the solution information from the Hilbert
subspace of the spin system to the one of the internal motion of the
single-atom system. The second step is within the Hilbert subspace of the
single-atom system and it is to transfer the solution information from the
Hilbert subspace of the atomic internal motion to the one of the atomic COM
motion. This two-step scheme for the solution-information transfer is chosen
mainly based on the following three points. $\left( i\right) $ It is usually
easier to achieve a strong interaction between the spin system and the
atomic internal motion and it also can be easily realized for a strong
interaction between the internal motion and the COM motion within a
single-atom system. Both the strong interactions can make the
solution-information transfer easy and fast. $\left( ii\right) $ The
interaction between the atomic internal and COM motions is needed for the
unitary selective manipulation of the atomic COM motion, because the unitary
selective manipulation (e.g., transition or excitation) among the
equal-space COM-motion energy levels needs the help from the discrete and
unequal-space internal-motion energy levels. $\left( iii\right) $ The atomic
COM motion usually could be more fragile and sensitive to the effect of its
environment than the atomic internal motion, where the effect could cause
decoherence of the atomic COM motion. Thus, during the solution-information
transfer the atomic COM motion should avoid maximally a direct contact with
its environment (e.g., the spin system and the auxiliary quantum system and
so on). In this two-step scheme only the first step is involved in a direct
and necessary contact between the single-atom system and its environment.
Therefore, except at the first step the single-atom system can keep isolated
from its environment during the solution-information transfer. An isolated
quantum system obeys strictly the unitary quantum dynamics.\newline
\newline
\newline
{\Large 3.1. Existence of the }${\Large SIC}${\Large \ unitary propagators}

In this subsection it proves that theoretically the $SIC$ unitary propagator 
$U_{at}^{sic}(t_{m})$ in (3.10) can be constructed exactly by the equation
(3.6) up to a global phase factor. This really proves that theoretically
there exists a $SIC$ unitary propagator of (3.2) that can be exactly
constructed by the basic equation (3.4) up to a global phase factor. Here
the number-state relation between quantum number and eigenfunction that is
obtained from the eigenvalue equation in quantum mechanics plays an
important role in showing that there exists the $SIC$\ unitary propagator.

For convenience consider first a $d-$qubit spin system which consists of $d$
non-interacting spins with $I=1/2$ and each spin of which acts as a qubit.
In theory the spin (or qubit) number $d$ may be infinitely large and hence
the spin system may have infinitely many eigenbases. Here the
quantum-mechanical knowledge for spin angular momentum is simply introduced.
According to the spin angular momentum theory in quantum mechanics [$%
18,19,10 $] one usually chooses the representation defined by both the
square spin operator ($\mathbf{I}^{2}$) and the $z-$component spin operator (%
$I_{z})$, which commute with each other, to investigate the spin angular
momentum properties for a single spin. (The spin angular-momentum operator $%
\mathbf{J} $ is equal to the spin operator $\mathbf{I}$ in unit $\hslash ,$
and the latter is related to the Pauli spin operator $\mathbf{\sigma }$ by $%
\mathbf{I}=\frac{1}{2}\mathbf{\sigma }$ for a single spin with $I=1/2.$) For
a multiple-spin system such as the $d-$qubit spin system one usually may
choose simply the representation defined by the commuting spin operator set $%
\{\mathbf{I}_{i}^{2},I_{iz}\}$ $(i=1,2,...,d)$ of all the $d$ spins of the
spin system. Suppose now that $\left\vert I_{i},m_{i}\right\rangle $ is the
common eigenbase of both the spin operators $\mathbf{I}_{i}^{2}$ and $I_{iz}$
of the $i-$th spin of the spin system. Then the tensor-product base $%
\left\vert I_{d},m_{d}\right\rangle ...\left\vert I_{2},m_{2}\right\rangle
\left\vert I_{1},m_{1}\right\rangle \equiv
\tbigotimes\nolimits_{l=1}^{d}\left\vert I_{d-l+1},m_{d-l+1}\right\rangle ,$
which is the tensor (or direct) product of all the $d$ single-spin
eigenbases $\{\left\vert I_{i},m_{i}\right\rangle \}$ with $i=1,2,...,d$, is
a common eigenbase of the commuting spin operator set $\{\mathbf{I}%
_{i}^{2},I_{iz}\}.$ For convenience hereafter denote $\left\vert
m_{i}\right\rangle $ as $\left\vert I_{i},m_{i}\right\rangle $ briefly. Now
the eigenvalue equation for the $i-$th spin operator $I_{iz}$ is given by [$%
10$]: $I_{iz}\left\vert m_{i}\right\rangle =m_{i}\left\vert
m_{i}\right\rangle ,$ where $m_{i}=\pm 1/2$ is the spin magnetic quantum
number and $\left\vert m_{i}\right\rangle $ is its associated spin
eigenbase. Then the total $z-$component spin operator $I_{z}=\tsum%
\nolimits_{i=1}^{d}I_{iz}$ of the spin system has the eigenvalue equation: $%
I_{z}\left\vert m_{d}\right\rangle ...\left\vert m_{2}\right\rangle
\left\vert m_{1}\right\rangle =M\left\vert m_{d}\right\rangle ...\left\vert
m_{2}\right\rangle \left\vert m_{1}\right\rangle ,$ where the eigenvalue,
i.e., the total spin magnetic quantum number $M,$ is given by $%
M=\tsum\nolimits_{i=1}^{d}m_{i}$ and its associated eigenbase is just the
tensor-product base $\left\vert m_{d}\right\rangle ...\left\vert
m_{2}\right\rangle \left\vert m_{1}\right\rangle $. The tensor-product base
thus is an eigenbase of the total spin operator $I_{z}.$ Moreover, it is
also a common eigenbase of the $z-$component spin-spin interaction operators 
$\{I_{iz}I_{jz}\}$. Therefore, as usual these $z-$component spin operators
and spin-spin interaction operators, i.e., $I_{iz},$ $I_{iz}I_{jz},$ etc.$,$
are diagonal operators in the spin angular momentum theory.

The $i-$th spin of the $d-$qubit spin system has the spin Hamiltonian $%
H_{i}=\Omega _{i}I_{iz},$ when the spin system is in an external field,
e.g., an external magnetic field with $z-$direction. Obviously the spin
Hamiltonian of the spin system is given by $H=\tsum\nolimits_{i=1}^{d}H_{i}.$
It is obvious that the single-spin eigenbase $\left\vert m_{i}\right\rangle $
is also an energy eigenbase of the spin Hamiltonian $H_{i}.$ Thus, the $i-$%
th spin has the energy eigenbasis set $\{\left\vert m_{i}\right\rangle \}$
which consists of only two orthonormal eigenbases $\left\vert
1/2\right\rangle $ and $\left\vert -1/2\right\rangle .$ It can be found that
the tensor-product base $\left\vert m_{d}\right\rangle ...\left\vert
m_{2}\right\rangle \left\vert m_{1}\right\rangle $ is an energy eigenbase of
the spin Hamiltonian $H$ of the spin system. Thus, the $d-$qubit spin system
owns the tensor-product energy eigenbasis set $\{\left\vert
m_{d}\right\rangle ...\left\vert m_{2}\right\rangle \left\vert
m_{1}\right\rangle \}.$ The energy-eigenvalue equation for the spin
Hamiltonian $H$ is given by 
\begin{equation*}
H\left\vert m_{d}\right\rangle ...\left\vert m_{2}\right\rangle \left\vert
m_{1}\right\rangle =\left( \tsum\nolimits_{i=1}^{d}\Omega _{i}m_{i}\right)
\left\vert m_{d}\right\rangle ...\left\vert m_{2}\right\rangle \left\vert
m_{1}\right\rangle .
\end{equation*}%
When $d$ is infinitely large, the eigenbase number $2^{d}$ of the energy
eigenbasis set of the $d-$qubit spin system is infinitely large. Once all
the energy eigenvalues and the energy eigenbasis set of the spin system are
obtained, the time evolution process of the spin system may be evaluated
conveniently by using the energy eigenfunction expansion. Finally, it is
pointed out that the above quantum-mechanical treatment for the $d-$qubit
non-interacting quantum spin system can be applied as well to a $d-$qubit
interacting quantum spin system whose spin-spin interactions are given by $%
\{A_{kl}I_{kz}I_{lz}\}$ with interacting constants $\{A_{kl}\}.$

Now by starting from the basic $SIC$ unitary operators of (3.1) one can
easily construct the $SIC$ unitary diagonal propagator of the $d-$qubit spin
system with $d=\infty $: 
\begin{equation*}
U_{\infty }^{sic}\left( a_{m}^{s}\right) =\exp (-ia_{m}^{s}\theta _{m\infty
}I_{\infty z})\tbigotimes ...
\end{equation*}%
\begin{equation}
\tbigotimes \exp (-ia_{m}^{s}\theta _{m2}I_{2z})\tbigotimes \exp
(-ia_{m}^{s}\theta _{m1}I_{1z})  \tag{3.12}
\end{equation}%
where the spin number $d=\infty $ is determined from $d=\log _{2}L$ for
given eigenbase number $L=2^{d}\rightarrow \infty $ of the energy eigenbasis
set of the spin system. Here each spin of the spin system can evolve
independently, and these rotation angles $\left\{ \theta _{mi}\right\} $ in
(3.12) can be set independently. If the $U_{\infty }^{sic}\left(
a_{m}^{s}\right) $ of (3.12) is written in the exponential operator form $%
U_{\infty }^{sic}\left( a_{m}^{s}\right) =\exp \left( -ia_{m}^{s}H_{\infty
}t_{m}/\hslash \right) ,$ then the spin Hamiltonian $H_{\infty }=\left(
t_{m}/\hslash \right) ^{-1}\sum_{l=1}^{\infty }\theta _{ml}I_{lz}.$ It can
be seen from the preceding paragraph that the tensor-product bases $%
\{\left\vert m_{d}\right\rangle ...\left\vert m_{2}\right\rangle \left\vert
m_{1}\right\rangle \}$ with $d=\infty $ are also the energy eigenbases of
the spin Hamiltonian $H_{\infty }.$ Evidently the $SIC$ unitary propagator $%
U_{\infty }^{sic}\left( a_{m}^{s}\right) $ is diagonal in the energy
representation.

How to build the $SIC$ unitary propagator of (3.12) out of the basic $SIC$
unitary operators of (3.1)? Suppose that there are $d$ spin subsystems to
form a composite spin system through tensor product. Each spin subsystem
(e.g., the $l-$th spin subsystem with $1\leq l\leq d$) is used independently
to prepare a $z-$component basic $SIC$ unitary operator $%
U_{lz}^{sic}(a_{m}^{s},\theta _{m}^{l})=\exp (-ia_{m}^{s}\theta
_{m}^{l}I_{mz}^{l})$ of (3.1) through the search-space dynamical reduction.
Here $I_{mz}^{l}$ is the $z-$component spin operator of the $m-$th spin in
the $l-$th spin subsystem. Then after the search-space dynamical reduction
there are $d$ basic $SIC$ unitary operators $\{U_{lz}^{sic}(a_{m}^{s},\theta
_{m}^{l}),$ $1\leq l\leq d\}$ of the composite spin system. Now these $d$
spins $\{I_{m\mu }^{l},$ $1\leq l\leq d\}$, each of which belongs to one
different spin subsystem, may constitute a $d-$qubit spin subsystem of the
composite spin system. Then the effective $SIC$ unitary propagator of the $%
d- $qubit spin subsystem which may be expressed as $U_{eff}^{sic}\left(
a_{m}^{s}\right)
=\tbigotimes\nolimits_{l=1}^{d}U_{lz}^{sic}(a_{m}^{s},\theta _{m}^{l})$ can
be generated in tensor-product form by these $d$ basic $SIC$ unitary
operators $\{U_{lz}^{sic}(a_{m}^{s},\theta _{m}^{l})\}$ prepared previously.
Suppose further that there is an auxiliary $d-$qubit spin ($I=1/2$) system
that does not belong to the composite spin system. Then one may perform a $%
QM $ unitary transformation\footnote{%
For example, the $QM$ unitary transformation may be realized by $%
I_{lz}=R_{kl}I_{kz}R_{kl}^{+}$ with the unitary operator $R_{kl}=\exp \left(
+i\pi I_{kx}I_{ly}\right) \exp \left( -i\pi I_{ky}I_{lx}\right) ,$ where the
indices $k$ and $l$ may be chosen suitably.}: $I_{mz}^{l}\rightarrow I_{lz},$
$l=1,2,...,d;$ here the spin operator $I_{lz}$ belongs to the $l-$th spin ($%
I=1/2$) of the auxiliary $d-$qubit spin system, while $I_{mz}^{l}$ belongs
to the $m-$th spin ($I=1/2$) of the $l-$th spin subsystem of the composite
spin system. After the $QM$ unitary transformation the effective $SIC$
unitary propagator $U_{eff}^{sic}\left( a_{m}^{s}\right) $ that works in the
composite spin system is transformed to the one (denoted by $%
U_{d}^{sic}\left( a_{m}^{s}\right) $ here) which works in the auxiliary $d-$%
qubit spin system. If now one sets $\theta _{m}^{l}=\theta _{ml}$, then the $%
SIC$ unitary propagator$\ U_{d}^{sic}\left( a_{m}^{s}\right) $ with $d=\log
_{2}L\rightarrow \infty $ is just the one of (3.12) up to tensor-product
order.

Below it is described theoretically how a suitable $SIC$ unitary propagator $%
U_{\infty }^{sic}\left( a_{m}^{s}\right) $ of (3.12) of the $d-$qubit spin
system with $d=\infty $ is unitarily transformed to the desired $SIC$\
unitary propagator $U_{at}^{sic}(t_{m})$ in (3.10) of the single-atom
system. This is the main step of the joint construction method in the
present work. Note that both the $SIC$\ unitary propagators $U_{\infty
}^{sic}\left( a_{m}^{s}\right) $ and $U_{at}^{sic}(t_{m})$ are the
duality-character unitary operators. Then the $QM$ unitary transformation
that changes the $U_{\infty }^{sic}\left( a_{m}^{s}\right) $ to the $%
U_{at}^{sic}(t_{m})$ must be performed simultaneously in both the physical
Hilbert space and its corresponding math Hilbert space of the composite
quantum system whose component subsystems include the $d-$qubit spin system
with $d=\infty $ and the single-atom system. Of course, here the $QM$\
unitary-transformational operator on the physical Hilbert space overlaps
completely with the one on the math Hilbert space. The solution-information (%
$a_{m}^{s}$) transfer is deterministic and obeys the information
conservation law during the $QM$\ unitary transformation. Therefore, during
the $QM$\ unitary transformation any quantum state that contains the
information ($a_{m}^{s}$) of the real solution to the unstructured search
problem always remains in the physical Hilbert space, while any state that
contains the information ($a_{m}^{s}$) of the candidate solution different
from the real solution is always in the math Hilbert space. Due to this
point below the theoretical construction of the $QM$\ unitary transformation
is performed in the physical Hilbert space of the composite quantum system
alone. But it is available as well in the corresponding math Hilbert space.

It is known from the preceding paragraphs that all the tensor-product energy
eigenbases $\{\left\vert m_{d}\right\rangle ...\left\vert m_{2}\right\rangle
\left\vert m_{1}\right\rangle \}$ form a complete orthonormal basis set of
the $d-$qubit spin system, even if the spin number $d$ is infinitely large.
Now suppose that $\left\vert \Psi _{\infty }\right\rangle $ is arbitrary
initial state of the time evolution process governed by the $SIC$ unitary
propagator of (3.12) in the $d-$qubit spin system with $2^{d}=\infty $. Then
according to the energy eigenfunction expansion principle an arbitrary state
such as $\left\vert \Psi _{\infty }\right\rangle $ of the $d-$qubit spin
system with $2^{d}=\infty $ can be expanded in the complete orthonormal
energy eigenbasis set $\left\{ \left\vert k\right\rangle \right\} $ of the
spin system:%
\begin{equation}
\left\vert \Psi _{\infty }\right\rangle =\sum_{k=0}^{\infty }C_{k}\left\vert
k\right\rangle  \tag{3.13}
\end{equation}%
where the tensor-product energy eigenbase $\left\vert k\right\rangle
=\left\vert m_{\infty }\right\rangle ...\left\vert m_{i}\right\rangle
...\left\vert m_{2}\right\rangle \left\vert m_{1}\right\rangle $ and $%
k=...+k_{i-1}\times 2^{i-1}+...+k_{1}\times 2^{1}+k_{0}\times 2^{0}$ with $%
k_{i-1}=\left( -m_{i}+\frac{1}{2}\right) =0$ or $1$ for $i=1,2,...,\infty .$
This expansion series is infinite, but it is always convergent [$13$]. With
the help of the expression (3.12) of the $SIC$ unitary propagator $U_{\infty
}^{sic}\left( a_{m}^{s}\right) ,$ the tensor-product eigenbase $\left\vert
k\right\rangle =\left\vert m_{\infty }\right\rangle ...\left\vert
m_{l}\right\rangle ...\left\vert m_{2}\right\rangle \left\vert
m_{1}\right\rangle ,$ and the eigenvalue equation $I_{lz}\left\vert
m_{l}\right\rangle =m_{l}\left\vert m_{l}\right\rangle $ with $m_{l}=\pm
1/2, $ one may set up the eigenvalue equation for the $U_{\infty
}^{sic}\left( a_{m}^{s}\right) $ of (3.12). This eigenvalue equation is
given by%
\begin{equation}
U_{\infty }^{sic}\left( a_{m}^{s}\right) \left\vert k\right\rangle =\exp
\left( -ia_{m}^{s}\sum_{l=1}^{\infty }\theta _{ml}m_{l}\right) \left\vert
k\right\rangle .  \tag{3.14}
\end{equation}%
Now the rotation angles $\left\{ \theta _{ml}\right\} $ in (3.12) or (3.14)
are chosen as $\theta _{ml}=-\alpha _{m}\times 2^{l-1}$ for $%
l=1,2,...,\infty $ and here $\alpha _{m}$ is a real constant. Then by using
the relations $k_{l-1}=\left( -m_{l}+\frac{1}{2}\right) $ and $%
k=\sum_{l=1}^{\infty }k_{l-1}\times 2^{l-1}=0,1,...,\infty $ the eigenvalue
in (3.14), i.e., the $SIC$\ complex exponential function may be rewritten as%
\begin{equation}
\exp \left( -ia_{m}^{s}\sum_{l=1}^{\infty }\theta _{ml}m_{l}\right)
=\tprod\limits_{l=1}^{\infty }\exp \left( i2^{l-2}\alpha
_{m}a_{m}^{s}\right) \exp \left( -i\alpha _{m}a_{m}^{s}k\right) .  \tag{3.15}
\end{equation}%
Let $U_{\infty }^{sic}\left( a_{m}^{s},\alpha _{m}\right) =U_{\infty
}^{sic}\left( a_{m}^{s}\right) $ when $\theta _{ml}=-\alpha _{m}\times
2^{l-1}$ for $l\geq 1$ in (3.12). By using the infinite expansion series
(3.13) of the initial state $\left\vert \Psi _{\infty }\right\rangle ,$ the
eigenvalue equation (3.14), and the relation (3.15) the time evolution
process of the $d-$qubit spin system with $2^{d}=\infty $ governed by the $%
SIC$\ unitary propagator $U_{\infty }^{sic}\left( a_{m}^{s},\alpha
_{m}\right) $ then is expressed as%
\begin{equation}
U_{\infty }^{sic}\left( a_{m}^{s},\alpha _{m}\right) \left\vert \Psi
_{\infty }\right\rangle =\tprod\limits_{l=1}^{\infty }\exp \left(
i2^{l-2}\alpha _{m}a_{m}^{s}\right) \sum_{k=0}^{\infty }C_{k}\exp \left(
-i\alpha _{m}a_{m}^{s}k\right) \left\vert k\right\rangle  \tag{3.16}
\end{equation}%
Here the infinite product term on the RH side is a global phase factor,
although it carries the solution information. Evidently the time evolution
process (3.16) is also a QUANSDAM (or UNIDYSLOCK) process.

Before the $QM$ unitary transformation that changes the $U_{\infty
}^{sic}\left( a_{m}^{s}\right) $ of (3.12) to the desired one, i.e., the $%
U_{at}^{sic}(t_{m})$ in (3.10) is constructed concretely, the
quantum-mechanical knowledge relevant to the unitary transformation between
different eigenbasis sets is introduced simply. This (basis) transformation
theory in quantum mechanics [$10$] is theoretical basis to realize the
solution-information transfer mentioned above and build the desired $SIC$
unitary propagator in (3.10). Suppose that there are two different
eigenbasis sets in the Hilbert space of some quantum system. Both the
eigenbasis sets each have the same number of eigenbases and the number may
be infinitely large. For simplicity, here assume that one (orthonormal)
eigenbasis set is generated by the COM-motion Hamiltonian $H_{c}$ of the
single-atom system\footnote{%
Here the Hamiltonian $H_{c}$ may be replaced with any appropriate
Hamiltonian of the single-atom system.} and is denoted as $\left\{
u_{k}\left( x\right) \right\} ,$ and another is generated by the above spin
Hamiltonian of the $d-$qubit spin system with $2^{d}=\infty $ and is given
by the tensor-product eigenbasis set $\left\{ \left\vert k\right\rangle
\right\} $ in (3.13). As a typical example, here the eigenbasis set $\left\{
u_{k}\left( x\right) \right\} $ may be given by (3.7b), which is the
COM-motion energy eigenbasis set of a single atom in external harmonic
potential field. Both the eigenbasis sets $\left\{ \left\vert k\right\rangle
\right\} $ and $\left\{ u_{k}\left( x\right) \right\} $ each contain the
same number of eigenbases and the number is infinitely large. Then quantum
mechanics [$10$] tells ones that there is a unitary transformation $W$ (or $%
W^{+}$) such that the eigenbasis set $\left\{ \left\vert k\right\rangle
\right\} $ can be transformed to $\left\{ u_{k}\left( x\right) \right\} $
(or vice versa) in the Hilbert space. In a simple form the $QM$ unitary
transformation $W$ may be taken as 
\begin{equation*}
W:\left\vert k\right\rangle \rightarrow u_{k}\left( x\right) ,\text{ }%
k=0,1,2,...,\infty .
\end{equation*}%
With the help of this $QM$ unitary transformation it can prove in theory
below that the $QM$ unitary transformation on the $SIC$ unitary propagator $%
U_{\infty }^{ic}\left( a_{m}^{s}\right) $ of (3.12), i.e., $WU_{\infty
}^{sic}\left( a_{m}^{s},\alpha _{m}\right) W^{+},$ is equal to the $SIC$
unitary propagator $U_{c}^{sic}(t_{m})=\exp \left[ -ia_{m}^{s}H_{c}t_{m}/%
\hslash \right] $ up to a global phase factor, where the eigenvalue equation
of the Hamiltonian $H_{c}$ is given by $H_{c}u_{k}\left( x\right)
=E_{k}^{c}u_{k}\left( x\right) $ with $E_{k}^{c}=ak+b.$ This is a simple
case. More complex cases will be considered later. It may be better to let
this theoretical proof be based on the eigenfunction expansion principle.
This principle shows that arbitrary state $|\Psi _{\infty }\rangle $ of the
Hilbert space may be expanded in terms of the eigenbases $\left\{
u_{k}\left( x\right) \right\} ,$ that is, $|\Psi _{\infty }\rangle
=\sum_{k=0}^{\infty }B_{k}u_{k}\left( x\right) ,$ as shown by (3.8), and it
can be expanded as well in the tensor-product eigenbasis set $\left\{
\left\vert k\right\rangle \right\} $, as shown by (3.13). Which one of the
two equivalent eigenfunction expansions is more convenient in the
theoretical proof, it will be used below.

Now the theoretical proof is given as follows. Suppose first that both the $%
U_{\infty }^{sic}\left( a_{m}^{s},\alpha _{m}\right) $ and $%
U_{c}^{sic}(t_{m})$ work on (i.e., act on) the same Hilbert space to which
the state $|\Psi _{\infty }\rangle $ belongs. Consider the simple case that
the energy eigenvalue $E_{k}^{c}$ of the Hamiltonian $H_{c}$\ associated
with the eigenbase $u_{k}\left( x\right) $ is a linear function of the
quantum number $k$, i.e., $E_{k}^{c}=ak+b.$ On the one hand, by applying $%
WU_{\infty }^{sic}\left( a_{m}^{s},\alpha _{m}\right) W^{+}$ to arbitrary
state $|\Psi _{\infty }\rangle $ with the above infinite eigenfunction
expansion series like (3.8) one obtains%
\begin{equation}
WU_{\infty }^{sic}\left( a_{m}^{s},\alpha _{m}\right) W^{+}|\Psi _{\infty
}\rangle =\sum_{k=0}^{\infty }B_{k}WU_{\infty }^{sic}\left( a_{m}^{s},\alpha
_{m}\right) \left\vert k\right\rangle  \tag{3.17}
\end{equation}%
where $W^{+}u_{k}\left( x\right) =\left\vert k\right\rangle $ is already
used. Now using the eigenvalue equation (3.14), the relation (3.15), and $%
u_{k}\left( x\right) =W\left\vert k\right\rangle $ the time evolution
process of (3.17) may be reduced to the form%
\begin{equation*}
WU_{\infty }^{sic}\left( a_{m}^{s},\alpha _{m}\right) W^{+}|\Psi _{\infty
}\rangle =\tprod\limits_{l=1}^{\infty }\exp \left( i2^{l-2}\alpha
_{m}a_{m}^{s}\right)
\end{equation*}%
\begin{equation}
\times \sum_{k=0}^{\infty }B_{k}\exp \left( -i\alpha _{m}a_{m}^{s}k\right)
u_{k}\left( x\right) .  \tag{3.18}
\end{equation}%
On the other hand, by applying directly $U_{c}^{sic}\left( t_{m}\right)
=\exp \left[ -ia_{m}^{s}H_{c}t_{m}/\hslash \right] $ to the same state $%
|\Psi _{\infty }\rangle $ and noticing that $E_{k}^{c}=ak+b$ one obtains the
time evolution process:%
\begin{equation}
U_{c}^{sic}\left( t_{m}\right) |\Psi _{\infty }\rangle =\exp \left[
-ia_{m}^{s}bt_{m}/\hslash \right] \sum_{k=0}^{\infty }B_{k}\exp \left[
-i\left( at_{m}/\hslash \right) a_{m}^{s}k\right] u_{k}\left( x\right) 
\tag{3.19}
\end{equation}%
where the eigenvalue equation $U_{c}^{sic}\left( t_{m}\right) u_{k}\left(
x\right) =\exp \left[ -ia_{m}^{s}\left( ak+b\right) t_{m}/\hslash \right]
u_{k}\left( x\right) $ is already used. Let $\alpha _{m}$ in (3.18) be $%
\alpha _{m}=\left( at_{m}/\hslash \right) .$ Then by substituting (3.19)
into (3.18) one obtains%
\begin{equation*}
WU_{\infty }^{sic}\left( a_{m}^{s},\alpha _{m}\right) W^{+}|\Psi _{\infty
}\rangle =\exp \left[ i\Phi _{1}\left( a_{m}^{s}\right) \right]
U_{c}^{sic}(t_{m})|\Psi _{\infty }\rangle
\end{equation*}%
with the global phase factor: 
\begin{equation*}
\exp \left[ i\Phi _{1}\left( a_{m}^{s}\right) \right] =\tprod%
\nolimits_{l=1}^{\infty }\exp \left[ i2^{l-2}\left( at_{m}/\hslash \right)
a_{m}^{s}\right] \exp \left[ ia_{m}^{s}bt_{m}/\hslash \right] .
\end{equation*}%
Notice that the state $\left\vert \Psi _{\infty }\right\rangle $ is
arbitrary. One can conclude that the $SIC$ unitary propagator $%
U_{c}^{sic}\left( t_{m}\right) $ can be exactly written as%
\begin{equation}
U_{c}^{sic}(t_{m})=\exp \left[ -i\Phi _{1}\left( a_{m}^{s}\right) \right]
WU_{\infty }^{sic}\left( a_{m}^{s},\alpha _{m}\right) W^{+}.  \tag{3.20}
\end{equation}%
Indeed, by the $QM$\ unitary transformation $W$ the $U_{\infty }^{sic}\left(
a_{m}^{s},\alpha _{m}\right) $ of (3.12) is transformed to $%
U_{c}^{sic}(t_{m})=\exp \left[ -ia_{m}^{s}H_{c}t_{m}/\hslash \right] $ with
the energy eigenvalue $E_{k}^{c}=ak+b$ of the Hamiltonian $H_{c}$ up to a
global phase factor.

Next consider a slightly complex case that the energy eigenvalue $E_{k}^{c}$
associated with the eigenbase $u_{k}\left( x\right) $ is a quadratic
function of the quantum number $k$, i.e., $E_{k}^{c}=ak^{2}+bk+c.$ In this
case, beside the $SIC$ unitary diagonal propagator $U_{\infty }^{sic}\left(
a_{m}^{s}\right) $ of (3.12), one needs to construct a new $SIC$ unitary
diagonal propagator:%
\begin{equation}
U_{2\infty }^{sic}\left( a_{m}^{s}\right) =\tprod\limits_{j>l=1}^{\infty
}\exp \left( -ia_{m}^{s}\theta _{jl}2I_{jz}I_{lz}\right)  \tag{3.21}
\end{equation}%
where the interaction parameter $\theta _{jl}$ can be set independently and $%
\{2I_{jz}I_{lz}\}$ are the $z-$component spin-spin interaction operators. As
usual, here the product operator $I_{jz}I_{lz}$ stands for the tensor (or
direct) product $(I_{jz}\tbigotimes I_{lz})$ of the two $z-$component
Cartesian single-spin operators $I_{jz}$ and $I_{lz}.$ It is diagonal. Then
it can prove that the eigenvalue equation for the $SIC$ unitary diagonal
propagator of (3.21) is given by%
\begin{equation}
U_{2\infty }^{sic}\left( a_{m}^{s}\right) \left\vert k\right\rangle =\exp
\left( -ia_{m}^{s}\tsum\limits_{j>l=1}^{\infty }2\theta
_{jl}m_{j}m_{l}\right) \left\vert k\right\rangle .  \tag{3.22}
\end{equation}%
Here the tensor-product base $\left\vert k\right\rangle $ is still given in
(3.13) and it is also an eigenbase of the spin interaction Hamiltonian of
the $SIC$ unitary propagator $U_{2\infty }^{sic}\left( a_{m}^{s}\right) $.

In the representation of the tensor-product basis set $\left\{ \left\vert
k\right\rangle \right\} $ in (3.13) (or the energy representation) the basic 
$SIC$ unitary operator $U_{z}^{sic}(a_{m}^{s},\theta _{m})$ of (3.1) is
diagonal and so are the two $SIC$\ unitary propagators $U_{\infty
}^{sic}\left( a_{m}^{s}\right) $ of (3.12) and $U_{2\infty }^{sic}\left(
a_{m}^{s}\right) $ of (3.21). This greatly simplifies the construction of
these two $SIC$ unitary propagators. It can be seen from (3.12) and (3.21)
that the $U_{\infty }^{sic}\left( a_{m}^{s}\right) $ consists of the
single-spin basic $SIC$ unitary diagonal operators $\{\exp (-i\theta
_{ml}a_{m}^{s}I_{lz})\}$ of (3.1), while the $U_{2\infty }^{sic}\left(
a_{m}^{s}\right) $ is built out of the two-spin $SIC$ unitary diagonal
operators $\left\{ \exp \left( -i\theta _{kl}a_{m}^{s}2I_{kz}I_{lz}\right)
\right\} ,$ each of which may be generated by making an appropriate $QM$
unitary transformation on the basic $SIC$ unitary diagonal operator $%
U_{z}^{sic}(a_{m}^{s},\theta _{m})\footnote{%
For example, the $QM$ unitary transformation may be realized by $\exp \left(
-i\theta _{kl}a_{m}^{s}2I_{kz}I_{lz}\right) =V_{kl}\exp \left( -i\theta
_{kl}a_{m}^{s}I_{kz}\right) V_{kl}^{+}$ with the $QM$ unitary operator $%
V_{kl}=\exp \left( +i\pi I_{ky}I_{lz}\right) \exp \left( -i\frac{\pi }{2}%
I_{ky}\right) $ and $\exp \left( -i\theta _{kl}a_{m}^{s}I_{mz}\right)
=R_{km}\exp \left( -i\theta _{kl}a_{m}^{s}I_{kz}\right) R_{km}^{+}$ with the 
$QM$\ unitary operator $R_{km}=\exp \left( +i\pi I_{kx}I_{my}\right) \exp
\left( -i\pi I_{ky}I_{mx}\right) $}\ $(See also the general theoretical
treatment below). All these $SIC$\ unitary diagonal operators and
propagators commute with each other. This is really important for convenient
construction of the $SIC$ unitary propagator $U_{c}^{sic}\left( t_{m}\right)
=\exp \left[ -ia_{m}^{s}H_{c}t_{m}/\hslash \right] $ in which the
Hamiltonian $H_{c}$ has the energy eigenvalue $E_{k}^{c}=ak^{2}+bk+c.$

Let $U_{2\infty }^{sic}\left( a_{m}^{s},\beta _{m}\right) =U_{2\infty
}^{sic}\left( a_{m}^{s}\right) $ when the parameter $\theta _{jl}$ is taken
as $\theta _{jl}=\beta _{m}\times 2^{j+l-2}$ in (3.21). Then construct the $%
SIC$ unitary diagonal propagator $U_{T}^{sic}\left( a_{m}^{s},\beta
_{m}\right) :$%
\begin{equation}
U_{T}^{sic}\left( a_{m}^{s},\beta _{m}\right) =U_{2\infty }^{sic}\left(
a_{m}^{s},\beta _{m}\right) \tprod\limits_{j=1}^{\infty }U_{\infty
}^{sic}\left( a_{m}^{s},2^{j-1}\beta _{m}\right)  \tag{3.23}
\end{equation}%
where $U_{\infty }^{sic}\left( a_{m}^{s},2^{j-1}\beta _{m}\right) $ may be
prepared directly from (3.12) by replacing $\alpha _{m}$ in $U_{\infty
}^{sic}(a_{m}^{s},\alpha _{m})$ with $2^{j-1}\beta _{m}.$ It can prove that
the $SIC$ unitary diagonal propagator $U_{T}^{sic}\left( a_{m}^{s},\beta
_{m}\right) $ acting on the tensor-product base $\left\vert k\right\rangle $
can generate the $SIC$ complex exponential function $\exp \left( -i\beta
_{m}a_{m}^{s}k^{2}\right) $ whose exponent is proportional to square of the
quantum number $k,$%
\begin{equation}
U_{T}^{sic}\left( a_{m}^{s},\beta _{m}\right) \left\vert k\right\rangle
=\exp \left[ i\Phi _{2}\left( a_{m}^{s}\right) \right] \exp \left( -i\beta
_{m}a_{m}^{s}k^{2}\right) \left\vert k\right\rangle ,  \tag{3.24}
\end{equation}%
where the global phase factor $\exp \left[ i\Phi _{2}\left( a_{m}^{s}\right) %
\right] =\tprod\nolimits_{j\geq l=1}^{\infty }\exp \left( i2^{j+l-3}\beta
_{m}a_{m}^{s}\right) .$

The theoretical proof for (3.24) is carried out below. By using the
relations $k_{l-1}=\left( -m_{l}+\frac{1}{2}\right) $ and $%
k=\sum_{l=1}^{\infty }k_{l-1}\times 2^{l-1}$ one can re-write $\exp \left(
-i\beta _{m}a_{m}^{s}k^{2}\right) $ in (3.24) in the form%
\begin{equation*}
\exp \left( -i\beta _{m}a_{m}^{s}k^{2}\right) =\exp \left( -i\beta
_{m}a_{m}^{s}\sum_{j>l=1}^{\infty }m_{j}m_{l}\times 2^{j+l-1}\right)
\end{equation*}%
\begin{equation}
\times \tprod\limits_{j=1}^{\infty }\exp \left( -i2^{j-1}\beta
_{m}a_{m}^{s}k\right) \times \tprod\limits_{j>l=1}^{\infty }\exp \left(
+i\beta _{m}a_{m}^{s}2^{j+l-3}\right) .  \tag{3.25}
\end{equation}%
Now let the parameter $\theta _{jl}=\beta _{m}\times 2^{j+l-2}$ in (3.22)
and $\alpha _{m}=\beta _{m}$ in (3.15). Then by acting the $%
U_{T}^{sic}\left( a_{m}^{s},\beta _{m}\right) $ of (3.23) on the
tensor-product base $\left\vert k\right\rangle $ one can obtain the
following eigenvalue equation:%
\begin{equation*}
U_{T}^{sic}\left( a_{m}^{s},\beta _{m}\right) \left\vert k\right\rangle
=\tprod\limits_{j=1}^{\infty }\tprod\limits_{l=1}^{\infty }\exp \left(
i2^{j+l-3}\beta _{m}a_{m}^{s}\right) \times \tprod\limits_{j=1}^{\infty
}\exp \left( -i2^{j-1}\beta _{m}a_{m}^{s}k\right)
\end{equation*}%
\begin{equation}
\times \exp \left( -i\beta _{m}a_{m}^{s}\sum_{j>l=1}^{\infty
}m_{j}m_{l}\times 2^{j+l-1}\right) \left\vert k\right\rangle  \tag{3.26}
\end{equation}%
where the equations (3.14), (3.15), and (3.22) are already used. By
substituting (3.25) into (3.26) one obtains%
\begin{equation*}
U_{T}^{sic}\left( a_{m}^{s},\beta _{m}\right) \left\vert k\right\rangle
=\tprod\limits_{j=1}^{\infty }\tprod\limits_{l=1}^{\infty }\exp \left(
i2^{j+l-3}\beta _{m}a_{m}^{s}\right)
\end{equation*}%
\begin{equation*}
\times \tprod\limits_{j>l=1}^{\infty }\exp \left( -i2^{j+l-3}\beta
_{m}a_{m}^{s}\right) \exp \left( -i\beta _{m}a_{m}^{s}k^{2}\right)
\left\vert k\right\rangle .
\end{equation*}%
A simple calculation shows that this equation is just (3.24).

With the help of the $QM$ unitary transformation $W$ it can prove that $%
WU_{T}^{sic}\left( a_{m}^{s},\beta _{m}\right) U_{\infty }^{sic}\left(
a_{m}^{s},\alpha _{m}\right) W^{+}$ is equal to $U_{c}^{sic}(t_{m})=\exp %
\left[ -ia_{m}^{s}H_{c}t_{m}/\hslash \right] $ up to a global phase factor,
where the eigenvalue equation for the Hamiltonian $H_{c}$ is given by $%
H_{c}u_{k}\left( x\right) =E_{k}^{c}u_{k}\left( x\right) $ with $%
E_{k}^{c}=ak^{2}+bk+c$. On the one hand, by acting $WU_{T}^{sic}\left(
a_{m}^{s},\beta _{m}\right) U_{\infty }^{sic}\left( a_{m}^{s},\alpha
_{m}\right) W^{+}$ on arbitrary state $|\Psi _{\infty }\rangle
=\sum_{k=0}^{\infty }B_{k}u_{k}\left( x\right) $ one obtains%
\begin{equation*}
WU_{T}^{sic}\left( a_{m}^{s},\beta _{m}\right) U_{\infty }^{sic}\left(
a_{m}^{s},\alpha _{m}\right) W^{+}|\Psi _{\infty }\rangle
=\tprod\limits_{l=1}^{\infty }\exp \left( i2^{l-2}\alpha _{m}a_{m}^{s}\right)
\end{equation*}%
\begin{equation}
\times \exp \left[ i\Phi _{2}\left( a_{m}^{s}\right) \right]
\sum_{k=0}^{\infty }B_{k}\exp \left( -i\beta _{m}a_{m}^{s}k^{2}\right) \exp
\left( -i\alpha _{m}a_{m}^{s}k\right) u_{k}\left( x\right)  \tag{3.27}
\end{equation}%
where use has already been made of the equations (3.14), (3.15), and (3.24)
as well as the unitary transformation $u_{k}\left( x\right) =W\left\vert
k\right\rangle $ and its inverse. On the other hand, by applying directly $%
U_{c}^{sic}(t_{m})=\exp \left[ -ia_{m}^{s}H_{c}t_{m}/\hslash \right] $ to
the state $|\Psi _{\infty }\rangle $ and noticing that $%
E_{k}^{c}=ak^{2}+bk+c $ one can find that%
\begin{equation}
U_{c}^{sic}\left( t_{m}\right) |\Psi _{\infty }\rangle =\sum_{k=0}^{\infty
}B_{k}\exp \left[ -ia_{m}^{s}\left( ak^{2}+bk+c\right) t_{m}/\hslash \right]
u_{k}\left( x\right)  \tag{3.28}
\end{equation}%
Let $\beta _{m}=at_{m}/\hslash $ and $\alpha _{m}=bt_{m}/\hslash $ in
(3.27). Then by substituting (3.28) into (3.27) it can be found that 
\begin{equation*}
WU_{T}^{sic}\left( a_{m}^{s},\beta _{m}\right) U_{\infty }^{sic}\left(
a_{m}^{s},\alpha _{m}\right) W^{+}|\Psi _{\infty }\rangle =\exp \left[ i\Phi
_{3}\left( a_{m}^{s}\right) \right] U_{c}^{sic}\left( t_{m}\right) |\Psi
_{\infty }\rangle
\end{equation*}%
where $\exp \left[ i\Phi _{3}\left( a_{m}^{s}\right) \right] $ is a global
phase factor. Notice that the state $|\Psi _{\infty }\rangle $ is arbitrary.
One therefore concludes that the $QM$ unitary transformation $W$ on the $SIC$
unitary diagonal propagator $U_{T}^{sic}\left( a_{m}^{s},\beta _{m}\right)
U_{\infty }^{sic}\left( a_{m}^{s},\alpha _{m}\right) ,\ $i.e., $%
WU_{T}^{sic}\left( a_{m}^{s},\beta _{m}\right) U_{\infty }^{sic}\left(
a_{m}^{s},\alpha _{m}\right) W^{+},$ is equal to $U_{c}^{sic}\left(
t_{m}\right) =\exp \left[ -ia_{m}^{s}H_{c}t_{m}/\hslash \right] $ with the
energy eigenvalue $E_{k}^{c}=ak^{2}+bk+c$ of the Hamiltonian $H_{c}$ up to a
global phase factor.

In the previous paragraphs two simple and important cases are theoretically
treated in detail that the energy eigenvalues are a linear and a quadratic
function of the quantum number $k$, respectively. This theoretical treatment
is divided into the two parts: the first part is the construction of the $%
SIC $\ unitary diagonal propagators and the second part is related to the $%
QM $ unitary transformation $W$, where $W$ is defined formally and its
existence is shown in quantum mechanics. This theoretical treatment shows
that there exist the $SIC$ unitary propagators $\{\exp \left[
-ia_{m}^{s}H_{c}t_{m}/\hslash \right] \}$ in which the discrete eigenspectra
of the Hamiltonian $H_{c}$ are the linear and quadratic functions of the
quantum number $k$, respectively.

In unitary quantum dynamics the theoretical treatment above can be
generalized in the framework of the multiple-quantum operator algebra space [%
$6$]. This generalized theoretical treatment is still divided into the two
parts. For the first part only the key points are first described simply,
while in the second part a detailed description for how to construct
explicitly the $QM$\ unitary transformation $W$ is given later. The\ $SIC$
unitary propagators that are constructed in the first part of the above
theoretical treatment are diagonal operators and so are the Hamiltonians of
these $SIC$\ unitary propagators. In accordance with Ref. [$6$], any unitary
diagonal operator that is generated by a diagonal Hamiltonian may be
constructed alone in the \textit{lo}ngitudinal \textit{m}agnetization and 
\textit{s}pin \textit{o}rder ($LOMSO$) operator subspace (i.e., the closed
diagonal operator subspace) of the multiple-quantum operator algebra space.
The characteristic property for the closed $LOMSO$ operator subspace is that
any pair of members of the subspace commute with each other. Therefore, in
the first part the key point for the general theoretical treatment based on
the multiple-quantum operator algebra space is that any $SIC$ unitary
diagonal operator (or propagator) is constructed in the closed $LOMSO$
operator subspace alone [$3$]. Given a complete set of the base operators of
the closed $LOMSO$ operator subspace, any diagonal Hamiltonian may be
linearly expanded in terms of the base operators. Here, in addition to the
unity operator, the base operators of the $LOMSO$ operator subspace for an $%
n-$qubit spin system ($n=1,2,...,\infty $) may be chosen as these $z-$%
component single-spin operators $\{I_{kz}\}$ and multiple-spin interaction
operators \{$2I_{kz}I_{lz}$, $4I_{kz}I_{lz}I_{mz}$, ...\} of the spin
system. And as usual these operators are called the longitudinal
magnetization and spin order product base operators $\{B_{k}^{z}\}$ in the
multiple-quantum operator algebra space [$6$]. Now for every $LOMSO$ base
operator $B_{k}^{z}$ one may generate a $SIC$ unitary diagonal propagator
(or operator) $R_{zk}^{sic}\left( \theta _{mk}\right) =\exp \left( -i\theta
_{mk}a_{m}^{s}B_{k}^{z}\right) $ [$3$], which corresponds to the $QM$\
elementary propagator $R_{zk}\left( \theta _{mk}\right) =\exp \left(
-i\theta _{mk}B_{k}^{z}\right) $. It can prove that any $SIC$ unitary
diagonal propagator $R_{zk}^{sic}\left( \theta _{mk}\right) $ is unitarily
equivalent to the $z-$component basic $SIC$ unitary operator $%
U_{z}^{sic}(a_{m}^{s},\theta _{m})$ of (3.1) in the sense that the $%
R_{zk}^{sic}\left( \theta _{mk}\right) $ can be transformed to the $%
U_{z}^{sic}(a_{m}^{s},\theta _{m})$ by a $QM$ unitary transformation. The $%
QM $ unitary transformation is non-diagonal except the identical
transformation. Consequently, though the $R_{zk}^{sic}\left( \theta
_{mk}\right) $ belongs to the closed $LOMSO$ operator subspace alone, its
preparation by starting from the basic $SIC$ unitary operators needs to be
performed in the multiple-quantum operator algebra space [$7,24,6$]. The
theoretical proof is simply described as follows. It is similar to that one
of Eq.(10) in Ref. [$24$]. An arbitrary $LOMSO$\ base operator $B_{k}^{z}$
may be written as $2^{l}I_{k_{1}z}I_{k_{2}z}...I_{k_{l}z}I_{k_{l+1}z}\equiv
2^{l}I_{k_{1}z}\tbigotimes I_{k_{2}z}\tbigotimes ...\tbigotimes
I_{k_{l}z}\tbigotimes I_{k_{l+1}z}$ with the spin indices $%
k_{1},k_{2},...,k_{l+1}$ $=1,2,...,n$ and $0\leq l<n.$ Then there is the
recursive relation for the $SIC$\ unitary propagator $R_{zk}^{sic}\left(
\theta _{mk}\right) :$%
\begin{equation*}
\exp \left( -i\theta
_{mk}a_{m}^{s}2^{l}I_{k_{1}z}I_{k_{2}z}...I_{k_{l}z}I_{k_{l+1}z}\right)
\end{equation*}%
\begin{equation}
=V_{k_{l}}\exp \left( -i\theta
_{mk}a_{m}^{s}2^{l-1}I_{k_{1}z}I_{k_{2}z}...I_{k_{l-1}z}I_{k_{l+1}z}\right)
V_{k_{l}}^{+},\text{ }n>l\geq 1,  \tag{3.29}
\end{equation}%
where the $QM$ non-diagonal unitary operator $V_{k_{l}}$ is given by%
\begin{equation*}
V_{k_{l}}=\exp \left( +i\pi I_{k_{l}z}I_{k_{l+1}y}\right) \exp \left( -i%
\frac{\pi }{2}I_{k_{l+1}y}\right)
\end{equation*}%
By repeating to use the recursive relation (3.29) while keeping the spin
index $k_{l+1}$ unchanged it can prove that the $R_{zk}^{sic}\left( \theta
_{mk}\right) $ can be transformed to the basic $SIC$ unitary operator $\exp
\left( -i\theta _{mk}a_{m}^{s}I_{k_{l+1}z}\right) $ by the $QM$\ unitary
transformation. Of course, $\exp \left( -i\theta
_{mk}a_{m}^{s}I_{k_{l+1}z}\right) $ also may be further transformed to the
desired one, e.g., the basic $SIC$\ unitary operator $\exp \left( -i\theta
_{mk}a_{m}^{s}I_{mz}\right) ,$ by making the $QM$ unitary transformation: $%
I_{k_{l+1}z}\rightarrow I_{mz}$. On the basis of (3.29) any $SIC\ $unitary
diagonal propagator $R_{zk}^{sic}\left( \theta _{mk}\right) $ can be
prepared by starting from the basic $SIC$ unitary operators. There is also
the $QM$ recursive relation (See Eq. (10) in Ref. [$24$]) corresponding to
the $SIC$ recursive relation (3.29). Finally any $SIC$ unitary propagator
that is generated by a diagonal Hamiltonian may be constructed alone in the $%
LOMSO$ operator subspace by starting from these $SIC$ unitary diagonal
operators $\{R_{zk}^{sic}\left( \theta _{mk}\right) \}$. As the special
cases, it may be thought that in the two simple cases above the $SIC$\
unitary diagonal propagators of (3.12) and (3.21) each also are
theoretically constructed in the closed $LOMSO$ operator subspace alone.
Note that in the two simple cases above the energy eigenvalues (i.e., $%
E_{k}^{c}$) are a linear and a quadratic function of the quantum number $k$,
respectively. If now the energy eigenvalue is any polynomial function of the
quantum number $k,$ then in this more general case the construction for the
corresponding $SIC$\ unitary diagonal propagator may be evidently performed
as well in the closed $LOMSO$ operator subspace alone.

In the second part of the above theoretical treatment the key important
problem, i.e., how the $QM$ unitary transformation $W$ is constructed
explicitly, has not yet been settled. In unitary quantum dynamics this
problem can be conveniently settled in the framework of the multiple-quantum
operator algebra space [$7,24,6$]. It must be pointed out that it tends to
be hard to construct explicitly the $QM$ unitary operator $W$ due to that
the basis set $\left\{ u_{k}\left( x\right) \right\} $ is not independent of
the tensor-product basis set $\left\{ \left\vert k\right\rangle \right\} $
in the same Hilbert space under study above. This hard problem can be
avoided, as shown below. Suppose that the original Hilbert space under study
is replaced with a composite Hilbert that is formed by the two Hilbert
spaces (denoted as $S_{1}$ and $S_{2}$) by tensor product. This means that
there are at least two component Hilbert subspaces ($S_{1}$ and $S_{2}$) in
the composite Hilbert space under study below. Here the original Hilbert
space, the component Hilbert subspace $S_{1}$ and $S_{2}$ each have the same
size and may be infinitely large. Suppose further that the $SIC$ unitary
diagonal propagators $U_{\infty }^{sic}\left( a_{m}^{s},\alpha _{m}\right) $
of (3.12) and $U_{T}^{sic}\left( a_{m}^{s},\beta _{m}\right) $ of (3.23)
each work on the component Hilbert subspace $S_{1}$ which belongs to the $d-$%
qubit spin system with $2^{d}=\infty $ alone, while the above $%
U_{c}^{sic}(t_{m})=\exp \left[ -ia_{m}^{s}H_{c}t_{m}/\hslash \right] $ works
on the component Hilbert subspace $S_{2},$ which may belong to a single-atom
system alone. If now the two basis sets $\left\{ \left\vert k\right\rangle
\right\} $ and $\left\{ u_{k}\left( x\right) \right\} $ are respectively
chosen by these two component Hilbert subspaces $S_{1}$ and $S_{2}$, then
thing could become simple for the explicit construction of the $QM$ unitary
operator $W$ on the composite Hilbert space $S_{1}\tbigotimes S_{2}$.
Correspondingly the original $QM$\ unitary operator $W$ that works on the
original Hilbert space is changed to the new one that works on the composite
Hilbert space $S_{1}\tbigotimes S_{2}$, and it will be seen later that the
latter may be constructed conveniently in the composite Hilbert space.

In principle the total composite Hilbert space under study below contains
the composite Hilbert space $S_{1}\tbigotimes S_{2}$ as its component
Hilbert subspace. Here only the composite Hilbert space $S_{1}\tbigotimes
S_{2}$ is directly treated theoretically, and any other component Hilbert
subspace of the total Hilbert space is not explicitly considered as it does
not affect the present theoretical treatment. Note that the basis sets $%
\left\{ \left\vert k\right\rangle \right\} $ and $\left\{ u_{k}\left(
x\right) \right\} $ are chosen by the two component Hilbert subspaces $S_{1}$
and $S_{2},$ respectively. Then in the composite Hilbert space $%
S_{1}\tbigotimes S_{2}$ they may be replaced with the basis subsets $\left\{
\left\vert k\right\rangle \tbigotimes \left\vert u_{0}\right\rangle \right\} 
$ and $\left\{ \left\vert 0\right\rangle \tbigotimes \left\vert
u_{k}\right\rangle \right\} ,$ respectively, here $\left\vert
u_{k}\right\rangle \equiv u_{k}\left( x\right) $ for $k=0,1,...,\infty .$
Correspondingly the component Hilbert subspaces $S_{1}$ and $S_{2}$ may be
replaced with the subspaces $S_{1}\tbigotimes \left\vert u_{0}\right\rangle $
and $\left\vert 0\right\rangle \tbigotimes S_{2}$ of the composite Hilbert
space $S_{1}\tbigotimes S_{2},$ respectively. Evidently the basis subsets $%
\left\{ \left\vert k\right\rangle \tbigotimes \left\vert u_{0}\right\rangle
\right\} $ and $\left\{ \left\vert 0\right\rangle \tbigotimes \left\vert
u_{k}\right\rangle \right\} $ are the complete basis sets of the subspaces $%
S_{1}\tbigotimes \left\vert u_{0}\right\rangle $ and $\left\vert
0\right\rangle \tbigotimes S_{2},$ respectively. Now the original unitary
transformation $W$ may be replaced with the following new one on the
composite Hilbert space $S_{1}\tbigotimes S_{2}$:%
\begin{equation*}
W_{12}:\left\vert k\right\rangle \tbigotimes \left\vert u_{0}\right\rangle
\rightarrow \left\vert 0\right\rangle \tbigotimes \left\vert
u_{k}\right\rangle ,\text{ }k=0,1,2,...,\infty .
\end{equation*}%
It is emphasized that $W_{12}$ is also a purely $QM$ unitary operator. It
can be found that both the basis subsets $\left\{ \left\vert k\right\rangle
\tbigotimes \left\vert u_{0}\right\rangle \right\} $ and $\left\{ \left\vert
0\right\rangle \tbigotimes \left\vert u_{k}\right\rangle \right\} $ are
orthogonal to each other, if the initialized tensor-product base $\left\vert
0\right\rangle \tbigotimes \left\vert u_{0}\right\rangle $ is omitted from
any one of the two basis subsets. This property is important to construct
conveniently the $QM$\ unitary operator $W_{12}$ in the composite Hilbert
space $S_{1}\tbigotimes S_{2}.$ The $QM$ unitary operator $W_{12}$ may be
used to transfer the solution information from the subspace $%
S_{1}\tbigotimes \left\vert u_{0}\right\rangle $ to $\left\vert
0\right\rangle \tbigotimes S_{2}$ in the composite Hilbert space $%
S_{1}\tbigotimes S_{2}$ or, equivalently speaking, from the component
Hilbert subspace $S_{1}$ to $S_{2}.$

The concrete construction for the $QM$\ unitary operator $W_{12}$ may be
conveniently carried out in the framework of the multiple-quantum operator
algebra space [$7,24,6$]. In analogous to the Pauli spin operators ($I_{\mu
}=\sigma _{\mu }/2$ with $\mu =x,y,z$) of a single-spin$-1/2$ system, the
Hermitian pseudospin operators in a multi-level quantum system may be
defined by (See, for example, Refs. [$7,15$])%
\begin{equation}
\left\{ 
\begin{array}{c}
Q_{x}^{KL}=\frac{1}{2}(|K\rangle \langle L|+|L\rangle \langle K|) \\ 
Q_{y}^{KL}=\frac{1}{2i}(|K\rangle \langle L|-|L\rangle \langle K|) \\ 
Q_{z}^{KL}=\frac{1}{2}(|K\rangle \langle K|-|L\rangle \langle L|)%
\end{array}%
\right.  \tag{3.30}
\end{equation}%
where $|K\rangle $ and $|L\rangle $ are any pair of energy levels of the
quantum system, which may form a two-level system in concept. Just like the
Pauli spin operators $\{I_{\mu }\}$ they satisfy the usual cyclic
commutation relations: $[Q_{\newline
\mu }^{KL},Q_{\nu }^{KL}]=iQ_{\lambda }^{KL}$ $\left( \mu ,\nu ,\lambda
=x,y,z\right) $. One important property of them is that the pseudospin
operator $Q_{\mu }^{KL}$ selectively acts on only the two-level subspace $%
\left\{ |K\rangle ,|L\rangle \right\} $ of the Hilbert space of the quantum
system. It can prove that all the independent $z-$component pseudospin
operators $\{Q_{z}^{KL}\}$ (plus the unity operator) form a complete set of
the base operators of the $LOMSO$ operator subspace of the multiple-quantum
operator algebra space of the quantum system. It also can prove that all
these independent pseudospin operators $\{Q_{\mu }^{KL}\}$ (plus the unity
operator) may form a complete set of the base operators of the
multiple-quantum operator algebra space.

The pseudospin operators defined by (3.30) are general in any quantum
system, but they could reflect less the symmetrical structure and property
of the quantum system. A further characterization (and classification) for
the pseudospin operators of a multi-level spin system may lead them to the
so-called selective multiple-quantum-transition operators. The latter could
reflect more the symmetrical structure and property of a spin system such as
an $n-$qubit spin system at least for the multiple-quantum-transition
symmetrical structure and property [$7,24,6$]. The
multiple-quantum-transition concept is original from the NMR spectroscopy [$%
25$].

In analogous to the unitary spin rotation operator $\exp \left( -i\theta
I_{\mu }\right) $ which is generated by the Pauli spin operator $I_{\mu }$ ($%
\mu =x,y,z$) the unitary pseudospin rotation operator $\exp \left( -i\theta
Q_{\mu }^{KL}\right) $ may be generated by the Hermitian pseudospin operator 
$Q_{\mu }^{KL}.$ The pseudospin rotation operator $\exp \left( -i\theta
Q_{\mu }^{KL}\right) $ ($\mu =x,y$) may induce the selective excitation or
transition between the energy levels $|K\rangle $ and $|L\rangle $ of the
multi-level quantum system. Experimentally the pseudospin rotation operators 
$\{\exp \left( -i\theta Q_{\mu }^{KL}\right) \}$ may be realized (directly
or indirectly) by the radiofrequency wave selective excitation [$28,25$] or
the laser light (or microwave) selective excitation [$28,29,17$]. Therefore,
they may be used to realize the $QM$ selective unitary transformation
between different energy eigenbases. As a typical example, the pseudospin
rotation operator $\exp \left( \pm i\pi Q_{x}^{KL}\right) $ may realize the
selective unitary transformations between the energy eigenbases $\left\vert
K\right\rangle $ and $\left\vert L\right\rangle $:%
\begin{equation}
\exp \left( \pm i\pi Q_{x}^{KL}\right) \left\vert K\right\rangle =\pm
i\left\vert L\right\rangle ,\text{ \ }\exp \left( \pm i\pi Q_{x}^{KL}\right)
\left\vert L\right\rangle =\pm i\left\vert K\right\rangle .  \tag{3.31}
\end{equation}%
These $QM$\ selective unitary transformations like (3.31) play an important
role in constructing conveniently the $QM$ unitary operator $W_{12}$.

In unitary quantum dynamics the present technique to construct the $QM$
unitary operator $W_{12}$ (and also the similar ones described below) mainly
follows the author's work [$7$]. Suppose that $\left\vert K\right\rangle $
is any energy eigenbase of the (non-interacting) composite quantum system
with the composite Hilbert space $S_{1}\tbigotimes S_{2}$ above. Evidently $%
\left\vert K\right\rangle $ may be taken as any tensor-product base $%
\left\vert k,u_{l}\right\rangle $ (i.e., $\left\vert k\right\rangle
\tbigotimes \left\vert u_{l}\right\rangle )$ of the composite Hilbert space $%
S_{1}\tbigotimes S_{2}.$ Now in (3.30) and (3.31) one may choose the energy
eigenbase $\left\vert K\right\rangle $ as $\left\vert k,u_{0}\right\rangle $
and $\left\vert L\right\rangle $ as $\left\vert 0,u_{k}\right\rangle ,\ $%
where both the energy eigenbases $\left\vert k,u_{0}\right\rangle $ and $%
\left\vert 0,u_{k}\right\rangle $ with $k\neq 0$ are orthogonal to each
other. Then the pseudospin operator $Q_{x}^{KL}$ of (3.30) may be rewritten
as%
\begin{equation}
Q_{x}^{k,u_{k}}=\frac{1}{2}(\left\vert k,u_{0}\right\rangle \left\langle
0,u_{k}\right\vert +\left\vert 0,u_{k}\right\rangle \left\langle
k,u_{0}\right\vert )  \tag{3.32}
\end{equation}%
with $\left\langle k,u_{0}\right\vert \equiv \left\langle k\right\vert
\tbigotimes \left\langle u_{0}\right\vert ,$ etc. In analogous way one may
re-write $Q_{y}^{KL}$ as $Q_{y}^{k,u_{k}}$ and $Q_{z}^{KL}\ $as $%
Q_{z}^{k,u_{k}}.$ There is an important property for the pseudospin
operators $\{Q_{\mu }^{k,u_{k}}\}$:%
\begin{equation}
\left[ Q_{\mu }^{k,u_{k}},Q_{\nu }^{l,u_{l}}\right] =0\text{ for }k\neq l%
\text{ and }\mu ,\nu =x,y,z  \tag{3.33}
\end{equation}%
This commutation relation says that both the pseudospin operators $Q_{\mu
}^{k,u_{k}}$ and $Q_{\nu }^{l,u_{l}}$ commute if $k\neq l,$ since there is
no overlapping between the two-level subspace of the operator $Q_{\mu
}^{k,u_{k}}$ and the one of the operator $Q_{\nu }^{l,u_{l}}.$ The
commutation relations of (3.33) play an important role in constructing
conveniently the $QM$ unitary operator $W_{12}.$ The $QM$ selective unitary
transformations of (3.31) between the two eigenbases $\left\vert
k,u_{0}\right\rangle $ $\left( \text{i.e., }\left\vert K\right\rangle
\right) $ and $\left\vert 0,u_{k}\right\rangle $ $\left( \text{i.e., }%
\left\vert L\right\rangle \right) $ caused by the pseudospin rotation
operator $\exp \left( \pm i\pi Q_{x}^{k,u_{k}}\right) $ then may be
re-written as%
\begin{equation}
\exp \left( \pm i\pi Q_{x}^{k,u_{k}}\right) \left\vert k,u_{0}\right\rangle
=\pm i\left\vert 0,u_{k}\right\rangle ,\text{ }\exp \left( \pm i\pi
Q_{x}^{k,u_{k}}\right) \left\vert 0,u_{k}\right\rangle =\pm i\left\vert
k,u_{0}\right\rangle  \tag{3.34a}
\end{equation}%
and moreover, there are the identical state transformations:%
\begin{equation}
\exp \left( \pm i\pi Q_{x}^{k,u_{k}}\right) \left\vert j,u_{0}\right\rangle
=\left\vert j,u_{0}\right\rangle ,\text{ }\exp \left( \pm i\pi
Q_{x}^{k,u_{k}}\right) \left\vert 0,u_{j}\right\rangle =\left\vert
0,u_{j}\right\rangle \text{ for }j\neq k  \tag{3.34b}
\end{equation}%
With the help of the commutation relations of (3.33), the $QM$\ selective
unitary transformations of (3.34a), and the identical state transformations
of (3.34b) one may construct explicitly the $QM$ unitary operator $W_{12}$
up to a global phase factor:%
\begin{equation}
W_{12}=\exp \left( -i\frac{\pi }{2}\left\vert 0,u_{0}\right\rangle
\left\langle 0,u_{0}\right\vert \right) \tprod\limits_{k=1}^{\infty }\exp
\left( -i\pi Q_{x}^{k,u_{k}}\right) .  \tag{3.35}
\end{equation}%
It is similar to a $QM$ subspace-selective unitary operator $[7]$. The $QM$\
unitary transformation $W_{12}$ can be proven simply below. By applying this
unitary operator $W_{12}$ to arbitrary basis state $\left\vert
j,u_{0}\right\rangle $ ($j>0$) and then using (3.33), (3.34a), and (3.34b)
one obtains%
\begin{equation*}
W_{12}\left\vert j,u_{0}\right\rangle =\exp \left( -i\frac{\pi }{2}%
\left\vert 0,u_{0}\right\rangle \left\langle 0,u_{0}\right\vert \right)
\tprod\limits_{k=1}^{\infty }\exp \left( -i\pi Q_{x}^{k,u_{k}}\right)
\left\vert j,u_{0}\right\rangle
\end{equation*}%
\begin{equation*}
=\exp \left( -i\pi Q_{x}^{j,u_{j}}\right) \left\vert j,u_{0}\right\rangle
=-i\left\vert 0,u_{j}\right\rangle
\end{equation*}%
and moreover, it can prove that $W_{12}\left\vert 0,u_{0}\right\rangle
=-i\left\vert 0,u_{0}\right\rangle .$ In analogous way one also obtains $%
W_{12}^{+}\left\vert j,u_{0}\right\rangle =i\left\vert 0,u_{j}\right\rangle $
for $j=0,1,...,\infty .$ Therefore, the $QM$ unitary transformation $W_{12}$
may be exactly written as%
\begin{equation}
W_{12}\left\vert j,u_{0}\right\rangle =-i\left\vert 0,u_{j}\right\rangle ,%
\text{ }W_{12}^{+}\left\vert j,u_{0}\right\rangle =i\left\vert
0,u_{j}\right\rangle ,\text{ for }j\geq 0.  \tag{3.36}
\end{equation}%
This unitary transformation $W_{12}$ is really the desired one, i.e., $%
W_{12}:\left\vert j,u_{0}\right\rangle \rightarrow \left\vert
0,u_{j}\right\rangle $ with $j\geq 0$ up to a global phase factor $(-i).$

The $QM$ subspace-selective unitary operator $W_{12}$ of (3.35) may be used
to realize the solution-information transfer from the subspace $%
S_{1}\tbigotimes \left\vert u_{0}\right\rangle $ to $\left\vert
0\right\rangle \tbigotimes S_{2}$ or equivalently from the component Hilbert
subspace $S_{1}$ to $S_{2}$ in the composite Hilbert space $S_{1}\tbigotimes
S_{2}.$ Suppose now that the solution information $\left( a_{m}^{s}\right) $
is initially loaded on the $SIC$ unitary propagator $U_{\infty }^{sic}\left(
a_{m}^{s},\alpha _{m}\right) $ of (3.12) which works on the component
Hilbert subspace $S_{1},$ and after the solution-information transfer it is
carried by the $SIC$ unitary propagator $U_{c}^{sic}(t_{m})=\exp \left[
-ia_{m}^{s}H_{c}t_{m}/\hslash \right] $ which works on the component Hilbert
subspace $S_{2},$ where the Hamiltonian $H_{c}$ owns the energy
eigenspectrum $\{E_{k}^{c}\}=\{ak+b\}$ and its associated eigenbasis set $%
\{u_{k}\left( x\right) \}.$ This is a simple case. Below it is treated in
detail. For other case a similar theoretical treatment is also available,
but it is not described here.

In an analogous way to proving (3.20) it can prove below that the explicit
expression for the $SIC$ unitary propagator $U_{c}^{sic}(t_{m})$ is
determined from%
\begin{equation}
U_{c}^{sic}(t_{m})=\exp \left[ -i\Phi _{1}\left( a_{m}^{s}\right) \right]
W_{12}U_{\infty }^{sic}\left( a_{m}^{s},\alpha _{m}\right) W_{12}^{+} 
\tag{3.37}
\end{equation}%
where the global phase factor $\exp \left[ -i\Phi _{1}\left(
a_{m}^{s}\right) \right] $ is still given in (3.20). The expression (3.37)
shows clearly that the $SIC$ unitary propagator $U_{c}^{sic}(t_{m})$ is
obtained from the $U_{\infty }^{sic}\left( a_{m}^{s},\alpha _{m}\right) $ by
the $QM$\ unitary transformation $W_{12}$ up to a global phase factor. As
shown in (3.12), the latter is generated by an infinite sequence of the
basic $SIC$ unitary operators. Therefore, this $U_{c}^{sic}(t_{m})$ is
really built out of these basic $SIC$ unitary operators. The
unitary-dynamical equation (3.37) works on the composite Hilbert space $%
S_{1}\tbigotimes S_{2}.$ But actually it exactly works on a much smaller
subspace $S_{1}\tbigotimes \left\vert u_{0}\right\rangle \cup \left\vert
0\right\rangle \tbigotimes S_{2}$ of the Hilbert space $S_{1}\tbigotimes
S_{2}.$ Hereafter denote $S_{12}\equiv S_{1}\tbigotimes \left\vert
u_{0}\right\rangle \cup \left\vert 0\right\rangle \tbigotimes S_{2}.$

The unitary-dynamical equation (3.37) shows that the solution-information
transfer from the subspace $S_{1}\tbigotimes \left\vert u_{0}\right\rangle $
to $\left\vert 0\right\rangle \tbigotimes S_{2}$ in the Hilbert space $%
S_{1}\tbigotimes S_{2}$ is realized by making the $QM$ subspace-selective
unitary transformation $W_{12}$ of (3.35) on the $SIC$\ unitary propagator $%
U_{\infty }^{sic}\left( a_{m}^{s},\alpha _{m}\right) .$

It will prove rigorously that the $SIC$ unitary propagator $%
U_{c}^{sic}\left( t_{m}\right) $ indeed obeys the unitary-dynamical equation
(3.37). As shown above, both $U_{c}^{sic}(t_{m})=\exp \left[
-ia_{m}^{s}H_{c}t_{m}/\hslash \right] $ and its Hamiltonian $H_{c}$ each act
on the component Hilbert subspace $S_{2}$ or equivalently on the subspace $%
\left\vert 0\right\rangle \tbigotimes S_{2}$ of the subspace $S_{12}$ of the
composite Hilbert space $S_{1}\tbigotimes S_{2}.$ Here the Hamiltonian $%
H_{c} $ owns the eigenvalue equation $H_{c}u_{k}\left( x\right)
=E_{k}^{c}u_{k}\left( x\right) $ with energy eigenvalue $E_{k}^{c}=ak+b$ and
its associated eigenbase $u_{k}\left( x\right) ,$ and the eigenbasis set $%
\{u_{k}\left( x\right) \}$ spans the component Hilbert subspace $S_{2}$. Now
in the energy eigenbasis set $\{u_{k}\left( x\right) \}$ arbitrary state $%
|\Psi _{c}\rangle $ of the component Hilbert subspace $S_{2}$ can be
expanded as $|\Psi _{c}\rangle =\sum_{k=0}^{\infty }B_{k}u_{k}\left(
x\right) ,$ which is like (3.8). Then correspondingly arbitrary state $%
\left\vert 0\right\rangle \tbigotimes |\Psi _{c}\rangle $ of the subspace $%
\left\vert 0\right\rangle \tbigotimes S_{2}$ can be expanded as%
\begin{equation}
\left\vert 0\right\rangle \tbigotimes |\Psi _{c}\rangle =\sum_{k=0}^{\infty
}B_{k}\left\vert 0,u_{k}\right\rangle .  \tag{3.38}
\end{equation}%
The theoretical proof for the equation (3.37) is described in detail as
follows. On the one hand, by applying $W_{12}U_{\infty }^{sic}\left(
a_{m}^{s},\alpha _{m}\right) W_{12}^{+}$ to the state $\left\vert
0\right\rangle \tbigotimes |\Psi _{c}\rangle $ with the infinite expansion
series (3.38) one obtains%
\begin{equation}
W_{12}U_{\infty }^{sic}\left( a_{m}^{s},\alpha _{m}\right)
W_{12}^{+}\left\vert 0\right\rangle \tbigotimes |\Psi _{c}\rangle
=W_{12}U_{\infty }^{sic}\left( a_{m}^{s},\alpha _{m}\right)
\sum_{k=0}^{\infty }iB_{k}\left\vert k,u_{0}\right\rangle  \tag{3.39}
\end{equation}%
where $W_{12}^{+}\left\vert 0,u_{k}\right\rangle =i\left\vert
k,u_{0}\right\rangle $ is already used, which is obtained from (3.36).
Notice that the $SIC$ unitary propagator $U_{\infty }^{sic}\left(
a_{m}^{s},\alpha _{m}\right) $ acts on the component Hilbert subspace $S_{1}$
of the composite Hilbert space $S_{1}\tbigotimes S_{2}$ alone. It can be
deduced from (3.14), (3.15), and (3.16) that the eigenvalue equation for the 
$U_{\infty }^{sic}\left( a_{m}^{s},\alpha _{m}\right) $ acting on any
tensor-product base $\left\vert k,u_{j}\right\rangle $ ($k,j=0,1,...,\infty $%
) of the composite Hilbert space $S_{1}\tbigotimes S_{2}$ is given by%
\begin{equation}
U_{\infty }^{sic}\left( a_{m}^{s},\alpha _{m}\right) \left\vert
k,u_{j}\right\rangle =\tprod\limits_{l=1}^{\infty }\exp \left(
i2^{l-2}\alpha _{m}a_{m}^{s}\right) \exp \left( -i\alpha
_{m}a_{m}^{s}k\right) \left\vert k,u_{j}\right\rangle .  \tag{3.40}
\end{equation}%
Now with the help of the eigenvalue equation (3.40) and $W_{12}\left\vert
k,u_{0}\right\rangle =-i\left\vert 0,u_{k}\right\rangle $ of (3.36) the
equation (3.39) is reduced to the form%
\begin{equation*}
W_{12}U_{\infty }^{sic}\left( a_{m}^{s},\alpha _{m}\right)
W_{12}^{+}\left\vert 0\right\rangle \tbigotimes |\Psi _{c}\rangle
\end{equation*}%
\begin{equation}
=\tprod\limits_{l=1}^{\infty }\exp \left( i2^{l-2}\alpha
_{m}a_{m}^{s}\right) \sum_{k=0}^{\infty }B_{k}\exp \left( -i\alpha
_{m}a_{m}^{s}k\right) \left\vert 0,u_{k}\right\rangle .  \tag{3.41}
\end{equation}%
This equation shows that $W_{12}U_{\infty }^{sic}\left( a_{m}^{s},\alpha
_{m}\right) W_{12}^{+}$ indeed works effectively on the subspace $\left\vert
0\right\rangle \tbigotimes S_{2}$ or equivalently on the component Hilbert
subspace $S_{2}.$ On the other hand, since both the $U_{c}^{sic}(t_{m})$ and 
$H_{c}$ each act on the component Hilbert subspace $S_{2},$ by applying
directly $U_{c}^{sic}\left( t_{m}\right) =\exp \left[ -ia_{m}^{s}H_{c}t_{m}/%
\hslash \right] $ to the state (3.38) one obtains%
\begin{equation*}
U_{c}^{sic}\left( t_{m}\right) \left\vert 0\right\rangle \tbigotimes |\Psi
_{c}\rangle =\exp \left( -ia_{m}^{s}\left( bt_{m}/\hslash \right) \right)
\end{equation*}%
\begin{equation}
\times \sum_{k=0}^{\infty }B_{k}\exp \left( -i\left( at_{m}/\hslash \right)
a_{m}^{s}k\right) \left\vert 0,u_{k}\right\rangle  \tag{3.42}
\end{equation}%
where the eigenvalue equation $H_{c}\left\vert 0,u_{k}\right\rangle
=E_{k}^{c}\left\vert 0,u_{k}\right\rangle $ with $E_{k}^{c}=ak+b$ is already
used, which is obtained from the eigenvalue equation $H_{c}u_{k}\left(
x\right) =E_{k}^{c}u_{k}\left( x\right) $ above. Let $\alpha _{m}=\left(
at_{m}/\hslash \right) $ in (3.42). Then by substituting (3.42) into (3.41)
it can be found that%
\begin{equation*}
W_{12}U_{\infty }^{sic}\left( a_{m}^{s},\alpha _{m}\right)
W_{12}^{+}\left\vert 0\right\rangle \tbigotimes |\Psi _{c}\rangle
\end{equation*}%
\begin{equation}
=\tprod\limits_{l=1}^{\infty }\exp \left( i2^{l-2}\alpha
_{m}a_{m}^{s}\right) \exp \left( ia_{m}^{s}\left( bt_{m}/\hslash \right)
\right) U_{c}^{sic}\left( t_{m}\right) \left\vert 0\right\rangle \tbigotimes
|\Psi _{c}\rangle  \tag{3.43}
\end{equation}%
Notice that the state $\left\vert 0\right\rangle \tbigotimes |\Psi
_{c}\rangle $ is arbitrary in the subspace $\left\vert 0\right\rangle
\tbigotimes S_{2}.$ Then the equation (3.43) leads directly to (3.37).

As pointed out previously, there is the hard problem for how to construct
explicitly the $QM$ unitary transformation $W$ in the preparation of the $%
SIC $ unitary propagator $U_{c}^{sic}(t_{m})$ by (3.20)$.$ Now the $%
U_{c}^{sic}(t_{m})$ also can be prepared by (3.37), where the $QM$ unitary
transformation $W_{12}$ can be easily constructed explicitly by (3.35). Here
the $QM$\ unitary operator $W_{12}$ corresponds to the $QM$ unitary operator$%
\ W.$ Therefore, the hard problem met in the first preparation for the $%
U_{c}^{sic}(t_{m})$ based on (3.20) is not met in the second one based on
(3.37). The first preparation employs the original Hilbert space whose size
is equal to the one of the component Hilbert space $S_{1}$ or $S_{2}$. In
contrast, the second preparation employs a much larger composite Hilbert
space $S_{1}\tbigotimes S_{2}.$

Evidently the solution information ($a_{m}^{s}$) may be further transferred
from the component Hilbert subspace $S_{2}$ to a third component Hilbert
subspace $S_{3},$ after it is transferred from the component Hilbert
subspace $S_{1}$ to $S_{2}$ through (3.37). Suppose that after the solution
information is transferred to the component Hilbert subspace $S_{3},$ the $%
SIC$ unitary propagator may be written as $V_{c}^{sic}(t_{v})=\exp \left[
-ia_{m}^{s}H_{c}^{v}t_{v}/\hslash \right] ,$ where the Hamiltonian $%
H_{c}^{v} $ works on the component Hilbert subspace $S_{3}.$ If now the
energy eigenspectrum $\{E_{k}^{v}\}$ of the Hamiltonian $H_{c}^{v}$ is the
same as the one ($\{E_{k}^{c}\}$) of the Hamiltonian $H_{c}$ of the $SIC$
unitary propagator $U_{c}^{sic}(t_{m})$ of (3.37), that is, $%
E_{k}^{v}=E_{k}^{c},$ then situation becomes simple. In this simple case
there always exists a $QM$ unitary transformation that transforms the
Hamiltonian $H_{c}$ to $H_{c}^{v}. $ Consequently it can prove that the $SIC$
unitary propagator $V_{c}^{sic}(t_{m})$ may be prepared by making a $QM$
unitary transformation $W_{23}$ on the $SIC$\ unitary propagator $%
U_{c}^{sic}(t_{m})$ of (3.37),%
\begin{equation}
V_{c}^{sic}(t_{m})=W_{23}U_{c}^{sic}(t_{m})W_{23}^{+}.  \tag{3.44}
\end{equation}%
Furthermore, if $E_{k}^{v}$ is equal to $E_{k}^{c}$ up to a constant
independent of the quantum number $k$, then situation is still simple. This
case means that the $SIC$ unitary propagator $V_{c}^{sic}(t_{m})$ has the
same eigenvalue set as the one of the $U_{c}^{sic}(t_{m})$ up to a global
phase factor. Theoretically this case still can be treated in analogous way
and the prepared $V_{c}^{sic}(t_{m})$ for this case is still given by (3.44)
up to a global phase factor. The theoretical demonstration for the
unitary-dynamical equation (3.44) may be carried out in an analogous way
that is used to prove the unitary-dynamical equation (3.37) above.

The solution-information transfer from the component Hilbert subspace $S_{1}$
to $S_{2}$ and then to $S_{3}$ is carried out in the composite Hilbert space 
$S_{1}\tbigotimes S_{2}\tbigotimes S_{3}.$ Here the solution-information
transfer from the component Hilbert subspace $S_{2}$ to $S_{3}$ is carried
out in the composite Hilbert subspace $S_{2}\tbigotimes S_{3}$ of the
Hilbert space $S_{1}\tbigotimes S_{2}\tbigotimes S_{3}.$ Then the $QM$
unitary operator $W_{23}$ in (3.44) works on only the Hilbert subspace $%
S_{2}\tbigotimes S_{3}.$ It is first constructed below. Note that the $SIC$
unitary propagator $V_{c}^{sic}(t_{m})$ of (3.44) and its Hamiltonian $%
H_{c}^{v}$ each effectively act on the component Hilbert subspace $S_{3}.$
Now suppose that the energy eigenvalue equation for the Hamiltonian $%
H_{c}^{v}$ is given by $H_{c}^{v}v_{k}\left( x\right) =E_{k}^{v}v_{k}\left(
x\right) ,$ where $E_{k}^{v}$ is the $k-$th energy eigenvalue and its
associated eigenbase is $v_{k}\left( x\right) $ $(\equiv \left\vert
v_{k}\right\rangle ).$ Then all the eigenbases $\{v_{k}\left( x\right) \}$
form an orthonormal basis set of the component Hilbert subspace $S_{3}.$ As
shown above, all the energy eigenbases $\{u_{k}\left( x\right) \}$ of the
Hamiltonian $H_{c}$ of the $U_{c}^{sic}(t_{m})$ in (3.44) form an
orthonormal basis set of the component Hilbert subspaces $S_{2}$. Then in
the composite Hilbert subspace $S_{2}\tbigotimes S_{3}$ both the basis sets $%
\{u_{k}\left( x\right) \}$ and $\{v_{k}\left( x\right) \}$ should be
respectively replaced with the basis subsets $\{\left\vert
u_{k}\right\rangle \tbigotimes \left\vert v_{0}\right\rangle \}$ and $%
\{\left\vert u_{0}\right\rangle \tbigotimes \left\vert v_{k}\right\rangle \}$%
, where $\left\vert u_{0}\right\rangle \tbigotimes \left\vert
v_{0}\right\rangle $ is the initialized state. The latter are the basis
subsets of the subspaces $S_{2}\tbigotimes \left\vert v_{0}\right\rangle $
and $\left\vert u_{0}\right\rangle \tbigotimes S_{3}$ of the composite
Hilbert subspace $S_{2}\tbigotimes S_{3},$ respectively. Both the basis
subsets $\{\left\vert u_{k}\right\rangle \tbigotimes \left\vert
v_{0}\right\rangle \}$ ($\equiv \left\vert u_{k},v_{0}\right\rangle $) and $%
\{\left\vert u_{0}\right\rangle \tbigotimes \left\vert v_{k}\right\rangle \}$
($\equiv \left\vert u_{0},v_{k}\right\rangle $) are orthogonal to each
other, if the initialized state $\left\vert u_{0},v_{0}\right\rangle $ is
omitted from any one of the two basis subsets. Then the $QM$ unitary
transformation $W_{23}$ may be defined by 
\begin{equation*}
W_{23}:\left\vert u_{k},v_{0}\right\rangle \rightarrow \left\vert
u_{0},v_{k}\right\rangle ,\text{ }k=0,1,2,...,\infty .
\end{equation*}%
The $QM$ unitary operator $W_{23}$ exactly works on the subspace $%
S_{23}=S_{2}\tbigotimes \left\vert v_{0}\right\rangle \cup \left\vert
u_{0}\right\rangle \tbigotimes S_{3}$ of the composite Hilbert subspace $%
S_{2}\tbigotimes S_{3}.$ It may be constructed in an analogous way that is
used to construct the $QM$ unitary operator $W_{12}$ above.

The $QM$ unitary operator $W_{23}$ may be constructed in the
(non-interacting) composite quantum system with the composite Hilbert
subspace $S_{2}\tbigotimes S_{3}.$ Suppose that $\left\vert K\right\rangle $
is any energy eigenbase of the composite quantum system. Evidently it may be
taken as any tensor-product base $\left\vert u_{k},v_{l}\right\rangle $ of
the composite Hilbert subspace $S_{2}\tbigotimes S_{3}.$ Then in accordance
with (3.30) one may set up the Hermitian pseudospin operator $%
Q_{x}^{u_{k},v_{k}}$ by%
\begin{equation}
Q_{x}^{u_{k},v_{k}}=\frac{1}{2}(\left\vert u_{k},v_{0}\right\rangle
\left\langle u_{0},v_{k}\right\vert +\left\vert u_{0},v_{k}\right\rangle
\left\langle u_{k},v_{0}\right\vert )  \tag{3.45}
\end{equation}%
and in analogous way one also may set up $Q_{y}^{u_{k},v_{k}}$ and $%
Q_{z}^{u_{k},v_{k}}.$ These pseudospin operators $\{Q_{\mu }^{u_{k},v_{k}}\}$
act on the composite Hilbert subspace $S_{2}\tbigotimes S_{3}.$ Just like
the pseudospin operators $\{Q_{\mu }^{k,u_{k}}\}$ defined by (3.32), etc.,
they satisfy the commutation relation like (3.33), 
\begin{equation}
\left[ Q_{\mu }^{u_{k},v_{k}},Q_{\lambda }^{u_{l},v_{l}}\right] =0\text{ for 
}k\neq l\text{ and }\mu ,\lambda =x,y,z.  \tag{3.46a}
\end{equation}%
The pseudospin rotation operator $\exp \left( -i\theta
Q_{x}^{u_{k},v_{k}}\right) $ that is generated by $Q_{x}^{u_{k},v_{k}}$ is
able to induce the selective energy-level transition $\left\vert
u_{k},v_{0}\right\rangle \rightarrow \left\vert u_{0},v_{k}\right\rangle $
(and vice versa). This selective energy-level transition may be used to
realize the $QM$ selective unitary transformations between the two energy
eigenbases $\left\vert u_{k},v_{0}\right\rangle $ and $\left\vert
u_{0},v_{k}\right\rangle $,%
\begin{equation}
\exp \left( \pm i\pi Q_{x}^{u_{k},v_{k}}\right) \left\vert
u_{k},v_{0}\right\rangle =\pm i\left\vert u_{0},v_{k}\right\rangle ,\text{ }%
\exp \left( \pm i\pi Q_{x}^{u_{k},v_{k}}\right) \left\vert
u_{0},v_{k}\right\rangle =\pm i\left\vert u_{k},v_{0}\right\rangle 
\tag{3.46b}
\end{equation}%
which are similar to (3.34a). There are also those identical state
transformations like (3.34b), 
\begin{equation}
\exp \left( \pm i\pi Q_{x}^{u_{k},v_{k}}\right) \left\vert
u_{j},v_{0}\right\rangle =\left\vert u_{j},v_{0}\right\rangle ,\exp \left(
\pm i\pi Q_{x}^{u_{k},v_{k}}\right) \left\vert u_{0},v_{j}\right\rangle
=\left\vert u_{0},v_{j}\right\rangle ,\text{ for }j\neq k.  \tag{3.46c}
\end{equation}%
With the help of the commutation relations of (3.46a), the $QM$ selective
unitary transformations of (3.46b), and the identical state transformations
of (3.46c) one may explicitly construct the $QM$ unitary operator $W_{23}$
up to a global phase factor by%
\begin{equation}
W_{23}=\exp \left( -i\frac{\pi }{2}\left\vert u_{0},v_{0}\right\rangle
\left\langle u_{0},v_{0}\right\vert \right) \tprod\limits_{k=1}^{\infty
}\exp \left( -i\pi Q_{x}^{u_{k},v_{k}}\right) .  \tag{3.47}
\end{equation}%
This $QM$ unitary operator $W_{23}$ also is similar to a $QM$
subspace-selective unitary operator [$7$], just like the $QM$\ unitary
operator $W_{12}$ of (3.35). By substituting this $QM$ unitary operator $%
W_{23}$ into (3.44) one may obtain the explicit expression for the $SIC$
unitary propagator $V_{c}^{sic}(t_{m}).$ The $QM$ unitary operator $W_{23}$
works exactly on the subspace $S_{23}$ of the Hilbert space $%
S_{2}\tbigotimes S_{3}$. Now by applying the unitary operator $W_{23}$ to
any tensor-product base $\left\vert u_{j},v_{0}\right\rangle $ $(j\geq 0)$
of the subspace $S_{23}$ one obtains, with the help of (3.46),%
\begin{equation*}
W_{23}\left\vert u_{j},v_{0}\right\rangle =\exp \left( -i\pi
Q_{x}^{u_{j},v_{j}}\right) \left\vert u_{j},v_{0}\right\rangle =-i\left\vert
u_{0},v_{j}\right\rangle
\end{equation*}%
In analogous way one also obtains $W_{23}^{+}\left\vert
u_{j},v_{0}\right\rangle =i\left\vert u_{0},v_{j}\right\rangle $ for $j\geq
0.$ Therefore, the $QM$ unitary operator $W_{23}$ of (3.47) acting on any
tensor-product base $\left\vert u_{j},v_{0}\right\rangle $ ($\left\vert
u_{0},v_{j}\right\rangle $) of the subspace $S_{23}$ may be exactly
described by%
\begin{equation}
W_{23}\left\vert u_{j},v_{0}\right\rangle =-i\left\vert
u_{0},v_{j}\right\rangle ,\text{ }W_{23}^{+}\left\vert
u_{j},v_{0}\right\rangle =i\left\vert u_{0},v_{j}\right\rangle ,\text{ for }%
j\geq 0.  \tag{3.48}
\end{equation}%
This unitary transformation $W_{23}$ is the desired one, i.e., $%
W_{23}:\left\vert u_{j},v_{0}\right\rangle \rightarrow \left\vert
u_{0},v_{j}\right\rangle $ for $j\geq 0$ up to a global phase factor $\left(
-i\right) .$

Below it can prove theoretically that the $SIC$\ unitary propagator $%
V_{c}^{sic}\left( t_{m}\right) =\exp \left[ -ia_{m}^{s}H_{c}^{v}t_{m}/%
\hslash \right] $ with energy eigenspectrum $\{E_{k}^{v}\}=\left\{
ak+b\right\} $ of the Hamiltonian $H_{c}^{v}$ is determined from (3.44),
that is, the $V_{c}^{sic}\left( t_{m}\right) $ obeys the unitary-dynamical
equation of (3.44). The $V_{c}^{sic}\left( t_{m}\right) $ exactly works on
the subspace:%
\begin{equation*}
S_{123}=S_{1}\tbigotimes \left\vert u_{0}\right\rangle \tbigotimes
\left\vert v_{0}\right\rangle \tbigcup \left\vert 0\right\rangle \tbigotimes
S_{2}\tbigotimes \left\vert v_{0}\right\rangle \tbigcup \left\vert
0\right\rangle \tbigotimes \left\vert u_{0}\right\rangle \tbigotimes S_{3},
\end{equation*}%
which is a subspace of the composite Hilbert space $S_{1}\tbigotimes
S_{2}\tbigotimes S_{3}.$ Moreover, it works effectively on the subspace $%
\left\vert 0\right\rangle \tbigotimes \left\vert u_{0}\right\rangle
\tbigotimes S_{3}$ of the subspace $S_{123}.$ Equivalently speaking, it
works effectively on the component Hilbert subspace $S_{3}.$ For convenience
hereafter denote $\{\left\vert k,u_{j},v_{l}\right\rangle \}$ as the
orthonormal tensor-product basis set $\{\left\vert k\right\rangle
\tbigotimes \left\vert u_{j}\right\rangle \tbigotimes |v_{l}\rangle \}$ of
the Hilbert space $S_{1}\tbigotimes S_{2}\tbigotimes S_{3}.$

First of all, by substituting the $U_{c}^{sic}(t_{m})$ of (3.37) into (3.44)
the $V_{c}^{sic}(t_{m})$ of (3.44) may be re-written as%
\begin{equation}
V_{c}^{sic}(t_{m})=\exp \left[ -i\Phi _{1}\left( a_{m}^{s}\right) \right]
W_{23}W_{12}U_{\infty }^{sic}\left( a_{m}^{s},\alpha _{m}\right)
W_{12}^{+}W_{23}^{+}.  \tag{3.49}
\end{equation}%
This unitary-dynamical equation is more convenient for the theoretical
proof. Later it is directly proven instead of the one of (3.44). Now in the
subspace $\left\vert 0\right\rangle \tbigotimes \left\vert
u_{0}\right\rangle \tbigotimes S_{3}$ arbitrary state $\left\vert
0\right\rangle \tbigotimes \left\vert u_{0}\right\rangle \tbigotimes |\Psi
_{v}\rangle $ may be expanded as%
\begin{equation}
\left\vert 0\right\rangle \tbigotimes \left\vert u_{0}\right\rangle
\tbigotimes |\Psi _{v}\rangle \equiv \left\vert 0,u_{0},\Psi
_{v}\right\rangle =\sum_{k=0}^{\infty }B_{k}\left\vert
0,u_{0},v_{k}\right\rangle .  \tag{3.50}
\end{equation}%
This infinite expansion series is convergent. It is directly obtained by
expanding arbitrary state $|\Psi _{v}\rangle $ of the component Hilbert
subspace $S_{3}$ in the complete energy eigenbasis set $\left\{ \left\vert
v_{k}\right\rangle \right\} $ of the Hamiltonian $H_{c}^{v}$. That is, it is
obtained from $|\Psi _{v}\rangle =\sum_{k=0}^{\infty }B_{k}\left\vert
v_{k}\right\rangle $, which is like (3.8).

Before the theoretical proof is described, these three different unitary
transformations of the $V_{c}^{sic}(t_{m})$ of (3.49) are set up in the
Hilbert space $S_{1}\tbigotimes S_{2}\tbigotimes S_{3}$ (or in its
subspace), which are the $SIC$\ unitary propagator $U_{\infty }^{sic}\left(
a_{m}^{s},\alpha _{m}\right) ,$ the $QM$\ unitary operator $W_{12}$ ($%
W_{12}^{+}$) and $W_{23}$ ($W_{23}^{+}$), respectively. Note that the $%
U_{\infty }^{sic}\left( a_{m}^{s},\alpha _{m}\right) $ of (3.12) works on
the component Hilbert space $S_{1}$ alone. Then it can be deduced from
(3.14), (3.15), and (3.16)\ that if the $U_{\infty }^{sic}\left(
a_{m}^{s},\alpha _{m}\right) $ acts on any tensor-product base $\left\vert
k,u_{i},v_{j}\right\rangle \ $with $k,i,j=0,1,...,\infty $ of the Hilbert
space $S_{1}\tbigotimes S_{2}\tbigotimes S_{3},$ then one may obtain the
eigenvalue equation for the $U_{\infty }^{sic}\left( a_{m}^{s},\alpha
_{m}\right) $ in the Hilbert space:%
\begin{equation*}
U_{\infty }^{sic}\left( a_{m}^{s},\alpha _{m}\right) \left\vert
k,u_{i},v_{j}\right\rangle =\tprod\limits_{l=1}^{\infty }\exp \left(
i2^{l-2}\alpha _{m}a_{m}^{s}\right)
\end{equation*}%
\begin{equation}
\times \exp \left( -i\alpha _{m}a_{m}^{s}k\right) \left\vert
k,u_{i},v_{j}\right\rangle  \tag{3.51}
\end{equation}%
The $QM$\ unitary operator $W_{12}$ is given by (3.35). It works on the
subspace $S_{12}$ of the component Hilbert subspace $S_{1}\tbigotimes S_{2}$
of the Hilbert space $S_{1}\tbigotimes S_{2}\tbigotimes S_{3}$ alone. If it
acts on any tensor-product base $\left\vert k,u_{0},v_{j}\right\rangle \
(\left\vert 0,u_{k},v_{j}\right\rangle )$ $(k,j=0,1,...,\infty )$ of the
subspace $S_{12}\tbigotimes S_{3}$ of the Hilbert space $S_{1}\tbigotimes
S_{2}\tbigotimes S_{3},$ then with the help of (3.36) it can be found that%
\begin{equation}
W_{12}\left\vert k,u_{0},v_{j}\right\rangle =-i\left\vert
0,u_{k},v_{j}\right\rangle ,\text{ }W_{12}^{+}\left\vert
k,u_{0},v_{j}\right\rangle =i\left\vert 0,u_{k},v_{j}\right\rangle 
\tag{3.52}
\end{equation}%
The $QM$\ unitary operator $W_{23}$ is given by (3.47). It acts on the
subspace $S_{23}$ of the component Hilbert subspace $S_{2}\tbigotimes S_{3}$
of the Hilbert space $S_{1}\tbigotimes S_{2}\tbigotimes S_{3}$ alone. Thus,
when it acts on any tensor-product base $\left\vert
k,u_{j},v_{0}\right\rangle $\ $(\left\vert k,u_{0},v_{j}\right\rangle )$ $%
(k,j=0,1,...,\infty )$ of the subspace $S_{1}\tbigotimes S_{23}$ of the
Hilbert space $S_{1}\tbigotimes S_{2}\tbigotimes S_{3},$ with the aid of
(3.48) one can set up the unitary transformation $W_{23}$ in the subspace $%
S_{1}\tbigotimes S_{23}:$%
\begin{equation}
W_{23}\left\vert k,u_{j},v_{0}\right\rangle =-i\left\vert
k,u_{0},v_{j}\right\rangle ,\text{ }W_{23}^{+}\left\vert
k,u_{j},v_{0}\right\rangle =i\left\vert k,u_{0},v_{j}\right\rangle 
\tag{3.53}
\end{equation}%
These three unitary transformations of (3.51), (3.52), and (3.53) are
further used for the theoretical proof of the unitary-dynamical equation
(3.49).

The theoretical proof is described as follows. On the one hand, by acting $%
W_{23}W_{12}U_{\infty }^{sic}\left( a_{m}^{s},\alpha _{m}\right)
W_{12}^{+}W_{23}^{+}$ of (3.49) on arbitrary state (3.50) of the subspace $%
\left\vert 0\right\rangle \tbigotimes \left\vert u_{0}\right\rangle
\tbigotimes S_{3}$ one obtains%
\begin{equation*}
W_{23}W_{12}U_{\infty }^{sic}\left( a_{m}^{s},\alpha _{m}\right)
W_{12}^{+}W_{23}^{+}\left\vert 0,u_{0},\Psi _{v}\right\rangle
\end{equation*}%
\begin{equation}
=W_{23}W_{12}U_{\infty }^{sic}\left( a_{m}^{s},\alpha _{m}\right)
\sum_{k=0}^{\infty }B_{k}\left( +i\right) \left( +i\right) \left\vert
k,u_{0},v_{0}\right\rangle  \tag{3.54}
\end{equation}%
where $W_{23}^{+}\left\vert 0,u_{0},v_{k}\right\rangle =i\left\vert
0,u_{k},v_{0}\right\rangle $ obtained from (3.53) and $W_{12}^{+}\left\vert
0,u_{k},v_{0}\right\rangle =i\left\vert k,u_{0},v_{0}\right\rangle $
obtained from (3.52) are already used. Then by using further the eigenvalue
equation (3.51) of the $U_{\infty }^{sic}\left( a_{m}^{s},\alpha _{m}\right) 
$ it can be found from (3.54) that%
\begin{equation*}
W_{23}W_{12}U_{\infty }^{sic}\left( a_{m}^{s},\alpha _{m}\right)
W_{12}^{+}W_{23}^{+}\left\vert 0,u_{0},\Psi _{v}\right\rangle
\end{equation*}%
\begin{equation}
=\tprod\limits_{l=1}^{\infty }\exp \left( i2^{l-2}\alpha
_{m}a_{m}^{s}\right) W_{23}W_{12}\sum_{k=0}^{\infty }B_{k}\left( +i\right)
\left( +i\right) \exp \left( -i\alpha _{m}a_{m}^{s}k\right) \left\vert
k,u_{0},v_{0}\right\rangle \newline
.  \tag{3.55}
\end{equation}%
With the help of $W_{12}\left\vert k,u_{0},v_{0}\right\rangle =-i\left\vert
0,u_{k},v_{0}\right\rangle $ of (3.52) and $W_{23}\left\vert
0,u_{k},v_{0}\right\rangle =-i\left\vert 0,u_{0},v_{k}\right\rangle $ of
(3.53) the equation (3.55) is finally reduced to the form%
\begin{equation*}
W_{23}W_{12}U_{\infty }^{sic}\left( a_{m}^{s},\alpha _{m}\right)
W_{12}^{+}W_{23}^{+}\left\vert 0,u_{0},\Psi _{v}\right\rangle
\end{equation*}%
\begin{equation}
=\tprod\limits_{l=1}^{\infty }\exp \left( i2^{l-2}\alpha
_{m}a_{m}^{s}\right) \sum_{k=0}^{\infty }B_{k}\exp \left( -i\alpha
_{m}a_{m}^{s}k\right) \left\vert 0,u_{0},v_{k}\right\rangle  \tag{3.56}
\end{equation}%
On the other hand, the $SIC$ unitary propagator $V_{c}^{sic}(t_{m})=\exp %
\left[ -ia_{m}^{s}H_{c}^{v}t_{m}/\hslash \right] $ effectively acts on the
subspace $\left\vert 0\right\rangle \tbigotimes \left\vert
u_{0}\right\rangle \tbigotimes S_{3}$ (or equivalently the component Hilbert
subspace $S_{3}$) of the Hilbert space $S_{1}\tbigotimes S_{2}\tbigotimes
S_{3}.$ Thus, by acting the $V_{c}^{sic}(t_{m})$ on arbitrary state (3.50)
of the subspace $\left\vert 0\right\rangle \tbigotimes \left\vert
u_{0}\right\rangle \tbigotimes S_{3}$ one obtains%
\begin{equation*}
V_{c}^{sic}(t_{m})\left\vert 0,u_{0},\Psi _{v}\right\rangle =\exp \left[
-ia_{m}^{s}\left( bt_{m}/\hslash \right) \right]
\end{equation*}%
\begin{equation}
\times \sum_{k=0}^{\infty }B_{k}\exp \left[ -i\left( at_{m}/\hslash \right)
a_{m}^{s}k\right] \left\vert 0,u_{0},v_{k}\right\rangle  \tag{3.57}
\end{equation}%
where the eigenvalue equation $V_{c}^{sic}(t_{m})\left\vert
0,u_{0},v_{k}\right\rangle =\exp \left[ -ia_{m}^{s}E_{k}^{v}t_{m}/\hslash %
\right] \left\vert 0,u_{0},v_{k}\right\rangle $ with $E_{k}^{v}=ak+b$ is
already used. Let $\alpha _{m}=\left( at_{m}/\hslash \right) $ in (3.57).
Then by substituting (3.57) into (3.56) one finds that%
\begin{equation*}
W_{23}W_{12}U_{\infty }^{sic}\left( a_{m}^{s},\alpha _{m}\right)
W_{12}^{+}W_{23}^{+}\left\vert 0,u_{0},\Psi _{v}\right\rangle
\end{equation*}%
\begin{equation}
=\tprod\limits_{l=1}^{\infty }\exp \left( i2^{l-2}\alpha
_{m}a_{m}^{s}\right) \exp \left[ ia_{m}^{s}\left( bt_{m}/\hslash \right) %
\right] V_{c}^{sic}(t_{m})\left\vert 0,u_{0},\Psi _{v}\right\rangle 
\tag{3.58}
\end{equation}%
Note that the state $\left\vert 0,u_{0},\Psi _{v}\right\rangle $ is
arbitrary in the subspace $\left\vert 0\right\rangle \tbigotimes \left\vert
u_{0}\right\rangle \tbigotimes S_{3}$. Then the equation (3.58) leads
directly to the equation (3.49).

It is theoretically described in detail above how the joint construction
method to prepare a desired $SIC$ unitary propagator, which is based on the
equation (3.6) and the energy eigenfunction expansion, is performed in the
framework of the multiple-quantum operator algebra space. In this quite
general theoretical work of the joint construction method the
solution-information transfer is realized unitarily from one component
Hilbert subspace to another so that the desired $SIC$ unitary propagator is
prepared. Now this general theoretical work may be applied to the concrete
case that the solution information is transferred from the component Hilbert
subspace of the $d-$qubit spin system with $2^{d}=\infty $ to the one of the
COM motion of a single-atom system so that the desired $SIC$ unitary
propagator $U_{at}^{sic}(t_{m})$ in (3.10) can be prepared. As pointed out
above, this solution-information transfer consists of the two steps, that
is, the first step is to transfer the solution information from the
component Hilbert subspace of the spin system to the one of the atomic
internal motion and the second step from the component Hilbert subspace of
the atomic internal motion to the one of the atomic COM motion in the
single-atom system.

Initially the solution information $\left( a_{m}^{s}\right) $ is loaded on
the $SIC$ unitary diagonal propagator $U_{\infty }^{sic}\left(
a_{m}^{s},\alpha _{m}\right) $ which is generated by the sequence (3.12) of
the basic $SIC$ unitary operators and works on the component Hilbert
subspace of the $d-$qubit spin system with $2^{d}=\infty .$ This component
Hilbert subspace acts as the component Hilbert subspace $S_{1}$ above. Then
the first step is to make a $QM$ unitary transformation on the $U_{\infty
}^{sic}\left( a_{m}^{s},\alpha _{m}\right) $ so that the solution
information carried by the $U_{\infty }^{sic}\left( a_{m}^{s},\alpha
_{m}\right) $ is transferred to an appropriate Hilbert subspace of the
internal motion of the single-atom system which acts as the component
Hilbert subspace $S_{2}$ above. Suppose here that $\left\{ \left\vert \psi
_{k}\right\rangle \right\} $ is a complete set (or subset) of the energy
eigenbases of the atomic internal motion and its eigenbase number is equal
to that one (i.e., $2^{d}$) of the Hilbert space of the $d-$qubit spin
system with $2^{d}=\infty $. Then $\left\{ \left\vert \psi _{k}\right\rangle
\right\} $ may form a Hilbert subspace of the atomic internal motion and can
act as the component Hilbert subspace $S_{2}$. For example, $\left\{
\left\vert \psi _{k}\right\rangle \right\} $ could be taken as the complete
set of the radial wavefunctions $\left\{ R_{n0}\left( r\right) \right\} $
(the angular-momentum quantum number $l=0$) for a hydrogen-like atom [$10$].
Now consider the simple case that the Hamiltonian $H_{c}$ of the $SIC$
unitary propagator $U_{c}^{sic}\left( t_{m}\right) =\exp \left[
-ia_{m}^{s}H_{c}t_{m}/\hslash \right] $ in the above general theoretical
work has the energy eigenspectrum $\{E_{k}^{c}\}=\{ak+b\}$ and its
associated eigenbasis set $\left\{ \left\vert \psi _{k}\right\rangle
\right\} $. That is, the Hamiltonian $H_{c}$ owns the eigenvalue equation: $%
H_{c}\left\vert \psi _{k}\right\rangle =E_{k}^{c}\left\vert \psi
_{k}\right\rangle .$ Obviously, here $\left\{ \left\vert \psi
_{k}\right\rangle \right\} $ is just taken as the eigenbasis set $\left\{
\left\vert u_{k}\right\rangle \right\} $ in the above general theoretical
work. This Hamiltonian $H_{c}$ could be different from the natural
Hamiltonian of the internal motion of the single-atom system. But
theoretically it may be considered as an effective Hamiltonian of the atomic
internal motion. According to the general theoretical work above there is a $%
QM$\ unitary transformation $W_{12}$ such that the $SIC$ unitary propagator $%
U_{c}^{sic}(t_{m}),$ which works on the component Hilbert subspace $\left\{
\left\vert \psi _{k}\right\rangle \right\} $, can be generated by making the
unitary transformation $W_{12}$ on the $U_{\infty }^{sic}\left(
a_{m}^{s},\alpha _{m}\right) .$ Here the $U_{c}^{sic}(t_{m})$ is determined
from (3.37) and $W_{12}$ from (3.35) in which the eigenbasis set $\left\{
\left\vert u_{k}\right\rangle \right\} $ is replaced with $\left\{
\left\vert \psi _{k}\right\rangle \right\} .$ Now the solution information
is carried by the $U_{c}^{sic}(t_{m}).$ This really means that it is already
transferred to the component Hilbert subspace $\left\{ \left\vert \psi
_{k}\right\rangle \right\} $ of the atomic internal motion.

Now the second step is to transfer further the solution information that is
carried by the $U_{c}^{sic}(t_{m})$ of (3.37) from the component Hilbert
subspace $\left\{ \left\vert \psi _{k}\right\rangle \right\} $ of the atomic
internal motion to the one of the atomic COM motion which acts as the
component Hilbert subspace $S_{3}$ above. Here consider the simple case that
the Hamiltonian $H_{c}^{v}$ of the $SIC$ unitary propagator $%
V_{c}^{sic}(t_{m})=\exp \left[ -ia_{m}^{s}H_{c}^{v}t_{m}/\hslash \right] ,$
which acts on the component Hilbert subspace ($S_{3}$) of the atomic COM
motion, has the energy eigenspectrum $\{E_{k}^{v}\}=\{ak+b\}$ and its
associated eigenbasis set $\left\{ \left\vert v_{k}\right\rangle \right\} .$
A typical example is that the Hamiltonian $H_{c}^{v}$ is taken as the
COM-motion Hamiltonian of (3.7a) of the single-atom system with external
harmonic potential field. Then one has $E_{k}^{v}=\left( k+1/2\right)
\hslash \omega $ or $a=\hslash \omega $ and $b=\frac{1}{2}\hslash \omega ,$
and the energy eigenbasis set $\left\{ \left\vert v_{k}\right\rangle
\right\} $ is given by the $\left\{ u_{k}\left( x\right) \right\} $ of
(3.7b). One sees that the Hamiltonian $H_{c}^{v}$ has the same energy
eigenspectrum as the effective Hamiltonian $H_{c}$ of the atomic internal
motion. Hence it is known from the general theoretical work above that there
is a $QM$ unitary transformation $W_{23}$ such that the $SIC$ unitary
propagator $V_{c}^{sic}(t_{m})$ may be generated by making the unitary
transformation $W_{23}$ on the $U_{c}^{sic}(t_{m}).$ Here the $SIC$\ unitary
propagator $V_{c}^{sic}(t_{m})$ is determined from (3.44) and $W_{23}$ from
(3.47) in which $\left\{ \left\vert u_{k}\right\rangle \right\} $ is
replaced with $\left\{ \left\vert \psi _{k}\right\rangle \right\} .$ If now $%
\left\{ \left\vert v_{k}\right\rangle \right\} $ is set to the $\left\{
u_{k}\left( x\right) \right\} $ of (3.7b), then the parameters $a$ and $b$
are given by $a=\hslash \omega $ and $b=\frac{1}{2}\hslash \omega ,$ and in
this case the Hamiltonian $H_{c}^{v}$ is just the Hamiltonian of the COM
motion (with the oscillatory angular frequency $\omega $) of the single-atom
system with external harmonic potential field. One therefore obtains finally
the $SIC$\ unitary propagator $V_{c}^{sic}(t_{m})$ that corresponds to the $%
QM$ unitary propagator of the COM\ motion of the single-atom system with
external harmonic potential field. Now the solution information is carried
by the $V_{c}^{sic}(t_{m}),$ This indicates that it is already transferred
to the component Hilbert subspace of the COM motion of the single-atom
system.

In the single-atom system the $SIC$ unitary propagator $U_{at}^{sic}(t_{m})$
in (3.10) is different from the currently prepared $SIC$ unitary propagator $%
V_{c}^{sic}(t_{m})$ only in the $SIC$ unitary propagator $\exp \left[
-ia_{m}^{s}H_{a}^{\prime }t_{m}/\hslash \right] $ with $H_{a}^{\prime
}=E_{0}^{a}\left\vert \psi _{0}\right\rangle \left\langle \psi
_{0}\right\vert .$ Here the $V_{c}^{sic}(t_{m})$ is prepared by (3.44). If
both the $SIC$ unitary propagators $U_{at}^{sic}(t_{m})$ and $%
V_{c}^{sic}(t_{m})$ work effectively on the same Hilbert subspace $\left\{
\left\vert \psi _{0}\right\rangle \tbigotimes \left\vert v_{k}\right\rangle
\right\} $ of the single-atom system, then due to that $\exp \left[
-ia_{m}^{s}H_{a}^{\prime }t_{m}/\hslash \right] $ generates only a global
phase factor $\exp \left( -ia_{m}^{s}E_{0}^{a}t_{m}/\hslash \right) ,$ as
shown in (3.10), there is no essential difference between the $%
U_{at}^{sic}(t_{m})$ in (3.10) and the currently prepared $%
V_{c}^{sic}(t_{m}).$ Note that the $U_{at}^{sic}(t_{m})$ works on the
Hilbert subspace $\left\{ \left\vert \psi _{0}\right\rangle \tbigotimes
\left\vert v_{k}\right\rangle \right\} ,$ while the $V_{c}^{sic}(t_{m})$
works effectively on the Hilbert subspace $\left\{ \left\vert 0\right\rangle
\tbigotimes \left\vert \psi _{0}\right\rangle \tbigotimes \left\vert
v_{k}\right\rangle \right\} ,$ indicating that both the $U_{at}^{sic}(t_{m})$
and $V_{c}^{sic}(t_{m})$ work effectively on the same component Hilbert
subspace $\left\{ \left\vert \psi _{0}\right\rangle \tbigotimes \left\vert
v_{k}\right\rangle \right\} .$ One therefore concludes that in theory there
always exists the $SIC$ unitary propagator $U_{at}^{sic}(t_{m})$ in (3.10).

The conclusion is important in theory that there always exists the $SIC$
unitary propagator $U_{at}^{sic}(t_{m})$ in (3.10). However, there is no
practical value for quantum-computing speedup if one proves merely that
conclusion theoretically, at least because the $SIC$ unitary propagator $%
U_{at}^{sic}(t_{m})$ (i.e., the currently prepared $V_{c}^{sic}(t_{m})$)
that is constructed theoretically in the $d-$qubit spin system with $%
2^{d}=\infty $ consists of infinitely many basic $SIC$ unitary operators.

The general theoretical work set up in this subsection is the theoretical
basis for further constructing a practical $SIC$\ unitary propagator in next
subsection.\newline
\newline
\newline
{\Large 3.2. The\ practical }${\Large SIC}${\Large \ unitary propagators}

From the point of view of quantum-computing speedup only two special cases
deserve consideration, one of which is that the spin number $d$ of the $d-$%
qubit spin system takes a finite number or even $d\varpropto \log n,$ where $%
2^{n}$ is the dimensional size of the unstructured search space; and another
is $d=1$. The first case can lead to a good approximation to the $SIC$
unitary propagator $U_{at}^{sic}(t_{m})$ in (3.10) that is constructed
theoretically above, and the well-approximated $U_{at}^{sic}(t_{m})$ could
be realized in practice. It is discussed in detail below. The second case
(i.e., $d=1$) could be quite different from the first case. It will not be
discussed here.

Now the first case is discussed that the spin number $d>1$ is finite. As
shown in the general theoretical work in the previous subsection, the $SIC$
unitary propagator $U_{c}^{sic}(t_{m})$ of (3.37) is constructed
theoretically by using the $d-$qubit spin system with $d=\log
_{2}L\rightarrow \infty .$ Such a $SIC$ unitary propagator can not be useful
for quantum-computing speedup. The question is whether there exists a $SIC$
unitary propagator that is sufficiently approximate to the theoretical one,
i.e., the $U_{c}^{sic}(t_{m})$ of (3.37) and may be realized in a $d_{s}-$%
qubit spin system with finite spin number $d_{s}.$

As far as the solution information ($a_{m}^{s}$) is concerned, the
theoretical work in this subsection is carried out in the (total) physical
Hilbert space under study for the case that the solution information is of
the real solution to the unstructured search problem. It is available as
well for the case that the solution information is of any candidate solution
different from the real solution as long as the physical Hilbert space of
the $d_{s}-$qubit spin system with finite spin number $d_{s}$ is replaced
with its corresponding math Hilbert space and at the same time the physical
Hilbert space of any other quantum system such as the single-atom system
also is replaced with its corresponding math Hilbert space.

For simplicity, the following theoretical work is limited to a $SIC$ unitary
propagator that can be made sufficiently close to the theoretical one, $%
U_{c}^{sic}(t_{m})=\exp \left[ -ia_{m}^{s}H_{c}t_{m}/\hslash \right] ,$ in
which the Hamiltonian $H_{c}$ has the energy eigenspectrum $%
\{E_{k}^{c}\}=\{ak+b\}.$ The $SIC$ unitary propagator $U_{c}^{sic}(t_{m})$
is generated by (3.37) theoretically. Therefore, the unitary-dynamical
equation of (3.37) is theoretical basis to construct a $SIC$ unitary
propagator $U_{ap}^{sic}(t_{m})$ that can be made sufficiently close to the
theoretical one, i.e., the $U_{c}^{sic}(t_{m})$ of (3.37), and can be
realized in a $d_{s}-$qubit spin system with finite spin number $d_{s}$. It
is known above that on the right-hand side of (3.37) the $SIC$ unitary
propagator $U_{\infty }^{sic}\left( a_{m}^{s},\alpha _{m}\right) $ is given
by (3.12), and one sees from (3.12) that the $U_{\infty }^{sic}\left(
a_{m}^{s},\alpha _{m}\right) $ contains infinitely many basic $SIC$ unitary
operators. If now the $U_{\infty }^{sic}\left( a_{m}^{s},\alpha _{m}\right) $
is modified to be a limited sequence that contains only a finite number
(e.g., $d_{s}$) of the basic $SIC$ unitary operators, then one may obtain a
new $SIC$ unitary diagonal propagator:%
\begin{equation*}
U_{ap}^{sic}\left( a_{m}^{s},\alpha _{m}\right) =\exp (-ia_{m}^{s}\theta
_{md_{s}}I_{d_{s}z})\tbigotimes ...\tbigotimes \exp (-ia_{m}^{s}\theta
_{m2}I_{2z})
\end{equation*}%
\begin{equation}
\tbigotimes \exp (-ia_{m}^{s}\theta _{m1}I_{1z})  \tag{3.59}
\end{equation}%
where $\theta _{ml}=-\alpha _{m}\times 2^{l-1}$ ($l=1,2,...,d_{s}$) and $%
\alpha _{m}=at_{m}/\hslash .$ This $SIC$\ unitary propagator can be prepared
with the same way that is used to prepare the $U_{\infty }^{sic}\left(
a_{m}^{s},\alpha _{m}\right) $ of (3.12). Obviously the $U_{ap}^{sic}\left(
a_{m}^{s},\alpha _{m}\right) $ works exactly on the Hilbert space of the $%
d_{s}-$qubit spin system. It is also known from (3.35) that the $QM$\
unitary operator $W_{12}$ in (3.37) contains infinitely many unitary
pseudospin rotation operators $\{\exp \left( -i\pi Q_{x}^{k,u_{k}}\right) \}$%
. Notice that in the Hilbert space of the $d_{s}-$qubit spin system the
complete eigenbasis set $\{\left\vert k\right\rangle \}$ contains a finite
number (i.e., $2^{d_{s}}$) of energy eigenbases. When the $U_{\infty
}^{sic}\left( a_{m}^{s},\alpha _{m}\right) $ in (3.37) is replaced with the $%
U_{ap}^{sic}\left( a_{m}^{s},\alpha _{m}\right) $ of (3.59), correspondingly
in (3.37) the $QM$\ unitary operator $W_{12}$ which is given by (3.35)
should be replaced with the following $QM$\ unitary operator:%
\begin{equation}
W_{12}^{ap}=\exp \left( -i\frac{\pi }{2}\left\vert 0,u_{0}\right\rangle
\left\langle 0,u_{0}\right\vert \right)
\tprod\limits_{k=1}^{2^{d_{s}}-1}\exp \left( -i\pi Q_{x}^{k,u_{k}}\right) 
\tag{3.60}
\end{equation}%
This $QM$ unitary operator is a limited product of the $2^{d_{s}}-1$
pseudospin rotation operators $\{\exp \left( -i\pi Q_{x}^{k,u_{k}}\right) \}$
and one unitary diagonal operator generated by $\left\vert
0,u_{0}\right\rangle \left\langle 0,u_{0}\right\vert $. It is similar to a $%
QM$\ subspace-selective unitary operator [$7$]. Therefore, with the help of
the $U_{ap}^{sic}\left( a_{m}^{s},\alpha _{m}\right) $ of (3.59) and $%
W_{12}^{ap}$ of (3.60), correspondingly the $SIC$ unitary propagator $%
U_{c}^{sic}(t_{m})$ of (3.37) should be replaced with the following $SIC$
unitary propagator:%
\begin{equation}
U_{ap}^{sic}(t_{m})=\exp \left[ -i\Phi _{ap}\left( a_{m}^{s}\right) \right]
W_{12}^{ap}U_{ap}^{sic}\left( a_{m}^{s},\alpha _{m}\right) \left(
W_{12}^{ap}\right) ^{+}  \tag{3.61}
\end{equation}%
where $\exp \left[ -i\Phi _{ap}\left( a_{m}^{s}\right) \right] $ is a global
phase factor. This $SIC$ unitary propagator can be realized by using the $%
d_{s}-$qubit spin system with finite spin number $d_{s}.$ And this means
that its realization does not need to use the $d-$qubit spin system with $%
2^{d}=\infty $. It will prove below that it can be made sufficiently close
to the $U_{c}^{sic}(t_{m})$ of (3.37).

It is known in the general theoretical work in the subsection 3.1 that the $%
SIC$\ unitary propagator $U_{c}^{sic}(t_{m})$ of (3.37) works on the
component Hilbert subspace $S_{2}$ of the composite Hilbert space $%
S_{1}\tbigotimes S_{2}$. Here it can prove that the $SIC$ unitary propagator 
$U_{ap}^{sic}(t_{m})$ of (3.61) also works on the same component Hilbert
subspace $S_{2}.$

With the help of the expression (3.59) of the $U_{ap}^{sic}\left(
a_{m}^{s},\alpha _{m}\right) $ one may set up the eigenvalue equation for
the $U_{ap}^{ic}\left( a_{m}^{s},\alpha _{m}\right) $ acting on any
tensor-product base $\left\vert k\right\rangle $ of the Hilbert space of the 
$d_{s}-$qubit spin system. It is given explicitly by%
\begin{equation}
U_{ap}^{sic}\left( a_{m}^{s},\alpha _{m}\right) \left\vert k\right\rangle
=\exp \left[ i\varphi _{0}\left( a_{m}^{s}\right) \right] \exp \left(
-i\alpha _{m}a_{m}^{s}k\right) \left\vert k\right\rangle \text{ for }0\leq
k<2^{d_{s}}  \tag{3.62a}
\end{equation}%
where the global phase factor $\exp \left[ i\varphi _{0}\left(
a_{m}^{s}\right) \right] =\tprod\nolimits_{l=1}^{d_{s}}\exp \left(
i2^{l-2}\alpha _{m}a_{m}^{s}\right) .$ This eigenvalue equation is obtained
in an analogous way that is used to set up the eigenvalue equation for the $%
U_{\infty }^{sic}\left( a_{m}^{s},\alpha _{m}\right) $ of (3.12) which is
given by (3.14) and (3.15). For the theoretically calculational convenience,
it could be thought theoretically that the Hilbert space of the $d_{s}-$%
qubit spin system is a component Hilbert subspace of the Hilbert space of
the $d-$qubit spin system with $2^{d}=\infty .$ Consequently, though the $%
U_{ap}^{sic}\left( a_{m}^{s},\alpha _{m}\right) $ works exactly on the
Hilbert space of the $d_{s}-$qubit spin system, in theory it also\ works on
the whole Hilbert space of the $d-$qubit spin system with $2^{d}=\infty $
which acts as the component Hilbert subspace $S_{1}$ in the general
theoretical work in the subsection 3.1. And moreover, in theory one should
have $U_{ap}^{sic}\left( a_{m}^{s},\alpha _{m}\right) \left\vert
k\right\rangle =\exp \left[ i\phi _{k}\left( a_{m}^{s}\right) \right]
\left\vert k\right\rangle $ if $k\geq 2^{d_{s}},\ $where the phase factor $%
\exp \left[ i\phi _{k}\left( a_{m}^{s}\right) \right] $ is explicitly
calculated by replacing the $U_{ap}^{sic}\left( a_{m}^{s},\alpha _{m}\right) 
$ with $E\tbigotimes U_{ap}^{sic}\left( a_{m}^{s},\alpha _{m}\right) \ $with
the unity operator $E,$ and here $E=E_{d}\tbigotimes ...\tbigotimes
E_{d_{s}+2}\tbigotimes E_{d_{s}+1}$ and $E_{l}$ is the $2\times 2$ unity
operator (or matrix) for $d\geq l\geq d_{s}+1$ and $d=\log
_{2}2^{d}\rightarrow \infty $. The complete expression for the eigenvalue
equation for the $U_{ap}^{sic}\left( a_{m}^{s},\alpha _{m}\right) $ acting
on any eigenbase $\left\vert k\right\rangle $ of the component Hilbert space 
$S_{1}$ then may be written as%
\begin{equation}
U_{ap}^{sic}\left( a_{m}^{s},\alpha _{m}\right) \left\vert k\right\rangle
=\left\{ 
\begin{array}{c}
\exp \left[ i\varphi _{0}\left( a_{m}^{s}\right) \right] \exp \left(
-i\alpha _{m}a_{m}^{s}k\right) \left\vert k\right\rangle \\ 
\exp \left[ i\phi _{k}\left( a_{m}^{s}\right) \right] \left\vert
k\right\rangle%
\end{array}%
\begin{array}{c}
if \\ 
if%
\end{array}%
\begin{array}{c}
0\leq k<2^{d_{s}} \\ 
k\geq 2^{d_{s}}%
\end{array}%
\right.  \tag{3.62b}
\end{equation}%
Obviously, this eigenvalue equation contains the one of (3.62a). As an
application of (3.62b), when the $U_{ap}^{sic}\left( a_{m}^{s},\alpha
_{m}\right) $ acts on the composite Hilbert space $S_{1}\tbigotimes S_{2}$,
it acts on the component Hilbert subspace $S_{1}$ alone. Then in this
application the unitary transformation (3.62b) still holds by replacing any
eigenbase $\left\vert k\right\rangle $ in (3.62b) with the tensor-product
base $\left\vert k,u_{l}\right\rangle $ ($l\geq 0$) of the composite Hilbert
space.

As shown in the subsection 3.1, the $QM$ unitary operator $W_{12}$ of (3.35)
works on the subspace $S_{12}$ of the composite Hilbert space $%
S_{1}\tbigotimes S_{2}.$ Here it can be shown that the $QM$ unitary operator 
$W_{12}^{ap}$ of (3.60) works on the subspace $S_{d_{s}}\tbigotimes
\left\vert u_{0}\right\rangle \cup \left\vert 0\right\rangle \tbigotimes
S_{2},$ where $S_{d_{s}}$ is the Hilbert space of the $d_{s}-$qubit spin
system, and theoretically it also works on the subspace $S_{12},$ since in
theory $S_{d_{s}}$ is considered as a component Hilbert subspace of the
Hilbert space $S_{1}$. With the help of the expression (3.60) of the $%
W_{12}^{ap}$ one may set up the unitary transformation $W_{12}^{ap}$ in the
subspace $S_{12}$. Now by applying the $W_{12}^{ap}$ of (3.60) to any
tensor-product base $\left\vert k,u_{0}\right\rangle $ ($\left\vert
0,u_{k}\right\rangle $) of the subspace $S_{12}$ and then with the help of
the commutation relations of (3.33), the $QM$ selective unitary
transformations of (3.34a), and the identical state transformations of
(3.34b), it can prove that%
\begin{equation}
W_{12}^{ap}:\left\vert k,u_{0}\right\rangle \rightarrow -i\left\vert
0,u_{k}\right\rangle ,\text{ if }k=0,1,2,...,2^{d_{s}}-1,  \tag{3.63a}
\end{equation}%
\begin{equation}
\left( W_{12}^{ap}\right) ^{+}:\left\vert k,u_{0}\right\rangle \rightarrow
i\left\vert 0,u_{k}\right\rangle ,\text{ if }0\leq k<2^{d_{s}}  \tag{3.63b}
\end{equation}%
and moreover, 
\begin{equation}
W_{12}^{ap}\left\vert k,u_{0}\right\rangle =\left\vert k,u_{0}\right\rangle 
\text{ and }W_{12}^{ap}\left\vert 0,u_{k}\right\rangle =\left\vert
0,u_{k}\right\rangle ,\text{ if }k\geq 2^{d_{s}}.  \tag{3.63c}
\end{equation}%
If $k\geq 2^{d_{s}},$ then for a tensor-product base $\left\vert
k,u_{l}\right\rangle =\left\vert k\right\rangle \tbigotimes \left\vert
u_{l}\right\rangle $ with any $l\geq 0$ the base $\left\vert k\right\rangle $
does not belong to the Hilbert space of the $d_{s}-$qubit spin system.
However, it is thought in theory that the base $\left\vert k\right\rangle $
with $k\geq 2^{d_{s}}$ still belongs to the Hilbert space of the $d-$qubit
spin system with $2^{d}=\infty .$ Therefore, the unitary transformation $%
W_{12}^{ap}\left\vert k,u_{0}\right\rangle =\left\vert k,u_{0}\right\rangle $
of (3.63c) with $k\geq 2^{d_{s}}$ makes sense only in theory. Then the
unitary transformations (3.63a) and (3.63b) show that the unitary operator $%
W_{12}^{ap}$ really acts on the component Hilbert subspace which is the
Hilbert space of the $d_{s}-$qubit spin system. In (3.63c) the unitary
transformation $W_{12}^{ap}\left\vert k,u_{0}\right\rangle =\left\vert
k,u_{0}\right\rangle $ with $k\geq 2^{d_{s}}$ is equivalent to that one
replaces the $W_{12}^{ap}$ with the diagonal operator $\sum_{k=2^{d_{s}}}^{%
\infty }\left\vert k,u_{0}\right\rangle \left\langle k,u_{0}\right\vert $ to
calculate this unitary transformation in the theoretical case $k\geq
2^{d_{s}}.$ This is similar to the theoretical case $k\geq 2^{d_{s}}$ met in
the unitary transformation (3.62b) of the $U_{ap}^{sic}\left(
a_{m}^{s},\alpha _{m}\right) $ above. Such a theoretical case also may be
met later in this subsection and hereafter it is no longer explained
explicitly.

If the $SIC$ unitary propagator $U_{ap}^{sic}(t_{m})$ of (3.61) acts
effectively on the component Hilbert subspace $S_{2}$ of the composite
Hilbert space $S_{1}\tbigotimes S_{2},$ then for arbitrary state $|\Psi
_{c}\rangle $ of the subspace $S_{2}$ the state that is generated by acting
the $U_{ap}^{sic}(t_{m})$ on the state $|\Psi _{c}\rangle $ is still a state
of the same subspace $S_{2}.$ Conversely, if both the state $|\Psi
_{c}\rangle $ and the state generated by acting the $U_{ap}^{ic}(t_{m})$ on
the state $|\Psi _{c}\rangle $ are of the same subspace $S_{2},$ then the $%
U_{ap}^{sic}(t_{m})$ is a $SIC$\ unitary propagator that acts effectively on
the subspace $S_{2}.$ This can be proven below on the basis of the energy
eigenfunction expansion principle.

Here still denote $\left\vert 0\right\rangle \tbigotimes |\Psi _{c}\rangle $
as arbitrary state of the subspace $\left\vert 0\right\rangle \tbigotimes
S_{2}$ of the subspace $S_{12}$ of the Hilbert space $S_{1}\tbigotimes
S_{2}. $ Then it can be expanded as the infinite expansion series of (3.38).
Now by applying the $U_{ap}^{sic}(t_{m})$ of (3.61) to the state $\left\vert
0\right\rangle \tbigotimes |\Psi _{c}\rangle $ of (3.38) and then using the
unitary transformations (3.62) and (3.63) one obtains%
\begin{equation*}
U_{ap}^{sic}(t_{m})\left\vert 0\right\rangle \tbigotimes |\Psi _{c}\rangle
=\exp \left[ -i\Phi _{ap}\left( a_{m}^{s}\right) \right] \exp \left[
i\varphi _{0}\left( a_{m}^{s}\right) \right]
\end{equation*}%
\begin{equation}
\times \{\sum_{k=0}^{2^{d_{s}}-1}B_{k}\exp \left( -i\alpha
_{m}a_{m}^{s}k\right) \left\vert 0,u_{k}\right\rangle
+\sum_{k=2^{d_{s}}}^{\infty }B_{k}\left\vert 0,u_{k}\right\rangle \}. 
\tag{3.64}
\end{equation}%
Notice that the infinite expansion series of (3.38) is convergent. By
comparing (3.64) with (3.38) one sees that the infinite series of (3.64) is
evidently convergent and moreover, both the states $\left\vert
0\right\rangle \tbigotimes |\Psi _{c}\rangle $ and $U_{ap}^{sic}(t_{m})\left%
\vert 0\right\rangle \tbigotimes |\Psi _{c}\rangle $ belong to the same
subspace $\left\vert 0\right\rangle \tbigotimes S_{2}.$ Note that the state $%
\left\vert 0\right\rangle \tbigotimes |\Psi _{c}\rangle $ is arbitrary in
the subspace $\left\vert 0\right\rangle \tbigotimes S_{2}.$ Then the $%
U_{ap}^{sic}(t_{m})$ is either the one that acts on the subspace $\left\vert
0\right\rangle \tbigotimes S_{2}$ or the one that effectively acts on the
subspace $\left\vert 0\right\rangle \tbigotimes S_{2}$. As shown in (3.61),
the unitary operator $W_{12}^{ap}$ and its inverse $\left(
W_{12}^{ap}\right) ^{+},$ which are the product factors of the $%
U_{ap}^{sic}(t_{m})$ of (3.61), act on the subspace $S_{12}$ which is beyond
the subspace $\left\vert 0\right\rangle \tbigotimes S_{2}.$ Then the $%
U_{ap}^{sic}(t_{m})$ can be only the one that effectively acts on the
subspace $\left\vert 0\right\rangle \tbigotimes S_{2}$.

It will prove below that the $U_{ap}^{sic}(t_{m})$ of (3.61) can be made
sufficiently close to the theoretical one, i.e., the $U_{c}^{sic}(t_{m})$ of
(3.37). Because both the $U_{ap}^{sic}(t_{m})$ and $U_{c}^{sic}(t_{m})$ work
effectively on the same subspace $\left\vert 0\right\rangle \tbigotimes
S_{2},$ that the $U_{ap}^{sic}(t_{m})$ is sufficiently close to the $%
U_{c}^{sic}(t_{m})$ here means that when they each act on arbitrary state of
the subspace, both the generated states are sufficiently close to one
another. It needs to prove that every unitary-transformational step of the $%
U_{ap}^{sic}(t_{m})$ can be made sufficiently close to the counterpart of
the $U_{c}^{sic}(t_{m}).$ Both the $U_{ap}^{sic}(t_{m})$ and $%
U_{c}^{sic}(t_{m})$ work on the subspace $S_{12}$ and they work effectively
on the same subspace $\left\vert 0\right\rangle \tbigotimes S_{2}$ of the
subspace $S_{12}.$ Therefore, it must be on the subspace $S_{12}$ to
investigate whether every unitary-transformational step of the $%
U_{ap}^{sic}(t_{m})$ can be made sufficiently close to the counterpart of
the $U_{c}^{sic}(t_{m}).$ It can be deduced from (3.37) and (3.61) that the $%
U_{c}^{sic}(t_{m})$ and $U_{ap}^{sic}(t_{m})$ each can generate the three
unitary-transformational steps, when they each act on arbitrary state $%
\left\vert 0\right\rangle \tbigotimes |\Psi _{c}\rangle $ of the subspace $%
\left\vert 0\right\rangle \tbigotimes S_{2}$. For the $U_{ap}^{sic}(t_{m})$
of (3.61) these three steps are successively $\left( W_{12}^{ap}\right)
^{+}\left\vert 0\right\rangle \tbigotimes |\Psi _{c}\rangle ,$ $%
U_{ap}^{sic}\left( a_{m}^{s},\alpha _{m}\right) \left( W_{12}^{ap}\right)
^{+}\left\vert 0\right\rangle \tbigotimes |\Psi _{c}\rangle ,$ and $%
U_{ap}^{sic}(t_{m})\left\vert 0\right\rangle \tbigotimes |\Psi _{c}\rangle .$
The final states of these three steps are denoted as $\Psi
_{1}^{ap}=e^{i\varphi _{1}}\left( W_{12}^{ap}\right) ^{+}\left\vert
0\right\rangle \tbigotimes |\Psi _{c}\rangle ,$ $\Psi _{2}^{ap}=e^{i\varphi
_{2}}U_{ap}^{sic}\left( a_{m}^{s},\alpha _{m}\right) \left(
W_{12}^{ap}\right) ^{+}\left\vert 0\right\rangle \tbigotimes |\Psi
_{c}\rangle ,$ and $\Psi _{3}^{ap}=e^{i\varphi
_{3}}W_{12}^{ap}U_{ap}^{sic}\left( a_{m}^{s},\alpha _{m}\right) \left(
W_{12}^{ap}\right) ^{+}\left\vert 0\right\rangle \tbigotimes |\Psi
_{c}\rangle $ $($or $\bar{\Psi}_{3}^{ap}=e^{i\varphi _{3}^{\prime
}}U_{ap}^{sic}(t_{m})\left\vert 0\right\rangle \tbigotimes |\Psi _{c}\rangle
),$ respectively. Here the additional global phase factors $\{e^{i\varphi
_{l}}\}$ do not affect these three final states, but they can adjust
arbitrarily the global phase factors of these three final states,
respectively. For the $U_{c}^{sic}(t_{m})$ of (3.37) these three steps are
successively $\Psi _{1}=W_{12}^{+}\left\vert 0\right\rangle \tbigotimes
|\Psi _{c}\rangle ,$ $\Psi _{2}=U_{\infty }^{sic}\left( a_{m}^{s},\alpha
_{m}\right) W_{12}^{+}\left\vert 0\right\rangle \tbigotimes |\Psi
_{c}\rangle ,$ and $\Psi _{3}=W_{12}U_{\infty }^{sic}\left( a_{m}^{s},\alpha
_{m}\right) W_{12}^{+}\left\vert 0\right\rangle \tbigotimes |\Psi
_{c}\rangle $ $($or $\bar{\Psi}_{3}=U_{c}^{sic}(t_{m})\left\vert
0\right\rangle \tbigotimes |\Psi _{c}\rangle )$. Then correspondingly there
must be the three norms to measure completely whether the $%
U_{ap}^{sic}(t_{m})$ is sufficiently close to the $U_{c}^{sic}(t_{m})$ in
all these three steps. Each of these three norms may be defined by (See, for
example, Chapt. One in Ref. [$26$])%
\begin{equation}
Normk=\left\vert \left\vert \Psi _{k}^{ap}-\Psi _{k}\right\vert \right\vert ,%
\text{ }k=1,2,3.  \tag{3.65}
\end{equation}%
The additional global phase factors $\{e^{i\varphi _{k}}\}$ of these final
states $\{\Psi _{k}^{ap}\}$ are merely used to eliminate the effect of the
relevant global phase factors on these three norms $Norm1,$ $Norm2,$ and $%
Norm3$ in calculation, respectively. Obviously, up to a global phase factor
the final state $\Psi _{1}^{ap}$ is sufficiently close to the final state $%
\Psi _{1}$ at the first unitary-transformational step, if the norm $Norm1$
is sufficiently close to zero. Likewise the final states $\Psi _{2}^{ap}$
and $\Psi _{3}^{ap}$ are sufficiently close to the final states $\Psi _{2}$
and $\Psi _{3}$ up to the global phase factors at the second and the third
step, respectively, if the norms $Norm2$ and $Norm3$ each are sufficiently
close to zero.

These three norms of (3.65) are explicitly calculated below. On the one
hand, by using the infinite series (3.38) of the state $\left\vert
0\right\rangle \tbigotimes |\Psi _{c}\rangle $ and the unitary
transformation $\left( W_{12}^{ap}\right) ^{+}$ obtained from (3.63) one can
calculate the final state $\Psi _{1}^{ap}$. It is given by%
\begin{equation}
\Psi _{1}^{ap}=\sum_{k=0}^{\infty }iB_{k}\left\vert k,u_{0}\right\rangle
-\sum_{k=2^{d_{s}}}^{\infty }iB_{k}\left\vert k,u_{0}\right\rangle
+\sum_{k=2^{d_{s}}}^{\infty }B_{k}\left\vert 0,u_{k}\right\rangle  \tag{3.66}
\end{equation}%
where the global phase factor $e^{i\varphi _{1}}$ is already set to $%
e^{i\varphi _{1}}=1.$ This shows that the final state $\Psi _{1}^{ap}$ is
indeed in the subspace $S_{12}.$ On the other hand, by applying directly the
unitary operator $\left( W_{12}\right) ^{+}$ which is obtained from (3.35)
to the infinite series (3.38) and then with the help of the unitary
transformations $\left( W_{12}\right) ^{+}\left\vert 0,u_{j}\right\rangle
=i\left\vert j,u_{0}\right\rangle $ ($j\geq 0$) which is obtained from
(3.36) one finds that the final state $\Psi _{1}$ is given by 
\begin{equation}
\Psi _{1}=\sum_{k=0}^{\infty }iB_{k}\left\vert k,u_{0}\right\rangle . 
\tag{3.67}
\end{equation}%
By comparing (3.66) with (3.67) one sees that on the RH side of (3.66) the
first infinite series exists and it is equal to the state of (3.67).
Furthermore it can prove that the last two infinite series on the RH side of
(3.66) each are convergent. Now by substituting (3.66) and (3.67) into
(3.65) with $k=1$ one can calculate the norm $Norm1$. It is found that $%
norm1 $ is bounded by%
\begin{equation}
Norm1=\left\Vert -\sum_{k=2^{d_{s}}}^{\infty }iB_{k}\left\vert
k,u_{0}\right\rangle +\sum_{k=2^{d_{s}}}^{\infty }B_{k}\left\vert
0,u_{k}\right\rangle \right\Vert \leq 2\sqrt{\sum_{k=2^{d_{s}}}^{\infty
}\left\vert B_{k}\right\vert ^{2}}  \tag{3.68}
\end{equation}%
In an analogous way the rest two norms $Norm2$ and $Norm3$ each can be
explicitly calculated. For the norm $Norm2$, on the one hand, one has%
\begin{equation*}
\Psi _{2}^{ap}=e^{i\varphi _{2}}\exp \left[ i\varphi _{0}\left(
a_{m}^{s}\right) \right] \{\sum_{k=0}^{\infty }iB_{k}\exp \left( -i\alpha
_{m}a_{m}^{s}k\right) \left\vert k,u_{0}\right\rangle
\end{equation*}%
\begin{equation}
-\sum_{k=2^{d_{s}}}^{\infty }iB_{k}\exp \left( -i\alpha
_{m}a_{m}^{s}k\right) \left\vert k,u_{0}\right\rangle
+\sum_{k=2^{d_{s}}}^{\infty }B_{k}\left\vert 0,u_{k}\right\rangle \} 
\tag{3.69}
\end{equation}%
This formula is obtained by applying the $U_{ap}^{sic}\left(
a_{m}^{s},\alpha _{m}\right) $ to the state (3.66) and then using the
unitary transformation of (3.62). On the other hand, by applying directly
the $U_{\infty }^{sic}\left( a_{m}^{s},\alpha _{m}\right) $ to the state
(3.67) and then using the eigenvalue equation (3.40) one obtains%
\begin{equation}
\Psi _{2}=\tprod\limits_{l=1}^{\infty }\exp \left( i2^{l-2}\alpha
_{m}a_{m}^{s}\right) \sum_{k=0}^{\infty }iB_{k}\exp \left( -i\alpha
_{m}a_{m}^{s}k\right) \left\vert k,u_{0}\right\rangle .  \tag{3.70}
\end{equation}%
In order to cancel the effect of the global phase factors of the two states $%
\Psi _{2}^{ap}$ and $\Psi _{2}$ on the norm $Norm2$ of (3.65) with $k=2$ in
calculation one may set suitably the global phase factor $e^{i\varphi _{2}}$
of the state $\Psi _{2}^{ap}$ in (3.69) such that these two states $\Psi
_{2}^{ap}$ and $\Psi _{2}$ achieve the same global phase factor. This can be
done by letting $e^{i\varphi _{2}}\exp \left[ i\varphi _{0}\left(
a_{m}^{s}\right) \right] =\tprod\nolimits_{l=1}^{\infty }\exp \left(
i2^{l-2}\alpha _{m}a_{m}^{s}\right) $. Of course, discarding directly the
two global phase factors in (3.69) and (3.70) also can cancel the effect.
Now by substituting (3.69) and (3.70) into (3.65) with $k=2$ the norm $norm2$
can be calculated. It is shown that $Norm2$ is bounded by%
\begin{equation*}
Norm2=\left\Vert -\sum_{k=2^{d_{s}}}^{\infty }iB_{k}\exp \left( -i\alpha
_{m}a_{m}^{s}k\right) \left\vert k,u_{0}\right\rangle
+\sum_{k=2^{d_{s}}}^{\infty }B_{k}\left\vert 0,u_{k}\right\rangle \right\Vert
\end{equation*}%
\begin{equation}
\leq 2\sqrt{\sum_{k=2^{d_{s}}}^{\infty }\left\vert B_{k}\right\vert ^{2}} 
\tag{3.71}
\end{equation}%
For calculation of the norm $Norm3$ one may employ the state (3.64), i.e., $%
U_{ap}^{sic}(t_{m})\left\vert 0\right\rangle \tbigotimes |\Psi _{c}\rangle $
and the state (3.41). The state (3.41) is just the state $\Psi _{3}$, while
the state $\bar{\Psi}_{3}^{ap}\ $is obtained by multiplying the state (3.64)
by the global phase factor $e^{i\varphi _{3}^{\prime }}.$ Now by
substituting the state $\Psi _{3}$ of (3.41) into (3.65) with $k=3$ and
replacing $\Psi _{3}^{ap}$ in (3.65) with the state $\bar{\Psi}_{3}^{ap}$
one obtains%
\begin{equation}
Norm3=\left\Vert \sum_{k=2^{d_{s}}}^{\infty }B_{k}\left( 1-\exp \left(
-i\alpha _{m}a_{m}^{s}k\right) \right) \left\vert 0,u_{k}\right\rangle
\right\Vert \leq 2\sqrt{\sum_{k=2^{d_{s}}}^{\infty }\left\vert
B_{k}\right\vert ^{2}}  \tag{3.72}
\end{equation}%
where $e^{i\varphi _{3}^{\prime }}\exp \left[ -i\Phi _{ap}\left(
a_{m}^{s}\right) \right] \exp \left[ i\varphi _{0}\left( a_{m}^{s}\right) %
\right] =\tprod\nolimits_{l=1}^{\infty }\exp \left( i2^{l-2}\alpha
_{m}a_{m}^{s}\right) $ is used to obtain the global phase factor $%
e^{i\varphi _{3}^{\prime }}.$ Now all these three norms of (3.65) or their
upper bounds are obtained. They are determined from (3.68), (3.71), and
(3.72), respectively.

After these three norms of (3.65) (or their upper bounds) are obtained, one
can further obtain the maximum one among these three norms (or its upper
bound). It can be found from (3.68), (3.71), and (3.72) that the maximum
norm is bounded by%
\begin{equation}
Normmax=\max \left( Norm1,Norm2,Norm3\right) \leq 2\sqrt{%
\sum_{k=2^{d_{s}}}^{\infty }\left\vert B_{k}\right\vert ^{2}}  \tag{3.73}
\end{equation}%
When the maximum norm $Normmax$ is sufficiently close to zero, every one of
these three norms of (3.65) is sufficiently close to zero automatically.
Notice that the initial state $\left\vert 0\right\rangle \tbigotimes |\Psi
_{c}\rangle $ is arbitrary. Then this means that the $SIC$ unitary
propagator $U_{ap}^{sic}(t_{m})$ of (3.61) is sufficiently close to the
theoretical one, i.e., the $U_{c}^{sic}(t_{m})$ of (3.37) up to a global
phase factor.

The upper bound of the maximum norm $Normmax$ that is determined from (3.73)
is closely related to convergence of the infinite eigenfunction-expansion
series (3.38) of the initial state $\left\vert 0\right\rangle \tbigotimes
|\Psi _{c}\rangle .$ Because convergence is a mathematical attribute, the
upper bound has nothing to do with any quantum effect of the initial state $%
\left\vert 0\right\rangle \tbigotimes |\Psi _{c}\rangle .$ Actually, it can
be found that the upper bound is determined completely by the residual terms
of the infinite expansion series (3.38) of the initial state (See below).

Suppose that the initial state $\left\vert 0\right\rangle \tbigotimes |\Psi
_{c}\rangle $ can be well approximated by the partial sum of the first $L$
terms in the infinite expansion series of (3.38). Here for convenience the
partial sum of all those residual terms with index $k\geq L$ in the infinite
expansion series of (3.38) is called the residual sum of the infinite
expansion series. Then it follows from (3.38) that the residual sum is given
by%
\begin{equation}
RES\left( L\right) =\left\vert 0\right\rangle \tbigotimes |\Psi _{c}\rangle
-\sum_{k=0}^{L-1}B_{k}\left\vert 0,u_{k}\right\rangle =\sum_{k=L}^{\infty
}B_{k}\left\vert 0,u_{k}\right\rangle  \tag{3.74}
\end{equation}%
and its norm is given by%
\begin{equation}
NRES\left( L\right) =\left\Vert \sum_{k=L}^{\infty }B_{k}\left\vert
0,u_{k}\right\rangle \right\Vert =\sqrt{\sum_{k=L}^{\infty }\left\vert
B_{k}\right\vert ^{2}}  \tag{3.75}
\end{equation}%
It can be seen from (3.73) and (3.75) that the maximum norm $Normmax$ is
bounded by%
\begin{equation}
Normmax\leq 2\times NRES\left( 2^{d_{s}}\right) .  \tag{3.76}
\end{equation}%
where the term number $L$ is set to $L=2^{d_{s}}$ in the norm $NRES(L)$
which is calculated from (3.75). This inequality shows clearly that the
upper bound of the maximum norm is solely determined by the norm of the
residual sum of the infinite expansion series (3.38) of the initial state $%
\left\vert 0\right\rangle \tbigotimes |\Psi _{c}\rangle $.

The infinite series (3.38) of the initial state $\left\vert 0\right\rangle
\tbigotimes |\Psi _{c}\rangle $ is really obtained directly from the
infinite expansion series $|\Psi _{c}\rangle =\sum_{k=0}^{\infty
}B_{k}u_{k}\left( x\right) $ of the state $|\Psi _{c}\rangle $ which is like
(3.8). Here the residual sum of the infinite series of the state $|\Psi
_{c}\rangle $ is given by $RES\left( L\right) =\sum_{k=L}^{\infty
}B_{k}u_{k}\left( x\right) .$ Its norm is $NRES\left( L\right) =\sqrt{%
\sum_{k=L}^{\infty }\left\vert B_{k}\right\vert ^{2}}.$ This norm is exactly
equal to the one of (3.75). Therefore, actually the upper bound of the
maximum norm $Normmax$ is solely determined directly by the norm of the
residual sum of the infinite series of the state $|\Psi _{c}\rangle .$

The energy eigenfunction expansion principle indicates that the infinite
expansion series of the state $|\Psi _{c}\rangle $ or the infinite series
(3.38) of the initial state $\left\vert 0\right\rangle \tbigotimes |\Psi
_{c}\rangle $ is always convergent. This shows that the residual sum $%
RES\left( L\right) $ $=\sum_{k=L}^{\infty }B_{k}u_{k}\left( x\right) $ or $%
RES(L)$ of (3.74) is sufficiently close to zero if the term number $L$ is
sufficiently large. Mathematically this means that for any prescribed real
number $\varepsilon >0$ there exists an integer $L>0$ such that the norm of
the residual sum $NRES\left( K\right) <\varepsilon $ when the integer $K>L.$
The energy eigenfunction expansion principle provides theoretical basis for
how to make the $U_{ap}^{sic}(t_{m})$ of (3.61) sufficiently close to the $%
U_{c}^{sic}(t_{m})$ of (3.37). Suppose that the maximum norm $Normmax$ is
required not to be more than a prescribed small value $\varepsilon _{1}>0,$
that is, $Normmax\leq \varepsilon _{1}.$ Then according to (3.76) one may
choose $NRES\left( 2^{d_{s}}\right) \leq \varepsilon _{1}/2=\varepsilon .$
Now the minimum term number $L=L_{\min }$ is determined from $NRES\left(
L_{\min }\right) \leq \varepsilon $. Once $L_{\min }$ is obtained, one may
determine the spin number $d_{s}$ such that $2^{d_{s}-1}<L_{\min }\leq
2^{d_{s}}.$ Let $K=2^{d_{s}},$ and evidently $K\geq L_{\min }.$ Then one has 
$NRES\left( 2^{d_{s}}\right) \leq \varepsilon .$ Therefore, this spin number 
$d_{s}$ can ensure $Normmax\leq \varepsilon _{1}.$ Obviously, here the term
number $2^{d_{s}}$ must not be less than the dimension of the reduced search
space. Once the $d_{s}-$qubit spin system is determined, one may construct
the $SIC$ unitary propagator $U_{ap}^{sic}(t_{m})$ from (3.61). Such
constructed $SIC$\ unitary propagator can be made sufficiently close to the
theoretical one, i.e., the $U_{c}^{sic}(t_{m})$ of (3.37) and can be
realized by using the $d_{s}-$qubit spin system with finite spin number $%
d_{s}.$

The computational complexity for the $SIC$ unitary propagator $%
U_{ap}^{sic}(t_{m})$ of (3.61) is dependent upon the one for the $SIC$
unitary propagator $U_{ap}^{sic}\left( a_{m}^{s},\alpha _{m}\right) $ and
the one for the $QM$\ subspace-selective unitary operator $W_{12}^{ap}$ (or $%
\left( W_{12}^{ap}\right) ^{+}$). As shown in (3.59), the $%
U_{ap}^{sic}\left( a_{m}^{s},\alpha _{m}\right) $ is a tensor product of the 
$d_{s}$ basic $SIC$ unitary diagonal operators of (3.1). As shown in (3.60),
the $QM$ unitary operator $W_{12}^{ap}$ (or $\left( W_{12}^{ap}\right) ^{+}$%
) is a product of the $2^{d_{s}}-1$ pseudospin rotation operators $\{\exp
\left( -i\pi Q_{x}^{k,u_{k}}\right) \}$ and one unitary diagonal operator.
Then the smaller the spin number $d_{s}$ of the $d_{s}-$qubit spin system,
the better the performance of the $U_{ap}^{sic}\left( a_{m}^{s},\alpha
_{m}\right) $ and also the $QM$ unitary operator $W_{12}^{ap}$ (or $\left(
W_{12}^{ap}\right) ^{+}$), and consequently the better the performance of
the $U_{ap}^{sic}(t_{m})$ of (3.61). This is on the one hand. On the other
hand, the spin number $d_{s}$ is closely related to convergence of the
infinite expansion series of the state $|\Psi _{c}\rangle .$ It is known
from the inequality $NRES\left( 2^{d_{s}}\right) \leq \varepsilon $ for any
prescribed real number $\varepsilon >0$ that if the infinite series of the
state $|\Psi _{c}\rangle $ is faster convergent, then the term number $L=$ $%
2^{d_{s}}$ for the first $L$ terms of the infinite series is smaller and
hence the spin number $d_{s}$ is smaller too. Consequently, if the infinite
series of the state $|\Psi _{c}\rangle $ is faster convergent, then the $SIC$
unitary propagator $U_{ap}^{sic}(t_{m})$ prepared by (3.61) has a better
performance in computational complexity.\newline

Below it further proves that the $SIC$ unitary propagator $%
V_{c}^{sic}(t_{m}) $ of (3.49) can be realized approximately by using a $%
d_{s}-$qubit spin system with finite spin number $d_{s}$. As shown in the
subsection 3.1, this $SIC$ unitary propagator can be constructed
theoretically by using a $d-$qubit spin system with $2^{d}=\infty .$
However, it does not have a practicable value. Below a $SIC$ unitary
propagator is constructed that can be made sufficiently close to the $%
V_{c}^{sic}(t_{m})$ of (3.49) and may be realized by using a $d_{s}-$qubit
spin system with finite spin number $d_{s}$. This construction starts from
the unitary-dynamical equation of (3.49).

First of all, the $SIC$ unitary propagator $U_{\infty }^{sic}\left(
a_{m}^{s},\alpha _{m}\right) $ in (3.49) is replaced with the $%
U_{ap}^{sic}\left( a_{m}^{s},\alpha _{m}\right) $ of (3.59). The former must
be realized by using a $d-$qubit spin system with $2^{d}=\infty $, while the
latter can be realized by using a $d_{s}-$qubit spin system with finite spin
number $d_{s}$. Then correspondingly the $QM$\ unitary operator $W_{12}$ in
(3.49) is replaced with the $W_{12}^{ap}$ of (3.60) and $W_{23}$ in (3.49)
with the $W_{23}^{ap}$ which will be set up below. Therefore, the $SIC$
unitary propagator $V_{c}^{sic}(t_{m})$ of (3.49) is replaced with the
following $SIC$ unitary propagator:%
\begin{equation}
V_{ap}^{sic}(t_{m})=\exp \left[ -i\Phi _{1}^{ap}\left( a_{m}^{s}\right) %
\right] W_{23}^{ap}W_{12}^{ap}U_{ap}^{sic}\left( a_{m}^{s},\alpha
_{m}\right) \left( W_{12}^{ap}\right) ^{+}\left( W_{23}^{ap}\right) ^{+}. 
\tag{3.77}
\end{equation}%
This $SIC$ unitary propagator can be realized by using a $d_{s}-$qubit spin
system with finite spin number $d_{s}$. Below it can prove that it can be
made sufficiently close to the theoretical one, i.e., the $%
V_{c}^{sic}(t_{m}) $ of (3.49).

The $SIC$ unitary propagator $V_{ap}^{sic}(t_{m})$ really works on the
composite Hilbert space $S_{d_{s}}\tbigotimes S_{2}\tbigotimes S_{3}.$
However, just like the $V_{c}^{sic}(t_{m})$ of (3.49), theoretically it also
works on the composite Hilbert space $S_{1}\tbigotimes S_{2}\tbigotimes
S_{3}.$ It will prove later that it works effectively on the subspace $%
\left\vert 0\right\rangle \tbigotimes \left\vert u_{0}\right\rangle
\tbigotimes S_{3}$ of the Hilbert space $S_{1}\tbigotimes S_{2}\tbigotimes
S_{3}$ or equivalently on the component Hilbert subspace $S_{3}$.

One needs first to set up these unitary transformations $W_{23}^{ap},$ $%
W_{12}^{ap},$ and $U_{ap}^{sic}\left( a_{m}^{s},\alpha _{m}\right) $ for the 
$SIC\ $unitary propagator $V_{ap}^{sic}(t_{m})$ of (3.77). The $SIC$\
unitary propagator $U_{ap}^{sic}\left( a_{m}^{s},\alpha _{m}\right) $ in
(3.77) is still given by (3.59). It works really on the Hilbert space $%
S_{d_{s}}$ of the $d_{s}-$qubit spin system and its eigenvalue equation is
still given by (3.62a). Theoretically it also works on the component Hilbert
subspace $S_{1}$ of the composite Hilbert spaces $S_{1}\tbigotimes S_{2}$
and $S_{1}\tbigotimes S_{2}\tbigotimes S_{3}.$ And the eigenvalue equation
for the $U_{ap}^{sic}\left( a_{m}^{s},\alpha _{m}\right) $ acting on any
eigenbase $\left\vert k\right\rangle $ of the component Hilbert subspace $%
S_{1}$ is still given by (3.62b). It can be deduced from (3.62b) that when
the $U_{ap}^{sic}\left( a_{m}^{s},\alpha _{m}\right) $ is applied to any
tensor-product base $\left\vert k,u_{i},v_{j}\right\rangle \ $with $%
k,i,j=0,1,...,\infty $ of the Hilbert space $S_{1}\tbigotimes
S_{2}\tbigotimes S_{3},$ its unitary transformation (or its eigenvalue
equation) in the Hilbert space may be written as%
\begin{equation*}
U_{ap}^{sic}\left( a_{m}^{s},\alpha _{m}\right) \left\vert
k,u_{i},v_{j}\right\rangle
\end{equation*}%
\begin{equation}
=\left\{ 
\begin{array}{c}
\exp \left[ i\varphi _{0}\left( a_{m}^{s}\right) \right] \exp \left(
-i\alpha _{m}a_{m}^{s}k\right) \left\vert k,u_{i},v_{j}\right\rangle \\ 
\exp \left[ i\phi _{k}\left( a_{m}^{s}\right) \right] \left\vert
k,u_{i},v_{j}\right\rangle%
\end{array}%
\begin{array}{c}
if \\ 
if%
\end{array}%
\begin{array}{c}
0\leq k<2^{d_{s}} \\ 
k\geq 2^{d_{s}}%
\end{array}%
\right.  \tag{3.78}
\end{equation}%
The $QM$ unitary operator $W_{12}^{ap}$ in (3.77) is still written as
(3.60). It works really on the subspace $S_{d_{s}}\tbigotimes \left\vert
u_{0}\right\rangle \cup \left\vert 0\right\rangle \tbigotimes S_{2},$ which
is a subspace of the subspace $S_{12}$ of the Hilbert space $%
S_{1}\tbigotimes S_{2}$, and theoretically it also works on the subspace $%
S_{12}.$ Now in the Hilbert space $S_{1}\tbigotimes S_{2}\tbigotimes S_{3}$
one may set up its unitary transformation. With the aid of (3.63) the
unitary transformation $W_{12}^{ap}$ on any tensor-product base $\left\vert
k,u_{i},v_{j}\right\rangle $ for $k,i,j=0,1,...,\infty $ of the subspace $%
S_{12}\tbigotimes S_{3}$ of the Hilbert space $S_{1}\tbigotimes
S_{2}\tbigotimes S_{3}$ may be written as%
\begin{equation}
W_{12}^{ap}\left\vert k,u_{0},v_{j}\right\rangle =-i\left\vert
0,u_{k},v_{j}\right\rangle ,\text{ if }0\leq k<2^{d_{s}}  \tag{3.79a}
\end{equation}%
\begin{equation}
\left( W_{12}^{ap}\right) ^{+}\left\vert k,u_{0},v_{j}\right\rangle
=i\left\vert 0,u_{k},v_{j}\right\rangle ,\text{ if }0\leq k<2^{d_{s}} 
\tag{3.79b}
\end{equation}%
\begin{equation}
W_{12}^{ap}\left\vert k,u_{0},v_{j}\right\rangle =\left\vert
k,u_{0},v_{j}\right\rangle \text{ and }W_{12}^{ap}\left\vert
0,u_{k},v_{j}\right\rangle =\left\vert 0,u_{k},v_{j}\right\rangle ,\text{ if 
}k\geq 2^{d_{s}}  \tag{3.79c}
\end{equation}%
where $\left\vert v_{j}\right\rangle $ is any energy eigenbase of the
component Hilbert subspace $S_{3}$. Notice that the subspace $%
S_{12}\tbigotimes S_{3}$ contains the subspace $S_{123}.$ Then the unitary
transformations (3.79) still can be applied in the subspace $S_{123}.$

The $QM$ unitary operator $W_{23}^{ap}$ in the $V_{ap}^{sic}(t_{m})$ of
(3.77) corresponds to the unitary operator $W_{23}$ of (3.47) in the $%
V_{c}^{sic}(t_{m})$ of (3.44). It may be written as%
\begin{equation}
W_{23}^{ap}=\exp \left( -i\frac{\pi }{2}\left\vert u_{0},v_{0}\right\rangle
\left\langle u_{0},v_{0}\right\vert \right)
\tprod\limits_{k=1}^{2^{d_{s}}-1}\exp \left( -i\pi
Q_{x}^{u_{k},v_{k}}\right) .  \tag{3.80}
\end{equation}%
This $QM$ unitary operator is a limited product of the $2^{d_{s}}-1$
pseudospin rotation operators $\{\exp \left( -i\pi
Q_{x}^{u_{k},v_{k}}\right) \}$ in addition to one unitary diagonal operator.
It is also similar to a $QM$ subspace-selective unitary operator [$7$]. Just
like the unitary operator $W_{23}$ of (3.47), it still works really on the
subspace $S_{23}$ of the Hilbert space $S_{2}\tbigotimes S_{3}.$ Now one may
set up the unitary transformation $W_{23}^{ap}$ on the subspace by applying
directly the $W_{23}^{ap}$ of (3.80) to any tensor-product base of the
subspace and then using the commutation relations of (3.46a), the $QM$
selective unitary transformations of (3.46b), and the identical state
transformations of (3.46c). This is similar to setting up the unitary
transformation $W_{23}$ of (3.48) in the subsection 3.1. It can prove that
the unitary transformation $W_{23}^{ap}$ on any tensor-product base $%
\left\vert u_{k},v_{0}\right\rangle $ ($\left\vert u_{0},v_{k}\right\rangle $%
) of the subspace $S_{23}$ is given by\newline
\begin{equation}
W_{23}^{ap}:\left\vert u_{k},v_{0}\right\rangle \rightarrow -i\left\vert
u_{0},v_{k}\right\rangle \text{ if }0\leq k<2^{d_{s}},  \tag{3.81a}
\end{equation}%
\begin{equation}
\left( W_{23}^{ap}\right) ^{+}:\left\vert u_{k},v_{0}\right\rangle
\rightarrow i\left\vert u_{0},v_{k}\right\rangle \text{ if }0\leq
k<2^{d_{s}},  \tag{3.81b}
\end{equation}%
and moreover,%
\begin{equation}
W_{23}^{ap}\left\vert u_{k},v_{0}\right\rangle =\left\vert
u_{k},v_{0}\right\rangle ,\text{ }W_{23}^{ap}\left\vert
u_{0},v_{k}\right\rangle =\left\vert u_{0},v_{k}\right\rangle ,\text{ if }%
k\geq 2^{d_{s}}.\newline
\tag{3.81c}
\end{equation}%
Here the tensor-product base $\left\vert u_{k},v_{0}\right\rangle $ (or $%
\left\vert u_{0},v_{k}\right\rangle $) still belongs to the Hilbert space $%
S_{2}\tbigotimes S_{3}$ of the composite quantum system under study even if $%
k\geq 2^{d_{s}}.$

Now consider further that the $QM$ unitary operator $W_{23}^{ap}$ of (3.80)
acts on any tensor-product base of the subspace $S_{123}$ of the Hilbert
space $S_{1}\tbigotimes S_{2}\tbigotimes S_{3}.$ It can be deduced from
(3.81) that the unitary transformation $W_{23}^{ap}$ on any tensor-product
base $\left\vert k,u_{i},v_{j}\right\rangle $ with $k,i,j=0,1,...,\infty $
of the subspace $S_{1}\tbigotimes S_{23}$ of the Hilbert space $%
S_{1}\tbigotimes S_{2}\tbigotimes S_{3}$ is given by%
\begin{equation}
W_{23}^{ap}\left\vert j,u_{k},v_{0}\right\rangle =-i\left\vert
j,u_{0},v_{k}\right\rangle ,\text{ if }0\leq k<2^{d_{s}}  \tag{3.82a}
\end{equation}%
\begin{equation}
\left( W_{23}^{ap}\right) ^{+}\left\vert j,u_{k},v_{0}\right\rangle
=i\left\vert j,u_{0},v_{k}\right\rangle ,\text{ if }0\leq k<2^{d_{s}} 
\tag{3.82b}
\end{equation}%
and moreover,%
\begin{equation}
W_{23}^{ap}\left\vert j,u_{k},v_{0}\right\rangle =\left\vert
j,u_{k},v_{0}\right\rangle ,\text{ }W_{23}^{ap}\left\vert
j,u_{0},v_{k}\right\rangle =\left\vert j,u_{0},v_{k}\right\rangle ,\text{ if 
}k\geq 2^{d_{s}}.  \tag{3.82c}
\end{equation}%
Here theoretically $\left\vert j\right\rangle $ $(j\geq 0)$ can be any
eigenbase of the component Hilbert subspace $S_{1}$, even if it belongs to
the Hilbert space $S_{d_{s}}$ of the $d_{s}-$qubit spin system when $0\leq
j<2^{d_{s}}.$ Note that the subspace $S_{1}\tbigotimes S_{23}$ contains the
subspace $S_{123}.$ Then the unitary transformations (3.82) are still
available in the subspace $S_{123}.$

These three unitary transformations (3.78), (3.79), and (3.82) are further
used to calculate theoretically the time evolution process governed by the $%
SIC$\ unitary propagator $V_{ap}^{sic}(t_{m})$ of (3.77) in the subspace $%
S_{123}$ and to show that the $V_{ap}^{sic}(t_{m})$ can be made sufficiently
close to the $V_{c}^{sic}(t_{m})$ of (3.49).

First it can prove that the $V_{ap}^{sic}(t_{m})$ of (3.77) works
effectively on the subspace $\left\vert 0\right\rangle \tbigotimes
\left\vert u_{0}\right\rangle \tbigotimes S_{3}$ of the Hilbert subspace $%
S_{123}.$ It is known from (3.57) that the $V_{c}^{sic}(t_{m})$ of (3.49)
works effectively on the subspace $\left\vert 0\right\rangle \tbigotimes
\left\vert u_{0}\right\rangle \tbigotimes S_{3}.$ Here still suppose that $%
\left\vert 0,u_{0},\Psi _{v}\right\rangle $ is arbitrary state of the
subspace $\left\vert 0\right\rangle \tbigotimes \left\vert
u_{0}\right\rangle \tbigotimes S_{3}$ and can be expanded as the infinite
series of (3.50). Now applying the $V_{ap}^{sic}(t_{m})$ to the infinite
series (3.50) of the state $\left\vert 0,u_{0},\Psi _{v}\right\rangle $ and
then using these three unitary transformations (3.78), (3.79), and (3.82)
one can calculate the time evolution process governed by the $%
V_{ap}^{sic}(t_{m})$. A detailed calculation shows that the time evolution
process is described by%
\begin{equation*}
V_{ap}^{sic}(t_{m})\left\vert 0,u_{0},\Psi _{v}\right\rangle =\exp \left[
-i\Phi _{1}^{ap}\left( a_{m}^{s}\right) \right] \exp \left[ i\varphi
_{0}\left( a_{m}^{s}\right) \right]
\end{equation*}%
\begin{equation}
\times \{\sum_{k=0}^{2^{d_{s}}-1}B_{k}\exp \left( -i\alpha
_{m}a_{m}^{s}k\right) \left\vert 0,u_{0},v_{k}\right\rangle
+\sum_{k=2^{d_{s}}}^{\infty }B_{k}\left\vert 0,u_{0},v_{k}\right\rangle \}. 
\tag{3.83}
\end{equation}%
By comparing the infinite series (3.83) with the infinite series (3.50) of
the state $\left\vert 0,u_{0},\Psi _{v}\right\rangle $ it can be deduced
that the infinite series (3.83) is convergent. Notice that the state $%
\left\vert 0,u_{0},\Psi _{v}\right\rangle $ is arbitrary in the subspace $%
\left\vert 0\right\rangle \tbigotimes \left\vert u_{0}\right\rangle
\tbigotimes S_{3}.$\ Then the equation (3.83) shows that the $%
V_{ap}^{sic}(t_{m})$ indeed acts effectively on the subspace $\left\vert
0\right\rangle \tbigotimes \left\vert u_{0}\right\rangle \tbigotimes S_{3}$
of the Hilbert subspace $S_{123}.$

The $SIC$\ unitary propagators $V_{c}^{sic}(t_{m})$ of (3.49) and the $%
V_{ap}^{sic}(t_{m})$ of (3.77) each have five unitary-transformational
steps, when they each act on arbitrary state (3.50) of the subspace $%
\left\vert 0\right\rangle \tbigotimes \left\vert u_{0}\right\rangle
\tbigotimes S_{3}$. In order to show that the $V_{ap}^{sic}(t_{m})$ can be
made sufficiently close to the $V_{c}^{sic}(t_{m})$ it needs to prove that
each of the five unitary-transformational steps of the $V_{ap}^{sic}(t_{m})$
can be made sufficiently close to the counterpart of the $V_{c}^{sic}(t_{m})$%
. These five steps each can be derived from (3.49) for the $%
V_{c}^{sic}(t_{m})$ and from (3.77) for the $V_{ap}^{sic}(t_{m})$,
respectively. For the $V_{ap}^{sic}(t_{m})$ of (3.77) these five steps are
given successively by $\tilde{\psi}_{1}^{ap}=\left( W_{23}^{ap}\right)
^{+}\left\vert 0,u_{0},\Psi _{v}\right\rangle ,$ $\tilde{\psi}%
_{2}^{ap}=\left( W_{12}^{ap}\right) ^{+}\tilde{\psi}_{1}^{ap},$ $\tilde{\psi}%
_{3}^{ap}=U_{ap}^{sic}\left( a_{m}^{s},\alpha _{m}\right) \tilde{\psi}%
_{2}^{ap},$ $\tilde{\psi}_{4}^{ap}=W_{12}^{ap}\tilde{\psi}_{3}^{ap},$ and $%
\tilde{\psi}_{5}^{ap}=W_{23}^{ap}\tilde{\psi}_{4}^{ap}.$ Let the state $%
\tilde{\Psi}_{k}^{ap}=e^{i\phi _{k}}\tilde{\psi}_{k}^{ap}$ for $k=1,2,3,4,5.$
Here the additional global phase factor $e^{i\phi _{k}}$ allows to be chosen
arbitrarily. Thus, both the states $\tilde{\Psi}_{k}^{ap}$ and $\tilde{\psi}%
_{k}^{ap}$ are the same one up to a global phase factor. For the $%
V_{c}^{sic}(t_{m})$ of (3.49) these five steps are given successively by $%
\tilde{\Psi}_{1}=W_{23}^{+}\left\vert 0,u_{0},\Psi _{v}\right\rangle ,$ $%
\tilde{\Psi}_{2}=W_{12}^{+}\tilde{\Psi}_{1},$ $\tilde{\Psi}_{3}=U_{\infty
}^{sic}\left( a_{m}^{s},\alpha _{m}\right) \tilde{\Psi}_{2},$ $\tilde{\Psi}%
_{4}=W_{12}\tilde{\Psi}_{3},$ and $\tilde{\Psi}_{5}=W_{23}\tilde{\Psi}_{4}.$
There are five norms which correspond respectively to these five steps, each
of which characterizes the difference between the $V_{ap}^{sic}(t_{m})$ and
the $V_{c}^{sic}(t_{m})$ at the corresponding step of these five steps.
These five norms are defined by%
\begin{equation*}
Norm\_k=\left\Vert \tilde{\Psi}_{k}^{ap}-\tilde{\Psi}_{k}\right\Vert ,\text{ 
}k=1,2,3,4,5.
\end{equation*}%
Here the additional global phase factor $e^{i\phi _{k}}$ of the state $%
\tilde{\Psi}_{k}^{ap}$ should be set suitably so that one can cancel the
effect of the global phase factors of the two states $\tilde{\Psi}_{k}^{ap}$
and $\tilde{\Psi}_{k}$ on the norm $Norm\_k$ in calculation.

These five norms $\{Norm\_k\}$ are calculated in detail below. By applying
respectively the unitary operators $\left( W_{23}^{ap}\right) ^{+}$ obtained
from (3.80) and the $W_{23}^{+}$ obtained from (3.47) to the state $%
\left\vert 0,u_{0},\Psi _{v}\right\rangle $ of (3.50) and then using
respectively the unitary transformations (3.82) and (3.53) one obtains%
\begin{equation}
\tilde{\psi}_{1}^{ap}=\sum_{k=0}^{2^{d_{s}}-1}iB_{k}\left\vert
0,u_{k},v_{0}\right\rangle +\sum_{k=2^{d_{s}}}^{\infty }B_{k}\left\vert
0,u_{0},v_{k}\right\rangle  \tag{3.84}
\end{equation}%
and%
\begin{equation}
\tilde{\Psi}_{1}=\sum_{k=0}^{\infty }iB_{k}\left\vert
0,u_{k},v_{0}\right\rangle .  \tag{3.85}
\end{equation}%
By substituting the two states (3.84) and (3.85) into the expression of the
norm $Norm\_1$ and then letting $e^{i\phi _{1}}=1$ one can calculate the $%
Norm\_1$ and its upper bound, and the calculated result is given by%
\begin{equation}
Norm\_1=\left\vert \left\vert -\sum_{k=2^{d_{s}}}^{\infty }iB_{k}\left\vert
0,u_{k},v_{0}\right\rangle +\sum_{k=2^{d_{s}}}^{\infty }B_{k}\left\vert
0,u_{0},v_{k}\right\rangle \right\vert \right\vert \leq 2\sqrt{%
\sum_{k=2^{d_{s}}}^{\infty }\left\vert B_{k}\right\vert ^{2}}  \tag{3.86}
\end{equation}%
Now the norm $Norm\_2$ is calculated as follows. By applying the $\left(
W_{12}^{ap}\right) ^{+}$ obtained from (3.60) to the state (3.84) and then
using the unitary transformation (3.79) one obtains%
\begin{equation}
\tilde{\psi}_{2}^{ap}=-\sum_{k=0}^{2^{d_{s}}-1}B_{k}\left\vert
k,u_{0},v_{0}\right\rangle +\sum_{k=2^{d_{s}}}^{\infty }iB_{k}\left\vert
0,u_{0},v_{k}\right\rangle  \tag{3.87}
\end{equation}%
Correspondingly, by applying the $W_{12}^{+}$ obtained from (3.35) to the
state (3.85) and then using the unitary transformation (3.52) one can find
that%
\begin{equation}
\tilde{\Psi}_{2}=-\sum_{k=0}^{\infty }B_{k}\left\vert
k,u_{0},v_{0}\right\rangle  \tag{3.88}
\end{equation}%
By substituting the two states (3.87) and (3.88) into the expression of the
norm $Norm\_2$ and then letting $e^{i\phi _{2}}=1$ one obtains%
\begin{equation}
Norm\_2=\left\Vert \sum_{k=2^{d_{s}}}^{\infty }B_{k}\left\vert
k,u_{0},v_{0}\right\rangle +\sum_{k=2^{d_{s}}}^{\infty }iB_{k}\left\vert
0,u_{0},v_{k}\right\rangle \right\Vert \leq 2\sqrt{\sum_{k=2^{d_{s}}}^{%
\infty }\left\vert B_{k}\right\vert ^{2}}  \tag{3.89}
\end{equation}%
The norm $Norm\_3$ can be calculated by using further the two states (3.87)
and (3.88). On the one hand, by applying the $U_{ap}^{sic}\left(
a_{m}^{s},\alpha _{m}\right) $ to the state (3.87) and then using the
unitary transformation (3.78) one can find that%
\begin{equation*}
\tilde{\psi}_{3}^{ap}=-\exp \left[ i\varphi _{0}\left( a_{m}^{s}\right) %
\right] \sum_{k=0}^{2^{d_{s}}-1}B_{k}\exp \left( -i\alpha
_{m}a_{m}^{s}k\right) \left\vert k,u_{0},v_{0}\right\rangle
\end{equation*}%
\begin{equation}
+\exp \left[ i\varphi _{0}\left( a_{m}^{s}\right) \right]
\sum_{k=2^{d_{s}}}^{\infty }iB_{k}\left\vert 0,u_{0},v_{k}\right\rangle 
\tag{3.90}
\end{equation}%
On the other hand, by applying the $U_{\infty }^{sic}\left( a_{m}^{s},\alpha
_{m}\right) $ to the state (3.88) and then using the unitary transformation
(3.51) one can obtain%
\begin{equation}
\tilde{\Psi}_{3}=-\tprod\limits_{l=1}^{\infty }\exp \left( i2^{l-2}\alpha
_{m}a_{m}^{s}\right) \times \sum_{k=0}^{\infty }B_{k}\exp \left( -i\alpha
_{m}a_{m}^{s}k\right) \left\vert k,u_{0},v_{0}\right\rangle .  \tag{3.91}
\end{equation}%
Now these two states (3.90) and (3.91) are substituted into the expression
of the norm $Norm\_3.$ Then the norm $Norm\_3$ and its upper bound can be
calculated by setting $e^{i\phi _{3}}\exp \left[ i\varphi _{0}\left(
a_{m}^{s}\right) \right] =\tprod\nolimits_{l=1}^{\infty }\exp \left(
i2^{l-2}\alpha _{m}a_{m}^{s}\right) $. It is found that%
\begin{equation*}
Norm\_3=\left\Vert \sum_{k=2^{d_{s}}}^{\infty }B_{k}\exp \left( -i\alpha
_{m}a_{m}^{s}k\right) \left\vert k,u_{0},v_{0}\right\rangle
+\sum_{k=2^{d_{s}}}^{\infty }iB_{k}\left\vert 0,u_{0},v_{k}\right\rangle
\right\Vert
\end{equation*}%
\begin{equation}
\leq 2\sqrt{\sum_{k=2^{d_{s}}}^{\infty }\left\vert B_{k}\right\vert ^{2}} 
\tag{3.92}
\end{equation}%
The norm $Norm\_4$ is calculated as follows. On the one hand, by acting the $%
W_{12}^{ap}$ of (3.60) on the state (3.90) and then using the unitary
transformation (3.79) one obtains%
\begin{equation*}
\tilde{\psi}_{4}^{ap}=\exp \left[ i\varphi _{0}\left( a_{m}^{s}\right) %
\right] \sum_{k=0}^{2^{d_{s}}-1}iB_{k}\exp \left( -i\alpha
_{m}a_{m}^{s}k\right) \left\vert 0,u_{k},v_{0}\right\rangle
\end{equation*}%
\begin{equation}
+\exp \left[ i\varphi _{0}\left( a_{m}^{s}\right) \right]
\sum_{k=2^{d_{s}}}^{\infty }B_{k}\left\vert 0,u_{0},v_{k}\right\rangle . 
\tag{3.93}
\end{equation}%
On the other hand, by acting the $W_{12}$ of (3.35) on the state (3.91) and
then using the unitary transformation (3.52) it can be found that%
\begin{equation}
\tilde{\Psi}_{4}=\tprod\limits_{l=1}^{\infty }\exp \left( i2^{l-2}\alpha
_{m}a_{m}^{s}\right) \sum_{k=0}^{\infty }iB_{k}\exp \left( -i\alpha
_{m}a_{m}^{s}k\right) \left\vert 0,u_{k},v_{0}\right\rangle .  \tag{3.94}
\end{equation}%
Both the states (3.93) and (3.94) are substituted into the expression of the
norm $Norm\_4.$ Then by setting $e^{i\phi _{4}}\exp \left[ i\varphi
_{0}\left( a_{m}^{s}\right) \right] =\tprod\nolimits_{l=1}^{\infty }\exp
\left( i2^{l-2}\alpha _{m}a_{m}^{s}\right) $ in the $Norm\_4$ one can find
that%
\begin{equation*}
Norm\_4=\left\Vert \sum_{k=2^{d_{s}}}^{\infty }B_{k}\left\vert
0,u_{0},v_{k}\right\rangle -\sum_{k=2^{d_{s}}}^{\infty }iB_{k}\exp \left(
-i\alpha _{m}a_{m}^{s}k\right) \left\vert 0,u_{k},v_{0}\right\rangle
\right\Vert
\end{equation*}%
\begin{equation}
\leq 2\sqrt{\sum_{k=2^{d_{s}}}^{\infty }\left\vert B_{k}\right\vert ^{2}} 
\tag{3.95}
\end{equation}%
\newline
Now let $\bar{\Psi}_{5}=V_{c}^{sic}(t_{m})\left\vert 0,u_{0},\Psi
_{v}\right\rangle ,$ $\bar{\psi}_{5}^{ap}=V_{ap}^{sic}(t_{m})\left\vert
0,u_{0},\Psi _{v}\right\rangle $ and $\bar{\Psi}_{5}^{ap}=e^{i\phi
_{5}^{\prime }}\bar{\psi}_{5}^{ap},$ here $e^{i\phi _{5}^{\prime }}$ is an
additional global phase factor. Then it can prove that the norm $Norm\_5$
may be rewritten as $Norm\_5=\left\Vert \bar{\Psi}_{5}^{ap}-\bar{\Psi}%
_{5}\right\Vert .$ Thus, it also may be calculated by using directly the
state $\bar{\psi}_{5}^{ap}$ of (3.83) and the state $\bar{\Psi}_{5}$ which
is equal to (3.57) with $\alpha _{m}=\left( at_{m}/\hslash \right) $. The
calculated result is%
\begin{equation}
Norm\_5=\left\Vert \sum_{k=2^{d_{s}}}^{\infty }B_{k}\left( 1-\exp \left(
-i\alpha _{m}a_{m}^{s}k\right) \right) \left\vert 0,u_{0},v_{k}\right\rangle
\right\Vert \leq 2\sqrt{\sum_{k=2^{d_{s}}}^{\infty }\left\vert
B_{k}\right\vert ^{2}}  \tag{3.96}
\end{equation}%
where $\exp \left[ -ia_{m}^{s}\left( bt_{m}/\hslash \right) \right]
=e^{i\phi _{5}^{\prime }}\exp \left[ -i\Phi _{1}^{ap}\left( a_{m}^{s}\right) %
\right] \exp \left[ i\varphi _{0}\left( a_{m}^{s}\right) \right] $ is used
to determine the global phase factor $e^{i\phi _{5}^{\prime }}.$

These five norms $\{Norm\_k\}$ with $k=1,2,3,4,5$ or their upper bounds are
determined from (3.86), (3.89), (3.92), (3.95), and (3.96), respectively. It
can be found that the maximum one among these five norms is bounded by%
\begin{equation}
Norm\_max=\max_{1\leq k\leq 5}\left( Norm\_k\right) \leq 2\sqrt{%
\sum_{k=2^{d_{s}}}^{\infty }\left\vert B_{k}\right\vert ^{2}}  \tag{3.97}
\end{equation}%
When the maximum norm is sufficiently close to zero, every one of these five
norms is certainly sufficiently close to zero too. Then in that case every
one of the five unitary-transformational steps of the $V_{ap}^{sic}(t_{m})$
of (3.77) is sufficiently close to its counterpart of the $%
V_{c}^{sic}(t_{m}) $ of (3.49).

It can be seen from (3.97) and (3.50) that the upper bound of the maximum
norm $Norm\_max$ is still controlled completely by the residual terms with
index $k\geq 2^{d_{s}}$ of the infinite series (3.50) of the initial state $%
\left\vert 0,u_{0},\Psi _{v}\right\rangle .$ Suppose that the state $%
\left\vert 0,u_{0},\Psi _{v}\right\rangle $ can be well approximated by the
partial sum of the first $L=2^{d_{s}}$ terms in the infinite series. Then
the residual sum for the infinite series (3.50) is written as 
\begin{equation*}
RES\left( L\right) =\left\vert 0\right\rangle \tbigotimes \left\vert
0\right\rangle \tbigotimes |\Psi _{v}\rangle
-\sum_{k=0}^{L-1}B_{k}\left\vert 0,0,v_{k}\right\rangle =\sum_{k=L}^{\infty
}B_{k}\left\vert 0,0,v_{k}\right\rangle
\end{equation*}%
and its norm is given by%
\begin{equation}
NRES\left( L\right) =\left\Vert \sum_{k=L}^{\infty }B_{k}\left\vert
0,0,v_{k}\right\rangle \right\Vert =\sqrt{\sum_{k=L}^{\infty }\left\vert
B_{k}\right\vert ^{2}}  \tag{3.98}
\end{equation}%
It can be seen from (3.97) and (3.98) that the maximum norm $Norm\_max$ is
bounded by%
\begin{equation}
Norm\_max\leq 2\times NRES\left( 2^{d_{s}}\right) .  \tag{3.99}
\end{equation}%
Consequently the maximum norm is fully controlled by the norm $NRES\left(
2^{d_{s}}\right) $ of the residual sum.

Actually the infinite series (3.50) of the initial state $\left\vert
0,u_{0},\Psi _{v}\right\rangle $ is obtained directly from the infinite
expansion series $|\Psi _{v}\rangle =\sum_{k=0}^{\infty }B_{k}\left\vert
v_{k}\right\rangle $ of the state $|\Psi _{v}\rangle $ which is similar to
(3.8). The residual sum of the infinite series of the state $|\Psi
_{v}\rangle $ is given by $RES\left( L\right) =\sum_{k=L}^{\infty
}B_{k}\left\vert v_{k}\right\rangle .$ Its norm is $NRES\left( L\right) =%
\sqrt{\sum_{k=L}^{\infty }\left\vert B_{k}\right\vert ^{2}}.$ This norm is
exactly equal to the one of (3.98). Therefore, the upper bound of the
maximum norm $Norm\_max$ is solely determined directly by the norm of the
residual sum of the infinite series of the state $|\Psi _{v}\rangle .$

According to the energy eigenfunction expansion principle the infinite
expansion series of the state $|\Psi _{v}\rangle $ or the infinite series
(3.50) of the initial state $\left\vert 0,u_{0},\Psi _{v}\right\rangle $ is
always convergent. Then the norm $NRES\left( L\right) $ of the residual sum
of the infinite series of the state $|\Psi _{v}\rangle $ or the one of
(3.50) can be made sufficiently close to zero if the term number $L$ is
chosen as a sufficiently large integer. Therefore, by the inequality (3.99)
one may control the upper bound of $Norm\_max$ via the norm $NRES\left(
L\right) $ with $L=2^{d_{s}}$ such that the $SIC$ unitary propagator $%
V_{ap}^{sic}(t_{m})$ of (3.77) is sufficiently close to the theoretical one,
i.e., the $V_{c}^{sic}(t_{m})$ of (3.49), and at the same time the $d_{s}-$%
qubit spin system with finite spin number $d_{s}$ can be used to realize the 
$V_{ap}^{sic}(t_{m}).$ Suppose that the desired $Norm\_max\leq \varepsilon
_{1},$ where $\varepsilon _{1}>0$ is a prescribed small value. Then
according to (3.99) one may set $NRES\left( 2^{d_{s}}\right) \leq
\varepsilon _{1}/2=\varepsilon $. Now using the equation (3.98) and the
relation $NRES\left( L_{\min }\right) \leq \varepsilon $ one may determine
the minimum term number $L_{\min }$. Then the spin number $d_{s}$ of the $%
d_{s}-$qubit spin system is determined such that $2^{d_{s}-1}<L_{\min }\leq
2^{d_{s}}.$ If now one lets the term number $K=2^{d_{s}},$ then evidently $%
K\geq L_{\min }$ and hence $NRES\left( K\right) \leq \varepsilon .$ Through
the term number $L=K=2^{d_{s}}$ this spin number $d_{s}$ can ensure the
desired $Norm\_max\leq \varepsilon _{1}.$ Here the term number $2^{d_{s}}$
must not be less than the dimension of the reduced search space. Now the $%
d_{s}-$qubit spin system is used to construct the $SIC$ unitary propagator $%
V_{ap}^{sic}(t_{m})$ from (3.77). Obviously, this $SIC$ unitary propagator
can be made sufficiently close to the theoretical one, i.e., the $%
V_{c}^{sic}(t_{m})$ of (3.49) and may be realized by using the $d_{s}-$qubit
spin system with finite spin number $d_{s}.$

In the subsection 3.1 above it proves that in theory there always exists the 
$SIC$ unitary propagator $U_{at}^{sic}(t_{m})$ in (3.10) of a single-atom
system in external harmonic potential field and moreover, this $SIC$ unitary
propagator may be prepared through the $SIC$ unitary propagator $%
V_{c}^{sic}(t_{m})$ of (3.49). The problem is that the $V_{c}^{sic}(t_{m})$
has no practical value due to that it is realized by using the $d-$qubit
spin system with $d=\infty .$ In this subsection it is shown that the $%
V_{c}^{sic}(t_{m})$ of (3.49) can be approximated sufficiently well by the $%
SIC$ unitary propagator $V_{ap}^{sic}(t_{m})$ of (3.77) and moreover, the
latter can be realized by using a $d_{s}-$qubit spin system with finite spin
number $d_{s}.$ This indicates that the $U_{at}^{sic}(t_{m})$ can be
prepared well approximately through the $V_{ap}^{sic}(t_{m})$ of (3.77) and
it can be realized well approximately by using the $d_{s}-$qubit spin system.

It can be shown that if the infinite expansion series of the state $|\Psi
_{v}\rangle $ is faster convergent, then the $SIC$ unitary propagator $%
V_{ap}^{sic}(t_{m})$ prepared by (3.77) has a better performance in
computational complexity.

A plenty of the energy eigenstates of the internal motion of the single-atom
system have been employed to construct the $SIC$ unitary propagator $%
U_{at}^{sic}(t_{m})$ in (3.10) or the $SIC$ unitary propagator $%
V_{ap}^{sic}(t_{m})$ of (3.77). The reason for this is that this makes the
above strict theoretical proofs simple and convenient. However, in practice
this can be made unnecessary.

The symmetric structure and property of a quantum system plays an important
role in simplifying the construction (or preparation) of the $SIC$ unitary
propagators in the quantum system. For example, the preparation for the $QM$
subspace-selective unitary operators $W_{12}^{ap}$ of (3.60) and $%
W_{23}^{ap} $ of (3.80) could be further simplified with the aid of the
multiple-quantum-transition symmetric structure and property [$7,6$].

It is expected that the joint construction method can be applied as well to
other quantum system in time and space and with discrete energy spectrum
than a single-atom system to construct a $SIC$ unitary propagator. \newline
\newline
\newline
{\Large 4. The }${\Large SIC}${\Large \ unitary propagators in the
coordinate representation}

In the preceding section it is described in detail how a $SIC$ unitary
propagator is prepared concretely in a single-atom system. When the $SIC$\
unitary propagators are further used to construct a powerful QUANSDAM
process in the quantum system, theoretically it is necessary to calculate
explicitly the sequence of the $SIC$\ unitary propagators of the QUANSDAM
process and the time evolution process when the sequence is applied to the
quantum system. As proposed in Ref. [$1$], the Green function method in
quantum mechanics [$10,14$] could be a general theoretical method better for
this theoretical calculation. However, in unitary quantum dynamics the Green
function method can become a general theoretical method to construct a
QUANSDAM (or UNIDYSLOCK) process, only if at the first step one is able to
calculate theoretically any $SIC$\ Green functions, i.e., any $SIC$ unitary
propagators in the coordinate representation, in a quantum system in time
and space.

In this section a general theory is established to treat theoretically and
calculate the $SIC$ Green functions of a quantum system in time and space.
It is mainly based on the conventional path integral technique (i.e., the
Feynman path integral technique [$14,20$]). The energy eigenfunction
expansion method also is an important component of the theory. Especially,
as typical applications of the theory, in this section the $SIC$\ quadratic
propagators of a quantum system such as a single-atom system are calculated
exactly and explicitly in the coordinate representation. The present work is
basic for further constructing an exponential QUANSDAM process in a quantum
system (e.g., a single-atom system) in future.

As shown in (3.2), a solution-information-carrying unitary propagator under
study may be defined by [$1$] 
\begin{equation}
U_{H}^{sic}(a_{m}^{s},T_{m})=\exp \left( -ia_{m}^{s}HT_{m}/\hslash \right) 
\tag{4.1}
\end{equation}%
where the $QM$ Hamiltonian $H$ is time-independent and $T_{m}$ is time
interval. This $SIC$\ unitary propagator owns the time-evolution-reversal
symmetry in the sense that $U_{H}^{sic}(+1,T_{m})=U_{H}^{sic}(-1,T_{m})^{+}.$
In a general case the Hamiltonian $H$ in (4.1) may be dependent upon the
double-valued logical number, i.e., $H=H(a_{m}^{s}),$ but this general case
will not be discussed below. In the coordinate representation the $SIC$
unitary propagator of (4.1) is written as%
\begin{equation}
G_{H}^{sic}(x_{b},t_{b};x_{a},t_{a})=\left\langle x_{b},t_{b}\right\vert
U_{H}^{sic}(a_{m}^{s},T_{m})\left\vert x_{a},t_{a}\right\rangle  \tag{4.2}
\end{equation}%
where $\left\vert x_{a},t_{a}\right\rangle $ $\left( \text{or }\left\vert
x_{b},t_{b}\right\rangle \right) $ is the initial (or the final) position or
coordinate state and the time interval $t_{b}-t_{a}=T_{m}$ if $a_{m}^{s}=+1$
and $t_{b}-t_{a}=-T_{m}$ if $a_{m}^{s}=-1$ or simply $%
t_{b}-t_{a}=a_{m}^{s}T_{m}$\ in value. The $SIC$\ unitary propagator in the
coordinate representation, i.e., $G_{H}^{sic}(x_{b},t_{b};x_{a},t_{a}),$
also may be simply called the $SIC$\ Green function \footnote{%
In quantum mechanics the propagator $U(T_{m})=\exp \left( -iHT_{m}/\hslash
\right) $ is usually called the (unitary) propagator in operator form. In
the coordinate representation it is given by $G(x_{b},t_{b};x_{a},t_{a})=%
\left\langle x_{b},t_{b}\right\vert U(T_{m})\left\vert
x_{a},t_{a}\right\rangle $ which is usually called the propagator in the
coordinate representation or the Green function in the time domain or simply
the Green function.}. Given the $SIC$\ unitary propagator of (4.1), the
present theoretical work in this section is to determine explicitly the $SIC$%
\ Green function of (4.2). Here it should be pointed out that the $QM$
Hamiltonian $H$ of the $SIC$\ unitary propagator of (4.1) may be the natural
one or the effective one of a quantum system in time and space. The present
theoretical work does not involve how to prepare and realize explicitly the $%
SIC$ unitary propagator in a quantum system. Therefore, the present
theoretical treatment or calculation is general for the $SIC$\ Green
function of (4.2).

Here it is emphasized that a\ $SIC$\ unitary propagator in the coordinate
representation still owns the dual character of quantum-computing speedup.
This is the same as the $SIC$ unitary propagator itself. This means that a $%
SIC$ Green function obeys both the unitary quantum dynamics and the
mathematical-logical principle of the unstructured search problem.

According to the Green function method in quantum mechanics [$10,14$] the
time evolution process governed by the $SIC$\ unitary propagator $%
U_{H}^{sic}(a_{m}^{s},T_{m})$ of (4.1) may be expressed as%
\begin{equation}
\Psi ^{sic}(x_{b},t_{b})=\int_{-\infty }^{\infty
}dx_{a}G_{H}^{sic}(x_{b},t_{b};x_{a},t_{a})\Psi ^{sic}(x_{a},t_{a}). 
\tag{4.3}
\end{equation}%
Here the $SIC$ Green function$\ G_{H}^{sic}(x_{b},t_{b};x_{a},t_{a})$ is
just the $SIC$ unitary propagator $U_{H}^{sic}(a_{m}^{s},T_{m})$ in the
coordinate representation and is given by (4.2), and the initial state $\Psi
^{sic}(x_{a},t_{a})$ may or may not contain the double-valued logical number 
$a_{m}^{s}.$ Once the $SIC$\ Green function $%
G_{H}^{sic}(x_{b},t_{b};x_{a},t_{a})$ is obtained from (4.2), by
substituting it into (4.3) one may calculate the time evolution process
(4.3). The time evolution process (4.3) is a unitary and deterministic
process and obeys the information conservation law.

A simplest $SIC$\ unitary propagator of (4.1) should be diagonal in the
coordinate representation from the perspective of the theoretical
calculation of the $SIC$\ Green function of (4.2). Then a $SIC$\ unitary
propagator is usually simplest if it is generated by the potential-energy
operator of a quantum system which is usually a function of the coordinate
operators. Though the $SIC$\ unitary propagator that is generated by the
kinetic-energy operator of a quantum system is not diagonal in the
coordinate representation, it is also simplest. These two types of simplest $%
SIC$\ unitary propagators may be explicitly given by%
\begin{equation}
\left\{ 
\begin{array}{c}
U_{\hat{V}}^{sic}(a_{m}^{s},T_{m})=\exp \left( -ia_{m}^{s}\hat{V}%
T_{m}/\hslash \right) \\ 
U_{\hat{T}}^{sic}(a_{m}^{s},T_{m})=\exp \left( -ia_{m}^{s}\hat{T}%
T_{m}/\hslash \right)%
\end{array}%
\right.  \tag{4.4}
\end{equation}%
where $\hat{T}=\frac{1}{2m}p^{2}$ and $\hat{V}=V\left( x\right) $ are the
kinetic and potential operators, respectively; $H=\hat{T}+\hat{V}$ is the
Hamiltonian of the quantum system; and $p$ and $x$ are the momentum and the
coordinate operator of the quantum system, respectively. These $QM$
operators $H,$ $\hat{T},$ and $\hat{V}\ $may be the natural ones or the
effective ones of the quantum system. Then the simplest $SIC$ Green
functions corresponding to the simplest $SIC$\ unitary propagators of (4.4)
may be exactly obtained by substituting (4.4) into (4.2), respectively.
However, a $SIC$\ unitary propagator in the coordinate representation is
generally complicated except for the simple or special cases. Many $SIC$\
Green functions of (4.2) may be exactly calculated just like the simplest $%
SIC$ Green functions, but most of (4.2) are difficult to be calculated
exactly.

Consider the special case that the $QM$ Hamiltonian in (4.1) is taken as the
potential operator $\hat{V}.$ In this case the $U_{H}^{sic}(a_{m}^{s},T_{m})$
of (4.1) is given by the $U_{\hat{V}}^{sic}(a_{m}^{s},T_{m})$ in (4.4).
Since the $SIC$\ unitary propagator $U_{\hat{V}}^{sic}(a_{m}^{s},T_{m})$ has
the eigenfunctions that are the coordinate states $\{|x\rangle \}$, that is,
it is diagonal, its corresponding $SIC$ Green function$\ G_{\hat{V}%
}^{sic}(x_{b},t_{b};x_{a},t_{a})$ is simplest and can be calculated exactly.
After substituting the $U_{\hat{V}}^{sic}(a_{m}^{s},t_{m})$ of (4.4) into
(4.2), this $G_{\hat{V}}^{sic}(x_{b},t_{b};x_{a},t_{a})$ may be simply
written as%
\begin{equation}
G_{\hat{V}}^{sic}(x_{b},t_{b};x_{a},t_{a})=\exp \left( -ia_{m}^{s}V\left(
x_{a}\right) T_{m}/\hslash \right) \delta \left( x_{b}-x_{a}\right) , 
\tag{4.5}
\end{equation}%
where the eigenvalue equation $U_{\hat{V}}^{sic}(a_{m}^{s},t_{m})\left\vert
x_{a},t_{a}\right\rangle =\exp \left( -ia_{m}^{s}V\left( x_{a}\right)
T_{m}/\hslash \right) \times $ $\left\vert x_{a},t_{a}\right\rangle $ is
already used. It can be seen that in the determination of the $G_{\hat{V}%
}^{sic}(x_{b},t_{b};$ $x_{a},t_{a})$ one need not consider explicitly the
double-valued logical number $a_{m}^{s}=\pm 1.$ Generally, if the
eigenfunctions of the Hamiltonian $H$ of the $U_{H}^{sic}(a_{m}^{s},T_{m})$
of (4.1) are the coordinate states, then the corresponding $SIC$\ Green
function $G_{H}^{sic}(x_{b},t_{b};x_{a},t_{a})$ of (4.2) also may be
expressed as (4.5) where the eigenvalue $V\left( x_{a}\right) $ of the
potential operator $\hat{V}$ is replaced with the eigenvalue $H\left(
x_{a}\right) $ of the Hamiltonian $H$.

The situation becomes generally complicated in the determination of the $SIC$
Green function if the eigenfunctions of the Hamiltonian $H$ are not the
coordinate states, but for the simple or special cases one still may
calculate exactly the $SIC$\ Green functions. Consider the simplest case
that the Hamiltonian $H$ is taken as the kinetic operator $\hat{T}$, that
is, $H=\hat{T}=\frac{1}{2m}p^{2}.$ As a typical example, this Hamiltonian
may be used to describe the free center-of-mass (COM) motion of a single
atom with mass $m$ in one-dimensional coordinate space. In this simple case
one has $U_{H}^{sic}(a_{m}^{s},T_{m})=U_{\hat{T}}^{sic}(a_{m}^{s},T_{m})$
which is given in (4.4). Evidently the Hamiltonian $\hat{T}$ does not have
the coordinate eigenfunctions. Now given the Hamiltonian $\hat{T}$ one wants
to determine the $SIC$\ unitary propagator $U_{\hat{T}%
}^{sic}(a_{m}^{s},T_{m})$ in the coordinate representation. The $SIC$ Green
function $G_{\hat{T}}^{sic}(x_{b},t_{b};x_{a},t_{a})$ corresponding to the $%
U_{\hat{T}}^{sic}(a_{m}^{s},T_{m})$ may be written as, after the identical
operator $\int_{-\infty }^{\infty }dp\left\vert p\right\rangle \left\langle
p\right\vert $ of the momentum eigenbases $\left\{ \left\vert p\right\rangle
\right\} $ is inserted into (4.2),%
\begin{equation}
G_{\hat{T}}^{sic}(x_{b},t_{b};x_{a},t_{a})=\int_{-\infty }^{\infty
}dp\left\langle x_{b},t_{b}\right\vert U_{\hat{T}}^{sic}(a_{m}^{s},T_{m})%
\left\vert p\right\rangle \left\langle p|x_{a},t_{a}\right\rangle . 
\tag{4.6}
\end{equation}%
Here the momentum eigenbase $\left\vert p\right\rangle $ also is an
eigenbase of the kinetic operator $\hat{T},$ that is, $\hat{T}\left\vert
p\right\rangle =\frac{1}{2m}p^{2}\left\vert p\right\rangle .$ Therefore, the
eigenvalue equation for the $U_{\hat{T}}^{sic}(a_{m}^{s},T_{m})$ is given by%
\begin{equation}
U_{\hat{T}}^{sic}(a_{m}^{s},T_{m})\left\vert p\right\rangle =\exp \left( -i%
\frac{p^{2}}{2m\hslash }\left( a_{m}^{s}T_{m}\right) \right) \left\vert
p\right\rangle .  \tag{4.7}
\end{equation}%
In (4.6) one also has the eigenfunctions $\left\langle
p|x_{a},t_{a}\right\rangle =\frac{1}{\sqrt{2\pi \hslash }}\exp \left(
-ipx_{a}/\hslash \right) $ and $\left\langle x_{b},t_{b}|p\right\rangle =%
\frac{1}{\sqrt{2\pi \hslash }}\exp \left( ipx_{b}/\hslash \right) .$ Now by
substituting the eigenvalue equation (4.7) and the eigenfunctions $%
\left\langle x_{b},t_{b}|p\right\rangle $ and $\left\langle
p|x_{a},t_{a}\right\rangle $ into (4.6) the $G_{\hat{T}}^{sic}(x_{b},t_{b};$ 
$x_{a},t_{a})$ is reduced to the form%
\begin{equation}
G_{\hat{T}}^{sic}(x_{b},t_{b};x_{a},t_{a})=\frac{1}{2\pi \hslash }%
\int_{-\infty }^{\infty }dp\exp \left( -i\frac{p^{2}}{2m\hslash }\left(
a_{m}^{s}T_{m}\right) +i\left( x_{b}-x_{a}\right) p/\hslash \right) . 
\tag{4.8}
\end{equation}%
With the help of the Gaussian integration formula: $\int_{-\infty }^{\infty
}dp\exp \left( -ap^{2}+bp\right) =\sqrt{\frac{\pi }{a}}\exp \left( \frac{%
b^{2}}{4a}\right) $ the formula (4.8) is further reduced to the form%
\begin{equation}
G_{\hat{T}}^{sic}(x_{b},t_{b};x_{a},t_{a})=\sqrt{\frac{m}{i2\pi \hslash
\left( a_{m}^{s}T_{m}\right) }}\exp \left( \frac{im}{2\hslash }\frac{\left(
x_{b}-x_{a}\right) ^{2}}{\left( a_{m}^{s}T_{m}\right) }\right)  \tag{4.9}
\end{equation}%
This is the final expression for the $SIC$\ unitary propagator $U_{\hat{T}%
}^{sic}(a_{m}^{s},T_{m})$ in (4.4) in the coordinate representation.

The $SIC$ unitary propagator of (4.1) that is generated by a $QM$ quadratic
Hamiltonian is called a $SIC$\ unitary quadratic propagator. Here the $QM$
quadratic Hamiltonian is a quadratic function of the coordinate operators
and/or their canonically conjugate momentum operators. Actually, the $U_{%
\hat{T}}^{sic}(a_{m}^{s},T_{m})$ in (4.4) with the kinetic operator $\hat{T}=%
\frac{1}{2m}p^{2}$ is a $SIC$\ quadratic propagator. If the potential
operator $\hat{V}$ of the $U_{\hat{V}}^{sic}(a_{m}^{s},T_{m})$ in (4.4) is
taken as $\hat{V}_{c}=\frac{1}{2}m\omega ^{2}x^{2},$ then the $U_{\hat{V}%
_{c}}^{sic}(a_{m}^{s},T_{m})$ is also a $SIC$ quadratic propagator. These
two $SIC$ quadratic propagators are very important in the construction of a
powerful QUANSDAM process. Beside these two $SIC$ quadratic propagators a
third $SIC$ quadratic propagator that is simple and also very important is
given by%
\begin{equation}
U_{H_{c}}^{sic}(a_{m}^{s},T_{m})=\exp \left[ -ia_{m}^{s}H_{c}T_{m}/\hslash %
\right]  \tag{4.10}
\end{equation}%
with the $QM$ quadratic Hamiltonian $H_{c}=\hat{T}+\hat{V}_{c}.$ As a
typical example, quantum\ mechanically this Hamiltonian $H_{c}$ may be the
natural one that describes the harmonic COM motion of a single atom in
external harmonic potential field in one-dimensional coordinate space (See
(3.7a) above). As shown in the section 3, this $SIC$ quadratic propagator
can be prepared and realized effectively in a single-atom system. Evidently
it does not own the coordinate eigenfunctions and hence it is not diagonal
in the coordinate representation. Below one wants to determine the explicit
expression of the $SIC$ quadratic Green function corresponding to the $SIC$
quadratic propagator of (4.10).

First of all, one may employ the energy eigenfunction expansion principle [$%
10,14,13$] to determine the $SIC$\ quadratic Green function corresponding to
(4.10). Given the Hamiltonian $H_{c}$ (See (3.7a) above) quantum
mechanically [$10$] one may solve the energy eigenvalue equation $%
H_{c}u_{k}\left( x\right) =E_{k}^{c}u_{k}\left( x\right) $ and as shown in
the subsection 3.1, the energy eigenvalue $E_{k}^{c}$ is given by $%
E_{k}^{c}=\left( k+1/2\right) \hslash \omega $ and its associated energy
eigenbase $u_{k}\left( x\right) $ is given by (3.7b). The energy eigenbases $%
\{u_{k}\left( x\right) \}$ satisfy the orthogonal and normalized relations [$%
10$]:%
\begin{equation}
\int_{-\infty }^{\infty }dxu_{k}^{\ast }\left( x\right) u_{l}\left( x\right)
=\delta _{kl}  \tag{4.11}
\end{equation}%
and%
\begin{equation}
\sum_{k=0}^{\infty }u_{k}^{\ast }\left( x\right) u_{k}\left( y\right)
=\delta \left( x-y\right) .  \tag{4.12}
\end{equation}%
Now with the help of the complete orthonormal set of the eigenbases $%
\{u_{k}\left( x\right) \}$ and the eigenvalues $\{E_{k}^{c}\}$ the $SIC$\
quadratic propagator of (4.10) in the coordinate representation may be
expressed as%
\begin{equation}
G_{H_{c}}^{sic}(x_{b},t_{b};x_{a},t_{a})=\sum_{k=0}^{\infty }u_{k}^{\ast
}\left( x_{a}\right) u_{k}\left( x_{b}\right) \exp \left(
-ia_{m}^{s}E_{k}^{c}T_{m}/\hslash \right)  \tag{4.13}
\end{equation}%
This expression is obtained by first inserting the identical operator $%
\sum_{k=0}^{\infty }|u_{k}\rangle \langle u_{k}|$ into (4.2), where $\left\{
|u_{k}\rangle \right\} $ is the complete orthonormal set of the eigenvectors
(or eigenbases) of the Hamiltonian $H_{c},$ then using the eigenvalue
equation:%
\begin{equation}
U_{H_{c}}^{sic}(a_{m}^{s},T_{m})|u_{k}\rangle =\exp \left(
-ia_{m}^{s}E_{k}^{c}T_{m}/\hslash \right) |u_{k}\rangle ,  \tag{4.14}
\end{equation}%
and finally using $u_{k}\left( x_{b}\right) =\left\langle
x_{b},t_{b}|u_{k}\right\rangle $ and $u_{k}^{\ast }\left( x_{a}\right)
=\left\langle u_{k}|x_{a},t_{a}\right\rangle .$ The expression (4.13) needs
to be further calculated. By substituting the eigenfunction $u_{k}\left(
x\right) $ given in (3.7b) in the subsection 3.1 and the eigenvalue $%
E_{k}^{c}=\left( k+1/2\right) \hslash \omega $ into (4.13) and then letting $%
s=\exp \left( -ia_{m}^{s}\omega T_{m}\right) ,$ $\theta _{s}=\omega
a_{m}^{s}T_{m},$ and $\xi =\alpha x,$ the expression (4.13) may be reduced
to the form%
\begin{equation*}
G_{H_{c}}^{sic}(x_{b},t_{b};x_{a},t_{a})=\left( \frac{\alpha }{\sqrt{\pi }}%
\right) \exp \left( -i\frac{1}{2}\theta _{s}\right) \exp \left( -\frac{1}{2}%
\left( \xi _{a}^{2}+\xi _{b}^{2}\right) \right)
\end{equation*}%
\begin{equation}
\times \sum_{k=0}^{\infty }\frac{H_{k}\left( \xi _{a}\right) H_{k}\left( \xi
_{b}\right) }{2^{k}k!}s^{k}  \tag{4.15}
\end{equation}%
This formula can be further simplified by using the Mehler formula [$21,22$]
for the Hermitian polynomials $\left\{ H_{k}\left( \xi \right) \right\} $.

The Mehler formula for the real Hermitian polynomials $\left\{ H_{k}\left(
\xi \right) \right\} $ may be written as [$21$]%
\begin{equation}
\sum_{k=0}^{\infty }\frac{H_{k}\left( x\right) H_{k}\left( y\right) }{2^{k}k!%
}s^{k}\newline
=\frac{1}{\sqrt{1-s^{2}}}\exp \left[ -\frac{s^{2}\left( x^{2}+y^{2}\right)
-2sxy}{1-s^{2}}\right]  \tag{4.16}
\end{equation}%
where the parameter $s$ may be complex and $\left\vert s\right\vert <1$.
With the help of the Gaussian integration one also can prove the Mehler
formula [$22$]. The Mehler formula exists if $\left\vert s\right\vert <1,$
because the infinite power series of the parameter $s$ on the left-hand side
of (4.16) is convergent if $\left\vert s\right\vert <1.$ If both $x$ and $y$
are real in (4.16), then the infinite power series converges on the
unit-circle in the $s-$plane except at points $s=\pm 1.$ If $s=\pm 1,$ the
infinite power series may be divergent [$21$]. However, it can prove that
for the case that $s=\pm 1$ and both $x$ and $y$ are real the Mehler formula
of (4.16) can be reduced to the orthonormal relation (4.12) of the
eigenbases $\{u_{k}\left( x\right) \}$. Therefore, when both $x$ and $y$ are
real, the Mehler formula of (4.16) can be used if $\left\vert s\right\vert
\leq 1.$

Now by using the Mehler formula (4.16) one may exactly calculate the
infinite power series on the right-hand side of (4.15). It is known that the
variable $\xi =\alpha x$ is real, that is, both $\xi _{a}=\alpha x_{a}$ and $%
\xi _{b}=\alpha x_{b}$ are real in (4.15). In (4.15) one has $\left\vert
s\right\vert =1.$ Then with the help of the Mehler formula (4.16) with $%
x=\xi _{a},$ $y=\xi _{b},$ and $s=\exp \left( -ia_{m}^{s}\omega T_{m}\right) 
$ the expression (4.15) can be reduced to the form%
\begin{equation*}
G_{H_{c}}^{sic}(x_{b},t_{b};x_{a},t_{a})=\frac{1}{\sqrt{1-s^{2}}}\left( 
\frac{\alpha }{\sqrt{\pi }}\right) \exp \left( -i\frac{1}{2}\theta
_{s}\right)
\end{equation*}%
\begin{equation}
\times \exp \left[ -\frac{1}{2}\xi _{a}^{2}-\frac{1}{2}\xi _{b}^{2}-\frac{%
s^{2}\left( \xi _{a}^{2}+\xi _{b}^{2}\right) -2s\xi _{a}\xi _{b}}{1-s^{2}}%
\right]  \tag{4.17}
\end{equation}%
\newline
A direct calculation shows that%
\begin{equation*}
\frac{1}{2}\xi _{a}^{2}+\frac{1}{2}\xi _{b}^{2}+\frac{s^{2}\left( \xi
_{a}^{2}+\xi _{b}^{2}\right) -2s\xi _{a}\xi _{b}}{1-s^{2}}
\end{equation*}%
\begin{equation}
=\left( \frac{m\omega }{i2\hslash \sin \theta _{s}}\right) \left[ \left(
x_{a}^{2}+x_{b}^{2}\right) \allowbreak \cos \theta _{s}-2x_{a}x_{b}\right] 
\tag{4.18}
\end{equation}%
and%
\begin{equation}
\frac{1}{\sqrt{1-s^{2}}}\left( \frac{\alpha }{\sqrt{\pi }}\right) \exp
\left( -i\frac{1}{2}\theta _{s}\right) =\sqrt{\frac{m\omega }{i2\pi \hslash
\sin \theta _{s}}}  \tag{4.19}
\end{equation}%
By substituting (4.18) and (4.19) into (4.17) the expression (4.17) is
further written in the form%
\begin{equation*}
G_{H_{c}}^{sic}(x_{b},t_{b};x_{a},t_{a})=\sqrt{\frac{m\omega }{i2\pi \hslash
\sin \left( \omega a_{m}^{s}T_{m}\right) }}
\end{equation*}%
\begin{equation}
\times \exp \left[ \frac{im\omega }{2\hslash \sin \left( \omega
a_{m}^{s}T_{m}\right) }\left( \left( x_{a}^{2}+x_{b}^{2}\right) \allowbreak
\cos \left( \omega a_{m}^{s}T_{m}\right) -2x_{a}x_{b}\right) \right] 
\tag{4.20}
\end{equation}%
This is the final expression for the $SIC$ unitary propagator $%
U_{H_{c}}^{sic}(a_{m}^{s},T_{m})$ of (4.10) in the coordinate representation.

Actually the two $SIC$ Green functions of (4.9) and (4.20) are obtained
exactly by using the energy eigenfunction expansion method in quantum
mechanics [$10,14,13$], respectively. These are the two typical applications
of the method. This general method to calculate a $SIC$ Green function is
simply described as follows. Suppose that the energy eigenvalue equation for
the $QM$\ Hamiltonian $H$ of the $U_{H}^{sic}(a_{m}^{s},T_{m})$ of (4.1) is
given by $H\left\vert \psi _{k}\right\rangle =E_{k}\left\vert \psi
_{k}\right\rangle \ $or $H\psi _{k}\left( x\right) =E_{k}\psi _{k}\left(
x\right) $ in the coordinate representation. Here all the eigenbases $%
\{\left\vert \psi _{k}\right\rangle \}$ (or $\{\psi _{k}\left( x\right) \}$)
form a complete orthonormal basis set. Then the eigenvalue equation for the $%
U_{H}^{sic}(a_{m}^{s},T_{m})$ may be written as%
\begin{equation}
U_{H}^{sic}(a_{m}^{s},T_{m})\left\vert \psi _{k}\right\rangle =\exp \left(
-ia_{m}^{s}E_{k}T_{m}/\hslash \right) \left\vert \psi _{k}\right\rangle . 
\tag{4.21}
\end{equation}%
Now by inserting the identity operator $\sum_{k=0}^{\infty }\left\vert \psi
_{k}\right\rangle \left\langle \psi _{k}\right\vert $ of the energy
eigenbases (or eigenvectors) $\{\left\vert \psi _{k}\right\rangle \}$ into
(4.2) and then using the eigenvalue equation (4.21) one obtains%
\begin{equation}
G_{H}^{sic}(x_{b},t_{b};x_{a},t_{a})=\sum_{k=0}^{\infty }\exp \left(
-ia_{m}^{s}E_{k}T_{m}/\hslash \right) \psi _{k}^{\ast }\left( x_{a}\right)
\psi _{k}\left( x_{b}\right)  \tag{4.22}
\end{equation}%
where the orthonormal eigenfunctions $\psi _{k}\left( x_{b}\right)
=\left\langle x_{b},t_{b}|\psi _{k}\right\rangle $ and $\psi _{k}^{\ast
}\left( x_{a}\right) =\left\langle \psi _{k}|x_{a},t_{a}\right\rangle $ are
already used. Theoretically the formula (4.22) can be used to calculate any $%
SIC$\ Green function. However, the key point for the energy eigenfunction
expansion method to be successful is that one can calculate exactly and
conveniently the infinite series on the RH side of (4.22). The simple $SIC$
Green functions of (4.9) and (4.20) can be obtained exactly by this method,
because their infinite series (or integration) (4.8) and (4.15) can be
exactly calculated conveniently. But this needs to use some tricky. For a
general case this may not be always carried out conveniently.

By using explicitly the double-valued logical number $a_{m}^{s}=\pm 1$ the $%
SIC$\ unitary propagator of (4.1) may be written value-equivalently as%
\begin{equation}
U_{H}^{sic}(a_{m}^{s},T_{m})=\left\{ 
\begin{array}{c}
\exp \left( -iHT_{m}/\hslash \right) \text{ } \\ 
\exp \left( +iHT_{m}/\hslash \right)%
\end{array}%
\begin{array}{c}
if \\ 
\text{ }if%
\end{array}%
\begin{array}{c}
a_{m}^{s}=+1 \\ 
a_{m}^{s}=-1%
\end{array}%
\right.  \tag{4.23}
\end{equation}%
One sees from (4.23) that the $U_{H}^{sic}(a_{m}^{s},t_{m})$ with $%
a_{m}^{s}=+1$ is the inverse of the $U_{H}^{sic}(a_{m}^{s},t_{m})$ with $%
a_{m}^{s}=-1$ and vice versa. It is known from the mathematical-logical
principle of the unstructured search problem that the real solution to the
unstructured search problem exists, but it is not known which one of the two
cases $a_{m}^{s}=+1$ and $a_{m}^{s}=-1$ is assigned to the real solution
before the solution information $\left( a_{m}^{s}\right) $ is extracted
quantum-computingly. Equivalently speaking, any one of the two cases $%
a_{m}^{s}=\pm 1$ may be possibly considered as the one of the real solution
before the information $\left( a_{m}^{s}\right) $ of the real solution is
extracted. However, it is known from (4.23) that, given the Hamiltonian $H$,
in either case one always can calculate (exactly or approximately) the $SIC$
unitary propagator $U_{H}^{sic}(a_{m}^{s},T_{m})$ with $a_{m}^{s}=+1$ in the
coordinate representation, and likewise in either case one always can
calculate the $U_{H}^{sic}(a_{m}^{s},T_{m})$ with $a_{m}^{s}=-1$ in the
coordinate representation.

For convenience denote $G_{\pm
1}^{sic}(x_{b},t_{b};x_{a},t_{a})=\left\langle x_{b},t_{b}\right\vert
U_{H}^{sic}(\pm 1,T_{m})\left\vert x_{a},t_{a}\right\rangle $ as the $SIC$\
Green functions with $a_{m}^{s}=\pm 1$, respectively, here $U_{H}^{sic}(\pm
1,T_{m})$ is given by (4.23). Suppose now that the $%
G_{+1}^{sic}(x_{b},t_{b};x_{a},t_{a})$ is already calculated directly in
quantum mechanics. Then the $G_{-1}^{sic}(x_{b},t_{b};x_{a},t_{a})$ may be
calculated simply as follows. Actually, according to quantum mechanics one
has%
\begin{equation}
G_{-1}^{sic}(x_{b},t_{b};x_{a},t_{a})=G_{+1}^{sic}(x_{a},t_{a};x_{b},t_{b})^{\ast }.
\tag{4.24a}
\end{equation}%
The RH side of (4.24a) is the inverse of the $%
G_{+1}^{sic}(x_{b},t_{b};x_{a},t_{a}),$ while the LH side is just the $%
G_{-1}^{sic}(x_{b},t_{b};x_{a},t_{a})$. Thus, the $%
G_{-1}^{sic}(x_{b},t_{b};x_{a},t_{a})$ may be simply calculated from
(4.24a). Of course, it also can be directly calculated in quantum mechanics.
Similarly, the $G_{+1}^{sic}(x_{b},t_{b};x_{a},t_{a})$ may be simply
calculated by the formula:%
\begin{equation}
G_{+1}^{sic}(x_{b},t_{b};x_{a},t_{a})=G_{-1}^{sic}(x_{a},t_{a};x_{b},t_{b})^{\ast },
\tag{4.24b}
\end{equation}%
if the $G_{-1}^{sic}(x_{b},t_{b};x_{a},t_{a})$ is already calculated in
quantum mechanics.

Below the relation is set up between the time interval $T_{m}$ and both the
initial time $t_{a}$ and final time $t_{b}$ in the $SIC$\ Green function of
(4.2) or (4.22). Consider the setting that the time interval $T_{m}\geq 0$
in (4.2) or (4.22). It is known from (4.2) and (4.23) that the $%
G_{+1}^{sic}(x_{b},t_{b};x_{a},t_{a})$ is generated by the unitary
propagator $\exp \left( -iHT_{m}/\hslash \right) $ with $T_{m}\geq 0.$ Then
there is $T_{m}=t_{b}-t_{a}\geq 0.$ This corresponds to the case $%
a_{m}^{s}=+1$. On the other hand, it is also known from (4.2) and (4.23)
that the $G_{-1}^{sic}(x_{b},t_{b};x_{a},t_{a})$ is obtained from $\exp
\left( +iHT_{m}/\hslash \right) $ which is the inverse of $\exp \left(
-iHT_{m}/\hslash \right) .$ Therefore, there is $T_{m}=t_{a}-t_{b}\geq 0$ or 
$-T_{m}=t_{b}-t_{a}\leq 0.$ This corresponds to the case $a_{m}^{s}=-1$.
These results for the two cases $a_{m}^{s}=\pm 1$ together show that in the
present setting $T_{m}\geq 0$ there is the relation $%
a_{m}^{s}T_{m}=t_{b}-t_{a}$ in value. Next consider another setting that the
time interval $T_{m}<0$ in (4.2) or (4.22). It can be deduced from (4.2) and
(4.23) that the $G_{-1}^{sic}(x_{b},t_{b};x_{a},t_{a})$ is generated by the
unitary propagator $\exp \left( -iH\left\vert T_{m}\right\vert /\hslash
\right) $ with $\left\vert T_{m}\right\vert >0$. Then one has $\left\vert
T_{m}\right\vert =t_{b}-t_{a}>0$ or $-T_{m}=t_{b}-t_{a}>0.$ This corresponds
to the case $a_{m}^{s}=-1.$ On the other hand, it is also known from (4.2)
and (4.23) that the $G_{+1}^{sic}(x_{b},t_{b};x_{a},t_{a})$ is generated by $%
\exp \left( +iH\left\vert T_{m}\right\vert /\hslash \right) $ which is the
inverse of $\exp \left( -iH\left\vert T_{m}\right\vert /\hslash \right) .$
Thus, there is $\left\vert T_{m}\right\vert =t_{a}-t_{b}>0$ or $%
T_{m}=t_{b}-t_{a}<0.$ This corresponds to the case $a_{m}^{s}=+1.$ These
results for both the cases $a_{m}^{s}=\pm 1$ together show that in the
present setting $T_{m}<0$ there is the relation $a_{m}^{s}T_{m}=t_{b}-t_{a}$
in value. It therefore concludes that in either setting (i.e., either $%
T_{m}\geq 0$ or $T_{m}<0)$ there is always the relation $%
a_{m}^{s}T_{m}=t_{b}-t_{a}$ in value between the time interval $T_{m}$ and
both the initial time $t_{a}$ and final time $t_{b}$ in the $SIC$ Green
function of (4.2) or (4.22).

Now it is proposed to combine the energy eigenfunction expansion method with
the Feynman\ path integral technique [$14$] to determine a more general $SIC$
Green function. This joint method to determine exactly a $SIC$ Green
function is described below.

Since the Hamiltonian of the $U_{H}^{sic}(a_{m}^{s},T_{m})$ of (4.1) is
time-independent, both the $SIC$ unitary propagator $%
U_{H}^{sic}(a_{m}^{s},T_{m})$ and its $SIC$ Green function $%
G_{H}^{sic}(x_{b},t_{b};x_{a},t_{a})$ (or $G_{\pm
1}^{sic}(x_{b},t_{b};x_{a},t_{a}))$ of (4.2) are not dependent upon the
initial (or final) time $t_{a}$ $\left( \text{or }t_{b}\right) $ separately
but dependent upon the time interval $t_{b}-t_{a}.$ This is the same as a $%
QM $\ unitary propagator with time-independent Hamiltonian and its $QM$
Green function. The $QM$\ unitary propagator $U_{H}(T_{m})=\exp \left(
-iHT_{m}/\hslash \right) $ corresponds to the $SIC$ unitary propagator $%
U_{H}^{sic}(a_{m}^{s},T_{m})$. Its $QM$\ Green function $%
G_{H}(x_{b},t_{b};x_{a},t_{a})=\left\langle x_{b},t_{b}\right\vert
U_{H}(T_{m})\left\vert x_{a},t_{a}\right\rangle $ corresponds to the $SIC$
Green function $G_{H}^{sic}(x_{b},t_{b};x_{a},t_{a})$. According to the
energy eigenfunction expansion method this $QM$\ Green function may be
formally expressed as [$10,14$]%
\begin{equation}
G_{H}(x_{b},t_{b};x_{a},t_{a})=\sum_{k=0}^{\infty }\exp \left(
-iE_{k}T_{m}/\hslash \right) \psi _{k}^{\ast }\left( x_{a}\right) \psi
_{k}\left( x_{b}\right) .  \tag{4.25}
\end{equation}%
One sees that this $QM$\ Green function $G_{H}(x_{b},t_{b};x_{a},t_{a})$
corresponds to the $SIC$ Green function $%
G_{H}^{sic}(x_{b},t_{b};x_{a},t_{a}) $ of (4.22). It is known from (4.22)
and (4.25) that the $G_{H}(x_{b},t_{b};x_{a},t_{a})$ is a function of the
time interval $t_{b}-t_{a}=T_{m},$ while the $%
G_{H}^{sic}(x_{b},t_{b};x_{a},t_{a})$ is a function of the time interval $%
t_{b}-t_{a}=a_{m}^{s}T_{m}.$ By comparing (4.25) with (4.22) it can be found
that if one replaces the time interval $T_{m}$ in the $%
G_{H}(x_{b},t_{b};x_{a},t_{a})$ of (4.25) with the time interval $%
a_{m}^{s}T_{m},$ then this $G_{H}(x_{b},t_{b};x_{a},t_{a})$ is exactly
changed to the $SIC$ Green function $G_{H}^{sic}(x_{b},t_{b};x_{a},t_{a})$
of (4.22). Suppose now that one can explicitly determine the exact
expression of the $QM$ Green function of (4.25) in quantum mechanics. Then
in the exact expression one replaces the time interval $T_{m}$ with $%
a_{m}^{s}T_{m}.$ Such replacement, i.e., $T_{m}\rightarrow a_{m}^{s}T_{m},$
leads to that the exact expression of the $QM$ Green function $%
G_{H}(x_{b},t_{b};x_{a},t_{a})$ of (4.25) is exactly changed to the one of
the $SIC$ Green function $G_{H}^{sic}(x_{b},t_{b};x_{a},t_{a})$ of (4.22).

The correspondent relation between the $SIC$ Green function of (4.22) and
the $QM$\ one of (4.25) may be employed to calculate a general $SIC$ Green
function of (4.22), if the $QM$\ Green function of (4.25) can be exactly
obtained in quantum mechanics. Below on the basis of the correspondent
relation the joint method based on both the energy eigenfunction expansion
method and the\ Feynman path integration technique is proposed to determine
exactly a more general $SIC$ Green function than those of (4.9) and (4.20).
The Feynman path integration technique [$14,20$] has been well developed in
quantum mechanics. It provides a formulation of quantum mechanics that is
different from the original Heisenberg matrix mechanics and the Schr\"{o}%
dinger wave mechanics. It is able to treat theoretically any quantum systems
in time and space in unitary quantum dynamics. Many $QM$ Green functions for
the simple and special quantum systems can be exactly determined by the path
integration technique, but most $QM$\ Green functions can not be exactly
obtained for complicated quantum systems. Now consider any quantum system
whose $QM$ Green function can be exactly determined by the path integration
technique. For such quantum system one can calculate exactly its $SIC$ Green
function by the joint method. Here for convenience the Hamiltonian of the
quantum system is limited to be time-independent. At the first step of the
joint method one calculates exactly the $QM$ Green function of the quantum
system by using the path integration technique. Denote $T_{m}$ as the time
interval between the final and the initial time in the $QM$ Green function.
Evidently this $QM$\ Green function is a function of the time interval $%
T_{m} $. At the second step one makes the time-interval replacement, $%
T_{m}\rightarrow a_{m}^{s}T_{m},$ in the exact expression of the $QM$ Green
function. Then it can be deduced from the correspondent relation that after
the time-interval replacement this $QM$ Green function is exactly changed to
the $SIC$\ Green function of the same quantum system. One therefore obtains
the exact expression of the $SIC$\ Green function that is correspondent to
the $QM$\ Green function.

In quantum mechanics there are a number of quantum systems whose $QM$ Green
functions can be calculated exactly by the path integration technique, and
there also may be the $QM$ Green functions of a quantum system with
different Hamiltonians which may be calculated exactly by the path
integration technique. In principle the $SIC$ Green functions corresponding
to all these exact $QM$\ Green functions can be exactly determined by the
joint method. As typical examples, below several applications of the joint
method are described to the exact calculation of the $SIC$ Green functions
in one-dimensional coordinate space in a single-atom system. The first two
applications are to re-calculate the $SIC$\ Green functions of (4.9) and
(4.20) by the joint method, respectively. As shown in (4.9) and (4.20),
these two $SIC$ Green functions are already obtained exactly by the energy
eigenfunction expansion method alone.

The $SIC$ Green function of (4.9) is generated by the quadratic Hamiltonian $%
\hat{T}=\frac{1}{2m}p^{2},$ while the one of (4.20) is generated by the
quadratic Hamiltonian$\ H_{c}$ with the expression (3.7a). Here, as shown
above, for a single-atom system (e.g., a hydrogen-like atom) $p$ and $x$ may
stand for the momentum and coordinate of the COM motion of the single atom,
respectively, and they are also a pair of canonical momentum and coordinate.
For a general quantum system in time and space $p$ and $x$ may be treated as
a pair of canonical momentum and coordinate (See the paragraph with (4.36)
in detail below). In the Feynman path integration technique the $QM$ Green
function of a quantum system tends to be characterized by the Lagrangian $L$
of the quantum system and its corresponding action $S=\int Ldt.$ With the
help of the relation between the Hamiltonian $H$ and the Lagrangian $L$ in a
quantum system (See (4.36) with one degree of freedom below) it can be found
that the quadratic Lagrangians corresponding to the Hamiltonians $\hat{T}$
and $H_{c}$ are given by $L_{0}=\frac{1}{2}m(\frac{dx}{dt})^{2}\equiv \frac{1%
}{2}m\dot{x}^{2}$ and $L_{c}=\frac{1}{2}m\dot{x}^{2}-\frac{1}{2}m\omega
^{2}x^{2},$ respectively.

The $QM$ unitary quadratic propagator that is generated by the kinetic
operator $\hat{T}$ is given by $U_{\hat{T}}(T_{m})=\exp \left( -i\hat{T}%
T_{m}/\hslash \right) ,$ while the one that is generated by the Hamiltonian$%
\ H_{c}$ is given by $U_{H_{c}}(T_{m})=\exp \left( -iH_{c}T_{m}/\hslash
\right) .$ The former corresponds to the $SIC$ quadratic propagator $U_{\hat{%
T}}^{sic}(a_{m}^{s},T_{m})$ in (4.4), while the latter corresponds to the $%
SIC\ $quadratic propagator $U_{H_{c}}^{sic}(a_{m}^{s},T_{m})$ of (4.10). It
can prove by the path integration technique that the $QM$\ Green function
that is generated by the quadratic Lagrangian $L_{0}$ or correspondent to
the $U_{\hat{T}}(T_{m})$ can be calculated exactly and may be written in the
form [$14,20$]%
\begin{equation}
G_{\hat{T}}(x_{b},t_{b};x_{a},t_{a})=\sqrt{\frac{m}{i2\pi \hslash T_{m}}}%
\exp \left( i\frac{m}{2\hslash }\frac{\left( x_{b}-x_{a}\right) ^{2}}{T_{m}}%
\right) ,  \tag{4.26}
\end{equation}%
and similarly the $QM$ Green function that is generated by the quadratic
Lagrangian $L_{c}$ or correspondent to the $U_{H_{c}}(T_{m})$ also can be
exactly calculated and may be written in the form [$14,20$]%
\begin{equation*}
G_{H_{c}}(x_{b},t_{b};x_{a},t_{a})=\sqrt{\frac{m\omega }{i2\pi \hslash \sin
\left( \omega T_{m}\right) }}
\end{equation*}%
\begin{equation}
\times \exp \left[ \frac{im\omega }{2\hslash \sin \left( \omega T_{m}\right) 
}\left( \left( x_{a}^{2}+x_{b}^{2}\right) \allowbreak \cos \left( \omega
T_{m}\right) -2x_{a}x_{b}\right) \right]  \tag{4.27}
\end{equation}%
It can be seen that both the $G_{\hat{T}}(x_{b},t_{b};x_{a},t_{a})$ and $%
G_{H_{c}}(x_{b},t_{b};x_{a},t_{a})$ are the functions of the time interval $%
t_{b}-t_{a}=T_{m}.$ The work in this paragraph is the first step of the
joint method.

Now the second step of the joint method can be easily carried out. In the
exact expression (4.26) of the $QM$\ Green function $G_{\hat{T}%
}(x_{b},t_{b};x_{a},t_{a})$ one makes the time-interval replacement $%
T_{m}\rightarrow a_{m}^{s}T_{m}.$ Then it can be found that the $G_{\hat{T}%
}(x_{b},t_{b};x_{a},t_{a})$ is exactly changed to the $SIC$ Green function $%
G_{\hat{T}}^{sic}(x_{b},t_{b};x_{a},t_{a})$ of (4.9). One thus obtains the
exact expression of the $SIC$ Green function $G_{\hat{T}%
}^{sic}(x_{b},t_{b};x_{a},t_{a})$ that is generated by the quadratic
Lagrangian $L_{0}$ and corresponds to the $QM$ Green function $G_{\hat{T}%
}(x_{b},t_{b};x_{a},t_{a}).$ This example confirms that the $SIC$ Green
function $G_{\hat{T}}^{sic}(x_{b},t_{b};x_{a},t_{a})$ that is exactly
determined by the joint method is completely consistent with the one of
(4.9) that is exactly obtained by the energy eigenfunction expansion method
alone. Similarly, if in the exact expression (4.27) of the $%
G_{H_{c}}(x_{b},t_{b};x_{a},t_{a})$ one makes the time-interval replacement $%
T_{m}\rightarrow a_{m}^{s}T_{m},$ then it can be found that the $QM$ Green
function $G_{H_{c}}(x_{b},t_{b};x_{a},t_{a})$ is exactly changed to the $SIC$
Green function $G_{H_{c}}^{sic}(x_{b},t_{b};x_{a},t_{a})$ of (4.20). The
exact expression for the $SIC$\ Green function that is generated by the
quadratic Lagrangian $L_{c}$ thus is obtained by the joint method. This is
the second example that confirms that both the joint method and the energy
eigenfunction expansion method can achieve the same result. These two
examples (or applications) therefore confirm that, just like the energy
eigenfunction expansion method, the joint method can obtain the correct
result in the exact calculation of the $SIC$\ unitary propagators in the
coordinate representation.

As a third application, consider the $SIC$ quadratic propagator $%
U_{H_{d}}^{sic}(a_{m}^{s},T_{m})$ $=\exp \left[ -ia_{m}^{s}H_{d}T_{m}/%
\hslash \right] ,$ where the time-independent $QM$ quadratic Hamiltonian $%
H_{d}$ is given by $H_{d}=\frac{1}{2m}p^{2}+\frac{1}{2}m\omega ^{2}x^{2}+fx.$
As shown by (4.36) below, the quadratic Lagrangian $L_{d}$ corresponding to
the Hamiltonian $H_{d}$ is given by $L_{d}=\frac{1}{2}m\dot{x}^{2}-\frac{1}{2%
}m\omega ^{2}x^{2}-fx.$ It is also time-independent. Now one wants to
determine the $SIC$ quadratic propagator $U_{H_{d}}^{sic}(a_{m}^{s},T_{m})$
in the coordinate representation, i.e., $%
G_{H_{d}}^{sic}(x_{b},t_{b};x_{a},t_{a}),$ by the joint method. Given the $%
QM $ quadratic Lagrangian $L_{d}$ (or the Hamiltonian $H_{d}$) one first
determines the $QM$ quadratic propagator $U_{H_{d}}(T_{m})=\exp \left[
-iH_{d}T_{m}/\hslash \right] $ in the coordinate representation, i.e., $%
G_{H_{d}}(x_{b},t_{b};x_{a},t_{a}),$ by the path integration technique. Here
the $QM$\ Green function $G_{H_{d}}(x_{b},t_{b};x_{a},t_{a})$ corresponds to
the $SIC$ Green function $G_{H_{d}}^{sic}(x_{b},t_{b};x_{a},t_{a}).$ The
exact expression of the$\ G_{H_{d}}(x_{b},t_{b};x_{a},t_{a})$ is already
obtained by the\ Feynman path integration technique [$14,20$]. Evidently the 
$G_{H_{d}}(x_{b},t_{b};x_{a},t_{a})$ is a function of the time interval $%
T_{m}.$ This is the first step of the joint method. At the second step in
the exact expression of the $G_{H_{d}}(x_{b},t_{b};x_{a},t_{a})$ one makes
the time-interval replacement $T_{m}\rightarrow a_{m}^{s}T_{m}.$ This
replacement results in that the $QM$ Green function $%
G_{H_{d}}(x_{b},t_{b};x_{a},t_{a})$ is exactly changed to the $SIC$ Green
function $G_{H_{d}}^{sic}(x_{b},t_{b};x_{a},t_{a}).$ One thus obtains the
exact expression of the $G_{H_{d}}^{sic}(x_{b},t_{b};x_{a},t_{a})$ which may
be explicitly written as%
\begin{equation*}
G_{H_{d}}^{sic}(x_{b},t_{b};x_{a},t_{a})=\sqrt{\frac{m\omega }{i2\pi \hslash
\sin (\omega a_{m}^{s}T_{m})}}
\end{equation*}%
\begin{equation*}
\times \exp \{i\frac{m\omega }{2\hslash \sin (\omega a_{m}^{s}T_{m})}%
[(x_{b}^{2}+x_{a}^{2})\cos (\omega a_{m}^{s}T_{m})-2x_{b}x_{a}]
\end{equation*}%
\begin{equation*}
+i\frac{f}{\hslash \omega }\frac{\cos \left( \omega a_{m}^{s}T_{m}\right) -1%
}{\sin \left( \omega a_{m}^{s}T_{m}\right) }\left( x_{b}+x_{a}\right)
\end{equation*}%
\begin{equation}
+i\allowbreak \frac{f^{2}}{2m\hslash \omega ^{3}}\frac{2\cos \left( \omega
a_{m}^{s}T_{m}\right) +\left( \omega a_{m}^{s}T_{m}\right) \sin \left(
\omega a_{m}^{s}T_{m}\right) -2}{\sin \left( \omega a_{m}^{s}T_{m}\right) }\}
\tag{4.28}
\end{equation}%
where the time interval $t_{b}-t_{a}=a_{m}^{s}T_{m}$. This is just the exact
expression of the $SIC$ quadratic propagator $%
U_{H_{d}}^{sic}(a_{m}^{s},T_{m})$ in the coordinate representation.

As a fourth application of the joint method, here consider a general $SIC$
quadratic propagator. A\ general $SIC$ quadratic propagator may be written
as $U_{H_{q}}^{sic}(a_{m}^{s},T_{m})=\exp \left[ -ia_{m}^{s}H_{q}T_{m}/%
\hslash \right] ,$ where the time-independent $QM$ quadratic Hamiltonian $%
H_{q}$ may be formally written as%
\begin{equation}
H_{q}\equiv H_{q}\left( p,x\right) =\frac{1}{2m}p^{2}+\frac{1}{2}b\left(
xp+px\right) +cx^{2}+dp+fx+g  \tag{4.29}
\end{equation}%
and its corresponding time-independent quadratic Lagrangian $L_{q},$ which
is obtained from (4.36) (See below), is given by%
\begin{equation}
L_{q}\equiv L_{q}\left( \dot{x},x\right) =\allowbreak \frac{1}{2}m\dot{x}%
^{2}\allowbreak -b_{L}x\dot{x}-c_{L}x^{2}-d_{L}\dot{x}-f_{L}x-g_{L} 
\tag{4.30}
\end{equation}%
where $b_{L}=mb,$ $c_{L}=c-\frac{1}{2}mb^{2},$ $d_{L}=md,$ $f_{L}=f-mbd,$
and $g_{L}=g-\frac{1}{2}md^{2}.$ With the help of the relation (4.36) below
the Hamiltonian $H_{q}$ can be changed to the corresponding Lagrangian $%
L_{q} $ and vice versa. Generally the action $S_{q}$ of the Lagrangian $%
L_{q} $ is given by $S_{q}=\int L_{q}dt.$ The $QM$ quadratic propagator that
is generated by the quadratic Hamiltonian $H_{q}$ is given by $%
U_{H_{q}}(T_{m})=\exp \left( -iH_{q}T_{m}/\hslash \right) .$ It corresponds
to the $SIC$ quadratic propagator $U_{H_{q}}^{sic}(a_{m}^{s},T_{m}).$

Quantum mechanically it can prove [$14,20$] that the $QM$ quadratic Green
function that is generated by the $U_{H_{q}}(T_{m})$ or by the quadratic
Lagrangian $L_{q}$ of (4.30) can be exactly determined by the path
integration technique and may be written in a general form [$20$]%
\begin{equation}
G_{H_{q}}(x_{b},t_{b};x_{a},t_{a})=\sqrt{\frac{m}{i2\pi \hslash }(-T_{ab})}%
\exp \left( \frac{i}{\hslash }S_{q}(x_{b},t_{b};x_{a},t_{a})\right) 
\tag{4.31}
\end{equation}%
where the $QM$ action $S_{q}(x_{b},t_{b};x_{a},t_{a})$ may be formally
expressed as%
\begin{equation}
S_{q}(x_{b},t_{b};x_{a},t_{a})=\frac{1}{2}m\left(
T_{bb}x_{b}^{2}+T_{ab}2x_{a}x_{b}+T_{aa}x_{a}^{2}\right)
+x_{a}B_{a}+x_{b}B_{b}+B_{0}.  \tag{4.32}
\end{equation}%
In the $QM$ Green function of (4.31) and the $QM$ action of (4.32) these
parameters $T_{aa},$ $T_{ab},$ $T_{bb},$ $B_{a},$ $B_{b},$ and $B_{0}$ each
are a real function of the time interval $T_{m},$ $e.g.,$ $%
T_{ab}=T_{ab}\left( T_{m}\right) ,$ $B_{a}=B_{a}\left( T_{m}\right) ,$ etc.,
and they each are independent of the coordinates $x_{b}$ and $x_{a}.$ This
is the first step of the joint method that one first obtains the exact
expression (4.31) of the $QM$ Green function of the quadratic Lagrangian $%
L_{q}$ by the path integration technique. The next step is that in the exact
expression (4.31) of the $QM$\ Green function $%
G_{H_{q}}(x_{b},t_{b};x_{a},t_{a})$ one makes the time-interval replacement,
i.e., $T_{m}\rightarrow a_{m}^{s}T_{m}.$ It can be seen from (4.31) and
(4.32) that this time-interval replacement is equivalent to that all these
parameters in (4.31) and (4.32) are made the replacement of the
time-interval variable: $T_{\lambda }\left( T_{m}\right) \rightarrow
T_{\lambda }\left( a_{m}^{s}T_{m}\right) $ for $\lambda =aa,$ $ab,$ $bb$ and 
$B_{\mu }\left( T_{m}\right) \rightarrow B_{\mu }\left(
a_{m}^{s}T_{m}\right) $ for $\mu =a,$ $b,$ $0.$ After the time-interval
replacement the $QM$ Green function $G_{H_{q}}(x_{b},t_{b};x_{a},t_{a})$ is
exactly changed to the $SIC$\ Green function $%
G_{H_{q}}^{sic}(x_{b},t_{b};x_{a},t_{a})$ that is generated by the $SIC$
quadratic propagator $U_{H_{q}}^{sic}(a_{m}^{s},t_{m})$ or by the quadratic
Lagrangian $L_{q}$. One thus obtains the exact expression of the $SIC$\
Green function $G_{H_{q}}^{sic}(x_{b},t_{b};x_{a},t_{a})$ which may be
formally written as%
\begin{equation}
G_{H_{q}}^{sic}(x_{b},t_{b};x_{a},t_{a})=\sqrt{\frac{m}{i2\pi \hslash }%
(-S_{ab})}\exp \left( \frac{i}{\hslash }S_{q}^{sic}(x_{b},t_{b};x_{a},t_{a})%
\right)  \tag{4.33}
\end{equation}%
and here the $SIC$ action $S_{q}^{sic}(x_{b},t_{b};x_{a},t_{a})$ may be
formally expressed as%
\begin{equation}
S_{q}^{sic}(x_{b},t_{b};x_{a},t_{a})=\frac{1}{2}m\left(
S_{bb}x_{b}^{2}+S_{ab}2x_{a}x_{b}+S_{aa}x_{a}^{2}\right)
+x_{a}Q_{a}+x_{b}Q_{b}+\Theta _{0},  \tag{4.34}
\end{equation}%
where these $SIC$ parameters $S_{aa},$ $S_{ab},$ $S_{bb},$ $Q_{a},$ $Q_{b},$
and $\Theta _{0}$ are the real functions of the time-interval variable $%
t_{b}-t_{a}=a_{m}^{s}T_{m}$ and they are explicitly given by $S_{\lambda
}=T_{\lambda }\left( a_{m}^{s}T_{m}\right) $ for $\lambda =aa,$ $ab,$ $bb;$ $%
Q_{\mu }=B_{\mu }\left( a_{m}^{s}T_{m}\right) $ for $\mu =a,$ $b;$ and $%
\Theta _{0}=B_{0}\left( a_{m}^{s}T_{m}\right) .$

It is known from (4.32) and (4.34) that both the $QM$ action $%
S_{q}(x_{b},t_{b};x_{a},t_{a})$ and the $SIC$ action $%
S_{q}^{sic}(x_{b},t_{b};x_{a},t_{a})$ each are a quadratic function of the
coordinates $x_{a}$ and $x_{b}.$ This is the characteristic feature of a $%
QM\ $quadratic Green function and also a $SIC$ quadratic Green function. Due
to this characteristic feature the formal expression (4.33) is general for a 
$SIC$\ quadratic Green function, even if the quadratic Lagrangian $L_{q}$ of
(4.30) (or the Hamiltonian $H_{q}$ of (4.29)) carries the solution
information ($a_{m}^{s}$), that is, $L_{q}=L_{q}\left( a_{m}^{s}\right) $
(or $H_{q}=H_{q}\left( a_{m}^{s}\right) $). It can prove that the previous $%
SIC$ quadratic Green functions of (4.9), (4.20), and (4.28) are the special
cases of the general $SIC$ quadratic Green function of (4.33).

The final application of the joint method is to determine the $SIC$\ Green
function with a time-independent $QM$\ non-quadratic Lagrangian (or
Hamiltonian). This non-quadratic Hamiltonian is given by $H_{s}=\frac{1}{2m}%
p^{2}+V\left( x\right) $ and the corresponding non-quadratic Lagrangian is $%
L_{s}=\frac{1}{2m}p^{2}-V\left( x\right) $ (See also (4.36) below), where
the potential energy $V\left( x\right) $ is given by $V\left( x\right) =0$
if $0<x<a$ and $V\left( x\right) =\infty $ if $0<x$ or $x>a.$ In the present
example this Hamiltonian describes the COM motion of a single atom in
external infinite square potential well with walls at $x=0$ and $x=a$. The $%
QM$ unitary propagator with the Hamiltonian $H_{s}$ is given by $\exp \left(
-iH_{s}T_{m}/\hslash \right) ,$ and here its corresponding $QM$\ Green
function with the Lagrangian $L_{s}$ is denoted as $%
G_{H_{s}}(x_{b},t_{b};x_{a},t_{a}).$ It can prove that by the path
integration technique the $QM$ Green function $%
G_{H_{s}}(x_{b},t_{b};x_{a},t_{a})$ can be exactly determined [$20,23$], and
it is a function of the time interval $T_{m}$. Now in the exact expression [$%
20,23$] of the $G_{H_{s}}(x_{b},t_{b};x_{a},t_{a})$ one makes the
time-interval replacement, i.e., $T_{m}\rightarrow a_{m}^{s}T_{m}.$ This
time-interval replacement leads to that the $QM$\ Green function $%
G_{H_{s}}(x_{b},t_{b};x_{a},t_{a})$ is exactly changed to the $SIC$ Green
function $G_{H_{s}}^{sic}(x_{b},t_{b};x_{a},t_{a})$ that is generated by the 
$SIC$\ unitary propagator $\exp \left( -ia_{m}^{s}H_{s}T_{m}/\hslash \right) 
$ or by the Lagrangian $L_{s}.$ Therefore, one obtains the exact expression
of the $SIC$\ Green function:%
\begin{equation*}
G_{H_{s}}^{sic}(x_{b},t_{b};x_{a},t_{a})=\sqrt{\frac{m}{i2\pi \hslash \left(
a_{m}^{s}T_{m}\right) }}
\end{equation*}%
\begin{equation}
\times \sum_{n=-\infty }^{\infty }\left[ \exp \left( i\frac{m}{2\hslash }%
\frac{\left( x_{b}-x_{a}+2na\right) ^{2}}{\left( a_{m}^{s}T_{m}\right) }%
\right) -\exp \left( i\frac{m}{2\hslash }\frac{\left( x_{b}+x_{a}+2na\right)
^{2}}{\left( a_{m}^{s}T_{m}\right) }\right) \right] .  \tag{4.35}
\end{equation}%
This is just the $SIC$\ unitary propagator $\exp \left(
-ia_{m}^{s}H_{s}T_{m}/\hslash \right) $ in the coordinate representation.
This $SIC$ Green function can not be exactly reduced to the formal
expression (4.33) of a general $SIC$ quadratic Green function. It is a $SIC$
non-quadratic Green function.

In the above theoretical treatments based on the joint method the path
integration technique has been indirectly employed to calculate exactly a $%
SIC$ unitary propagator in the coordinate representation. It is well known
in quantum mechanics that most of $QM$ Green functions can not be exactly
determined by the path integration technique. Since the joint method is
usually limited to determine an exact $SIC$\ Green function which
corresponds to an exact $QM$ Green function and it is used to treat a $SIC$
unitary propagator that is generated by a time-independent Hamiltonian, it
is a limited method. Below it is simply described how the path integration
technique is directly used to calculate (exactly or approximately) a general 
$SIC$\ Green function.

In classical dynamics a physical system may be described by a set of
canonical dynamical variables, i.e., the canonical momentum ($p_{i}$) and
coordinate ($q_{i}$) variables $\{q_{i},p_{i}\}$. If the degree of freedom
of the system is $g,$ then the set contains $2g$ variables $\{q_{i},p_{i}\}$%
. Both the Hamiltonian $H$ and the Lagrangian $L$ of the system are the
functions of the canonical momentum and coordinate variables $%
\{q_{i},p_{i}\} $, respectively. That is, $H=H\left( \mathbf{q},\mathbf{p}%
,t\right) $ and $L=L\left( \mathbf{q},\mathbf{\dot{q}},t\right) ,$ where $%
\mathbf{q=\{}q_{i}\mathbf{\}}$ and $\mathbf{p}=\{p_{i}\}$. Then according to
classical dynamics \footnote{%
See, for example, D. T. Greenwood, \textit{Classical dynamics}, Chapt. 4,
Dover edition, Prentice-Hall, Inc. (1977)} there is the relation between the
Hamiltonian $H$ and the Lagrangian $L:$ 
\begin{equation}
H\left( \mathbf{q},\mathbf{p},t\right) =\sum_{i=1}^{g}p_{i}\dot{q}%
_{i}-L\left( \mathbf{q},\mathbf{\dot{q}},t\right)  \tag{4.36}
\end{equation}%
With the help of this relation the Hamiltonian $H$ can be changed to the
Lagrangian $L$ and vice versa. For the quantum system corresponding to the
classical physical system above the relation (4.36) still holds between the
Hamiltonian $H$ and the Lagrangian $L$ in quantum mechanics. However, in
quantum mechanics $\mathbf{q=\{}q_{i}\mathbf{\}}$ and $\mathbf{p}=\{p_{i}\}$
in (4.36) are coordinate and momentum operators, respectively. Of course, in
the quantization transition from the classical system to the quantum system
these canonical coordinate and momentum variables$\ \{q_{i},p_{i}\}$ should
be expressed in Cartesian coordinates [$10$]. In quantum mechanics a unitary
propagator is usually generated with a Hamiltonian ($H$). In the Feynman
path integration technique a Green function is usually characterized with a
Lagrangian ($L$).

In the definition (4.1) the Hamiltonian used to define a $SIC$\ unitary
propagator is time-independent. Does the $SIC$ unitary propagator make sense
if the Hamiltonian is time-dependent? The present theoretical work dose not
involve the concrete preparation and realization of a $SIC$ unitary
propagator. This provides convenience to define theoretically a $SIC$\
unitary propagator in a quantum system in time and space whose Hamiltonian
is time-dependent. It also could become convenient for general theoretical
treatment or calculation of a $SIC$ Green function in the framework of the
path integration technique. Theoretically a $SIC$ unitary propagator may be
defined in any quantum system in time and space whose Hamiltonian may or may
not be time-dependent. Now a $SIC$\ unitary propagator, which is generated
by a time-dependent Hamiltonian $H\left( t\right) ,$ may be defined by%
\begin{equation*}
U_{H}^{sic}(a_{m}^{s},t_{0}+a_{m}^{s}T_{m},t_{0})
\end{equation*}%
\begin{equation*}
=\lim_{N\rightarrow \infty }\exp \left[ -\frac{i}{\hslash }a_{m}^{s}H\left(
t_{0}+a_{m}^{s}T_{m}-\frac{1}{2}\left( a_{m}^{s}T_{m}/N\right) \right)
\left( T_{m}/N\right) \right]
\end{equation*}%
\begin{equation*}
\times \exp \left[ -\frac{i}{\hslash }a_{m}^{s}H\left( t_{0}+a_{m}^{s}T_{m}-%
\frac{3}{2}\left( a_{m}^{s}T_{m}/N\right) \right) \left( T_{m}/N\right) %
\right]
\end{equation*}%
\begin{equation*}
\times ...\times \exp \left[ -\frac{i}{\hslash }a_{m}^{s}H\left( t_{0}+\frac{%
3}{2}\left( a_{m}^{s}T_{m}/N\right) \right) \left( T_{m}/N\right) \right]
\end{equation*}%
\begin{equation}
\times \exp \left[ -\frac{i}{\hslash }a_{m}^{s}H\left( t_{0}+\frac{1}{2}%
\left( a_{m}^{s}T_{m}/N\right) \right) \left( T_{m}/N\right) \right] 
\tag{4.37}
\end{equation}%
where $t_{0}+a_{m}^{s}T_{m}$ and $t_{0}$ are the final and the initial time
of the quantum system, respectively. It can be found that when the
Hamiltonian $H\left( t\right) $ is time-independent, the $SIC$\ unitary
propagator of (4.37) is reduced to the one of (4.1). The $SIC$ unitary
propagator defined by (4.37) may correspond to the following $QM$\ unitary
propagator:%
\begin{equation*}
U_{H}(t_{0}+T_{m},t_{0})=\lim_{N\rightarrow \infty }\exp \left[ -\frac{i}{%
\hslash }H\left( t_{0}+T_{m}-\frac{1}{2}\left( T_{m}/N\right) \right) \left(
T_{m}/N\right) \right]
\end{equation*}%
\begin{equation*}
\times \exp \left[ -\frac{i}{\hslash }H\left( t_{0}+T_{m}-\frac{3}{2}\left(
T_{m}/N\right) \right) \left( T_{m}/N\right) \right]
\end{equation*}%
\begin{equation*}
\times ...\times \exp \left[ -\frac{i}{\hslash }H\left( t_{0}+\frac{3}{2}%
\left( T_{m}/N\right) \right) \left( T_{m}/N\right) \right]
\end{equation*}%
\begin{equation}
\times \exp \left[ -\frac{i}{\hslash }H\left( t_{0}+\frac{1}{2}\left(
T_{m}/N\right) \right) \left( T_{m}/N\right) \right]  \tag{4.38}
\end{equation}%
\newline
If one replaces the time interval $T_{m}$ in the $QM$ unitary propagator of
(4.38) with the time interval $a_{m}^{s}T_{m},$ then the $QM$ unitary
propagator is exactly changed to the $SIC$ unitary propagator of (4.37).

The main advantage for the definition (4.37) of a $SIC$ unitary propagator $%
U_{H}^{sic}(a_{m}^{s},t_{0}+a_{m}^{s}T_{m},t_{0})$ is that the $%
U_{H}^{sic}(a_{m}^{s},t_{0}+a_{m}^{s}T_{m},t_{0})$ is governed simply and
directly by a single time-dependent Hamiltonian $H(t)$ and that the $%
U_{H}^{sic}(a_{m}^{s},t_{0}+a_{m}^{s}T_{m},t_{0})$ with $a_{m}^{s}=+1$ is
the inverse of the $U_{H}^{sic}(a_{m}^{s},t_{0}+a_{m}^{s}T_{m},t_{0})$ with $%
a_{m}^{s}=-1$ and vice versa, that is, 
\begin{equation}
U_{H}^{sic}(+1,t_{0}+T_{m},t_{0})=U_{H}^{sic}(-1,t_{0},t_{0}+T_{m})^{+}. 
\tag{4.39}
\end{equation}%
This relation may be called the global time-evolution-reversal symmetry.
However, such definition as (4.37) of a $SIC$ unitary propagator is usually
complicated, since it is involved in the piecewise time-independent
Hamiltonians which may be dependent upon the double-valued logical number $%
a_{m}^{s}.$ Consequently it usually could be complicated to construct and
realize the $SIC$ unitary propagator in a quantum system.

A simpler definition for a $SIC$\ unitary propagator generated by a
time-dependent Hamiltonian $H(t)$ may be given by%
\begin{equation*}
U_{H}^{sic}(a_{m}^{s},t_{0}+a_{m}^{s}T_{m},t_{0})=\lim_{N\rightarrow \infty
}\exp \left[ -\frac{i}{\hslash }a_{m}^{s}H\left( t_{0}+T_{m}-\frac{1}{2}%
\left( T_{m}/N\right) \right) \left( T_{m}/N\right) \right]
\end{equation*}%
\begin{equation*}
\times \exp \left[ -\frac{i}{\hslash }a_{m}^{s}H\left( t_{0}+T_{m}-\frac{3}{2%
}\left( T_{m}/N\right) \right) \left( T_{m}/N\right) \right]
\end{equation*}%
\begin{equation*}
\times ...\times \exp \left[ -\frac{i}{\hslash }a_{m}^{s}H\left( t_{0}+\frac{%
3}{2}\left( T_{m}/N\right) \right) \left( T_{m}/N\right) \right]
\end{equation*}%
\begin{equation}
\times \exp \left[ -\frac{i}{\hslash }a_{m}^{s}H\left( t_{0}+\frac{1}{2}%
\left( T_{m}/N\right) \right) \left( T_{m}/N\right) \right]  \tag{4.40}
\end{equation}%
This $SIC$ unitary propagator is obtained by combining the definition (4.1)
of the $SIC$ unitary propagator generated by a time-independent Hamiltonian
with the $QM$\ unitary propagator of (4.38) that is generated by a
time-dependent Hamiltonian $H(t)$. The $SIC$\ unitary propagator of (4.40)
is different from the one of (4.37) in that the piecewise time-independent
Hamiltonians in (4.40) are independent of the double-valued logical number $%
a_{m}^{s}$ (here suppose that the initial time $t_{0}$ is independent of $%
a_{m}^{s}$).

The $SIC$\ unitary propagator of (4.40) does not own the global
time-evolution-reversal symmetry of (4.39), but it owns the local
time-evolution-reversal symmetry, because in the short time interval $%
T_{m}/N $ every $SIC$ exponential propagator $%
U_{H_{j}}^{sic}(a_{m}^{s},T_{m}/N)$ in (4.40) satisfies%
\begin{equation}
U_{H_{j}}^{sic}(+1,T_{m}/N)=U_{H_{j}}^{sic}(-1,T_{m}/N)^{+},\text{ }%
j=1,2,...,N  \tag{4.41}
\end{equation}%
Here, as defined by (4.1), $U_{H_{j}}^{sic}(a_{m}^{s},T_{m}/N)=\exp \left[ -%
\frac{i}{\hslash }a_{m}^{s}H_{j}T_{m}/N\right] ,$ where the piecewise
time-independent Hamiltonian $H_{j}$ is given by%
\begin{equation*}
H_{j}=H\left( t_{0}+T_{m}-\left( j-1/2\right) \left( T_{m}/N\right) \right) ,
\end{equation*}%
as shown in (4.40).

It could be thought that the $SIC$ unitary propagator of (4.40) is
effectively generated by the time-dependent Hamiltonian $a_{m}^{s}H(t).$
When $a_{m}^{s}=+1,$ the Hamiltonian $a_{m}^{s}H(t)$ is the natural one
(i.e., $H(t)$) of the quantum system. However, when $a_{m}^{s}=-1,$ the
Hamiltonian $a_{m}^{s}H(t)=-H\left( t\right) $ is not the natural one and it
may be considered as the effective time-dependent Hamiltonian of the quantum
system. Given the effective time-dependent Hamiltonian $a_{m}^{s}H(t)$ one
may set up the time-dependent $SIC$\ Schr\"{o}dinger equation (See Appendix
A) to describe the $SIC$ unitary propagator of (4.40).

As shown in (4.2), corresponding to the $SIC$\ unitary propagator of (4.37)
or (4.40) the $SIC$ Green function may be re-expressed as $%
G_{H}^{sic}(x_{b},t_{b};x_{a},t_{a})=\left\langle x_{b},t_{b}\right\vert
U_{H}^{sic}(a_{m}^{s},t_{0}+a_{m}^{s}T_{m},t_{0})\left\vert
x_{a},t_{a}\right\rangle ,$ where the initial time $t_{a}=t_{0}$ and in
value the final time $t_{b}$ is obtained from $a_{m}^{s}T_{m}=t_{b}-t_{0}.$

The $SIC$\ unitary propagators of (4.37) and (4.40) and their corresponding $%
SIC$\ Green functions each still own the dual character of quantum-computing
speedup. They obey both the unitary quantum dynamics and the
mathematical-logical principle of the unstructured search problem. This
means that one needs to employ both the physical Hilbert space and the math
Hilbert space to describe completely the quantum-computing speedup process
that is governed by the $SIC$\ unitary propagator of (4.37) or (4.40). This
quantum-computing speedup process is unitary and deterministic. It obeys the
information conservation principle.

The following theoretical treatment is devoted to setting up a standard
path-integration formula for a $SIC$ Green function. It is limited to
one-dimensional coordinate space. It is based on the Feynman path
integration technique [$14,20$]. Suppose that a quantum system in time and
space has the time-independent Hamiltonian $H=\hat{T}+\hat{V},$ where $\hat{T%
}=\frac{1}{2m}p^{2}$ and $\hat{V}=V\left( x\right) $ are the kinetic and
potential energy operators, respectively. According to the definition (4.1)
the $SIC$ unitary propagator of the quantum system then may be written as $%
U_{H}^{sic}(a_{m}^{s},T_{m})=\exp \left[ -ia_{m}^{s}\left( \hat{T}+\hat{V}%
\right) T_{m}/\hslash \right] .$ With the help of the Trotter product
formula the $U_{H}^{sic}(a_{m}^{s},T_{m})$ may be written as%
\begin{equation*}
U_{H}^{sic}(a_{m}^{s},T_{m})=\lim_{N\rightarrow \infty }\left[ \exp \left( -%
\frac{i}{\hslash }a_{m}^{s}\hat{T}T_{m}/N\right) \exp \left( -\frac{i}{%
\hslash }a_{m}^{s}\hat{V}T_{m}/N\right) \right] ^{N}
\end{equation*}%
Then the corresponding $SIC$ Green function of (4.2) is given by%
\begin{equation*}
G_{H}^{sic}(x_{b},t_{b};x_{a},t_{a})
\end{equation*}%
\begin{equation}
=\lim_{N\rightarrow \infty }\left\langle x_{b},t_{b}\right\vert \left[ \exp
\left( -\frac{i}{\hslash }a_{m}^{s}\hat{T}T_{m}/N\right) \exp \left( -\frac{i%
}{\hslash }a_{m}^{s}\hat{V}T_{m}/N\right) \right] ^{N}\left\vert
x_{a},t_{a}\right\rangle  \tag{4.42}
\end{equation}%
By inserting the $N-1$ coordinate identity operators $\{\int
dx_{j}\left\vert x_{j}\right\rangle \left\langle x_{j}\right\vert \}$ for $%
j=1,2,...,N-1$ into (4.42) one obtains%
\begin{equation*}
G_{H}^{sic}(x_{b},t_{b};x_{a},t_{a})=\lim_{N\rightarrow \infty }\int
dx_{1}dx_{2}...dx_{N-1}
\end{equation*}%
\begin{equation}
\times \underset{j=0}{\overset{N-1}{\dprod }}\left\langle x_{j+1}\right\vert
\exp \left( -\frac{i}{\hslash }a_{m}^{s}\hat{T}\varepsilon \right) \exp
\left( -\frac{i}{\hslash }a_{m}^{s}\hat{V}\varepsilon \right) \left\vert
x_{j}\right\rangle  \tag{4.43}
\end{equation}%
where $\varepsilon =T_{m}/N,$ $x_{0}=x_{a},$ and $x_{N}=x_{b}.$ There is the
eigenvalue equation $\exp \left( -ia_{m}^{s}\hat{V}\varepsilon /\hslash
\right) \left\vert x_{j}\right\rangle =$ $\exp \left( -ia_{m}^{s}V\left(
x_{j}\right) \varepsilon /\hslash \right) \left\vert x_{j}\right\rangle $.
There is also the expression of the matrix element: 
\begin{equation*}
\left\langle x_{j+1}\right\vert \exp \left( -ia_{m}^{s}\hat{T}\varepsilon
/\hslash \right) \left\vert x_{j}\right\rangle =\sqrt{\frac{m}{i2\pi \hslash
\left( a_{m}^{s}\varepsilon \right) }}\exp \left[ i\frac{m}{2\hslash }\left(
a_{m}^{s}\varepsilon \right) \left( \frac{x_{j+1}-x_{j}}{\varepsilon }%
\right) ^{2}\right] .
\end{equation*}%
This expression can be obtained by the way that is used to determine the $%
SIC $\ Green function of (4.9), here one needs to use the important property
of the double-valued logical number $a_{m}^{s},$ i.e., $\left(
a_{m}^{s}\right) ^{2}=1.$ Now with the help of the eigenvalue equation and
the expression one can reduce the $SIC$\ Green function of (4.43) to the form%
\begin{equation*}
G_{H}^{sic}(x_{b},t_{b};x_{a},t_{a})=\lim_{N\rightarrow \infty }\int
dx_{1}dx_{2}...dx_{N-1}\left( \sqrt{\frac{m}{i2\pi \hslash \left(
a_{m}^{s}\varepsilon \right) }}\right) ^{N}
\end{equation*}%
\begin{equation}
\times \exp \left[ \frac{i}{\hslash }\left( a_{m}^{s}\varepsilon \right)
\sum_{j=0}^{N-1}\left( \frac{1}{2}m\left( \frac{x_{j+1}-x_{j}}{\varepsilon }%
\right) ^{2}-V\left( x_{j}\right) \right) \right]  \tag{4.44}
\end{equation}%
This is a standard path-integration expression of the $SIC\ $Green function.
In the limit $N\rightarrow \infty $ or $\varepsilon =T_{m}/N\rightarrow 0$
the exponent (omitting $ia_{m}^{s}/\hslash $) in (4.44) may be written as%
\begin{equation}
\lim_{\varepsilon \rightarrow 0}\varepsilon \sum_{j=0}^{N-1}\left( \frac{1}{2%
}m\left( \frac{x_{j+1}-x_{j}}{\varepsilon }\right) ^{2}-V\left( x_{j}\right)
\right) =\int_{0}^{T_{m}}Ldt  \tag{4.45}
\end{equation}%
Here the $QM$ Lagrangian $L$ is given by 
\begin{equation}
L=\frac{1}{2}m\dot{x}^{2}-V\left( x\right) ,  \tag{4.46}
\end{equation}%
and the right side of (4.45) is just the $QM$ action $S:$%
\begin{equation}
S=\int_{t_{a}}^{t_{b}}Ldt.  \tag{4.47}
\end{equation}%
The factor $\left( \frac{m}{i2\pi \hslash \left( a_{m}^{s}\varepsilon
\right) }\right) ^{N/2}$ in (4.44) is a `normalization constant'. It ensures
that the $G_{H}^{sic}(x_{b},t_{b};x_{a},t_{a})$ of (4.44) is unitary. Now
the $G_{H}^{sic}(x_{b},t_{b};x_{a},t_{a})$ also may be formally written as%
\begin{equation}
G_{H}^{sic}(x_{b},t_{b};x_{a},t_{a})=\int_{x\left( t_{a}\right) }^{x\left(
t_{b}\right) }D\left[ x\left( t\right) \right] \exp \left( \frac{i}{\hslash }%
a_{m}^{s}\int_{t_{a}}^{t_{b}}Ldt\right)  \tag{4.48}
\end{equation}%
The path integration formula (4.44) can be easily generalized for the $SIC$\
unitary propagator of (4.40) that is generated by a time-dependent
Hamiltonian.

A check for correction of the path integral formula (4.44) may be carried
out by using (4.44) to calculate exactly the $SIC$\ Green function $G_{\hat{T%
}}^{sic}(x_{b},t_{b};x_{a},t_{a})$ that is generated by the kinetic operator 
$\hat{T}=\frac{1}{2m}p^{2}.$ By using the path integral formula (4.44) the $%
G_{\hat{T}}^{sic}(x_{b},t_{b};x_{a},t_{a})$ can be calculated exactly by the
way [$14,20$] that is used to calculate the corresponding $QM$\ Green
function $G_{\hat{T}}(x_{b},t_{b};x_{a},t_{a})$ (See also (4.26) above). An
important point is that one must repeat to use the important property $%
\left( a_{m}^{s}\right) ^{2}=1$ in the calculation. The detailed calculation
shows that the previous expression (4.9) of the $G_{\hat{T}%
}^{sic}(x_{b},t_{b};x_{a},t_{a}),$ which is obtained by the energy
eigenfunction expansion method alone,\ is exactly reproduced by the path
integral formula (4.44).

In the path integration technique most $SIC$\ Green functions can not be
exactly obtained. It could be feasible to employ the perturbation method to
calculate approximately a general $SIC$\ Green function. Suppose that the
total Hamiltonian of a quantum system in time and space is given by $%
H=H_{0}+H_{1},$ where $H_{0}$ and $H_{1}$ are the main term and the
perturbation term, respectively. According to the definition (4.1) the two $%
SIC$ unitary propagators $U^{sic}\left( t\right) =\exp \left(
-ia_{m}^{s}Ht/\hslash \right) $ and $U_{0}^{sic}\left( t\right) =\exp \left(
-ia_{m}^{s}H_{0}t/\hslash \right) $ are defined with the total Hamiltonian $%
H $ and the main Hamiltonian $H_{0},$ respectively. And the two
corresponding $SIC$\ Green functions $G_{H}^{sic}(x_{b},t_{b};x_{a},t_{a})$
and $G_{H_{0}}^{sic}(x_{b},t_{b};x_{a},t_{a})$ are obtained from (4.2),
respectively. Then the perturbation equation that these two $SIC$ Green
functions obey may be written as%
\begin{equation*}
G_{H}^{sic}(x_{b},t_{b};x_{a},0)=G_{H_{0}}^{sic}(x_{b},t_{b};x_{a},0)
\end{equation*}%
\begin{equation}
+\frac{1}{i\hslash }\int_{0}^{t}dt^{\prime }\int dx^{\prime
}G_{H_{0}}^{sic}(x_{b},t_{b};x^{\prime },t^{\prime })\left(
a_{m}^{s}H_{1}\left( x^{\prime }\right) \right) G_{H}^{sic}(x^{\prime
},t^{\prime };x_{a},0)  \tag{4.49}
\end{equation}%
\newline
where the initial time $t_{a}=0$ and the final time $%
t_{b}=t_{b}-t_{a}=a_{m}^{s}t$ in value. This is the $SIC$ perturbation
equation that a $SIC$\ Green function obeys. In principle it should be able
to derive this $SIC$ perturbation equation from the standard path
integration formula (4.44), but here there is not such a detailed
derivation. However, this $SIC$\ perturbation equation can be directly
obtained from the $SIC$ perturbation equation in operator form:%
\begin{equation}
U^{sic}\left( t\right) =U_{0}^{sic}\left( t\right) +\frac{1}{i\hslash }%
\int_{0}^{t}U_{0}^{sic}\left( t-t^{\prime }\right) \left(
a_{m}^{s}H_{1}\right) U^{sic}\left( t^{\prime }\right) dt^{\prime } 
\tag{4.50}
\end{equation}%
This is the $SIC$ perturbation equation that a $SIC$\ unitary propagator
obeys. It is derived in detail in Appendix A on the basis of the motion
equation in the interaction picture in quantum mechanics.

If the $SIC$\ Green function $G_{H_{0}}^{sic}(x_{b},t_{b};x_{a},0)$ in
(4.49) can be calculated conveniently by the path integration formula
(4.44), then by iterating the $SIC$ perturbation equation (4.49) one is able
to calculate the $SIC$\ Green function $G_{H}^{sic}(x_{b},t_{b};x_{a},0)$ to
any order approximation.

So far a single $SIC$ unitary propagator and its corresponding $SIC$\ Green
function have been treated explicitly. Generally a QUANSDAM (or UNIDYSLOCK)\
process is a sequence of many $SIC$\ unitary propagators. Therefore, it is
necessary to treat explicitly the product of two or more $SIC$\ unitary
propagators and its corresponding $SIC$ Green function.

A concatenate $SIC$\ unitary propagator is defined as the product of two or
more $SIC$ unitary propagators. Given two $SIC$ unitary propagators $%
U_{H_{1}}^{sic}(a_{m}^{s},T_{1})$ and $U_{H_{2}}^{sic}(a_{m}^{s},T_{2})$ the
concatenate $SIC$\ unitary propagator of them may be written as%
\begin{equation}
U_{H_{12}}^{sic}(a_{m}^{s},T_{12})=U_{H_{2}}^{sic}(a_{m}^{s},T_{2})U_{H_{1}}^{sic}(a_{m}^{s},T_{1})
\tag{4.51}
\end{equation}%
According to the definition (4.2) this concatenate $SIC$\ unitary propagator
is given by $G_{H_{12}}^{sic}\left( x_{b},t_{b};x_{a},t_{a}\right) \equiv
\left\langle
x_{b},t_{b}|U_{H_{12}}^{sic}(a_{m}^{s},T_{12})|x_{a},t_{a}\right\rangle $ in
the coordinate representation. And it can prove that%
\begin{equation}
G_{H_{12}}^{sic}\left( x_{b},t_{b};x_{a},t_{a}\right) =\int_{-\infty
}^{\infty }dx_{c}G_{H_{2}}^{sic}\left( x_{b},t_{b};x_{c},t_{c}\right)
G_{H_{1}}^{sic}\left( x_{c},t_{c};x_{a},t_{a}\right) .  \tag{4.52}
\end{equation}%
This integral formula of the product of a pair of $SIC$\ Green functions is
obtained by first substituting (4.51) into the definition of the $%
G_{H_{12}}^{sic}\left( x_{b},t_{b};x_{a},t_{a}\right) $ and then inserting
the operator identity $\int_{-\infty }^{\infty }dx_{c}\left\vert
x_{c},t_{c}\right\rangle \left\langle x_{c},t_{c}\right\vert $ of the
coordinate eigenbases $\left\{ \left\vert x_{c},t_{c}\right\rangle \right\} $
into between $U_{H_{2}}^{sic}(a_{m}^{s},T_{2})$ and $%
U_{H_{1}}^{sic}(a_{m}^{s},T_{1})$. If both the $%
U_{H_{1}}^{sic}(a_{m}^{s},T_{1})$ and $U_{H_{2}}^{sic}(a_{m}^{s},T_{2})$ in
(4.51) each are taken as the form of (4.1), then in (4.52) the time interval 
$t_{c}-t_{a}=a_{m}^{s}T_{1}$ and $t_{b}-t_{c}=a_{m}^{s}T_{2}.$ The integral
formula (4.52) may be called the integral product formula of the $SIC$ Green
functions. By the integral formula (4.52) one is able to determine the $SIC$%
\ Green function that is generated by the product of two $SIC$\ unitary
propagators. This integral product formula for a pair of $SIC$ Green
functions corresponds to the conventional $QM$\ one for a pair of $QM$\
Green functions [$14$].

If both the $U_{H_{1}}^{sic}(a_{m}^{s},T_{1})$ and $%
U_{H_{2}}^{sic}(a_{m}^{s},T_{2})$ in (4.51) are any $SIC$ quadratic
propagators, then it can prove that the concatenate $SIC$ unitary propagator 
$U_{H_{12}}^{sic}(a_{m}^{s},T_{12})$ of (4.51) is a $SIC$\ quadratic
propagator too. A detail proof is given below that is based on the $SIC$%
-Green-function integral product formula (4.52), the $SIC$\ quadratic Green
functions, and the Gaussian integration. The $SIC$\ Green functions for the $%
SIC$\ quadratic propagators $U_{H_{1}}^{sic}(a_{m}^{s},T_{1})$ and $%
U_{H_{2}}^{sic}(a_{m}^{s},T_{2})$ can be obtained from (4.33) above,
respectively. According to (4.33), formally both the $SIC$ quadratic Green
functions of the $U_{H_{1}}^{sic}(a_{m}^{s},T_{1})$ and $%
U_{H_{2}}^{sic}(a_{m}^{s},T_{2})$ may be written respectively as%
\begin{equation*}
G_{H_{1}}^{sic}\left( x_{c},t_{c};x_{a},t_{a}\right) =\sqrt{\frac{m}{i2\pi
\hslash }(-S_{ac})}
\end{equation*}%
\begin{equation}
\times \exp \left( \frac{i}{\hslash }\left[ \frac{1}{2}m\left(
S_{cc}x_{c}^{2}+S_{ac}2x_{a}x_{c}+S_{aa}x_{a}^{2}\right)
+x_{a}Q_{a}+x_{c}Q_{c}+\Theta _{0}\right] \right)  \tag{4.53}
\end{equation}%
and%
\begin{equation*}
G_{H_{2}}^{sic}(x_{b},t_{b};x_{c},t_{c})=\sqrt{\frac{m}{i2\pi \hslash }%
(-S_{cb}^{\prime })}
\end{equation*}%
\begin{equation}
\times \exp \left( \frac{i}{\hslash }\left[ \frac{1}{2}m\left(
S_{bb}^{\prime }x_{b}^{2}+S_{cb}^{\prime }2x_{c}x_{b}+S_{cc}^{\prime
}x_{c}^{2}\right) +x_{c}Q_{c}^{\prime }+x_{b}Q_{b}^{\prime }+\Theta
_{0}^{\prime }\right] \right)  \tag{4.54}
\end{equation}%
where all these real $SIC$ parameters $S_{aa},$ $S_{ac},$ $S_{cc},$ $Q_{a},$ 
$Q_{c},$ and $\Theta _{0};$ $S_{cc}^{\prime },$ $S_{cb}^{\prime },$ $%
S_{bb}^{\prime },$ $Q_{c}^{\prime },$ $Q_{b}^{\prime },$ and $\Theta
_{0}^{\prime }$ each may depend on the double-valued logical number $\left(
a_{m}^{s}=\pm 1\right) ,$ e.g., $S_{aa}=S_{aa}\left( a_{m}^{s}\right) ,$ etc$%
.$ By substituting (4.53) and (4.54) into the integral product formula
(4.52) one obtains%
\begin{equation*}
G_{H_{12}}^{sic}\left( x_{b},t_{b};x_{a},t_{a}\right) =\sqrt{\frac{m}{i2\pi
\hslash }(-S_{cb}^{\prime })}\sqrt{\frac{m}{i2\pi \hslash }(-S_{ac})}\exp
\left( \frac{i}{\hslash }\left( \Theta _{0}+\Theta _{0}^{\prime }\right)
\right)
\end{equation*}%
\begin{equation*}
\times \exp \left( \frac{i}{\hslash }\left[ \frac{1}{2}m\left(
S_{bb}^{\prime }x_{b}^{2}\right) +\frac{1}{2}m\left( S_{aa}x_{a}^{2}\right)
+x_{b}Q_{b}^{\prime }+x_{a}Q_{a}\right] \right)
\end{equation*}%
\begin{equation*}
\times \int_{-\infty }^{\infty }dx_{c}\exp \{\frac{i}{\hslash }[\frac{1}{2}%
m\left( S_{cb}^{\prime }2x_{c}x_{b}+S_{cc}^{\prime }x_{c}^{2}\right)
\end{equation*}%
\begin{equation}
+\frac{1}{2}m\left( S_{cc}x_{c}^{2}+S_{ac}2x_{a}x_{c}\right)
+x_{c}Q_{c}^{\prime }+x_{c}Q_{c}]\}  \tag{4.55}
\end{equation}%
The integration on the RH side of (4.55) is a Gaussian integration about the
variable $x_{c}$. For convenience this Gaussian integration is denoted as $%
G\left( x_{b},x_{a}\right) .$ It can be exactly calculated. The calculated
result is given by%
\begin{equation}
G\left( x_{b},x_{a}\right) =\int_{-\infty }^{\infty }dx_{c}\exp \left(
-ax_{c}^{2}+bx_{c}\right) =\sqrt{\frac{\pi }{a}}\exp \left( \frac{b^{2}}{4a}%
\right)  \tag{4.56}
\end{equation}%
where $a=-\frac{im}{2\hslash }\left( S_{cc}^{\prime }+S_{cc}\right) $ and $b=%
\frac{i}{\hslash }\left[ m\left( S_{cb}^{\prime }x_{b}+S_{ac}x_{a}\right)
+\left( Q_{c}^{\prime }+Q_{c}\right) \right] .$ Now by substituting (4.56)
into (4.55) the $SIC$ Green function of (4.55)\ can be reduced to the form%
\begin{equation*}
G_{H_{12}}^{sic}\left( x_{b},t_{b};x_{a},t_{a}\right) =\exp \left( \frac{i}{%
\hslash }\left( \Theta _{0}+\Theta _{0}^{\prime }\right) \right) \exp \left(
-\frac{i}{2\hslash m}\frac{\left( Q_{c}^{\prime }+Q_{c}\right) ^{2}}{\left(
S_{cc}^{\prime }+S_{cc}\right) }\right)
\end{equation*}%
\begin{equation*}
\times \sqrt{\frac{m}{i2\pi \hslash }(-S_{cb}^{\prime })}\sqrt{\frac{m}{%
i2\pi \hslash }(-S_{ac})}\sqrt{\frac{2i\pi \hslash }{m\left( S_{cc}^{\prime
}+S_{cc}\right) }}
\end{equation*}%
\begin{equation*}
\times \exp \left( \frac{i}{\hslash }\left[ \frac{1}{2}m\left(
S_{bb}^{\prime }x_{b}^{2}\right) +\frac{1}{2}m\left( S_{aa}x_{a}^{2}\right)
+x_{b}Q_{b}^{\prime }+x_{a}Q_{a}\right] \right)
\end{equation*}%
\begin{equation}
\times \exp \left( -\frac{i}{2\hslash m}\frac{\left[ m^{2}\left(
S_{cb}^{\prime }x_{b}+S_{ac}x_{a}\right) ^{2}+2m\left( Q_{c}^{\prime
}+Q_{c}\right) \left( S_{cb}^{\prime }x_{b}+S_{ac}x_{a}\right) \right] }{%
\left( S_{cc}^{\prime }+S_{cc}\right) }\right)  \tag{4.57}
\end{equation}%
By this expression it can be further shown that the $G_{H_{12}}^{sic}\left(
x_{b},t_{b};x_{a},t_{a}\right) $ of (4.57) is indeed a $SIC$ quadratic Green
function. Actually, the $G_{H_{12}}^{sic}\left(
x_{b},t_{b};x_{a},t_{a}\right) $ can be formally written as%
\begin{equation}
G_{H_{12}}^{sic}\left( x_{b},t_{b};x_{a},t_{a}\right) =\sqrt{\frac{m}{i2\pi
\hslash }(-S_{ab}^{\prime \prime })}\exp \left( \frac{i}{\hslash }%
S_{12}^{sic}\left( x_{b},t_{b};x_{a},t_{a}\right) \right) \newline
\tag{4.58}
\end{equation}%
where the $SIC$\ action $S_{12}^{sic}\left( x_{b},t_{b};x_{a},t_{a}\right) $
is a quadratic function of the coordinates $x_{b}$ and $x_{a}$ and may be
formally written as%
\begin{equation}
S_{12}^{sic}(x_{b},t_{b};x_{a},t_{a})=\frac{1}{2}m\left( S_{bb}^{\prime
\prime }x_{b}^{2}+S_{ab}^{\prime \prime }2x_{a}x_{b}+S_{aa}^{\prime \prime
}x_{a}^{2}\right) +x_{a}Q_{a}^{\prime \prime }+x_{b}Q_{b}^{\prime \prime
}+\Theta _{0}^{\prime \prime }  \tag{4.59}
\end{equation}%
All these real $SIC$ parameters in (4.58) and (4.59), i.e., $S_{aa}^{\prime
\prime },$ $S_{ab}^{\prime \prime },$ $S_{bb}^{\prime \prime },$ $%
Q_{a}^{\prime \prime },$ $Q_{b}^{\prime \prime },$ and $\Theta _{0}^{\prime
\prime },$ each can be explicitly obtained from (4.57). They generally
depend on the double-valued logical number $a_{m}^{s}.$ A detailed
calculation shows that they are explicitly given by%
\begin{equation}
S_{bb}^{\prime \prime }=S_{bb}^{\prime }-\frac{\left( S_{cb}^{\prime
}\right) ^{2}}{S_{cc}^{\prime }+S_{cc}},\text{ }S_{ab}^{\prime \prime }=-%
\frac{S_{cb}^{\prime }S_{ac}}{S_{cc}^{\prime }+S_{cc}},\text{ }%
S_{aa}^{\prime \prime }=S_{aa}-\frac{S_{ac}^{2}}{S_{cc}^{\prime }+S_{cc}}, 
\tag{4.60a}
\end{equation}%
\begin{equation}
Q_{b}^{\prime \prime }=Q_{b}^{\prime }-\frac{S_{cb}^{\prime }\left(
Q_{c}^{\prime }+Q_{c}\right) }{S_{cc}^{\prime }+S_{cc}},\text{ }%
Q_{a}^{\prime \prime }=Q_{a}-\frac{S_{ac}\left( Q_{c}^{\prime }+Q_{c}\right) 
}{S_{cc}^{\prime }+S_{cc}}\newline
,  \tag{4.60b}
\end{equation}%
\begin{equation}
\Theta _{0}^{\prime \prime }=\Theta _{0}^{\prime }+\Theta _{0}-\frac{1}{2m}%
\frac{\left( Q_{c}^{\prime }+Q_{c}\right) ^{2}}{S_{cc}^{\prime }+S_{cc}} 
\tag{4.60c}
\end{equation}%
As shown in (4.59), the $SIC$\ action $S_{12}^{sic}\left(
x_{b},t_{b};x_{a},t_{a}\right) $ in (4.58) is a quadratic function of the
coordinates $x_{b}$ and $x_{a}.$ This shows that the $G_{H_{12}}^{sic}\left(
x_{b},t_{b};x_{a},t_{a}\right) $ of (4.58) is a $SIC$\ quadratic Green
function. This further shows that the concatenate $SIC$ unitary propagator $%
U_{H_{12}}^{sic}(a_{m}^{s},T_{12})$ of (4.51), which generates the $%
G_{H_{12}}^{sic}\left( x_{b},t_{b};x_{a},t_{a}\right) $ of (4.58), is indeed
a $SIC$ quadratic propagator. It therefore concludes that the product of a
pair of $SIC$ quadratic propagators is still a $SIC$ quadratic propagator.
The only exception is that the $G_{H_{12}}^{sic}\left(
x_{b},t_{b};x_{a},t_{a}\right) $ of (4.58) may be reduced to a $QM$\
quadratic Green function up to a global phase factor, when both the physical
and the math Hilbert space to describe the $G_{H_{12}}^{sic}\left(
x_{b},t_{b};x_{a},t_{a}\right) $ are completely overlapping with each other.
The $SIC$ quadratic propagator $U_{H_{12}}^{sic}(a_{m}^{s},T_{12})$ of
(4.51) that generates the $G_{H_{12}}^{sic}(x_{b},t_{b};$ $x_{a},t_{a})$ of
(4.58) may not be written in the simple form of (4.1). Thus, it is a general 
$SIC$\ quadratic propagator. With the help of the same method as the above
one it can prove that the above conclusion that the product of a pair of $%
SIC $ quadratic propagators is a $SIC$ quadratic propagator is still
available for general $SIC$\ quadratic propagators.

It can seen from the above theoretical proof that the property that the
product of a pair of $SIC$ quadratic propagators is still a $SIC$ quadratic
propagator is obtained via the Gaussian integration of the product of a pair
of $SIC$\ quadratic Green functions. As shown in (4.53) and (4.54), these $%
SIC$ quadratic Green functions are the quadratic (or Gaussian) functions of
the coordinate variables ($x_{a},$ $x_{b},$ $x_{c}$). It is known that the
product of a pair of Gaussian functions is still a Gaussian function and the
Gaussian integration of the product is still a Gaussian function.\footnote{%
Suppose that $G_{1}\left( x,y,z\right) $ and $G_{2}\left( x,y,z\right) $ are
two Gaussian functions with variables $x,$ $y,$ and $z$. Then the product of
the two Gaussian functions, $G_{3}\left( x,y,z\right) =G_{1}\left(
x,y,z\right) G_{2}\left( x,y,z\right) ,$ is still a Gaussian function.
Moreover, the Gaussian integral of the product, $G\left( x,y\right)
=\int_{-\infty }^{\infty }dzG_{3}\left( x,y,z\right) ,$ is also a Gaussian
function of the variables $x$ and $y$. Similarly, $G\left( x,z\right) $ and $%
G\left( y,z\right) $ are also Gaussian functions.} Then the Gaussian
integration of the product of a pair of $SIC$\ quadratic Green functions is
still a Gaussian function and hence a $SIC$ quadratic Green function. This
is the first interpretation for the property. It also may be thought that
the property is original from the symmetrical structure and property of the
quantum system with a quadratic Hamiltonian, which is the fundamental
quantum-computing-speedup resource [$1,2$]. In the quantum-computing speedup
theory the symmetrical structure and property of the quantum system may be
completely characterized by the quantum symmetry group whose elements are
the quadratic unitary operators on the physical Hilbert space and also the
one whose elements are the quadratic unitary operators on the math Hilbert
space, here both the quantum symmetry groups are the same one in the present
case. This is the second interpretation. The second interpretation is
fundamental (See Refs. [$1,2$] and also the section one above), but the
first one really plays direct role in constructing theoretically a powerful
QUANSDAM process by the Green function method.

One $SIC$ unitary propagator may be transformed to another by making a $QM$
unitary transformation. This property can be applied as well to a $SIC$
quadratic propagator. As shown in the previous section 3, it is useful for
the preparation of the $SIC$ unitary propagators by starting from the basic $%
SIC$\ unitary operators. It is also important for those $SIC$ unitary
propagators that may not be prepared (or implemented) directly or
conveniently. As a typical example, consider the simplest$\ SIC$ quadratic
propagators $\exp \left( -ia_{m}^{s}\hat{V}_{c0}t_{m}/\hslash \right) 
\mathbf{\ }$with $\hat{V}_{c0}=\frac{1}{2}m\omega _{0}^{2}x^{2}\mathbf{\ }$%
and $\exp \left( -ia_{m}^{s}\hat{T}t_{m}/\hslash \right) $ with $\hat{T}=%
\frac{1}{2m}p^{2}$ to be prepared in a single-atom system. Both the $SIC$
quadratic propagators could not be prepared directly in a single-atom system
in the absence of external potential field, since such a single atom with
free COM motion has a continuous but not discrete COM-motion energy
spectrum. However, if they could be prepared conveniently in the single-atom
system with external harmonic potential field, then in principle they could
be effectively prepared in the same single-atom system in the absence of
external potential field. This may be done as follows: after (and before)
they each are prepared in the single-atom system with external harmonic
potential field, one turns off the external harmonic potential field
immediately. Now according to quantum mechanics the Heisenberg motion
equation for the COM motion of a single atom (e.g., a single ion [$17$]) in
external harmonic potential field may be written in the analytical solution
form (See, for example, Page 102 in Ref. [$26$])%
\begin{equation}
x\left( t\right) =\exp \left( iH_{c0}t/\hslash \right) x\exp \left(
-iH_{c0}t/\hslash \right) =x\cos \left( \omega _{0}t\right) +\frac{p}{%
m\omega _{0}}\sin \left( \omega _{0}t\right)  \tag{4.61a}
\end{equation}%
\begin{equation}
p\left( t\right) =\exp \left( iH_{c0}t/\hslash \right) p\exp \left(
-iH_{c0}t/\hslash \right) =p\cos \left( \omega _{0}t\right) -m\omega
_{0}x\sin \left( \omega _{0}t\right)  \tag{4.61b}
\end{equation}%
where the atomic COM Hamiltonian $H_{c0}$ is given by $H_{c0}=\hat{T}+\hat{V}%
_{c0}$ with the oscillatory angular frequency $\omega _{0}$. By setting $%
\omega _{0}t=\pi /2$ in the motion equation one obtains%
\begin{equation}
\exp \left( iH_{c0}t_{c}/\hslash \right) x\exp \left( -iH_{c0}t_{c}/\hslash
\right) =\frac{p}{m\omega _{0}}  \tag{4.62a}
\end{equation}%
\begin{equation}
\exp \left( iH_{c0}t_{c}/\hslash \right) p\exp \left( -iH_{c0}t_{c}/\hslash
\right) =-m\omega _{0}x  \tag{4.62b}
\end{equation}%
where $t=t_{c}=\frac{\pi }{2\omega _{0}}.$ Note that $\exp \left(
-iH_{c0}t_{c}/\hslash \right) $ is a $QM$ unitary operator. With the help of
the $QM$ unitary transformations of (4.62a) and (4.62b) one obtains the
following $QM$ unitary transformations between the two $SIC$ quadratic
propagators $\exp \left( -ia_{m}^{s}\hat{V}_{c0}t_{m}/\hslash \right) $ and $%
\exp \left( -ia_{m}^{s}\hat{T}t_{m}/\hslash \right) :$%
\begin{equation}
\exp \left( iH_{c0}t_{c}/\hslash \right) \exp \left( -ia_{m}^{s}\hat{V}%
_{c0}t_{m}/\hslash \right) \exp \left( -iH_{c0}t_{c}/\hslash \right) =\exp
\left( -ia_{m}^{s}\hat{T}t_{m}/\hslash \right)  \tag{4.63a}
\end{equation}%
\begin{equation}
\exp \left( iH_{c0}t_{c}/\hslash \right) \exp \left( -ia_{m}^{s}\hat{T}%
t_{m}/\hslash \right) \exp \left( -iH_{c0}t_{c}/\hslash \right) =\exp \left(
-ia_{m}^{s}\hat{V}_{c0}t_{m}/\hslash \right)  \tag{4.63b}
\end{equation}%
It can be seen from (4.63a) and (4.63b) that when one of the two $SIC$
quadratic propagators (or their corresponding $SIC$\ Green functions) can be
prepared or calculated conveniently, another also can be obtained
conveniently by the $QM$ unitary transformation of (4.63a) or (4.63b). 
\newline
\newline
\newline
{\Large 5. Discussion}

A quantum-computing speedup process owns the dual character that it obeys
both the unitary quantum dynamics and the mathematical-logical principle of
a computational problem to be solved. This is essentially different from a
conventional quantum computation (algorithm) [$32,33$] which is essentially
a purely quantum-physical process [$1$]. It takes into account the quantum
symmetry that is considered as the fundamental quantum-computing-speedup
resource. Both the dual character and the fundamental
quantum-computing-speedup resource are considered to be responsible for
achieving an essential quantum-computing speedup to solve a hard
computational problem in the quantum-computing speedup theory. Consequently
from the point of view of pure quantum mechanics both the unitary quantum
dynamics and the quantum symmetry are considered as the two pillars to build
an efficient quantum algorithm to solve a hard computational problem [$1,2$].

In contrast, it has been thought that the quantum-computational speedup for
a conventional quantum computation (algorithm) is achieved through the
quantum parallelism [$33$] based on the superposition of quantum states.
Then from the point of view of orthodox quantum mechanics it is essentially
based on the non-local effect of entanglement quantum states, and a
conventional quantum computation (algorithm) is indeterministic.

The largest difficulty facing the conventional quantum computation [$32,33$]
is perhaps that it is subjected to the square speedup limit on solving an
unstructured search problem, although most hard computational problems can
not be solved in an exponential quantum-computational speedup in the
conventional quantum computation.

In quantum mechanics there are two fundamentally different laws that govern
the time evolution process of a quantum system: the quantum measurement and
the unitary quantum dynamics [$13$]. The unitary time evolution process is
deterministic in the sense that in isolated quantum system state changes
deterministically in the process. In contrast, the quantum measurement is
indeterministic because in quantum system state changes in probability in
the quantum measurement. In orthodox quantum mechanics it is more reasonable
to say that the unitary quantum dynamics is computationally theoretical tool
than that it is fundamental physical law. The existing research works on the
non-local effect of entanglement quantum states in the past decades further
strengthen orthodox quantum mechanics. The conventional quantum computation [%
$32,33$] is based on orthodox quantum mechanics.

Is independent mathematical-logical principle of a computational problem
necessary in the quantum-computing speedup theory? It is essential and
necessary. If it was not independent of any physical laws, then ultimately
it would be contained by orthodox quantum mechanics and the
quantum-computing speedup theory would lose its own independence and
existence. Therefore, only when the mathematical-logical principle is taken
into account, can the determinacy of the unitary quantum dynamics become
substantial in a quantum-computing speedup process [$1$]. The
mathematical-logical principle of an unstructured search problem to be
solved was not considered as an independent principle, which is independent
of the unitary quantum dynamics and the quantum symmetry, to make
contribution to the quantum-searching speedup until the early 2001 [$3$].
According to the mathematical-logical principle of the unstructured search
problem the real solution to the search problem is completely determined in
the $HSSS$ quantum search process. The real solution state cannot be made a
random transition between the physical Hilbert space and the corresponding
math Hilbert space and it is always in the physical Hilbert space during the 
$HSSS$ quantum search process [$1$]. Suppose here that existence of the real
solution state reflects the reality in physics. Then it may be deduced from
these that if the reality is fundamental and universal, then so is the
determinism in physics.

The conventional quantum computation itself is not able to make certain for
whether the mathematical-logical principle of a computational problem is
obeyed or not in a conventional quantum computation (algorithm) to solve the
computational problem, because the conventional quantum computation
(algorithm) obeys the indeterministic (or probabilistic) quantum-physical
law, and indeterminism (or probability) is fundamental in orthodox quantum
mechanics and also in the conventional quantum computation. The conventional
quantum computation which tends to be unitary [$32$] contains the Bennett's
reversible computation [$11,12$] of classical physics.\footnote{%
This is due to that quantum physics contains classical physics.} Apparently,
from the viewpoint of the quantum-computing speedup theory [$1,2$] it seems
that the mathematical-logical principle of a computational problem should be
obeyed, when a conventional quantum computation (algorithm) solves the
problem. However, that is not so from the point of view of the
quantum-mechanical basis of the conventional quantum computation (i.e.,
orthodox quantum mechanics). A conventional quantum computation (algorithm)
is essentially a purely quantum-physical process [$1$]. Note that
indeterminism (or probability) is fundamental in orthodox quantum mechanics
and hence any quantum-physical process is fundamentally indeterministic. The
conventional quantum computation (algorithm) to solve the computational
problem therefore is also indeterministic. Consequently, whether the
mathematical-logical principle of the computational problem is obeyed is
indeterministic in the conventional quantum computation (algorithm).

A quantum-computing speedup process to solve a computational problem still
employs the quantum measurement to output the final computational result.
Then the mathematical-logical principle of the computational problem is
still obeyed during the quantum measurement. This leads to that there are
two fundamentally different ways (deterministic vs. indeterministic) to
describing the quantum measurement outputting the final computational
result, when a quantum-computing speedup process and a conventional quantum
computation (algorithm) solve the same computational problem, respectively.

As far as the $HSSS$ quantum search process is concerned, the real solution
to the unstructured search problem exists and can never be changed in the $%
HSSS$ quantum search process and especially during the quantum measurement,
because it is determined uniquely by the mathematical-logical principle of
the unstructured search problem. In order to show clearly the essential
difference between the $HSSS$ quantum search process and a conventional
quantum search algorithm in the quantum measurement to output the final
computational result, here consider the special case that the $HSSS$ quantum
search process simply consists of the duality-character oracle operations
(See the section 2 above) and the relevant $QM$ unitary operators, it does
not employ the search-space dynamical reduction, and it is simply performed
in an $n-$qubit quantum system only. On the one hand, suppose that after
this special $HSSS$ quantum search process is applied to the initial state
which is an expansion series of the usual computational bases of the quantum
system, just before the quantum measurement the quantum state in the
physical Hilbert space of the $n-$qubit quantum system is written as%
\begin{equation*}
\left\vert \Psi _{x_{0}}\right\rangle =B_{x_{0}}\left\vert
x_{0}\right\rangle +\sum_{x=0,x\neq x_{0}}^{N-1}B_{x}\left\vert
x\right\rangle
\end{equation*}%
where, as shown in the section 2, $\left\vert x_{0}\right\rangle $ is the
real solution state, and $\{\left\vert x\right\rangle \}$ which includes $%
\left\vert x_{0}\right\rangle $ is the usual computational basis set with
base number $N=2^{n}$. The coefficients in the final state $\left\vert \Psi
_{x_{0}}\right\rangle $ satisfy the normalization condition: $%
\sum_{x=0}^{N-1}\left\vert B_{x}\right\vert ^{2}=1.$ Particularly the
coefficient $B_{x_{0}}$ of the real solution state $\left\vert
x_{0}\right\rangle $ generally satisfies $\left\vert B_{x_{0}}\right\vert
^{2}\leq 1.$ On the other hand, consider that the conventional quantum
search algorithm is applied to the same initial state. Assume that after the
conventional quantum search algorithm is applied, just before the quantum
measurement the quantum state of the $n-$qubit quantum system is written as%
\begin{equation*}
\left\vert \Psi _{x_{0}^{\prime }}^{\prime }\right\rangle =B_{x_{0}^{\prime
}}^{\prime }\left\vert x_{0}^{\prime }\right\rangle +\sum_{x=0,x\neq
x_{0}^{\prime }}^{N-1}B_{x}^{\prime }\left\vert x\right\rangle
\end{equation*}%
where $\left\vert x_{0}^{\prime }\right\rangle $ is the marked state of the
conventional quantum search algorithm with $x_{0}^{\prime }=x_{0}$ to be
found, and $\{\left\vert x\right\rangle \}$ which includes $\left\vert
x_{0}^{\prime }\right\rangle $ is still the usual computational basis set.
The final state $\left\vert \Psi _{x_{0}^{\prime }}^{\prime }\right\rangle $
satisfies the normalization condition: $\sum_{x=0}^{N-1}\left\vert
B_{x}^{\prime }\right\vert ^{2}=1,$ and the coefficient $B_{x_{0}^{\prime
}}^{\prime }$ of the marked state $\left\vert x_{0}^{\prime }\right\rangle $
generally satisfies $\left\vert B_{x_{0}^{\prime }}^{\prime }\right\vert
^{2}\leq 1.$ Now by quantum measuring the final state $\left\vert \Psi
_{x_{0}}\right\rangle $ one is able to output the final computational result
for the $HSSS$ quantum search process. Likewise, by quantum measuring the
final state $\left\vert \Psi _{x_{0}^{\prime }}^{\prime }\right\rangle $ one
may obtain the final computational result of the conventional quantum search
algorithm. The quantum measurement is the final step for any quantum search
process (or algorithm) above. At this final step there is still the
essential difference between the $HSSS$ quantum search process and the
conventional quantum search algorithm. This can be shown below.

First of all, though both the final states $\left\vert \Psi
_{x_{0}}\right\rangle $ and $\left\vert \Psi _{x_{0}^{\prime }}^{\prime
}\right\rangle $ each are a finite series of the same $N$ usual
computational bases of the $n-$qubit quantum system, there is still their
essential difference. According to the theoretical analysis for the
duality-character oracle operations of the $HSSS$ quantum search process, as
shown in the section 2, it can prove theoretically that in the final state $%
\left\vert \Psi _{x_{0}}\right\rangle $ only the real solution state $%
\left\vert x_{0}\right\rangle $ is a member of the unstructured search
space, while any other computational base $\left\vert x\right\rangle $ than $%
\left\vert x_{0}\right\rangle $ is not. In contrast, in the final state $%
\left\vert \Psi _{x_{0}^{\prime }}^{\prime }\right\rangle $ of the
conventional quantum search algorithm the marked state $\left\vert
x_{0}^{\prime }\right\rangle $ is naturally a member of the unstructured
search space, but any other computational base $\left\vert x\right\rangle $
than $\left\vert x_{0}^{\prime }\right\rangle $ is also a member of the
unstructured search space. How can this essential difference between the
final states $\left\vert \Psi _{x_{0}}\right\rangle $ and $\left\vert \Psi
_{x_{0}^{\prime }}^{\prime }\right\rangle $ cause the deterministic vs.
indeterministic difference between the $HSSS$ quantum search process and the
conventional quantum search algorithm?

It seems that there is apparently no difference between the $HSSS$ quantum
search process and the conventional quantum search algorithm at the final
step if both the amplitudes $B_{x_{0}}$ in the final state $\left\vert \Psi
_{x_{0}}\right\rangle $ and $B_{x_{0}^{\prime }}^{\prime }$ in the final
state $\left\vert \Psi _{x_{0}^{\prime }}^{\prime }\right\rangle $ satisfy $%
\left\vert B_{x_{0}^{\prime }}^{\prime }\right\vert ^{2}=\left\vert
B_{x_{0}}\right\vert ^{2}=1,$ because for any one of both the quantum search
processes (or algorithms) the output state is the state $\left\vert
x_{0}\right\rangle $ ($x_{0}^{\prime }=x_{0}$) only and the final
computational result is the real solution ($x_{0}$) to the unstructured
search problem. While for the $HSSS$ quantum search process this final step
is evidently deterministic, apparently it seems to be also deterministic for
the conventional quantum search algorithm. However, once $\left\vert
B_{x_{0}^{\prime }}^{\prime }\right\vert ^{2},$ $\left\vert
B_{x_{0}}\right\vert ^{2}<1,$ the essential difference is exposed between
the $HSSS$ quantum search process and the conventional quantum search
algorithm. The essential difference is analyzed in detail in the following
two parts $(1)$ and $(2)$ for the conventional quantum search algorithm and
the $HSSS$ quantum search process, respectively.

$\left( 1\right) $ This part is devoted to the conventional quantum search
algorithm. Suppose that the probability $\left\vert B_{x_{0}^{\prime
}}^{\prime }\right\vert ^{2}$ of the marked state $\left\vert x_{0}^{\prime
}\right\rangle $ is the maximum one in the final state $\left\vert \Psi
_{x_{0}^{\prime }}^{\prime }\right\rangle $ and it is still less than one,
i.e., $\left\vert B_{x_{0}^{\prime }}^{\prime }\right\vert ^{2}<1.$
According to orthodox quantum mechanics each time after the quantum
measurement is performed the final state $\left\vert \Psi _{x_{0}^{\prime
}}^{\prime }\right\rangle $ is changed in probability to some computational
basis state (e.g., $\left\vert x_{0}^{\prime }\right\rangle $ or $\left\vert
x\right\rangle $). As shown above, this basis state of the final state $%
\left\vert \Psi _{x_{0}^{\prime }}^{\prime }\right\rangle $ is always a
member (or candidate solution state) of the unstructured search space,
regardless of the marked state $\left\vert x_{0}^{\prime }\right\rangle $ or
any other basis state $\left\vert x\right\rangle $. It is just the output
state of the conventional quantum search algorithm after performing the
quantum measurement one time, and hence it is considered as a
measurement-outputting real solution state to the unstructured search
problem. This is consistent with the `inverse' statement that the real
solution state may be possibly considered as any candidate solution state in
the final state $\left\vert \Psi _{x_{0}^{\prime }}^{\prime }\right\rangle $
before the real solution is found, and as stated in the section 2, this
`inverse' statement means the indeterministic description. Of course, due to 
$\left\vert B_{x_{0}^{\prime }}^{\prime }\right\vert ^{2}<1$ this
measurement-outputting real solution state may or may not be finally
assigned as the real solution state to the unstructured search problem.
However, by repeating many times to run the conventional quantum search
algorithm including the quantum measurement of the final state $\left\vert
\Psi _{x_{0}^{\prime }}^{\prime }\right\rangle $ one can obtain the
probability distribution for all the measurement-outputting real solution
states. Then with the help of this probability distribution one can find the
measurement-outputting real solution state with maximum probability, i.e.,
the marked state $\left\vert x_{0}^{\prime }\right\rangle $ with $%
x_{0}^{\prime }=x_{0}$. Finally this marked state, which is found by the
conventional quantum search algorithm, is assigned as the real solution
state to the unstructured search problem.

The detailed analysis above reveals the indeterministic property for the
conventional quantum search algorithm that the quantum measurement, which is
indeterministic and fundamental in orthodox quantum mechanics, determines
the real solution state to the unstructured search problem instead of the
mathematical-logical principle of the unstructured search problem. This
really means that while it is able to find the real solution to the
unstructured search problem, the conventional quantum search algorithm
itself can not make certain for whether the mathematical-logical principle
of the unstructured search problem is obeyed or not. This is consistent with
the previous general conclusion\ that the conventional quantum computation
itself is not able to make certain for whether the mathematical-logical
principle of a computational problem is obeyed or not in a conventional
quantum computation (algorithm) to solve the computational problem. The
reason for these comes from orthodox quantum mechanics. According to
orthodox quantum mechanics, not only whether the mathematical-logical
principle of the unstructured search problem is obeyed is uncertain in the
conventional quantum search algorithm, but also whether the state $%
\left\vert x_{0}\right\rangle $ can be assigned as the real solution state
to the unstructured search problem or not is also uncertain before the
quantum measurement.

$\left( 2\right) $ This part is devoted to the $HSSS$ quantum search
process. Consider first the basic case that the coefficient $B_{x_{0}}$ of
the real solution state $\left\vert x_{0}\right\rangle $ is given by $%
\left\vert B_{x_{0}}\right\vert ^{2}=1$ in the final state $\left\vert \Psi
_{x_{0}}\right\rangle .$ Now by quantum measurement the final state $%
\left\vert \Psi _{x_{0}}\right\rangle $ is changed deterministically to the
real solution state $\left\vert x_{0}\right\rangle .$ The latter state is
just the output state of the $HSSS$ quantum search process. Next consider
that the absolute coefficient $B_{x_{0}}$ is smaller than one or $\left\vert
B_{x_{0}}\right\vert ^{2}<1.$ For convenience here suppose that the term $%
B_{x_{0}}\left\vert x_{0}\right\rangle $ is the dominating term, while the
term $\sum_{x=0,x\neq x_{0}}^{N-1}B_{x}\left\vert x\right\rangle $ is small
term in the final state $\left\vert \Psi _{x_{0}}\right\rangle ,$ that is, $%
\left\vert B_{x_{0}}\right\vert ^{2}>\sum_{x=0,x\neq x_{0}}^{N-1}\left\vert
B_{x}\right\vert ^{2}.$ Then by quantum measurement the final state $%
\left\vert \Psi _{x_{0}}\right\rangle $ is changed to the real solution
state $\left\vert x_{0}\right\rangle $ or any other basis state $\left\vert
x\right\rangle $ than $\left\vert x_{0}\right\rangle $ in probability. As
shown above, in this case only the real solution state $\left\vert
x_{0}\right\rangle $ is a member of the unstructured search space, while any
other basis state $\left\vert x\right\rangle $ than $\left\vert
x_{0}\right\rangle $ is not. Thus, in the final state $\left\vert \Psi
_{x_{0}}\right\rangle $ any basis state $\left\vert x\right\rangle $ other
than $\left\vert x_{0}\right\rangle $ is not a candidate solution state to
the unstructured search problem. As shown in the section 2, this is original
from the duality-character oracle operation of the $HSSS$ quantum search
process. This is the substantial difference for the $HSSS$ quantum search
process from the conventional quantum search algorithm. Consequently, if by
the quantum measurement the final state $\left\vert \Psi
_{x_{0}}\right\rangle $ is changed to the basis state $\left\vert
x\right\rangle ,$ one cannot say that the $HSSS$ quantum search process
obtains a measurement-outputting real solution state. This also means that
assigning the basis state $\left\vert x\right\rangle $ to the real solution
state is not consistent with the mathematical-logical principle of the
unstructured search problem. The real solution to the unstructured search
problem has nothing to do with quantum measurement. It is uniquely
determined by the mathematical-logical principle of the unstructured search
problem. The mathematical-logical principle therefore must be obeyed in the
quantum measurement. Once this is taken into account, it is only reasonable
that the final state $\left\vert \Psi _{x_{0}}\right\rangle $ is changed to
the real solution state $\left\vert x_{0}\right\rangle $ by the quantum
measurement, although $\left\vert B_{x_{0}}\right\vert ^{2}<1.$ Therefore,
in theory the $HSSS$ quantum search process is still deterministic even if $%
\left\vert B_{x_{0}}\right\vert ^{2}<1.$ Now one needs to further consider
reasonably the case that the final state $\left\vert \Psi
_{x_{0}}\right\rangle $ is changed to the basis state $\left\vert
x\right\rangle $ by the quantum measurement. This is related to the small
term $\sum_{x=0,x\neq x_{0}}^{N-1}B_{x}\left\vert x\right\rangle $ of the
final state $\left\vert \Psi _{x_{0}}\right\rangle ,$ which is a finite
series of all the basis states $\{\left\vert x\right\rangle \}$ except the
real solution state $\left\vert x_{0}\right\rangle .$ Here it is not
discussed whether the term could make a contribution to the overlapping
integrations between the final states of the physical and math Hilbert
spaces. Equivalently speaking, the small term is always changed to a basis
state $\left\vert x\right\rangle $ ($\neq \left\vert x_{0}\right\rangle $)
by the quantum measurement. Thus, it does not make a positive contribution
to the determination of the real solution state by the quantum measurement.
It disturbs the determination of the real solution state. Then it is an
error source to determining the real solution state by the quantum
measurement for the $HSSS$ quantum search process. Theoretically, it
vanishes for the $HSSS$ quantum search process whose final state is only the
real solution state.

The quantum-computing speedup theory mainly deals with the quantum-computing
speedup and its mechanism. It involves the fundamental aspects of quantum
physics, i.e., unitarity, symmetry, reality, and determinism. \newline
\newline
\newline
{\Large References}\newline
1. X. Miao, \textit{The unitary dynamical state-locking process, the HSSS
quantum search process, and the quantum-computing speedup theory},
arXiv.org:1612.05969 [quant-ph] (2016)\newline
2. X. Miao, \textit{The universal quantum driving force to speed up a
quantum computation -- The unitary quantum dynamics}, arXiv:1105.3573
[quant-ph] (2011)\newline
3. X. Miao, \textit{Universal construction for the unsorted quantum search
algorithms, }https://arXiv.org/abs/quant-ph/0101126 (2001)\newline
4. Y. Lecerf, \textit{Machines} \textit{de} \textit{Turing} \textit{r\'{e}%
versibles}. \textit{R\'{e}cursive} \textit{insolubilit\'{e}} \textit{en} 
\textit{n} $\in $ \textit{N} \textit{de} \textit{l}$^{\prime }$ \textit{\'{e}%
quation} $\mathit{u=}\theta ^{\mathit{n}}\mathit{u}$, \textit{o\`{u} }$%
\theta $\textit{\ est un} $\mathit{\ll }$\textit{isomorphisme de codes}$%
\mathit{\gg }$, C. R. Acad. Sci. 257, 2597 (1963)\newline
5. X. Miao, \textit{Efficient dynamical reduction from the exponentially
large unstructured search space of a search problem to a polynomially small
subspace in an n-qubit spin system}, Unpublished work (2012)\newline
6. X. Miao, \textit{Multiple-quantum operator algebra spaces and description
for the unitary time evolution of multilevel spin systems}, Molec. Phys. 98,
625 (2000)\newline
7. X. Miao, \textit{Efficient multiple-quantum transition processes in an }$%
n-$\textit{qubit spin system}, http://arxiv.org/abs/quant-ph/0411046 (2004)%
\newline
8. L. K. Grover, \textit{Quantum mechanics helps in searching for a needle
in a haystack}, Phys. Rev. Lett. 79, 325 (1997)\newline
9. C. H. Bennett, E. Bernstein, G. Brassard, and U. Vazirani, \textit{%
Strengths \ and \ weaknesses \ of \ quantum\ \ computing}, \
http://arxiv.org/abs/quant-ph/ 9701001 (1997)\newline
10. L. I. Schiff, \textit{Quantum mechanics}, 3rd, McGraw-Hill book company,
New York, 1968\newline
11. C. H. Bennett, \textit{Logical reversibility of computation}, IBM J.
Res. Develop. 17, 525 (1973)\newline
12. E. Fredkin and T. Toffoli, \textit{Conservative logic}, Int. J. Theor.
Phys. 21, 219 (1982)\newline
13. J. von Neumann, \textit{Mathematical foundations of quantum mechanics},
(Translated by R. T. Beyer), Princeton University Press, 1955\newline
14. R. P. Feynman and A. R. Hibbs, \textit{Quantum mechanics and path
integrals}, McGraw-Hill, New York, 1965\newline
15. X. Miao, \textit{Unitary manipulation of a single atom in time and space
--- The spatially-selective and internal-state-selective triggering pulses},
http://arxiv.org/ abs/quant-ph/1309.3758 (2013)\newline
16. H. Dehmelt, \textit{Less is more: Experiments with an individual atomic
particle at rest in free space}, Am. J. Phys. 58, 17 (1990)\newline
17. D. J. Heinzen and D. J. Wineland, \textit{Quantum-limited cooling and
detection of radio-frequency oscillations by laser-cooled ions}, Phys. Rev.
A 42, 2977 (1990)\newline
18. A. R. Edmonds, \textit{Angular momentum in quantum mechanics}, 2nd edn,
Princeton University Press, Princeton (1974)\newline
19. M. E. Rose, \textit{Elementary theory of angular momentum}, Wiley, New
York (1957)\newline
20. L. S. Schulman, \textit{Techniques and applications of path integration}%
, Dover, New York (2005)\newline
21. E. Hille, \textit{A class of reciprocal functions}, Annals of
Mathematics, Second series, Vol. 27, 427 (1926)\newline
22. S. Thangavelu, \textit{Lectures on Hermite and Laguerre expansions},
Chapter one, Princeton University Press, Princeton (1993)\newline
23. M. Goodman, \textit{Path integral solution to the infinite square well},
Am. J. Phys. 49, 843 (1981)\newline
24. X. Miao, \textit{Universal construction of unitary transformation of
quantum computation with one- and two-body interactions},
http://arxiv.org/abs/quant-ph/0003068 (2000)\newline
25. R. R. Ernst, G. Bodenhausen, and A. Wokaun, \textit{Principles of
Nuclear Magnetic Resonance in One and Two Dimensions}, Oxford University
Press, Oxford (1987)\newline
26. T. F. Jordan, \textit{Linear operators for quantum mechanics}, Dover,
New York (2006)\newline
\textit{27. }R. Freeman\textit{, Spin Choreography, Spektrum, Oxford, 1997}%
\newline
28. L. Allen and J. H. Eberly, \textit{Optical resonance and two-level atoms}%
, Dover, New York, 1987\newline
29. K. Bergmann, H. Theuer, and B. W. Shore, \textit{Coherent population
transfer among quantum states of atoms and molecules}, Rev. Mod. Phys. 70,
1003 (1998)\newline
30. X. Miao, \textit{The STIRAP-based \ unitary \ decelerating \ and \
accelerating \ processes \ of a single free atom},
http://arxiv.org/abs/quant-ph/0707.0063 (2007)\newline
31. X. Miao, \textit{Unitarily manipulating in time and space a Gaussian
wave-packet motional state of a single atom in a quadratic potential field},
http: //arxiv.org/ abs/quant-ph/0708.2129 (2007)\newline
32. P. Benioff, \textit{Quantum mechanical Hamiltonian models of Turing
machines}, J. Statist. Phys. 29, 515 (1982)\newline
33.\ D. Deutsch, \textit{Quantum theory, the Church-Turing principle and the
universal quantum computer}, Proc. Roy. Soc. Lond. A 400, 96 (1985)\newline
34. X. Miao, \textit{The basic principles to construct a generalized
state-locking pulse field and simulate efficiently the reversible and
unitary halting protocol of a universal quantum computer},
http://arxiv.org/abs/quant-ph/0607144 (2006)\newline
\newline
\newline
{\Large Appendix A. The perturbation equations for a }$SIC${\Large \ unitary
propagator and a }$SIC${\Large \ Green function}

The $SIC$ perturbation equations are derived below. Suppose that the
time-independent Hamiltonian of a quantum system in time and space is given
by%
\begin{equation}
H=H_{0}+H_{1}\newline
\tag{A1}
\end{equation}%
where $H_{0}$ is the main term and $H_{1}$ is the perturbation term. With
the total Hamiltonian $H$ and the main Hamiltonian $H_{0}$ two $SIC$\
unitary propagators are defined by $U^{sic}\left( t\right) =\exp \left(
-ia_{m}^{s}Ht/\hslash \right) $ and $U_{0}^{sic}\left( t\right) =\exp \left(
-ia_{m}^{s}H_{0}t/\hslash \right) ,$ respectively. In the Schr\"{o}dinger
picture the time evolution process that is governed by the $SIC$\ unitary
propagator $U^{sic}\left( t\right) $ is written as%
\begin{equation}
\left\vert \Psi _{s}^{sic}\left( t\right) \right\rangle =U^{sic}\left(
t\right) \left\vert \Psi _{s}^{sic}\left( 0\right) \right\rangle  \tag{A2}
\end{equation}%
where $\left\vert \Psi _{s}^{sic}\left( t\right) \right\rangle $ is the $SIC$%
\ state at the time $t$ in the Schr\"{o}dinger picture and $\left\vert \Psi
_{s}^{sic}\left( 0\right) \right\rangle $ is the initial state, which may or
may not carry the solution information, at the initial time $t=0.$ The Schr%
\"{o}dinger equation that describes the time evolution process (A2) is given
by%
\begin{equation}
i\hslash \frac{d}{dt}\left\vert \Psi _{s}^{sic}\left( t\right) \right\rangle
=a_{m}^{s}H\left\vert \Psi _{s}^{sic}\left( t\right) \right\rangle  \tag{A3}
\end{equation}%
The $SIC$ Schr\"{o}dinger equation (A3) and the two $SIC$ unitary
propagators $U^{sic}\left( t\right) $ and $U_{0}^{sic}\left( t\right) $ are
theoretical basis to set up the perturbation equations that a $SIC$ unitary
propagator and a $SIC$\ Green function obey, respectively.

Generally, if the Hamiltonian $H$ is time-dependent, i.e., $H=H\left(
t\right) ,$ then the $SIC$ Schr\"{o}dinger equation (A3) is rewritten as%
\begin{equation*}
i\hslash \frac{d}{dt}\left\vert \Psi _{s}^{sic}\left( t\right) \right\rangle
=a_{m}^{s}H\left( t\right) \left\vert \Psi _{s}^{sic}\left( t\right)
\right\rangle
\end{equation*}%
This time-dependent $SIC$\ Schr\"{o}dinger equation also may be used to
define the $SIC$\ unitary propagator (4.40) that is generated by a
time-dependent Hamiltonian $H(t)$ in a quantum system in time and space.

Now the interaction picture is defined by%
\begin{equation}
\left\vert \Psi _{I}^{sic}\left( t\right) \right\rangle =U_{0}^{sic}\left(
t\right) ^{+}\left\vert \Psi _{s}^{sic}\left( t\right) \right\rangle 
\tag{A4}
\end{equation}%
\begin{equation}
H_{I}^{sic}\left( t\right) =U_{0}^{sic}\left( t\right)
^{+}H_{1}U_{0}^{sic}\left( t\right) \newline
\tag{A5}
\end{equation}%
where $\left\vert \Psi _{I}^{sic}\left( t\right) \right\rangle $ is the $SIC$%
\ state in the time $t$ in the interaction picture. It is clear that at the
initial time $t=0$ one has the initial state $\left\vert \Psi
_{I}^{sic}\left( 0\right) \right\rangle =\left\vert \Psi _{s}^{sic}\left(
0\right) \right\rangle .$ Here the initial state $\left\vert \Psi
_{I}^{sic}\left( 0\right) \right\rangle $ or $\left\vert \Psi
_{s}^{sic}\left( 0\right) \right\rangle $ can be arbitrary. The interaction
Hamiltonian $H_{I}^{sic}\left( t\right) $ carries the solution information,
although the perturbation term $H_{1}$ does not.

The motion equation for the state $\left\vert \Psi _{I}^{sic}\left( t\right)
\right\rangle $ in the interaction picture can be derived from (A4) and
(A5). It is given by%
\begin{equation}
i\hslash \frac{d}{dt}\left\vert \Psi _{I}^{sic}\left( t\right) \right\rangle
=a_{m}^{s}H_{I}^{sic}\left( t\right) \left\vert \Psi _{I}^{sic}\left(
t\right) \right\rangle  \tag{A6}
\end{equation}%
This equation has the formal solution:%
\begin{equation}
\left\vert \Psi _{I}^{sic}\left( t\right) \right\rangle =U_{I}^{sic}\left(
t\right) \left\vert \Psi _{I}^{sic}\left( 0\right) \right\rangle  \tag{A7}
\end{equation}%
where the $SIC$\ unitary propagator $U_{I}^{sic}\left( t\right) $ in the
interaction picture can be formally written as, as shown from the motion
equation (A10) below,%
\begin{equation}
U_{I}^{sic}\left( t\right) =\hat{T}\exp \left( -\frac{i}{\hslash }%
a_{m}^{s}\int_{0}^{t}H_{I}^{sic}\left( t^{\prime }\right) dt^{\prime }\right)
\tag{A8}
\end{equation}%
with the time-ordering operator $\hat{T}$. At the initial time $t=0$ one has 
$U_{I}^{sic}\left( 0\right) =E$ (the unity operator).

Now by using the equations (A2) and (A7) it can be obtained from the
definition (A4) of the interaction picture that $U_{I}^{sic}\left( t\right)
\left\vert \Psi _{I}^{sic}\left( 0\right) \right\rangle =U_{0}^{sic}\left(
t\right) ^{+}U^{sic}\left( t\right) \times $ $\left\vert \Psi
_{s}^{sic}\left( 0\right) \right\rangle .$ Note that the initial state $%
\left\vert \Psi _{I}^{sic}\left( 0\right) \right\rangle =\left\vert \Psi
_{s}^{sic}\left( 0\right) \right\rangle $ is arbitrary. Then one has%
\begin{equation}
U^{sic}\left( t\right) =U_{0}^{sic}\left( t\right) U_{I}^{sic}\left( t\right)
\tag{A9}
\end{equation}%
On the other hand, with the help of the equation (A7) and the fact that the
initial state $\left\vert \Psi _{I}^{sic}\left( 0\right) \right\rangle $ is
arbitrary the motion equation (A6) of the state $\left\vert \Psi
_{I}^{sic}\left( t\right) \right\rangle $ can be reduced to the following
motion equation of the $SIC$ unitary propagator $U_{I}^{sic}\left( t\right) $
in the interaction picture:%
\begin{equation}
i\hslash \frac{d}{dt}U_{I}^{sic}\left( t\right) =a_{m}^{s}H_{I}^{sic}\left(
t\right) U_{I}^{sic}\left( t\right)  \tag{A10}
\end{equation}%
with $U_{I}^{sic}\left( 0\right) =E$ at the initial time $t=0$. The formal
solution of the motion equation (A10) is just given by (A8).

By integrating the motion equation (A10) one obtains%
\begin{equation}
U_{I}^{sic}\left( t\right) =E+\frac{1}{i\hslash }a_{m}^{s}%
\int_{0}^{t}H_{I}^{sic}\left( t^{\prime }\right) U_{I}^{sic}\left( t^{\prime
}\right) dt^{\prime }  \tag{A11}
\end{equation}%
Then by using this equation and the equation (A9) one further obtains%
\begin{equation}
U^{sic}\left( t\right) =U_{0}^{sic}\left( t\right) +\frac{1}{i\hslash }%
a_{m}^{s}U_{0}^{sic}\left( t\right) \int_{0}^{t}H_{I}^{sic}\left( t^{\prime
}\right) U_{I}^{sic}\left( t^{\prime }\right) dt^{\prime }  \tag{A12}
\end{equation}%
It can prove by (A5) and (A9) that $H_{I}^{sic}\left( t^{\prime }\right)
U_{I}^{sic}\left( t^{\prime }\right) =U_{0}^{sic}\left( t^{\prime }\right)
^{+}H_{1}U^{sic}\left( t^{\prime }\right) .$ By using this relation the
equation (A12) can be rewritten in the form%
\begin{equation}
U^{sic}\left( t\right) =U_{0}^{sic}\left( t\right) +\frac{1}{i\hslash }%
\int_{0}^{t}U_{0}^{sic}\left( t-t^{\prime }\right) \left(
a_{m}^{s}H_{1}\right) U^{sic}\left( t^{\prime }\right) dt^{\prime } 
\tag{A13}
\end{equation}%
This is the $SIC$ perturbation equation that the $SIC$\ unitary propagators $%
U^{sic}\left( t\right) $ and $U_{0}^{sic}\left( t\right) $ obey. This
operator perturbation equation directly leads to that the $SIC$\ Green
functions $G_{H}^{sic}(x_{b},t_{b};x_{a},t_{a})$ (generated by $%
U^{sic}\left( t\right) $) and $G_{H_{0}}^{sic}(x_{b},t_{b};$ $x_{a},t_{a})$
(generated by $U_{0}^{sic}\left( t\right) $) obey the following equation:%
\begin{equation*}
G_{H}^{sic}(x_{b},t_{b};x_{a},0)=G_{H_{0}}^{sic}(x_{b},t_{b};x_{a},0)
\end{equation*}%
\begin{equation}
+\frac{1}{i\hslash }\int_{0}^{t}dt^{\prime }\int dx^{\prime
}G_{H_{0}}^{sic}(x_{b},t_{b};x^{\prime },t^{\prime })\left(
a_{m}^{s}H_{1}\left( x^{\prime }\right) \right) G_{H}^{sic}(x^{\prime
},t^{\prime };x_{a},0)  \tag{A14}
\end{equation}%
\newline
where the initial time $t_{a}=0$ and the final time $%
t_{b}=t_{b}-t_{a}=a_{m}^{s}t$ in value. This is the $SIC$ perturbation
equation that a $SIC$\ Green function obeys.

It can prove that if in (A13) one makes the replacements $U^{sic}\left( \tau
\right) \rightarrow U\left( \tau \right) ,$ $U_{0}^{sic}\left( \tau \right)
\rightarrow U_{0}\left( \tau \right) ,$ and $\left( a_{m}^{s}H_{1}\right)
\rightarrow H_{1},$ where $\tau =t,$ $t-t^{\prime },$ or $t^{\prime }$, then
the $SIC$ perturbation equation (A13) is exactly changed to the $QM$
perturbation equation that a $QM$ unitary propagator obeys. Here the $QM$\
unitary propagators $U\left( \tau \right) $ and $U_{0}\left( \tau \right) $
correspond to the $SIC$\ unitary propagators $U^{sic}\left( \tau \right) $
and $U_{0}^{sic}\left( \tau \right) ,$ respectively. Similarly it can prove
that if in (A14) one makes the replacements $%
G_{H}^{sic}(x_{b},t_{b};x_{a},t_{a})\rightarrow
G_{H}(x_{b},t_{b};x_{a},t_{a}),$ $G_{H_{0}}^{sic}(x_{b},t_{b};x_{a},t_{a})%
\rightarrow G_{H_{0}}(x_{b},t_{b};x_{a},t_{a}),$ and $a_{m}^{s}H_{1}\left(
x^{\prime }\right) \rightarrow H_{1}\left( x^{\prime }\right) ,$ then the $%
SIC$ perturbation equation (A14) is exactly changed to the $QM$ perturbation
equation that a $QM$\ Green function obeys [$10,14$]. \newline
\newline
\newline

\end{document}